\RequirePackage{lineno}
\documentclass[twocolumn,aps,prc,showpacs,superscriptaddress,floatfix]{revtex4-2}
\usepackage{rotating}
\usepackage{epsfig}
\usepackage[colorlinks=true,allcolors=blue]{hyperref}
\usepackage{lineno}

\usepackage{amsmath}

\usepackage{graphicx}
\usepackage[caption=false]{subfig}
\graphicspath{ {./}, {./image} }
\DeclareGraphicsExtensions{ .pdf, .png}

\usepackage[normalem]{ulem}

\usepackage[dvipsnames]{xcolor}

\newcommand{\snn}   {\sqrt{s_{_{NN}}}}
\newcommand{\avfd}  {\textsc{avfd}}
\newcommand{\ampt}  {\textsc{ampt}}
\newcommand{\epos}  {\textsc{epos\footnotesize{4}}}
\newcommand{\hydjet}{\textsc{hydjet\scriptsize{++}}}
\newcommand{\hijing}  {\textsc{hijing}}
\newcommand{\pythia}  {\textsc{pythia}}
\newcommand{\gevc}  {GeV/$c$}
\newcommand{\gevcc} {GeV/$c^2$}
\newcommand{\pt}    {p_{\rm T}}
\newcommand{\mt}    {m_{\rm T}}
\newcommand{\minv}  {m_{\rm inv}}
\newcommand{\ns}    {n_5/s}
\newcommand{\Ks}    {K_S}
\newcommand{\dg}    {\Delta\gamma}
\newcommand{\dgess} {\dg_{\textsc{ess}}}
\newcommand{\dgese} {\dg_{\textsc{ese}}}

\newcommand{\pair}  {{\rm pair}}
\newcommand{\qpair} {q_{2,\pair}}
\newcommand{\vpair} {v_{2,\pair}}
\newcommand{\single}{{\rm single}}
\newcommand{\vsing} {v_{2,\single}}
\newcommand{\qpoi}  {q_{2,\textsc{poi}}}
\newcommand{\qh}    {\hat{q}_{2}}
\newcommand{\qhpoi} {\hat{q}_{2,\textsc{poi}}}
\newcommand{\qhpair}{\hat{q}_{2,\pair}}
\newcommand{\res}   {{\rm res}}
\newcommand{\vres}  {v_{\rm 2,\res}}
\newcommand{\vrho}  {v_{\rm 2,\rho}}
\newcommand{\vobs}  {v_2^{\rm obs}}
\newcommand{\os}    {{\rm os}}
\newcommand{\sm}    {{\rm ss}}

\newcommand{\two}   {\{2\}}

\newcommand\mean[1]{\left\langle#1\right\rangle}

\begin{document}

\title{Investigating event-shape methods in the search for the chiral magnetic effect in relativistic heavy ion collisions}

\author{Han-Sheng Li}
\email{li3924@purdue.edu}
\affiliation{Department of Physics and Astronomy, Purdue University, West Lafayette, Indiana 47907, USA}
\author{Yicheng Feng}
\email{feng216@purdue.edu}
\affiliation{Department of Physics and Astronomy, Purdue University, West Lafayette, Indiana 47907, USA}
\author{Fuqiang Wang}
\email{fqwang@purdue.edu}
\affiliation{Department of Physics and Astronomy, Purdue University, West Lafayette, Indiana 47907, USA}

\begin{abstract}
The Chiral Magnetic Effect (CME) is a phenomenon in which electric charge is separated by a strong magnetic field from local domains of chirality imbalance and parity violation in quantum chromodynamics. The CME-sensitive observable, the charge-dependent three-point azimuthal correlator $\Delta\gamma$, is contaminated by a major physics background proportional to the particle's elliptic {\em flow} anisotropy $v_2$. Event-shape engineering (ESE) binning events in {\em dynamical} fluctuations of $v_2$ and event-shape selection (ESS) binning events in {\em statistical} fluctuations of $v_2$ are two methods to search for the CME by projecting $\dg$ to the {\em measured} anisotropy $v_2=0$ intercept. We conduct a systematic study of these two methods using physics models as well as toy model simulations. It is observed that the ESE method fulfills the general premise of measuring the CME but is statistically hungry. 
It is found that the intercept from the ESS method depends on the details of the event content, such as the mixtures of background-contributing sources, because of statistical fluctuations of  intertwining variables used in the method, and is thus not practically useful to measure the CME.
\end{abstract}


\maketitle

\section{Introduction}

It has been predicted by quantum chromodynamics (QCD) that vacuum fluctuations can result in gluon fields of nonzero topological charges in local metastable domains. Interactions of quarks with such gluon fields can cause chirality imbalance, breaking the parity and charge-parity symmetries in those domains. Such chirality imbalance, under a strong magnetic field, would result in charge separation, a phenomenon called the chiral magnetic effect (CME)~\cite{Kharzeev:1998kz,Kharzeev:2004ey,Kharzeev:2007jp,Fukushima:2008xe}. 

Ultra-strong magnetic fields are presumably produced in non-central relativistic heavy ion collisions~\cite{Skokov:2009qp,Deng:2012pc}. 
The magnetic field is on average perpendicular to the reaction plane (RP, the plane span by the beam and the impact parameter direction of the collision). 
The CME-induced charge separation is along the direction of the magnetic field, and may be conveniently quantified by the $a_1$ variable   in Fourier series of particle azimuthal distributions~\cite{Voloshin:1994mz,Voloshin:2004vk},
\begin{equation}
    dN_\pm/d\tilde\phi \propto 1 \pm 2a_1\sin\tilde\phi + 2v_2\cos2\tilde\phi + \cdots\,,
\end{equation}
where $\tilde\phi$ is the azimuthal angle of the particle momentum vector with respect to the RP and the subscript `$\pm$' indicates particle charge sign. 
The elliptic anisotropy $v_2$ harmonic is the leading modulation in particle distributions produced in relativistic heavy ion collisions~\cite{Heinz:2013th}.
Because of a vanishing mean $a_1$ due to random fluctuations, a commonly used observable is  the three-point correlator~\cite{Voloshin:2004vk},
\begin{equation}
    \gamma=\mean{\cos(\phi_\alpha+\phi_\beta-2\psi)}\,,
    \label{eq:g}
\end{equation}
where $\phi_\alpha$ and $\phi_\beta$ are the azimuthal angles of two particles of interest (POIs), and $\psi$ is that of the RP (or participant plane because of fluctuations in the collision overlap geometry~\cite{Alver:2006wh}).
To cancel charge-independent backgrounds, such as effects from global momentum conservation, the difference between opposite-sign (OS) and same-sign (SS) correlators is used~\cite{Abelev:2009ac,Abelev:2009ad},
\begin{equation}
    \Delta\gamma \equiv \gamma_\os - \gamma_\sm\,.
    \label{eq:dg}
\end{equation}
The CME signal presented in the $\Delta\gamma$ observable would then be $2a_1^2$.

Definite signals of the CME have not yet been observed~\cite{Abelev:2009ac,Abelev:2009ad,Abelev:2012pa,Kharzeev:2015znc,Zhao:2018ixy,Zhao:2019hta}. The major difficulty is the large contamination in $\Delta\gamma$ from mundane QCD backgrounds. Those backgrounds arise from genuine two-particle correlations, such as correlations between daughter particles from a resonance decay, or between particles from the same jet or back-to-back dijet, coupled with elliptic {\em flow} anisotropies of those background sources~\cite{Voloshin:2004vk,Wang:2009kd,Liao:2010nv,Bzdak:2010fd,Schlichting:2010qia,Pratt:2010zn}.
This background correlation can be schematically expressed as~\cite{Voloshin:2004vk,Zhao:2019hta,Zhao:2018skm,Zhao:2018pnk}
\begin{equation}
    \Delta\gamma_\res = \mean{\cos(\phi_\alpha+\phi_\beta-2\phi_\res)}\vres\,,
\end{equation}
where the subscript `res' stands generically for correlated two-particle clusters such as resonances  and jets, and $\vres\equiv\mean{\cos2(\phi_\res-\psi)}$ is the elliptic flow anisotropy of those background-contributing sources.

Large efforts have since been invested  to eliminate or mitigate these backgrounds~\cite{Adamczyk:2013kcb,Adamczyk:2014mzf,Khachatryan:2016got,STAR:2019xzd,ALICE:2020siw}, including innovative observables~\cite{Acharya:2017fau,Sirunyan:2017quh,Xu:2017qfs,Voloshin:2018qsm,STAR:2021pwb,Choudhury:2021jwd}.
One of the techniques is the event-shape engineering (ESE)~\cite{Schukraft:2012ah,Acharya:2017fau,Sirunyan:2017quh} method, grouping events into classes of different $v_2$ values relying on the {\em dynamical} fluctuations of $v_2$. A variation of this method is to select events according to the particle emission pattern, called the event-shape selection (ESS) method~\cite{Xu:2023elq}, relying on the {\em statistical} fluctuations of $v_2$. 
The main difference between the two methods is in the ways of  quantifying the event shape. The ESE method~\cite{Schukraft:2012ah} uses a variable calculated {\em not} from the POIs that are used for the physics measurement of $\dg$ and $v_2$, while the ESS method uses a variable calculated from the POIs~\cite{Xu:2023elq}. 
Each method analyzes $\dg$ as a function of the measured $v_2$ in events binned according to the corresponding event-shape variable and projects $\dg$ to $v_2=0$ to obtain the intercept, presumably more sensitive to CME signals than the overall $\dg$ measurement and less sensitive (or ideally insensitive) to backgrounds.

The ESE method is relatively straightforward to comprehend. The ESS method is, however, complex because the POIs are used in both event selection and physics measurements. 
In this paper, we conduct a systematic study of the ESS and ESE methods for the CME search using several physics models as well as toy models of varying ingredients, with the goal to elucidate what the projected intercept entails from each method. 

The focus of our study is placed on the effects of the event classifying variables on the intercept under various circumstances of the simulated events. We note, however, that the $v_2$ measured in experiments contains not only the flow anisotropy but also the so-called nonflow correlations, dependent on analysis methods. Nonflow correlations refer to those unrelated to the collision geometry common to the entire event, such as resonance decays and (di-)jet correlations~\cite{Borghini:2000cm,Borghini:2006yk,Wang:2008gp,Ollitrault:2009ie}. While we touch upon the nonflow issue in this article, a full and thorough discussion of nonflow effects on the ESE and ESS methods is outside the scope of the present study.

The rest of the paper is organized as follows. Section~\ref{sec:tech} describes the details of the ESE and ESS methods (and a related early method~\cite{Adamczyk:2013kcb}) used for the CME search. Section~\ref{sec:models} gives brief descriptions of the models used in this study. Section~\ref{sec:results} presents our model simulation results using ESS and ESE and discusses the findings. Finally, a summary is given in Section~\ref{sec:summary}. The appendix compiles all the $\dg$ vs.~$v_2$ plots from the simulations using the ESS and ESE methods.

\section{Event-Shape Techniques\label{sec:tech}}
In ESE~\cite{Schukraft:2012ah}, events from a narrow centrality bin are  grouped according to the ellipic flow vector magnitude $q_2$ calculated from particles (or other detected signals in  experiment like energy depositions in calorimeters) that are different from the POIs. 
The event-by-event $q_2$ quantity is generally defined as
\begin{eqnarray}
    q_2^2 &=& \frac{1}{N} 
    \left[ 
    \left( \sum_{i=1}^{N}\cos 2\phi_i \right )^2 + \left( \sum_{i=1}^{N}\sin 2\phi_i \right)^2
    \right] \nonumber\\
    &=& 1 + \frac{1}{N}\sum_{i\neq j} \cos 2(\phi_i-\phi_j) \,,
    \label{eq:q2}
\end{eqnarray}
where $N$ is the number of particles in the event within a given kinematic region (we use ``particles'' in our description for convenience without loss of generality).
One may use the normalized $\hat{q}_2$ quantity~\cite{Xu:2023elq} for ESE,
\begin{equation}
    \qh^2 \equiv q_2^2/\mean{q_2^2}\,.
    \label{eq:qh}
\end{equation}
The average of $q_2^2$ over all events, assuming Poisson fluctuations in multiplicity, is given by
\begin{equation}
\mean{q_2^2} \approx 1+Nv_{2,q}^2\two\,,
\label{eq:q2mean}
\end{equation}
where 
\begin{equation}
    v_2^2\two = \mean{\cos 2(\phi_1-\phi_2)}\,,
    \label{eq:v22}
\end{equation}
is the two-particle cumulant elliptic flow anisotropy ($\phi_1$ and $\phi_2$ are the azimuthal angles of the pair), and the subscript `$q$' in $v_{2,q}$ indicates that it is calculated from the particles used for computing $q_2^2$.

In  ESE, the $q_2^2$ is typically measured at forward and backward rapidities~\cite{Acharya:2017fau,Sirunyan:2017quh}, 
within a given narrow centrality range, and events are selected according to the measured value of $q_2^2$ to study $\dg$ as a function of $v_2$ typically in the midrapidity region.
The value of $q_2^2$ varies from event to event, and the variations are dominated by statistical fluctuations. 
These statistical fluctuations in $q_2^2$ at forward and backward rapidities are uncorrelated to those in $v_2$ (or any other quantities) at midrapidity, so the statistical fluctuation effect in $v_2$ is averaged to zero in those $q_2^2$-selected events.
There are, however, also dynamical fluctuations affecting the value of $q_2^2$. The main source of dynamical fluctuations is the initial geometry, which can fluctuate event-by-event within a given (narrow) centrality bin~\cite{Poskanzer:1998yz,Bhalerao:2006tp}, or even at a centrality precisely determined by a particular centrality measure. Even if the colliding geometry was precisely fixed, there can still be fluctuations in the subsequent ``hydrodynamic'' evolution and interactions of the collision system. Because the sources of these dynamical fluctuations are common to the entire collision event, they affect $q_2^2$ and $v_2$ (and any other quantities) of the same event. 
Thus, the event classes selected by $q_2$ (or equivalently $q_2^2$ or $\qh^2$) within a given narrow centrality range will have a different average $\mean{v_2}$ in the midrapidity region, or any other kinematic regions displaced from that for $q_2^2$ computation. 
In other words, because $q_2$ is {\em dynamically} correlated with $\mean{v_2}$, ESE is dividing events in groups of varying $\mean{v_2}$, which dynamically fluctuates from event to event. 
We note that since $\qh^2$ and $\mean{v_2}$ are usually computed in separate $\eta$ regions, longitudinal decorrelation of flow~\cite{Bozek:2010vz,Xiao:2012uw} can reduce the power of $\qh^2$ in selecting dynamical fluctuations in $\mean{v_2}$. However, this effect is small because the flow decorrelation has been measured to be at most a few percent~\cite{Adams:2019fpo,Yan:2023ugh}.

\begin{figure*}[hbt]
    \includegraphics[width=0.33\textwidth]{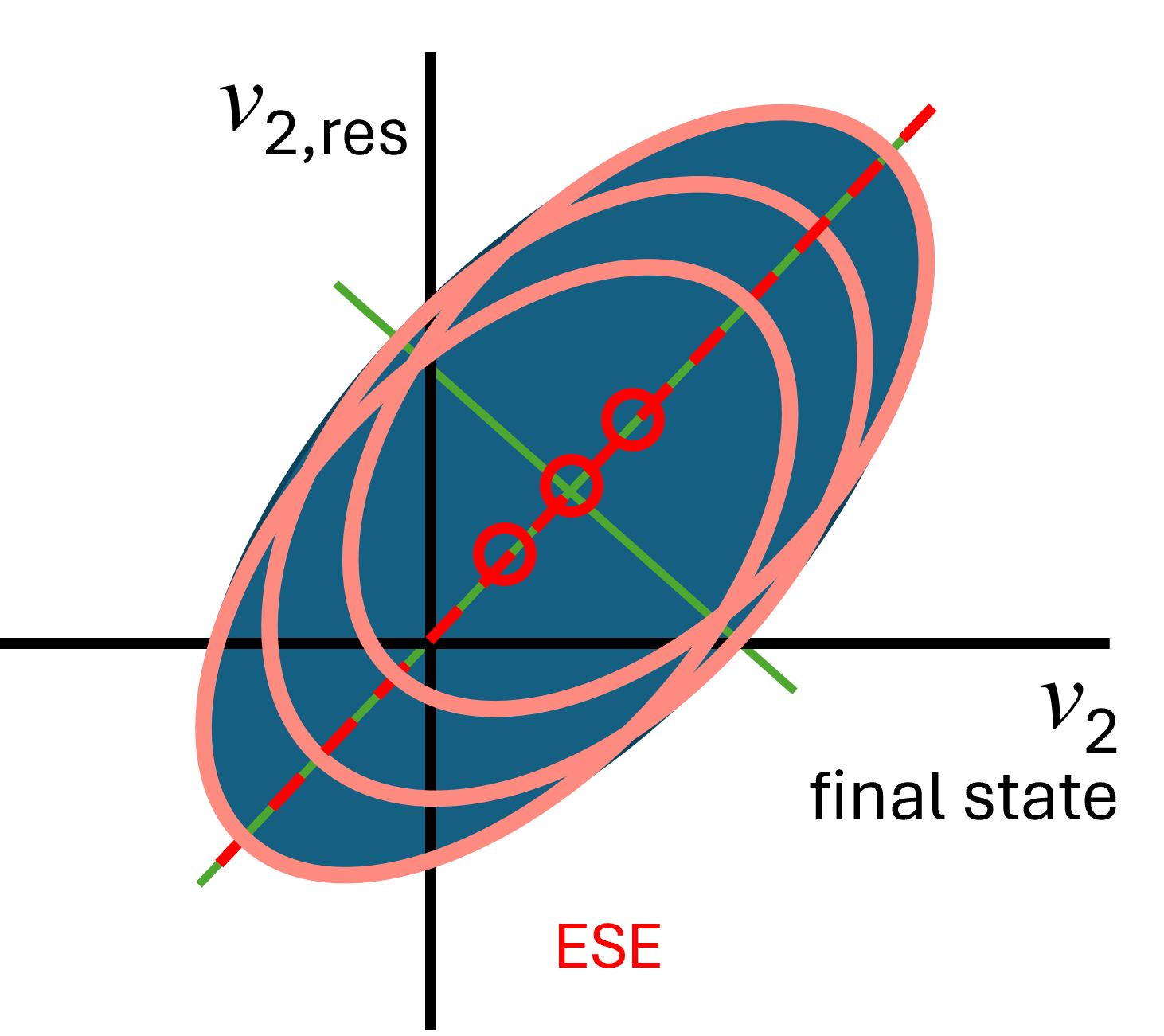}\hfill
    \includegraphics[width=0.33\textwidth]{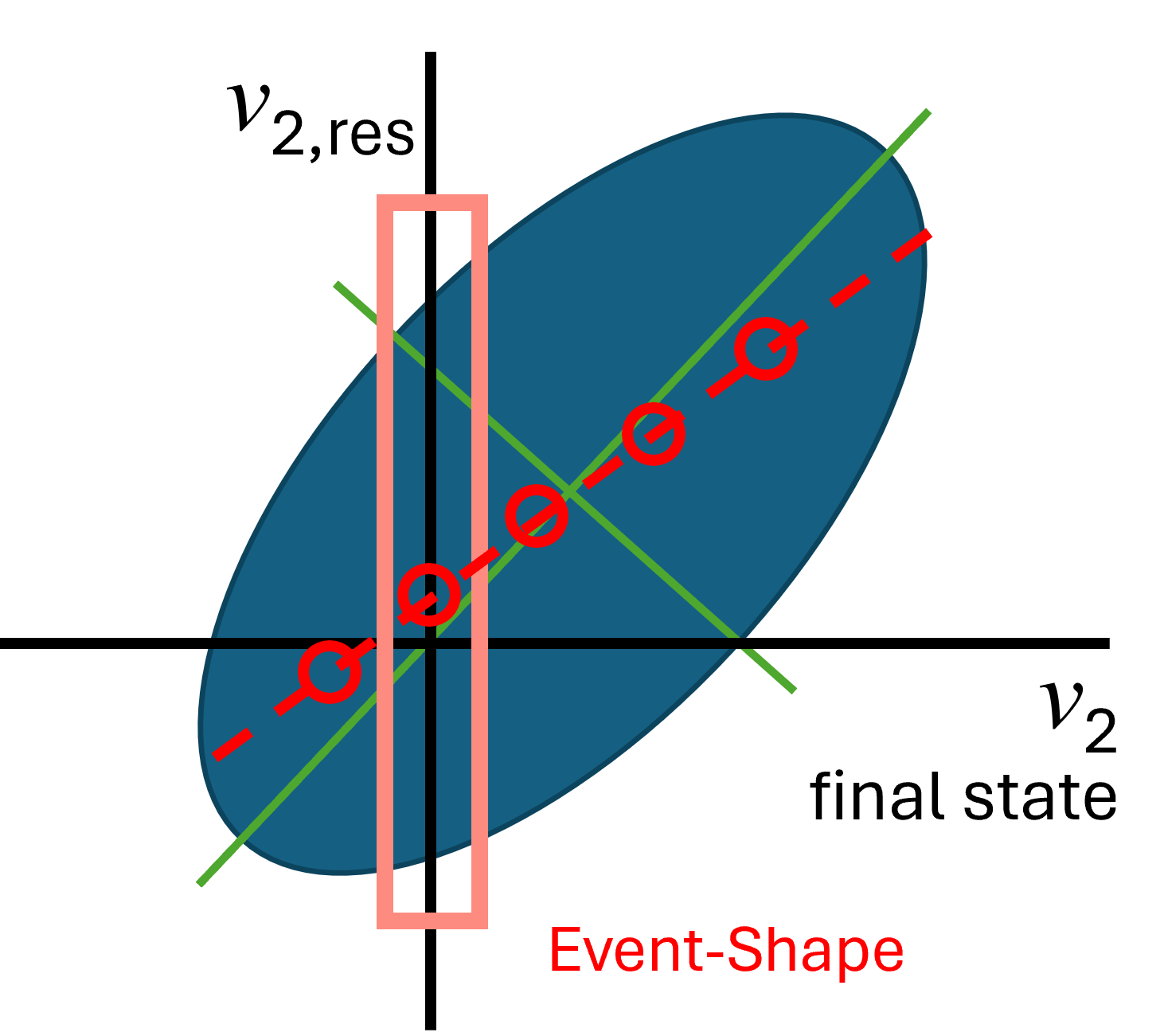}\hfill
    \includegraphics[width=0.33\textwidth]{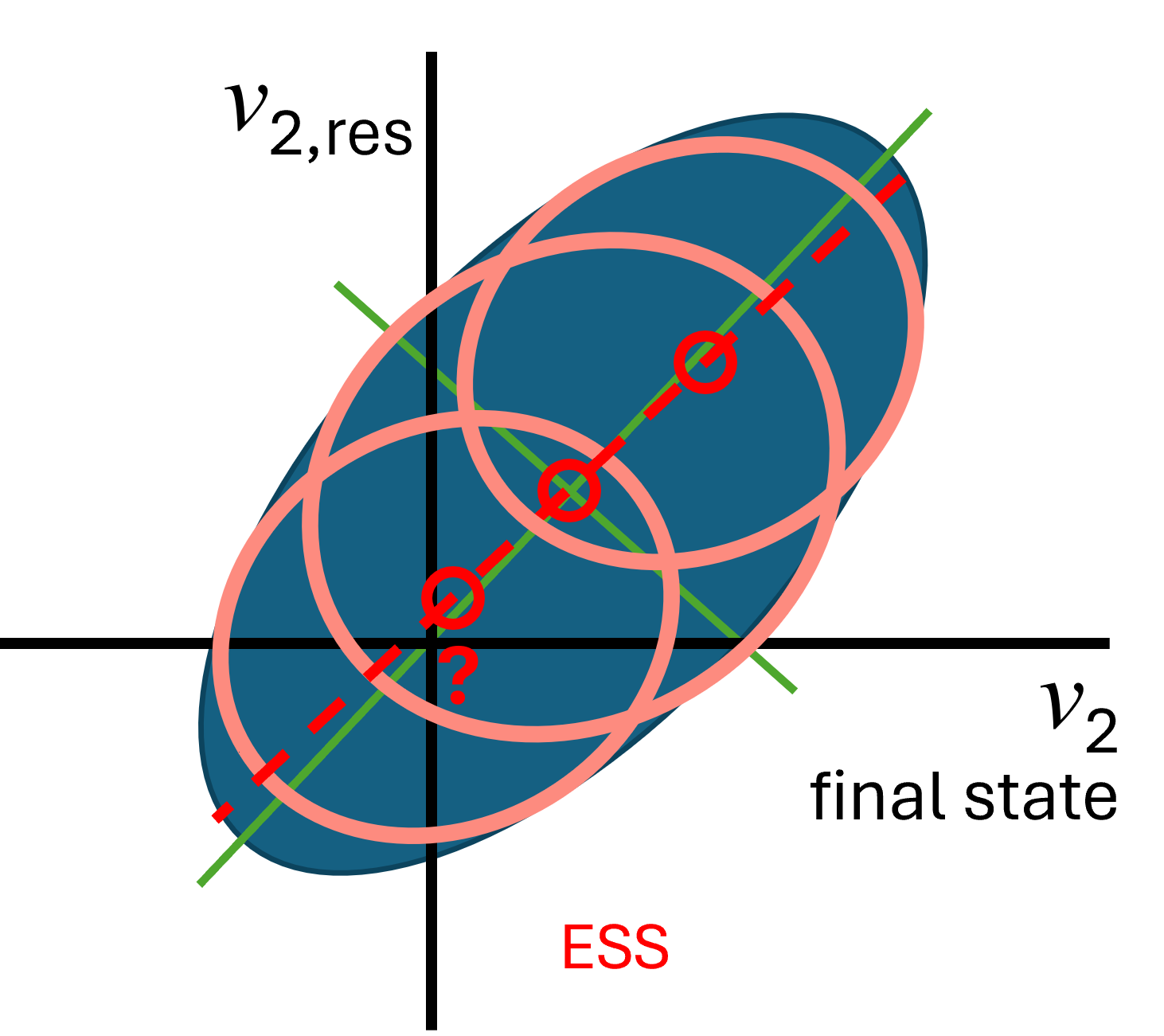}\hfill
    \caption{Cartoons illustrating the various event-shape methods in terms of the event-by-event resonance $\vres$ vs.~final-state single particle $v_2$ in events within a narrow centrality bin. Here $\vres$ and $v_2$ are taken to be the event-by-event elliptic shape variables, $\mean{\cos2(\phi-\psi)}$, with respect to the event plane $\psi$ from another kinematic region different from the particles of interest (POIs).
    (Left panel) Event-Shape Engineering (ESE) where events are selected according to $q_2$ in kinematic region displaced from that of POIs' (each $q_2$-class is indicated by the brown ellipse), in which the average $\mean{\vrho}$ is proportional to $\mean{v_2}$, unbiased, and the spread in $\mean{v_2}$ and $\mean{\vres}$ (indicated by the red circles) are due to {\em dynamical} fluctuations. 
    (Center penal) Event-Shape (ES) method where events are binned in the observed elliptic shape variable $\vobs$ of POIs (one bin is indicated by the brown rectangle), and the average $\vres$ in those events (indicated by the red circles) are strongly biased, not strictly proportional to $\vobs$ and with a positive $\vres$ value at $\vobs=0$. The wide range in $\vobs$ (including negative values) is mainly due to {\em statistical} fluctuations. 
    (Right panel) Event-Shape Selection (ESS) where events are selected according to $\qhpair^2$ of pairs of POIs (each $\qhpair^2$-class is indicated by the brown ellipse) in which the $\vres$ is unnecessarily proportional to $v_2$ (indicated by the red circles) due to complicated biases from the $\qhpair^2$ selection using the same POIs as for $v_2$. The wide spreads in $v_2$ and $\vres$ are mainly due to {\em statistical} fluctuations inherited in the ESS method involving the same POIs. The green cross in each cartoon indicates the overall averages $\mean{\mean{v_2}}$ and $\mean{\mean{\vres}}$ over the entire event sample.} 
    \label{cartoon}
\end{figure*}

Since the ESE method selects events according to dynamical fluctuations of the $\mean{v_2}$ of the POIs, decoupled from the $q_2^2$ computation, it is guaranteed that the average resonance $\mean{\vres}$, and generally the elliptic anisotropies of all elements in the event including all CME background sources, are proportional to $\mean{v_2}$. 
Note that the final-state $\mean{v_2}$ contains contributions not only from primordial particles, but also from all  decay products of resonances and clusters. 
Since the averages $\mean{\vres}$ and $\mean{v_2}$ in those $q_2$-selected event classes are all connected to the initial geometry, they are all proportional to each other. 
This is illustrated in the cartoon in Fig.~\ref{cartoon} (left), where the three ellipses illustrate event selections by $q_2$ and the hollow circles are the average $\mean{v_2}$ and $\mean{\vres}$ in each $q_2$-selected event class.
In the ESE analyses for CME searches~\cite{Acharya:2017fau,Sirunyan:2017quh}, one analyzes $\dg$ as a function of $\mean{v_2}$ in each event class and projects the $\dg$ measurement to zero $\mean{v_2}$. The ESE method guarantees that the elliptic flows of the background sources also vanish at zero $\mean{v_2}$.
In other words, a linear projection to $\mean{v_2}=0$ would be equivalent to a projection to $\mean{\vres}=0$, where the CME backgrounds are zero. 
Since all those events are from a narrow centrality bin where the CME signal does not vary much, the intercept of the linear projection is presumably the CME signal. 
Note that there may still be complications from nonflow effects~\cite{Feng:2021pgf} depending on how $\dg$ and $\mean{v_2}$ are measured, which is outside the scope of the present work.

Since the CME background comes from the elliptic anisotropies imprinted in the POIs, it is interesting to select events with an event-shape ellipticity $\vobs\equiv\mean{\cos2(\phi-\psi)}$ calculated by POIs themselves, where the event-plane (EP) azimuthal angle $\psi$ is determined in another kinematic region away from the POIs, and examine the $\dg$ observable as a function of $\vobs$. 
This method selects on the event-by-event elliptic-shape quantity (particle emissin pattern) $\vobs$ which can fluctuate even to negative values because of large statistical fluctuations, the intercept at $\vobs=0$, more sensitive to the CME, can be well determined from data. 
This event-shape (ES) analysis was performed by STAR~\cite{Adamczyk:2013kcb} and the intercept of the linear fit is consistent with zero with the then-available statistics. 
However, the backgrounds are not determined by $\vobs$ but by $\mean{\vres}$. 
Statistically fluctuated $\vobs=0$ does not necessarily guarantee $\mean{\vres}=0$. 
This is illustrated by the cartoon in Fig.~\ref{cartoon} (center). 
Although $\mean{\vres}$ is proportional to $\mean{v_2}$ with an unbiased event selection, the strongly (maximally) biased selection of events by $\vobs$ will obscure the relationship between the averages $\mean{\vres}$ and the $\vobs$ in each bin of $\vobs$ of those events, as indicated by the hollow box in Fig.~\ref{cartoon} (center). 
Because of the overall positive means $\mean{\vobs}=\mean{v_2}$ and $\mean{\vres}$ over the entire event sample, the linear projection will give a positive intercept of $\mean{\vres}$ at $\vobs=0$. 
In other words, with finite positive average $\mean{v_2}$ in heavy ion collisions, $\mean{\vres}>0$ at $\vobs=0$ with events binned in $\vobs$. This has been verified by model study~\cite{wang:2016iov}. 
Thus, positive residual background remains in the linear fit intercept in the analysis of $\dg$ as a function of $\vobs$~\cite{Adamczyk:2013kcb}, although the background is significantly reduced compared to that in the inclusive $\dg$ measurement.
Ideally, in the ES analysis, one would want to select events with the $\vobs$ values of all resonances (background sources) to be zero, however, such selection is experimentally insurmountable. 

It has been recently proposed by Xu {\em et al.}~\cite{Xu:2023elq} to select events according to the $\qhpoi^2$  of the POIs, and study the $\dg$ as a function of $v_2$ of the POIs. This is similar in spirit to the ES method above where the events are selected by $\vobs$ and the $\dg$ is examined as a function of the same $\vobs$ variable. 
Here, the selection variable $\qhpoi^2$ and the  $v_2$ variable are different, but related as they are computed by the same POIs. The $\qh^2$ variable is an event-by-event quantity connected to two-particle cumulant anisotropy (cf Eq.~\ref{eq:q2}), whereas $v_2$ can be measured with respect to $\psi$ from another kinematic region like in the ES method. (Of course $v_2$ can also be computed by the two-particle cumulant method, in which case it is more directly connected to $\qh^2$.) 
The variations in the event-by-event $\qhpoi^2$ quantity, within a given narrow centrality bin, are mainly from {\em statistical} fluctuations, 
just like the $\qh^2$ in the ESE method. However, unlike ESE, 
the $v_2$ variable in those $\qhpoi^2$-selected events by ESS are related to $\qhpoi^2$ itself and are calculated from the same POIs, so the variations in $v_2$ is also primarily of statistical nature. (Note, we simply use $v_2$ here, instead of the $\mean{v_2}$ for ESE, to indicate that the $v_2$ in ESS is of statistical nature although it is an ``average'' $v_2$ in events selected by $\qhpoi^2$.)
This technique selects on the statistical fluctuations in event shape, not on the dynamical fluctuations in $\mean{v_2}$ like in the ESE method, and it is referred to as the Event-Shape-Selection (ESS) method~\cite{Xu:2023elq}. 
The most important difference in the ESS method from the ESE method is that the $\qh^2$ is computed from the POIs in the former but in the later the momentum space in which to compute $\qh^2$ is displaced from that of the POIs'. 
In both approaches, one groups events (within a given narrow centrality bin) according to $\hat{q}_2^2$ and studies the $\dg$ variable as a function of the $v_2$ of the POIs.
It is worthy to emphasize again that, while the event selection variables in both methods are dominated by statistical fluctuations event-by-event, these statistical fluctuations are canceled in $v_2$ in the ESE method because the calculations of $\qh^2$ and $v_2$ use different sets of particles, but do not cancel (rather strongly correlated) in the $v_2$ in the ESS method because the calculations of $\qhpair^2$ and $v_2$ use the same POIs.

Several combinations of $\qhpoi^2$ and $v_2$ variables have been investigated in Ref.~\cite{Xu:2023elq}. The $\qpoi^2$ can be computed using single particle azimuthal angles of the POIs by Eq.~\ref{eq:q2}. It can also be defined by pairs of POIs similar to Eq.~\ref{eq:q2}; in this case, the $\qpair^2$ of pairs is calculated by substituting the $i^{\rm th}$ particle's $\phi_i$ by the azimuthal angle of the $i^{\rm th}$ pair in Eq.~\ref{eq:q2}, and the sum runs over all pairs of POIs in the event. 
The azimuthal angle of a pair is defined to be that of the total momentum of the two POIs (vector sum of their momenta).
The normalized $\qhpair^2$ is then
\begin{equation}
    \qhpair^2 \equiv \qpair^2/\mean{\qpair^2}\,.
    \label{eq:qhpair}
\end{equation}
where $\mean{\qpair^2} \approx 1 + N_\pair\vpair^2\two$.
Similarly, one can define  $v_2$ to be that of single particles of POIs or that of pairs of POIs; in the latter, one replace the azimuthal angle $\phi$ of a single particle by that of a particle pair in
\begin{equation}
    v_2 = \mean{\cos2(\phi-\psi)}\,.
    \label{eq:v2}
\end{equation}
One can study the $\dg$ as a function of $v_2$ of single POI particles (or $\vpair$ of pairs of POI particles) selecting events on $\qhpoi^2$ of single POIs (or $\qhpair^2$ of pairs of POIs). 
It is found by the Anomalous-Viscous Fluid Dynamics (\avfd) model study that, out of the four combinations,  it is the best to use the combination of pair $\qhpair^2$ and single $\vsing$~\cite{Xu:2023elq}, because the linear intercept of $\dg$ as a function of $\vsing$ of single POIs selecting events on $\qhpair^2$ of pairs of POIs reflects most closely  the true CME in \avfd.
However, since the same POIs are used for  $\qhpair^2$, $\vsing$ and $\dg$, self-correlations are present in the measurement the effects/biases of which are hard to discern.
It is unclear how much background is remaining in the ESS intercept, and whether the remaining background is positive or negative.
This is illustrated by the cartoon in Fig.~\ref{cartoon} (right).
In this work, we focus on the best combination of pair $\qhpair^2$ and single $\vsing$ in ESS as found in Ref.~\cite{Xu:2023elq}, and examine what possible biases may exist and the possible level of remaining background in the ESS intercept.

\section{Model setup and analysis details\label{sec:models}}

We investigate four physics models that have been used to simulate heavy ion collisions. Three of them are dynamical models (\avfd, \ampt, \epos) and the fourth is a parameterization model (\hydjet). The \avfd\ model is designed to study CME-related physics and has the capability to implement axial charge current and magnetic field. The other three models do not have those CME capabilities and are useful only for background studies, which is the main purpose of this work. 
The centralities of the models are determined by the charged hadron multiplicity within $|
\eta|<0.5$ as in the STAR experiments~\cite{Abelev:2008ab}. 

For further insights, we also investigate background behaviors with two versions of a toy model. The inputs to the toy models are parameterizations of the measured data.

\begin{figure*}
    \includegraphics[width=0.3\textwidth]{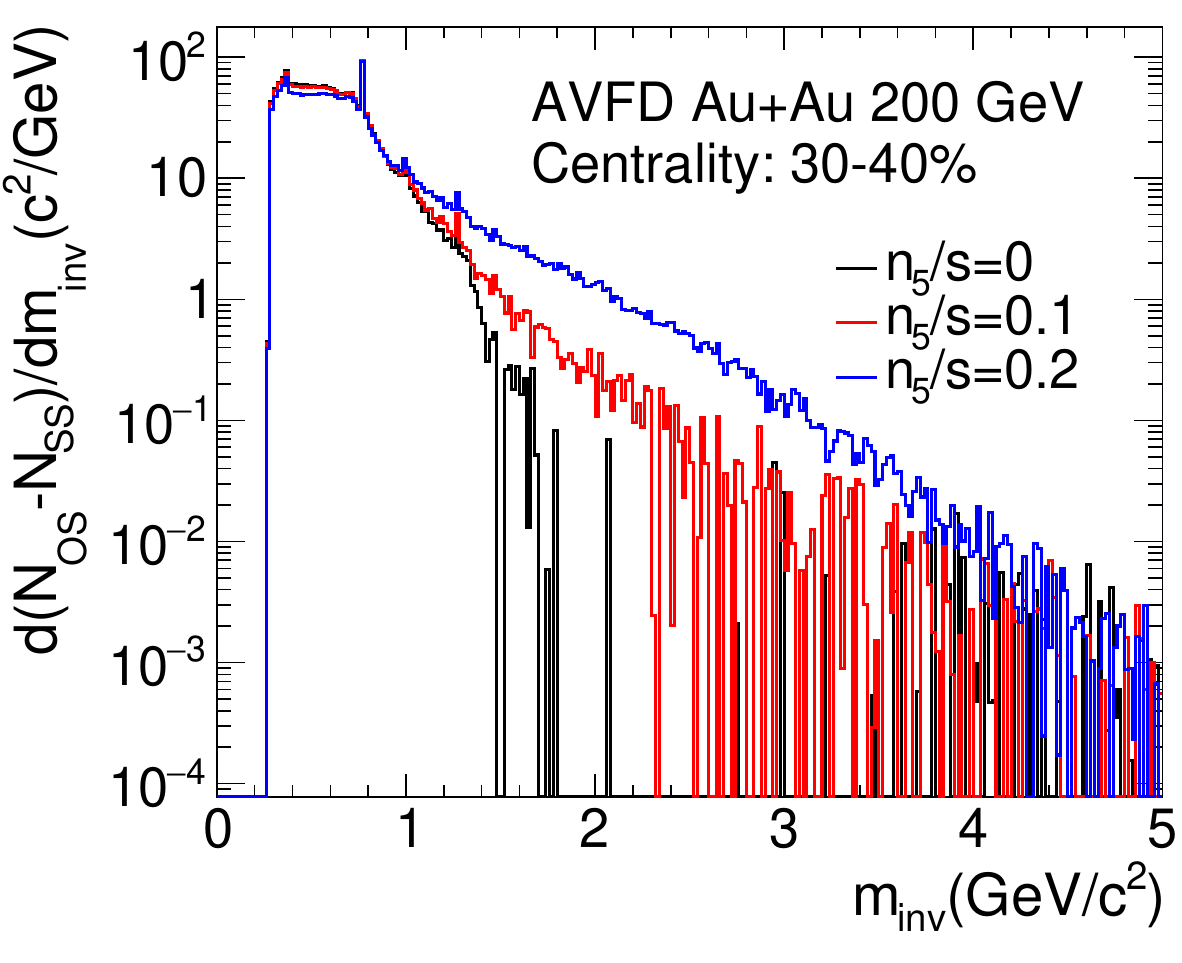}
    \includegraphics[width=0.3\textwidth]{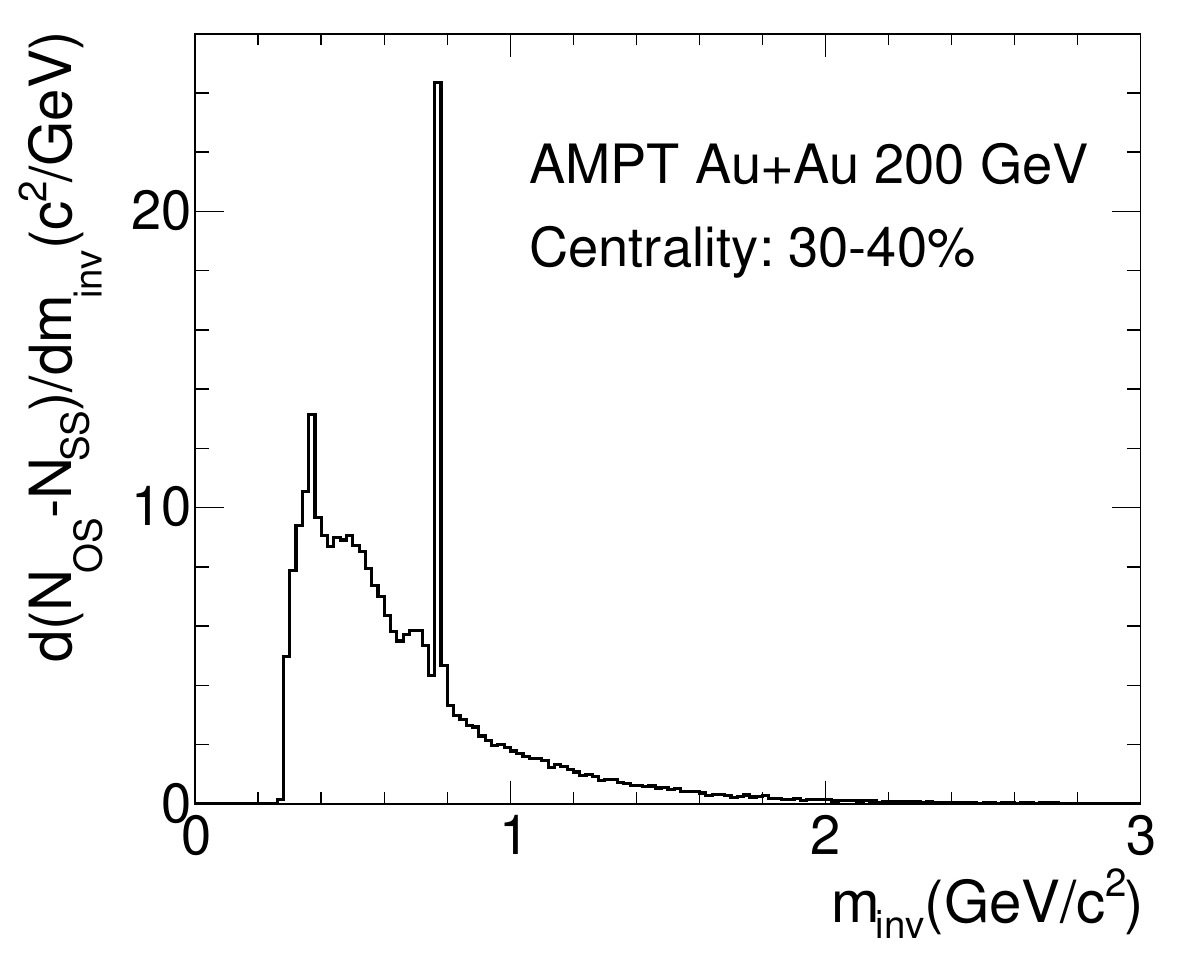}
    \includegraphics[width=0.3\textwidth]{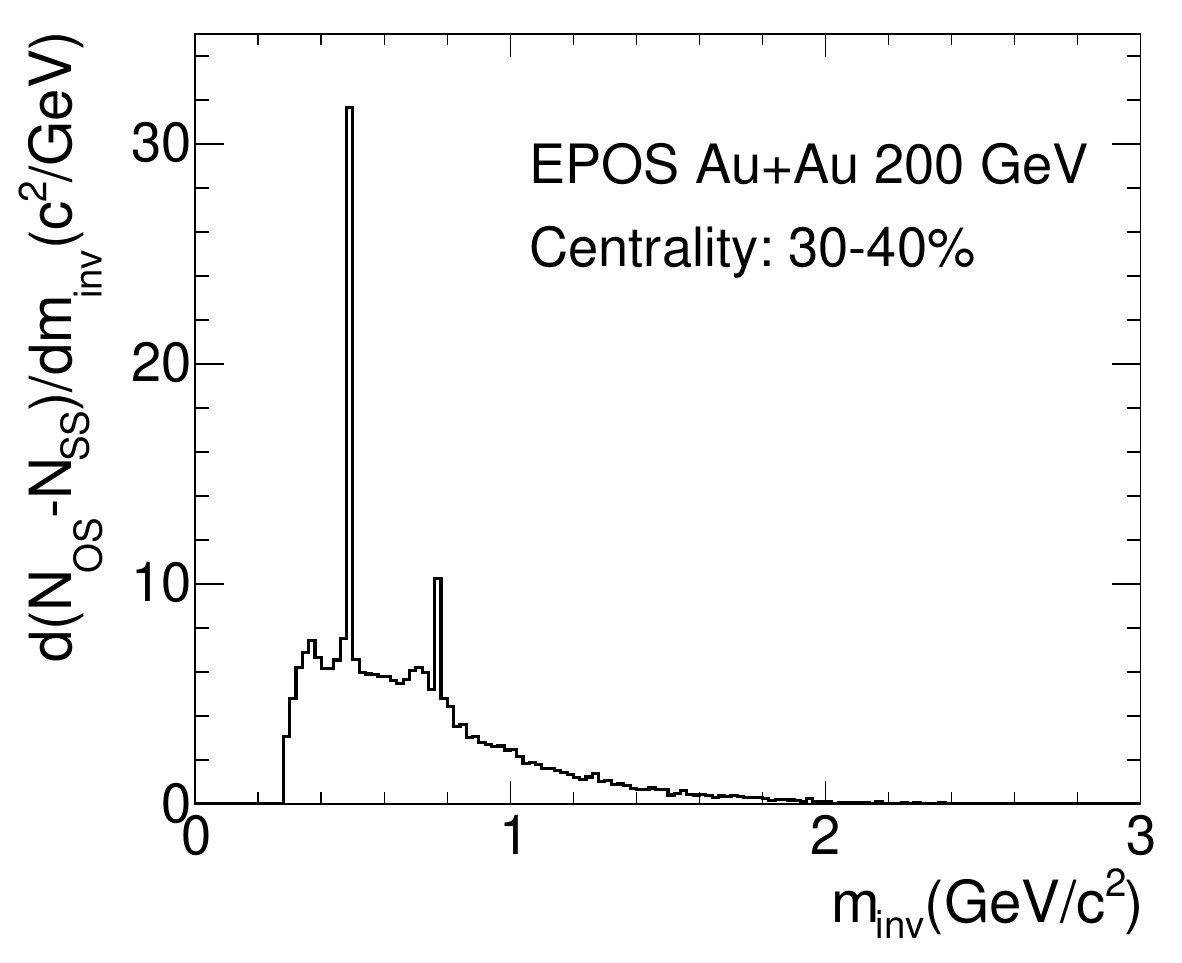}
    \includegraphics[width=0.3\textwidth]{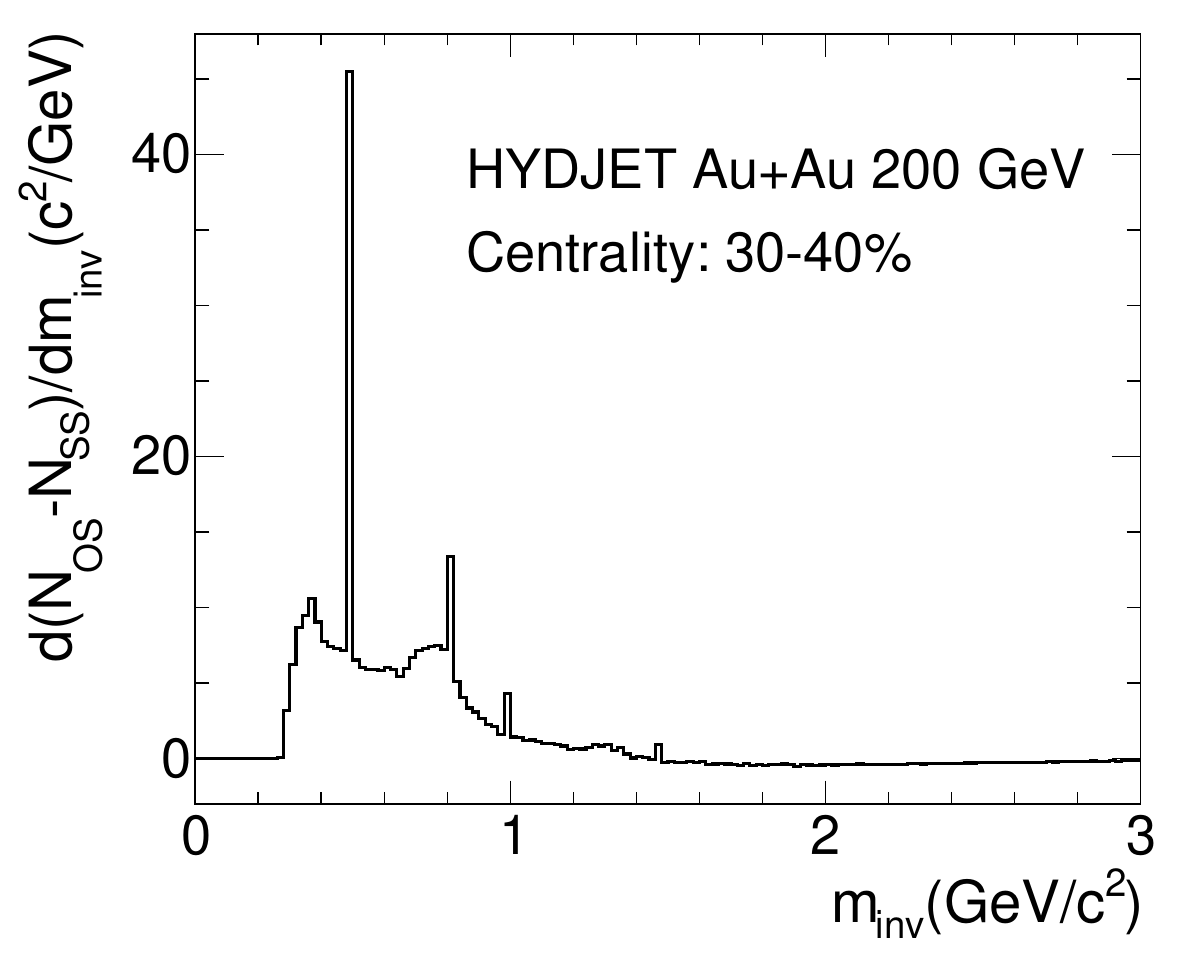}
    \includegraphics[width=0.3\textwidth]{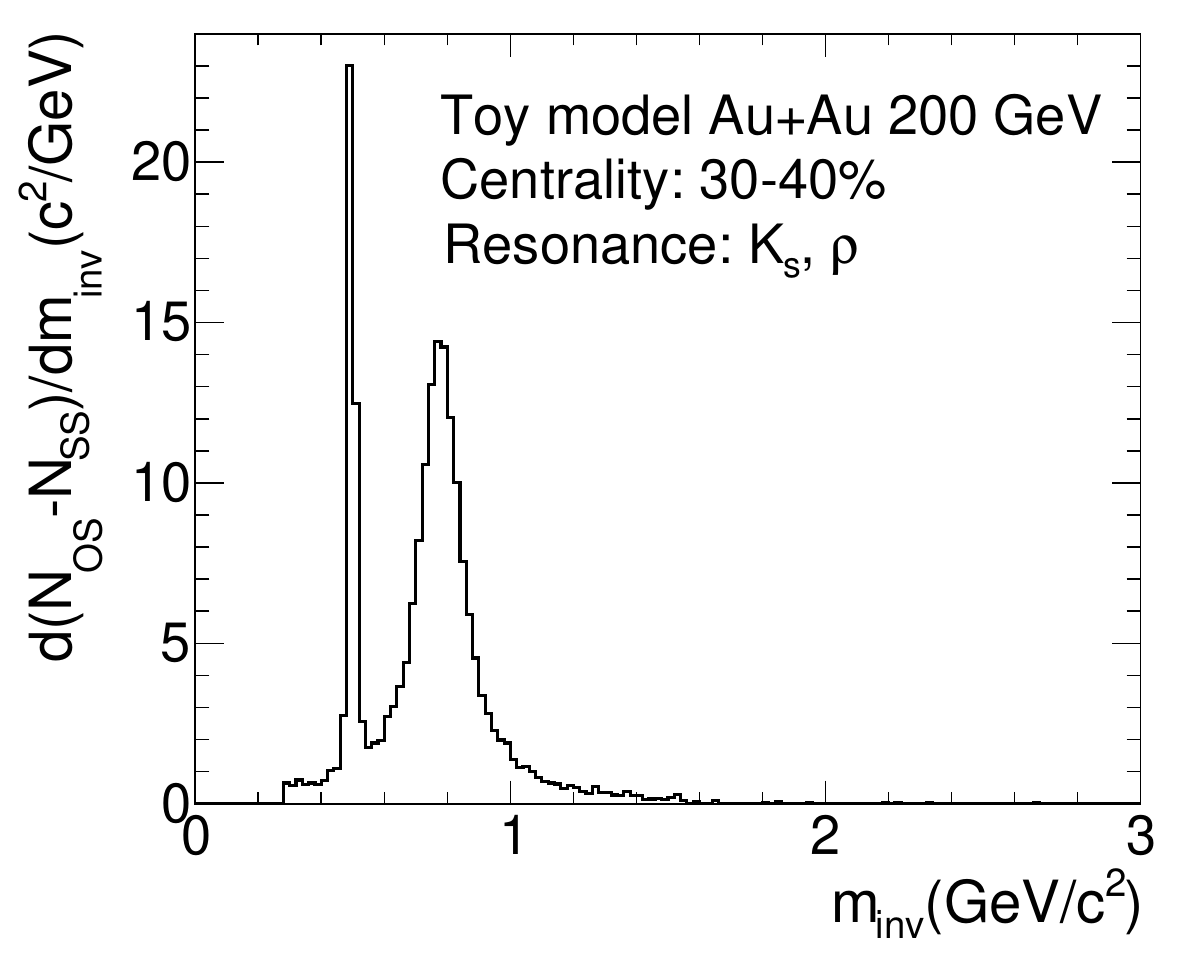}
    \includegraphics[width=0.3\textwidth]{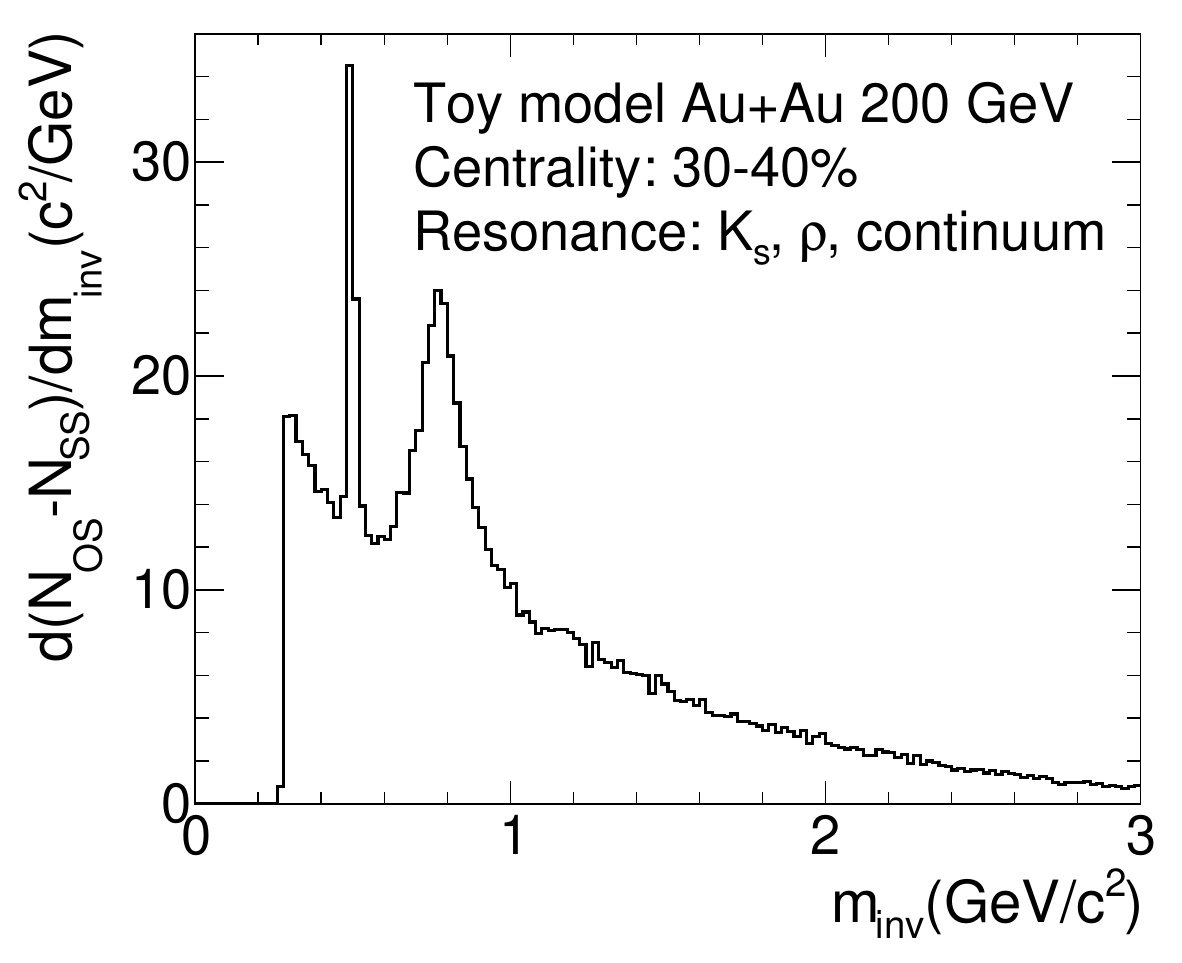}
    \caption{$\minv$ distributions. Shown are the per-event $\minv$ distribution of OS pairs subtracted by that of SS pairs from \avfd\ with axial current densities $n_5/s=0, 0.1$, and $0.2$ (upper left), \ampt\ (upper center), \epos\ (upper right), \hydjet\ (lower left), Toy Model I simulation including primordial pions, and $\Ks$ and $\rho^0$-resonance decay pions (lower center), and Toy Model II simulation including in addition decay pions from a mass continuum (lower right). Particles used to form the $\minv$ distributions are taken from all charged hadrons (defined to be $\pi^\pm$, $K^\pm$, $p$, and $\bar{p}$ in our study and treated as pions with pion mass) and are from the kinematic range of $|\eta|<1$ and $0.2<\pt<2$~\gevc. In the toy models all final-state particles are charged pions.} 
    \label{fig:minv}
\end{figure*}

\subsection{AVFD Model\label{sec:avfd}}
The \avfd\ model~\cite{Shi:2017cpu,Jiang:2016wve,Shi:2019wzi} is an anomalous fluid dynamics developed to describe the evolution of chiral fermion currents in the quark-gluon plasma (QGP) created in relativistic heavy ion collisions in addition to the normal simulations by VISHNU (a package combining the hydrodynamic evolution of the QGP with a microscopic hadronic cascade)~\cite{Song:2010aq}. 
The average magnetic field is calculated from the spectator protons, and it is nominally along the direction perpendicular to the RP. 
In the model, the magnetic field is quantified by event-by-event simulations~\cite{Bloczynski:2012en} taking into account the fluctuations of the magnetic field direction with respect to the RP.
A modest time evolution is assumed for the decreasing magnetic field with a typical lifetime comparable to the initial time of hydrodynamic evolution $\sim 0.6 \text{ fm}/c$.
The initial condition for the axial charge density ($n_5$) is dynamically generated in \avfd\ to be proportional to the entropy density ($s$), and the strength is set via the proportionality coefficient $\ns$. 
Three $\ns$ values are used in the simulations, $n_{5}/s$=0, 0.1, and 0.2. 
The QGP medium evolution is simulated by viscous hydrodynamics to describe the bulk background in heavy ion collisions, with transport parameters for the diffusion coefficient as well as the relaxation time. 
A final hadronic stage after the hydrodynamic freeze-out is included with hadronic re-scatterings and resonance decays.

Au+Au collisions are simulated by \avfd\ at $\snn=200$~GeV in~\cite{Choudhury:2021jwd}. In this study, we focus on the centrality range of 30--40\% as in~\cite{Choudhury:2021jwd}. 
Figure~\ref{fig:minv} (upper left) shows the difference of the $\minv$ distributions between OS and SS pairs for the three values of $\ns$. The CME implemented in \avfd\ affects the relative OS and SS pair $\minv$ distributions--the distribution flattens with larger $\ns$. This arises from contributions from back-to-back OS pairs (hence larger $\minv$) and near-side SS pairs (hence smaller $\minv$) as a result of the CME, resulting in enhancement at large $\minv$ and depletion at small $\minv$ for large $\ns$ compared to small one.
Several resonance peaks are apparent in the $\minv$ distributions: $\rho^0$, $f_0(980)$, and $f_2(1270)$. No intrinsic mass width is implemented in \avfd, so the resonances appear as $\delta$-functions. 
Weak-decay hadrons are treated as stable particles in \avfd, so no $\Ks$ peak is present in the $\minv$ distribution. 

\subsection{AMPT Model\label{sec:ampt}}
\ampt\ (A Multi-Phase Transport) is a parton transport model~\cite{Lin:2004en}. It consists of a fluctuating initial condition, parton elastic scatterings, quark coalescence for hadronization, and hadronic interactions. The initial condition of \ampt\ is based on the \hijing\ model~\cite{Gyulassy:1994ew}. The string melting version of \ampt, which we use for our study, converts these initial hadrons into their valence quarks and antiquarks~\cite{Lin:2001zk,Lin:2004en}. 
The (anti-)quarks further evolve via two-body elastic scatterings, treated with Zhang’s parton cascade~\cite{Zhang:1997ej} with a total parton scattering cross section of 3~mb. After parton scatterings cease, a simple quark coalescence model is applied to convert partons into hadrons~\cite{Lin:2004en}. Subsequent interactions of those formed hadrons are modeled by a hadron cascade, including meson-meson, meson-baryon, and baryon-baryon elastic and inelastic scatterings~\cite{Lin:2004en}. The model has been extensively tested to reproduce the $\pt$ spectra and flows of bulk particles.

Minimum bias (MB) Au+Au collisions are simulated by \ampt\ 
at $\snn=200$ and 27~GeV. We terminate the hadronic interactions at a cutoff time of 30~fm/$c$ (by model default).
Figure~\ref{fig:minv} (upper center) shows the difference of the $\minv$ distributions between OS and SS pairs for 30--40\% centrality, as an example.  
The $\rho^0$ peak is evident. No intrinsic mass width is implemented in \ampt, so the resonance peak is a  sharp $\delta$-function. Weak-decay hadrons are treated as stable particles in \ampt, so no $\Ks$ peak is present in the $\minv$ distribution. 

\subsection{EPOS4 Model\label{sec:epos}}
\epos~\cite{Werner:2023zvo} is a Monte Carlo model to simulate high-energy proton-proton and nucleus-nucleus collisions, among others. At high energies many (nucleonic and partonic) interactions happen simultaneously at once, not sequentially. Parallel multiparton interactions are modeled via the pomeron exchange mechanism and result in complex configurations composed of many strings~\cite{Werner:2023fne,Werner:2023mod}. The string decay products further interact, forming a medium treated by the core-corona picture followed by fluid dynamic evolution of the former and ordinary hadronic interactions in the latter~\cite{Werner:2023jps}. The \epos\ model has been extensively tested and tuned using proton-proton and heavy ion  data.

MB Au+Au collisions are  simulated by \epos\ (version 4.0.0) at $\snn=200$ and 27~GeV.
Figure~\ref{fig:minv} (upper right) shows the difference of the $\minv$ distributions between OS and SS pairs for 30--40\% centrality, as an example. 
The $\Ks$ and $\rho^0$ resonance peaks are evident. No intrinsic mass width is implemented in \epos, so the resonance peaks are  sharp $\delta$-functions. 
All $\Ks$ particles are decayed in \epos\ and their decay pions are treated as ``primordial'' pions. 

\subsection{HYDJET++ Model\label{sec:hydjet}}
\hydjet\ is an event generator to simulate heavy ion collisions by combining  two independent components, namely, the soft physics part and the hard physics part~\cite{Lokhtin:2012re,Bravina:2013xla}. 
The former is determined by thermal equilibrium where hadrons are produced on the hypersurface represented by a parameterization of relativistic hydrodynamics with given freeze-out conditions. 
Chemical and kinetic freeze-outs are separate, and hadronic resonances are decayed. 
The hard physics part starts with an initial parton configuration from \pythia~\cite{Sjostrand:2000wi}, lets partons rescatter traversing the hot and dense nuclear medium and losing energy via collisions and radiative gluon emission, and follows with parton hadronization and particle formation.
The model parameters are tuned to reproduce heavy ion  data 
on charged particle multiplicity, $\pt$-spectra and flow.

MB Au+Au collisions are generated by \hydjet\ (version 2.4) at $\snn=200$ and 27~GeV.
Figure~\ref{fig:minv} (lower left) shows the difference of the $\minv$ distributions between OS and SS pairs for 30--40\% centrality, as an example.
The $\Ks$, $\rho^0$, and $f_0(980)$ resonance peaks are evident. No intrinsic mass width is implemented in \hydjet\, so the resonance peaks are   sharp $\delta$-functions. 
All $\Ks$ particles are decayed in \hydjet\ and their decay pions are treated as ``primordial'' pions. 

\subsection{Toy Model I\label{sec:toy1}}
To gain insights, we have also generated particles using a toy model~\cite{wang:2016iov}. 
Toy models are useful and convenient for investigating the outcome of an analysis method given the known input to the model.

Two versions of the toy model are examined. 
In  Toy Model I, we include primordial pions and two resonances, $\Ks$ and $\rho^0$. The resonance mass is generated according to the Breit-Wignar distribution of proper mass and width~\cite{ParticleDataGroup:2014cgo}.
We generate particles uniformly within the rapidity range of $-1.5<y<1.5$, 
taking event-by-event particle multiplicities according to Poisson statistics about the means of three times the measured average pseudorapidity densities~\cite{STAR:2003vqj,Abelev:2008ab,STAR:2015tnn}.
The $\pt$ spectra are parameterized based on the measured data~\cite{PHENIX:2003qdj,STAR:2003jwm}, namely,
\begin{equation}
    \frac{dN}{d\pt} = 
    \begin{cases}
    \pt \left[\exp\left(\frac{\mt}{T_{\rm BE}}\right)-1\right]^{-1} & \text{for pions, and} \\
    \pt \exp\left(-\frac{\mt-m_0}{T}\right) & {\rm for \, resonances,}
    \end{cases}
    \label{eq:pt}
\end{equation}
where transverse momentum $\mt\equiv\sqrt{\pt^2+m_0^2}$ with $m_0$ being the rest mass of the corresponding particle. 
The particle or resonance $\pt$ is sampled from Eq.~\ref{eq:pt}.
At a given $\pt$, the average $\mean{v_2}$ of pion, $\Ks$, or $\rho^0$ meson is calculated based on the following parameterization of data motivated by the number-of-constituent-quark scaling~\cite{Dong:2004ve},
\begin{equation}
    \mean{v_2}(\pt) = n_q\left[\frac{a}{1+\exp\left(-\frac{(\mt-m_0)/n_q-b}{c}\right)}-d\right]\,,
\end{equation}
where $n_q$ is the number of constituent quarks for a given particle and, for our purpose, $n_q=2$.
A 40\% $v_2$ fluctuation is included about the parameterized $\mean{v_2}(\pt)$ event-by-event to obtain the final $v_2$ of the particle, based on which the azimuthal angle is generated about the fixed event plane at $\psi=0$.

The primordial pion, $\Ks$, and $\rho^0$ input information are taken from experimental measurements of Au+Au collisions at 200~GeV in the centrality range of 30--40\%, which we currently focus on.
All parameters used in the toy model can be found in Ref.~\cite{wang:2016iov}.
Figure~\ref{fig:minv} (lower center) shows the difference of the $\minv$ distributions between the OS and SS pairs from the generated toy-model data.  
We require that all resonances in our toy model (i.e.~$\Ks$ and $\rho^0$) to decay into a pair of charged pions. 
The decay is isotropic in the parent rest frame, and the decay pions are  properly boosted to the lab frame.
The $\Ks$ and $\rho^0$ resonance peaks are evident. 

As default only 2\% of the measured $\Ks$ abundance are generated in our toy model because this is the approximate fraction of $\Ks$ the decay daughters of which are both  reconstructed as primordial pions in the STAR experiment. 
We vary this fraction to investigate its effect on our results. 

\subsection{Toy Model II\label{sec:toy2}}
As shown in Fig.~\ref{fig:minv}, all physics models have a mass continuum in the $\minv$ distribution on which the resonance peaks reside. To mimic this, we add a mass continuum in our second version of the toy model. The mass distribution of the continuum is generated according to a probability $\propto e^{-\minv c^2/{\rm GeV}}$ between the two-pion mass threshold (0.28~\gevcc) and 5~\gevcc. Given a generated mass value, a Breit-Wignar distribution (mean at the mass value and width equal to the $\rho^0$ width~\cite{ParticleDataGroup:2014cgo}) is sampled to obtain the mass of the continuum ``resonance.''  All particles in the generated mass continuum are decayed into $\pi^+\pi^-$ isotropically in the parent rest frame. The decay pions are properly boosted to the lab frame.

\subsection{Analysis Details}\label{subsec:ana}
\begin{figure*}
    \includegraphics[width=0.3\textwidth]{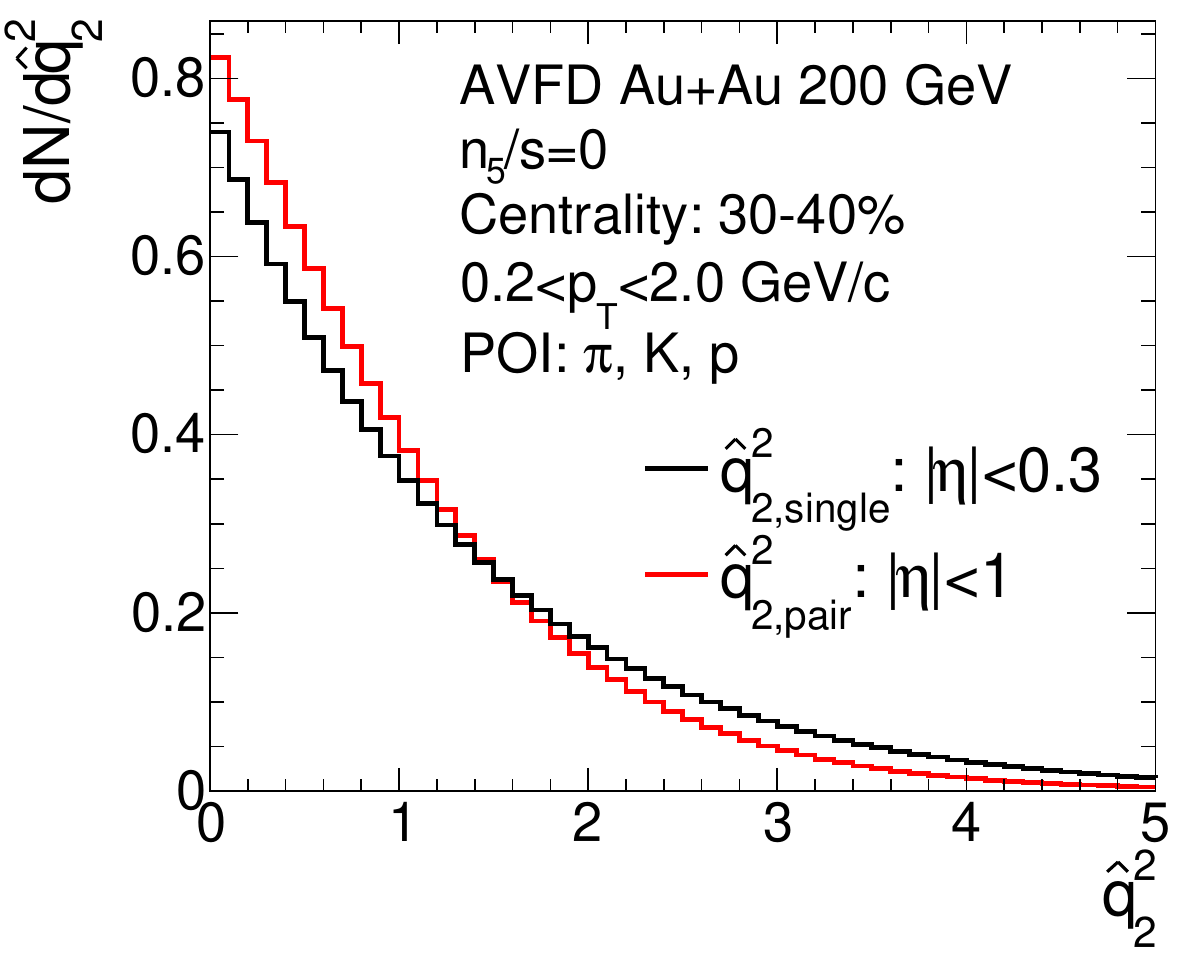}
    \includegraphics[width=0.3\textwidth]{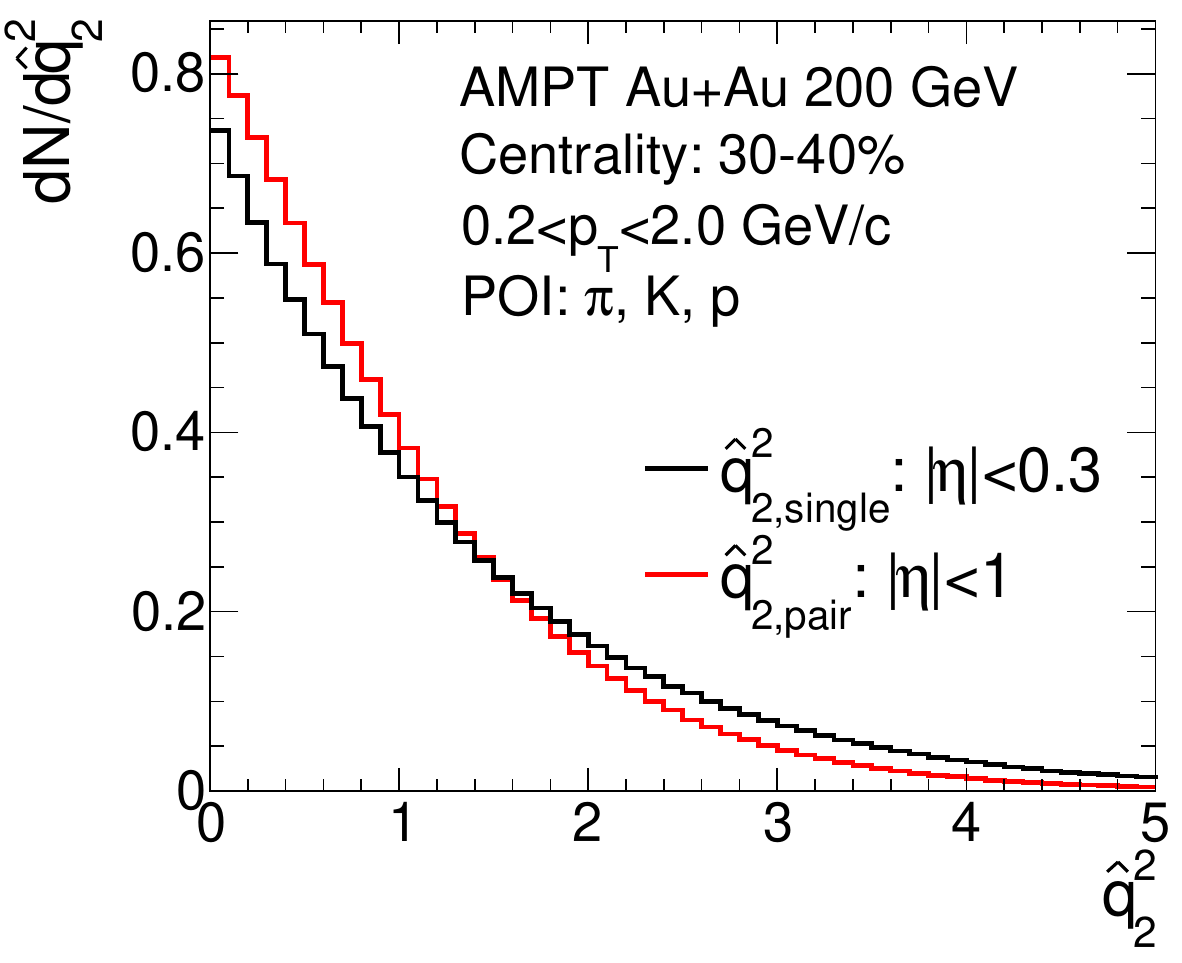}
    \includegraphics[width=0.3\textwidth]{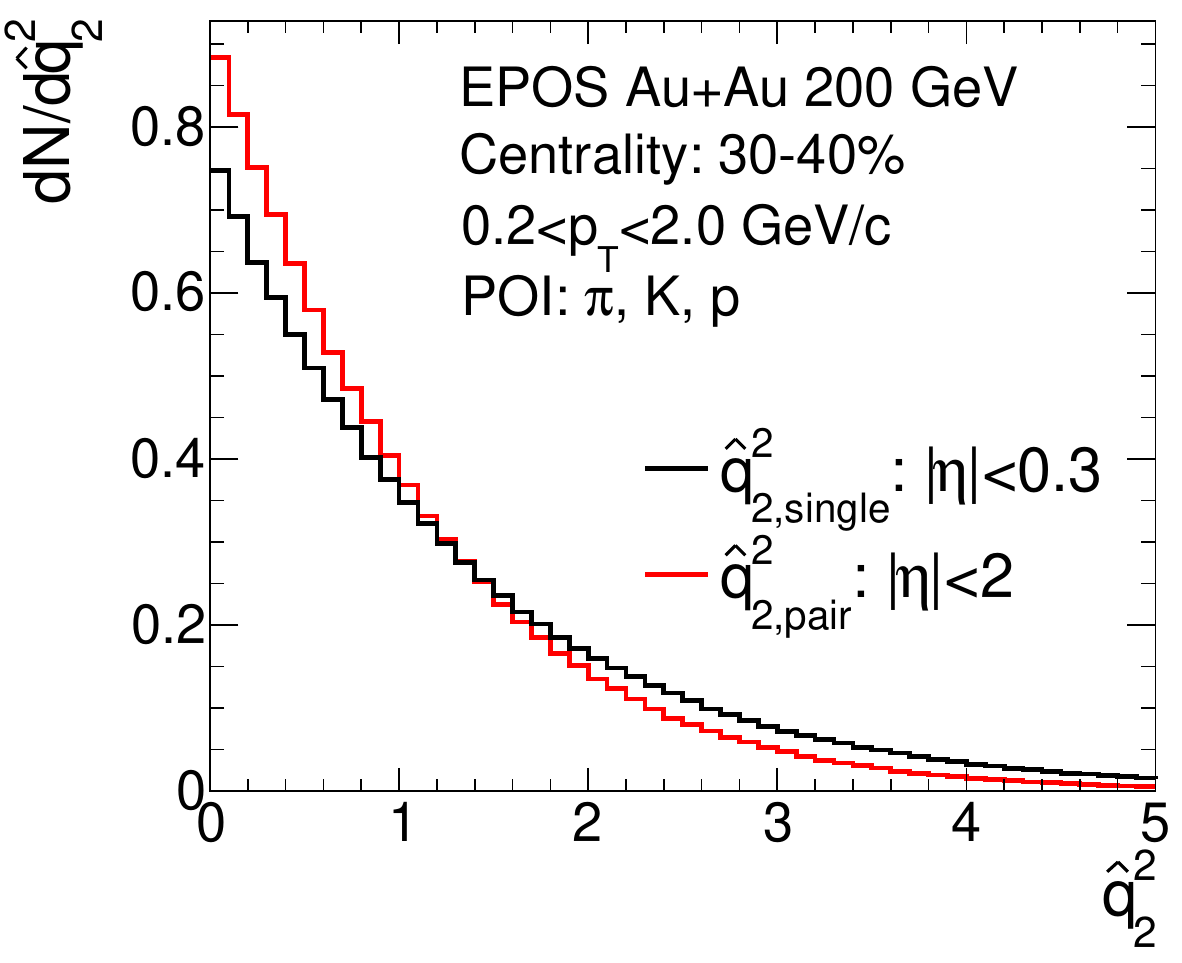}
    \includegraphics[width=0.3\textwidth]{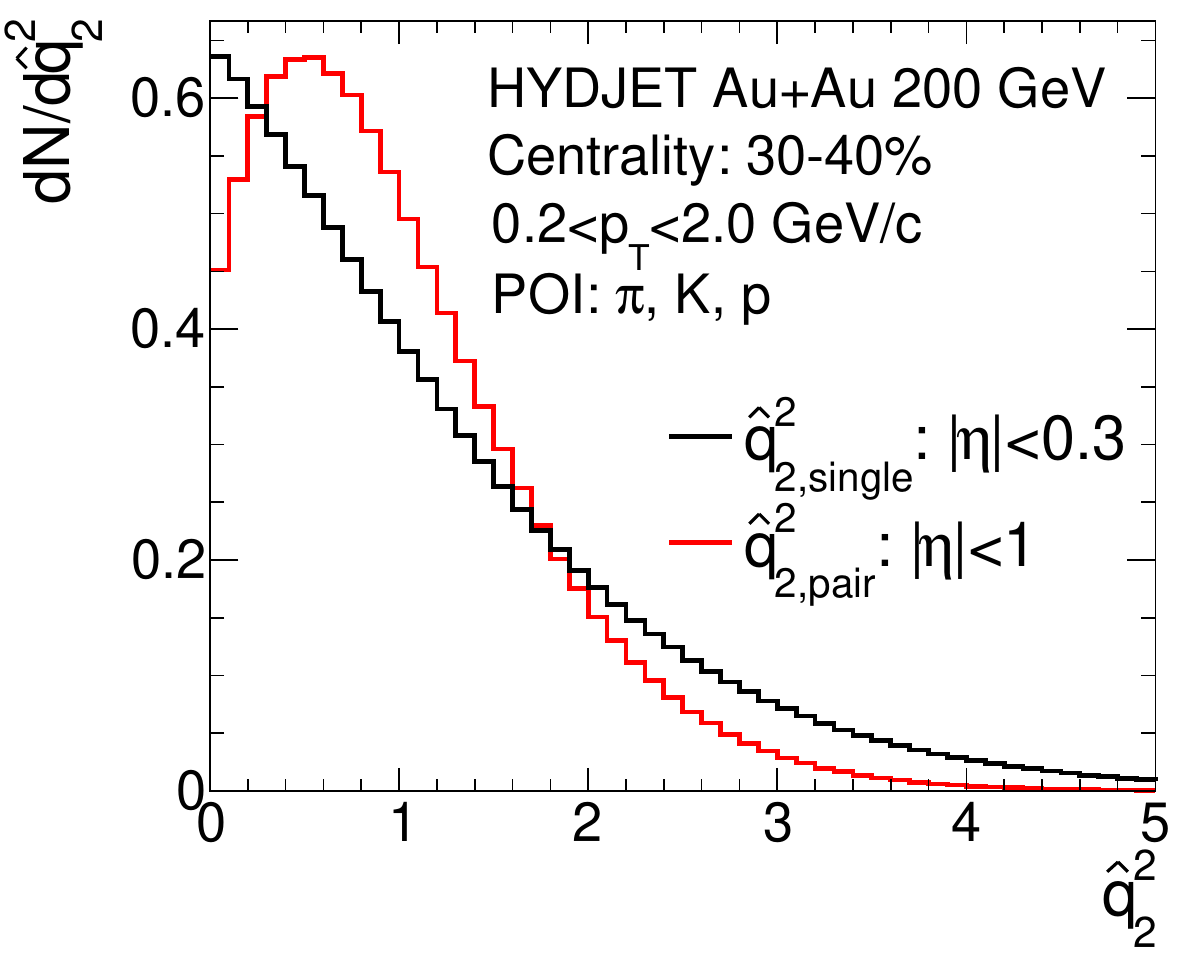}
    \includegraphics[width=0.3\textwidth]{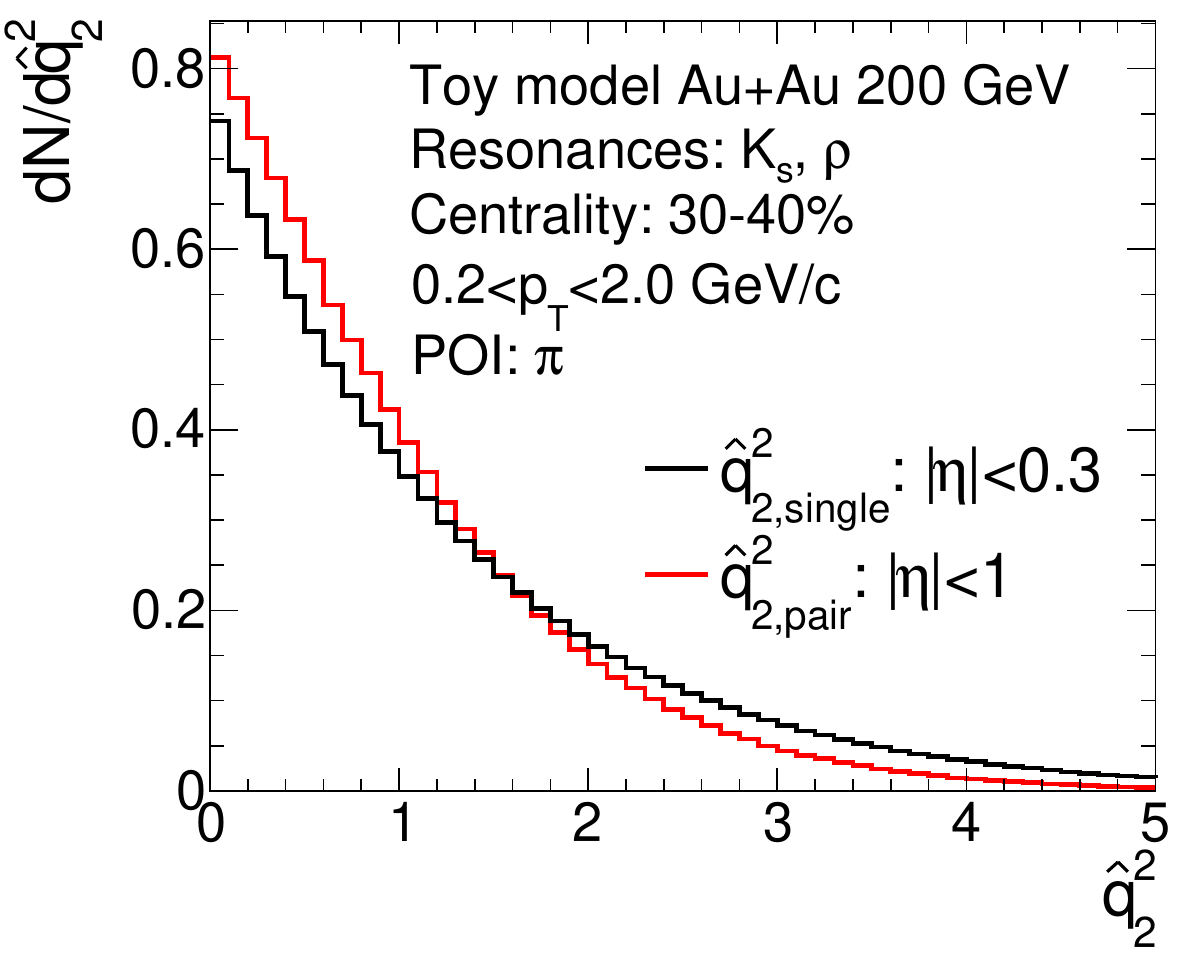}
    \includegraphics[width=0.3\textwidth]{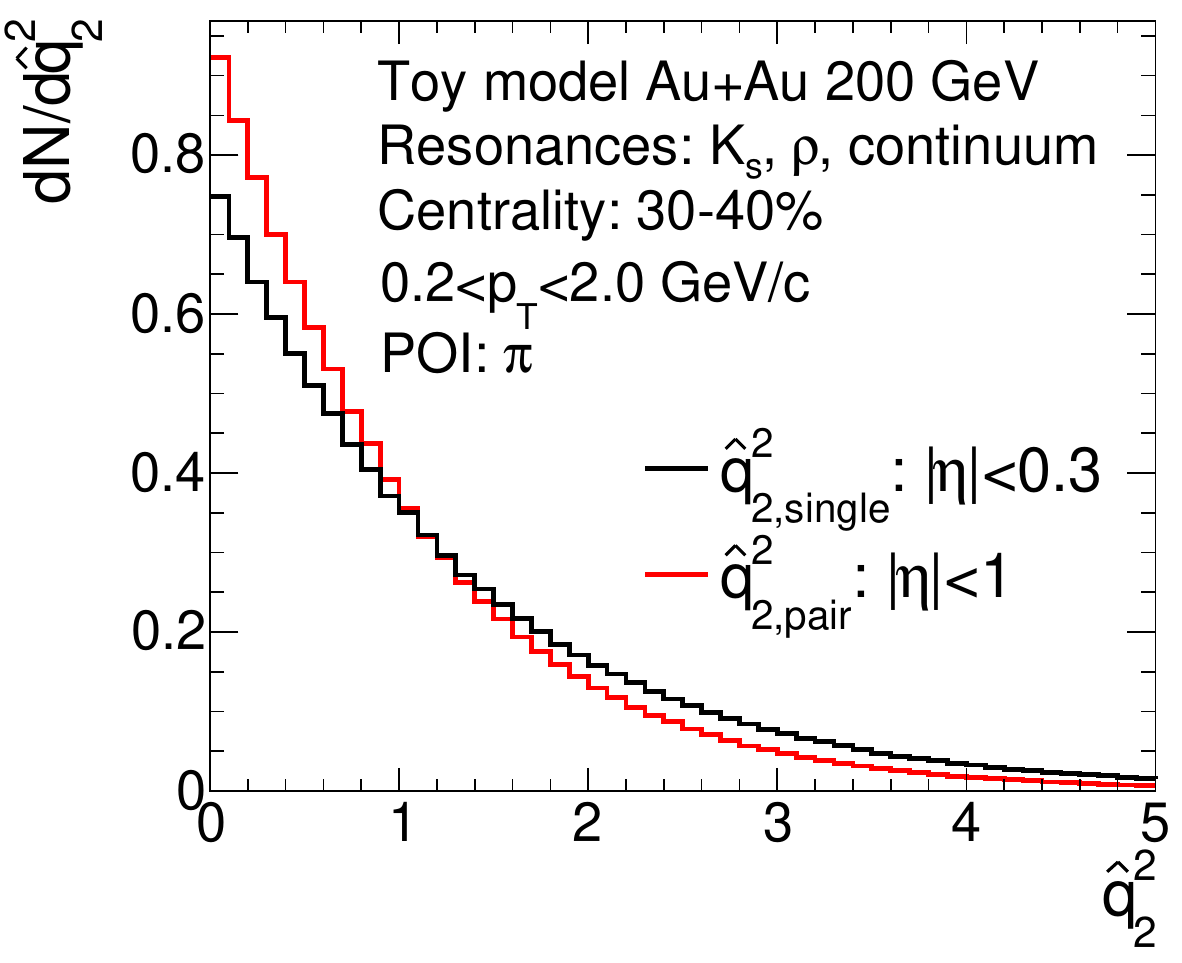}
    \caption{\label{fig:q2}$\qh^2$ distributions. Shown are the single $\qh^2$ and pair $\qhpair^2$ distributions of POIs from \avfd\ with $\ns=0$ (upper left), \ampt\ (upper center), \epos\ (upper right), \hydjet\ (lower left), Toy Model I (lower center), and Toy Model II (lower right). Shown in  all panels are for the 30--40\% centrality of Au+Au collisions, as examples. The $q_2^2$ (used in the ESE analysis) is calculated from particles in $|\eta|<0.3$, and the pair $\qpair^2$ (used in the ESS analysis) is calculated from particles within $|\eta|<1$, both with $0.2 < \pt < 2$~\gevc.}
\end{figure*}

POIs are assumed to be all charged hadrons, defined to be $\pi^\pm$, $K^\pm$, $p$ and $\bar{p}$ in our study, and are within the transverse momentum range of $0.2 < \pt < 2$~\gevc\ and a certain central pseudorapidity ($\eta$) range, typical of midrapidity detectors such as the STAR experiment~\cite{Ackermann:2002ad}.

In the ESS method, the POIs are taken to be within $|\eta|<1$ (or $|\eta|<2$). The event selection quantity, pair $\qhpair^2$, is  computed  by the two-particle cumulant method in Eqs.~\ref{eq:q2},\ref{eq:qhpair}, always using all particle pairs of the same sample of POIs. 

In the ESE method, the acceptance $|\eta|<1$ (or $|\eta|<2$) is divided into three subevents: 
\begin{itemize}
    \item west subevent: $-1$$<\eta<$$-0.3$ (or $-2$$<\eta<$$-0.3$),
    \item center subevent: $|\eta|<0.3$,
    \item east subevent: $0.3<\eta<1$ (or $0.3<\eta<2$).
\end{itemize}
The POIs are taken from either the west subevent or the east subevent, but not from both. The event selection quantity $\qh^2$ is computed by the two-particle cumulant method in Eqs.~\ref{eq:q2},\ref{eq:qh}, using particles from the center subevent, $|\eta|<0.3$ (or from a forward/backward region $3<|\eta|<4$).

Figure~\ref{fig:q2} shows the event-by-event $\qhpair^2$ and $\qh^2$ distributions from the ESS and ESE methods, respectively. 
The two distributions are similar for each of the models; the $\qh^2$ distribution ($|\eta|<0.3$) is broader than the $\qhpair^2$ distribution ($|\eta|<1$), presumably due to larger fluctuations in the smaller acceptance. 
We have checked that the distribution of the single particle $\qh^2$ calculated from $|\eta|<1$ is similar to the pair $\qhpair^2$ from the same acceptance.

\begin{figure*}
    \includegraphics[width=0.3\textwidth]{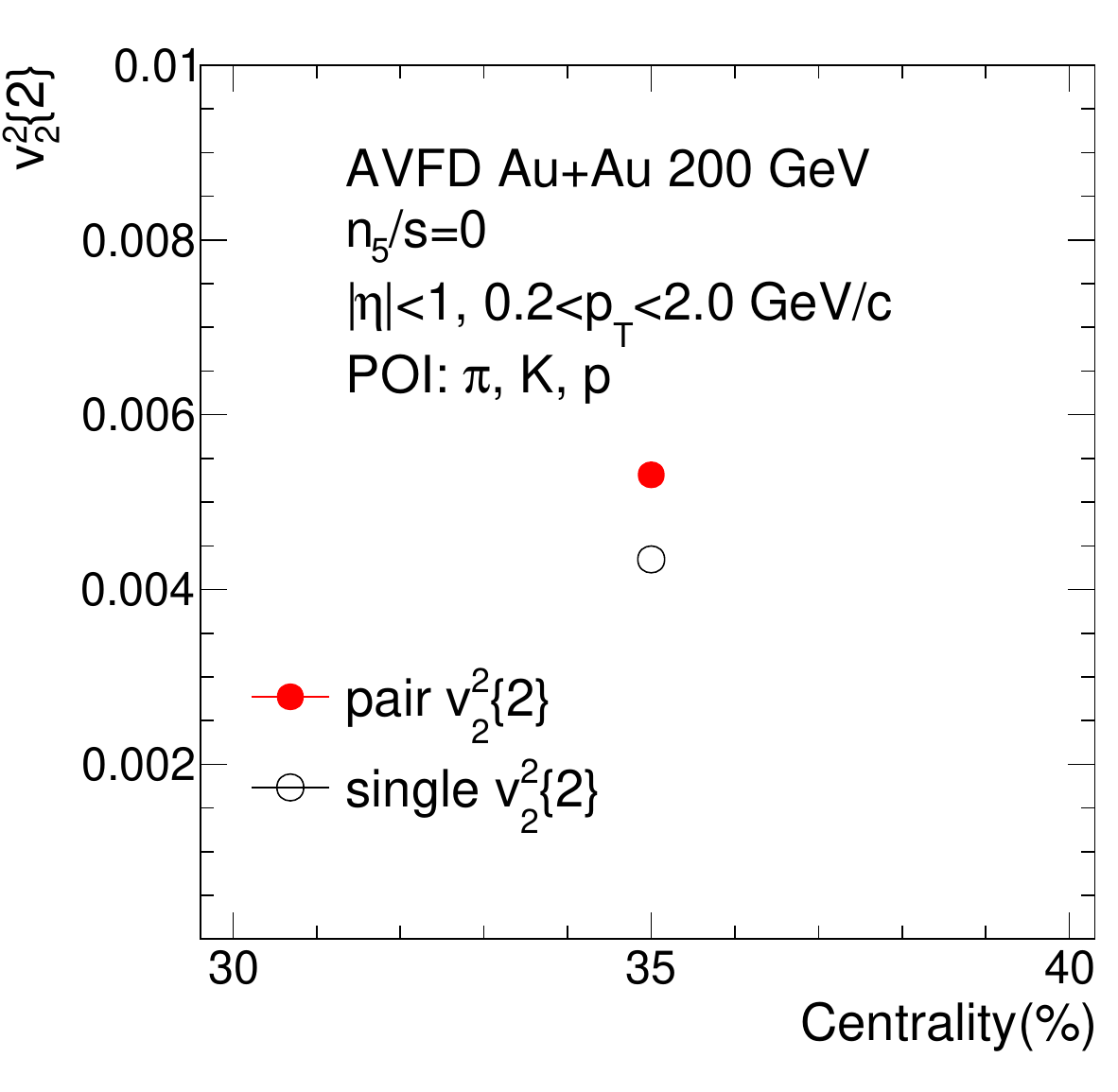}
    \includegraphics[width=0.3\textwidth]{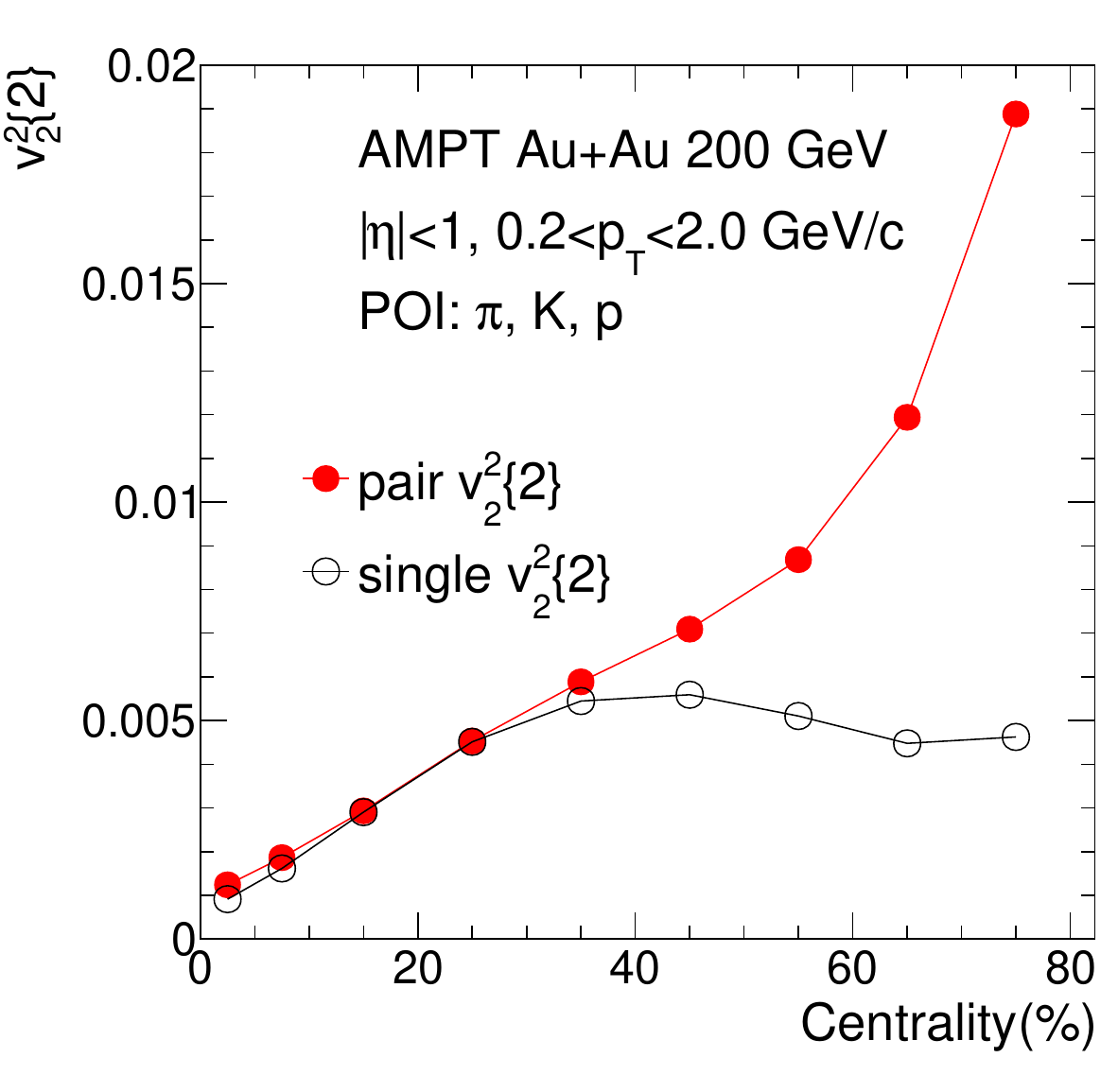}
    \includegraphics[width=0.3\textwidth]{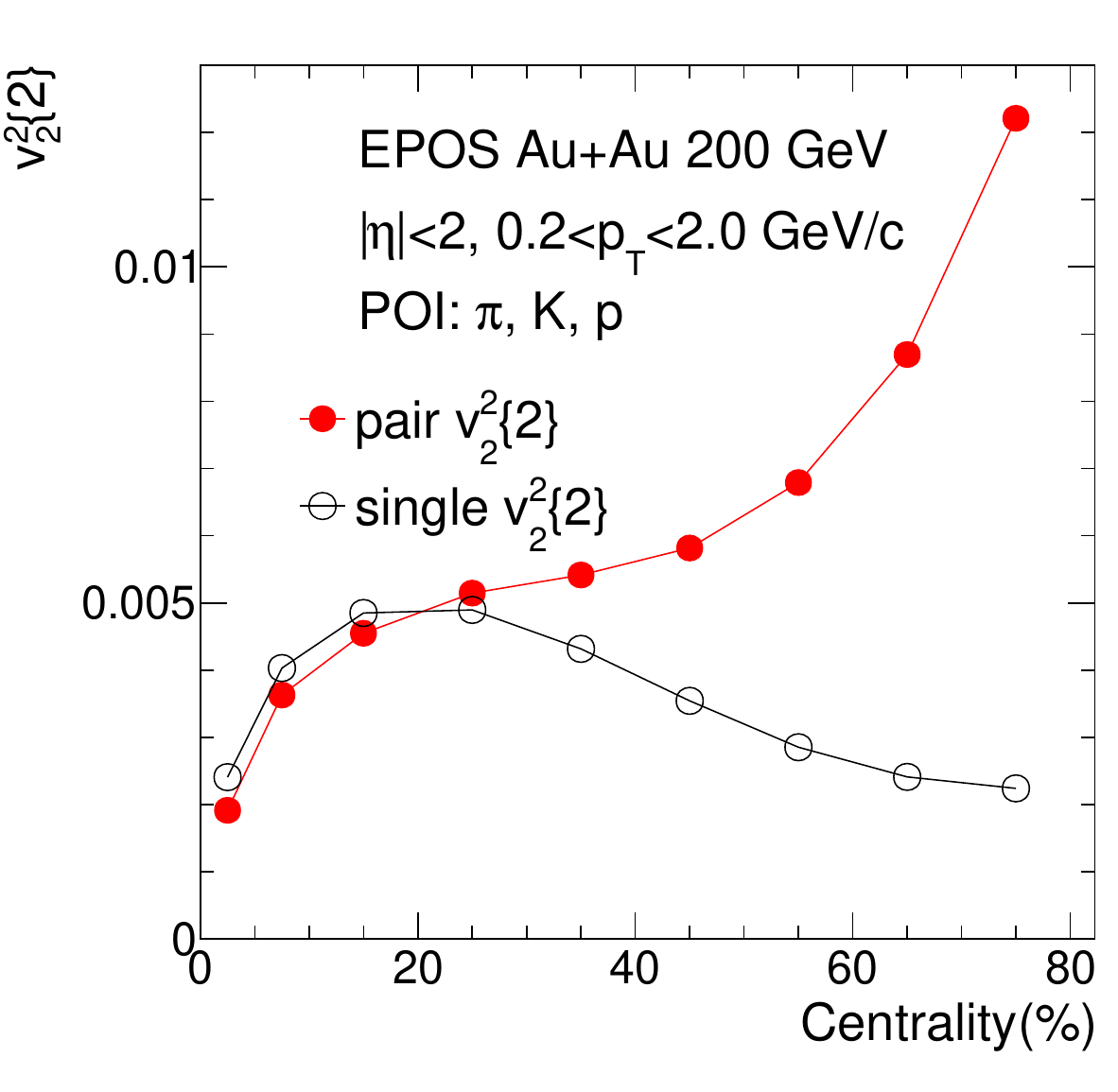}
    \includegraphics[width=0.3\textwidth]{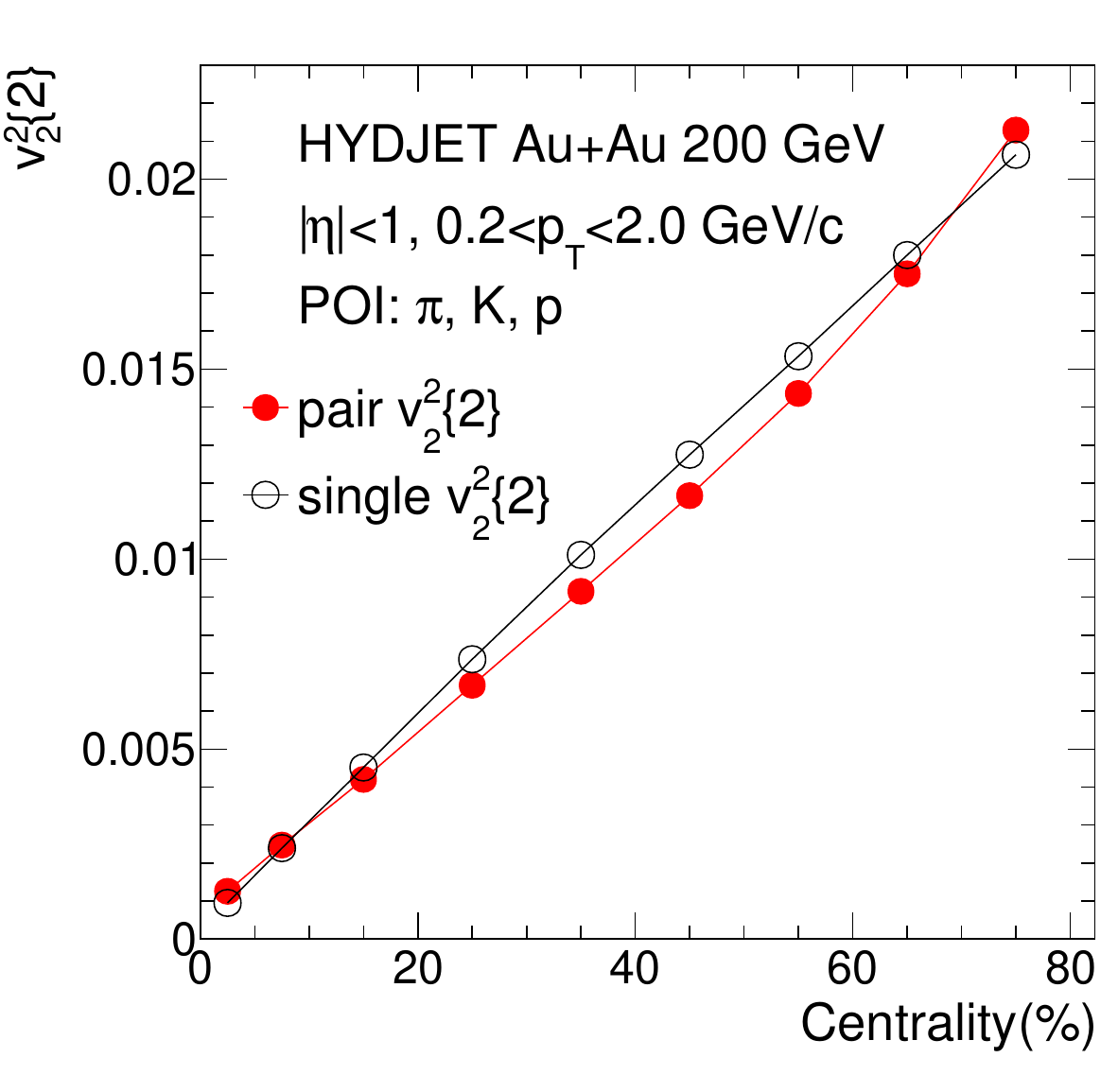}
    \includegraphics[width=0.3\textwidth]{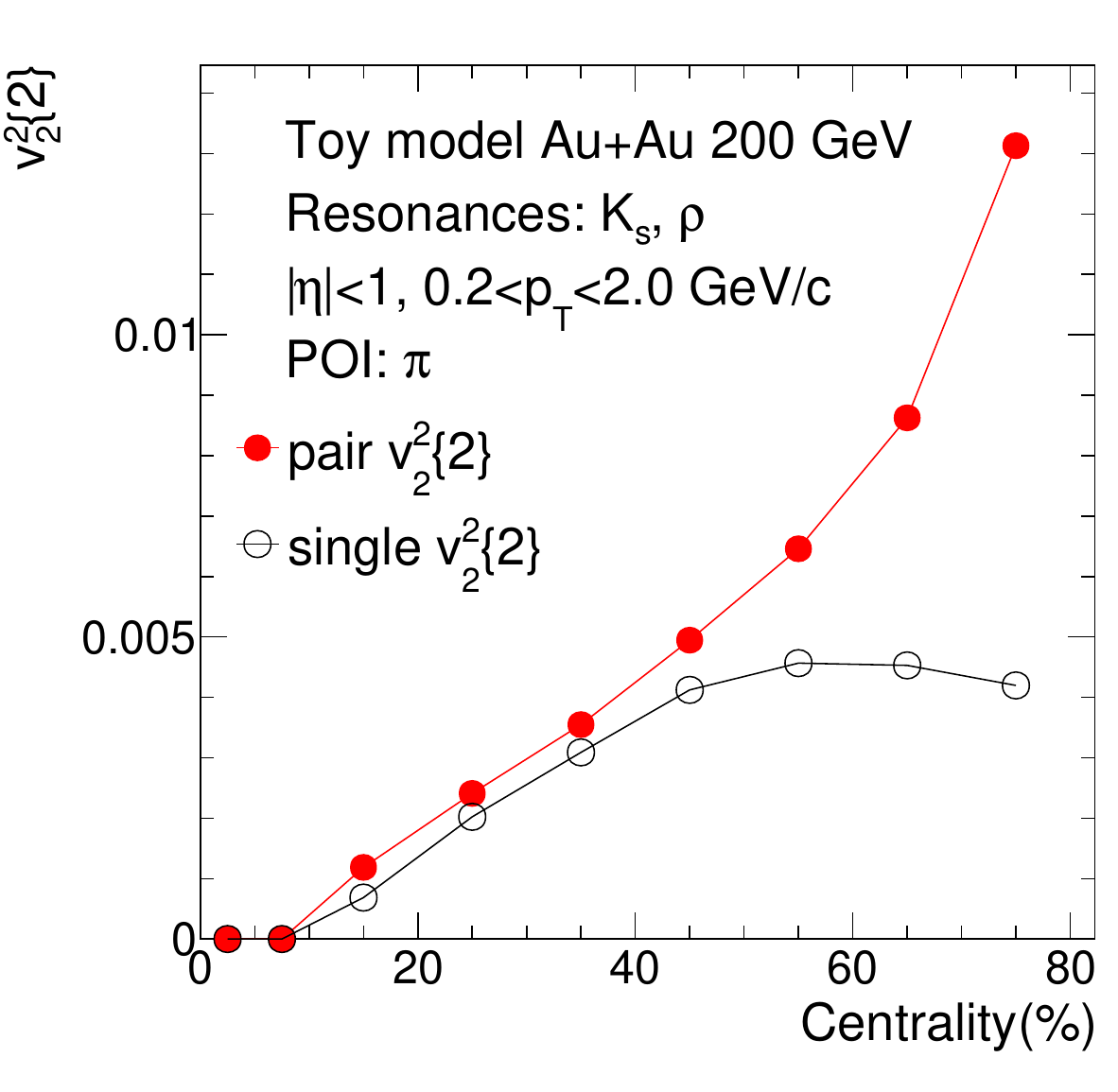}
    \includegraphics[width=0.3\textwidth]{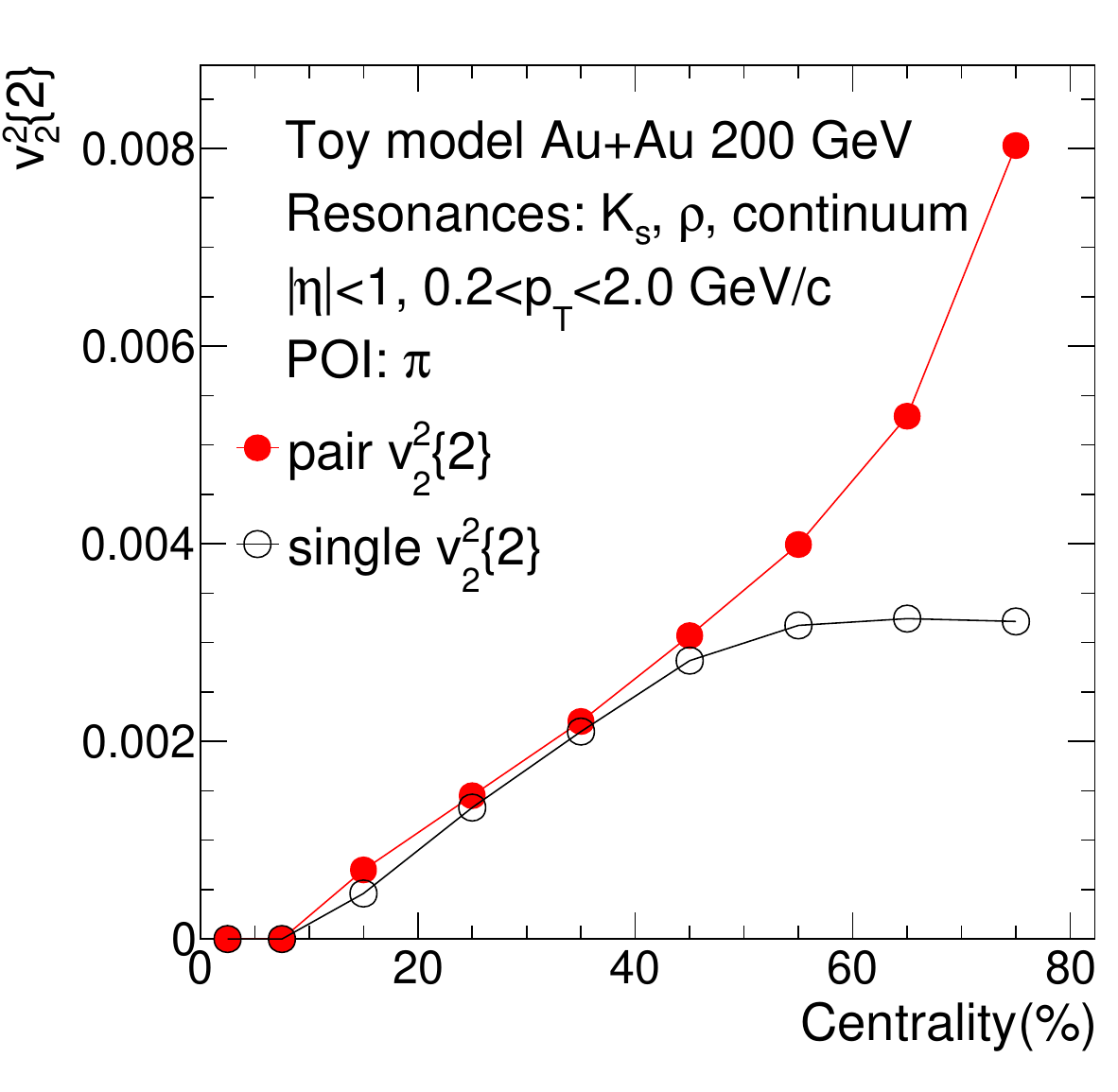}
    \caption{\label{fig:v2}Average $v_2^2\two$.     
     Shown are the two-particle cumulant $v_2^2\two$ of POIs and $\vpair^2\two$ of particle pairs of POIs as functions of centrality from \avfd\ for $\ns=0$ (upper left), \ampt\ (upper center), \epos\ (upper right), \hydjet\ (lower left), Toy Model I (lower center), and Toy Model II (lower right). 
    POIs are from acceptances $|\eta|<1$ (except for \epos, $|\eta|<2$) and $0.2 < \pt < 2$~\gevc. The large departure of $\vpair^2\two$ from $v_2^2\two$, both calculated by the cumulant method (Eq.~\ref{eq:v22}), is largely due to self-correlations where two pairs share a common particle. The issue is severe in peripheral collisions except for \hydjet\ where the average multiplicity is significantly larger than the rest models; for reference, for 30--40\% centrality of Au+Au, the average multiplicities within $|\eta|<0.5$ are 151, 169, 88, and 255 for \avfd, \ampt, \epos, and \hydjet, respectively.}
\end{figure*}

The $\qh^2$ variable calculated from single particles has been widely used. 
The $\qhpair^2$ variable calculated from particle pairs is relatively new~\cite{Xu:2023elq}.
As seen from Eq.~\ref{eq:q2mean}, the average $\mean{q_2^2}$ is related to the two-particle cumulant $v_2\two$. 
To gain insights, we calculate by Eq.~\ref{eq:v22} the $v_2\two$ of single POIs and the $\vpair\two$ of particle pairs of POIs; 
In the latter case, the $\phi_1$ and $\phi_2$ in Eq.~\ref{eq:v22} are the azimuthal angles of two pairs, each formed by two POIs.
The $\vpair\two$ consists of a component from pairing of ``random'' (hydrodynamic) particle pairs or a hydrodynamic pair with a resonance-decay pair, and a component from two-resonance $v_2\two$. The first component reflects, in a non-trivial way, the single particle $v_2$~\cite{Li:2024pue}. The last component is simply resonance (background contributing source) $v_2\two$.
Both $v_2\two$ and $\vpair\two$ contain nonflow effects, for example, from resonance decays (in the case of $\vpair\two$, the two pairs both contain daughters from the same resonance decay). 
Figure~\ref{fig:v2} shows $v_{2}\two$ (from 2 POIs) and $\vpair\two$ (from 2 pairs of POIs) as functions of centrality in different simulations of Au+Au collisions at $\snn = 200 \text{ GeV}$. 
They roughly agree at high multiplicity (\hydjet, or central collisions of other models), and $\vpair\two$ becomes much larger than $v_{2}\two$ at low multiplicity (peripheral collisions of models except for \hydjet). 
This discrepancy comes from the self-correlations where the two pairs share one common POI. 
To get a feeling of this effect, consider a simplified scenario with $\phi_{\pair} = (\phi_{1} + \phi_{2})/2$. 
For two pairs with indices $(1,3)$ and $(2,3)$ where particle 3 is shared, their correlation is $\cos 2[(\phi_{1}+\phi_{3})/2 - (\phi_{2}+\phi_{3})/2] = \cos(\phi_{1}-\phi_{2})$, and the fraction of this case in the $\vpair^{2}\two$ calculation is the triplet multiplicity over the quadruplet one, which is $\sim 1/N$. 
Therefore, besides the common flow component, $\vpair\two$ has an extra part $\sim\langle \cos(\phi_{1}-\phi_{2}) \rangle / N$, which  accounts qualitatively for its excess over $v_{2}\two$ when multiplicity is low. 

In the ESS method, events are divided into bins of $\qhpair^2$ of equal bin width 0.1. 
In the ESE method, events are grouped into five ranges in $q_2$, four equal size of width 0.5 from $q_2=0$ to 2, and the last range $q_2>2$. Note this is equivalent to grouping the events according to $\qh^2$ with corresponding divisions.

For the ESS analysis, in each event class according to $\qhpair^2$, we calculate the single particle $\vsing$ by Eq.~\ref{eq:v2} and the $\dg$ correlator by Eqs.~\ref{eq:g},\ref{eq:dg} using  the EP method. 
The EP azimuthal angle is taken to be $\psi=0$ as fixed in models. 
The experimental analogy is data analysis using the first-order event plane reconstructed from spectator neutrons in zero-degree calorimeters (ZDC)~\cite{Adler:2003sp}. 

Likewise, in each event class according to $q_2$ in the ESE analysis, we also use the known  EP of $\psi=0$ in models to calculate $\mean{v_2}$ by Eq.~\ref{eq:v2} and $\dg$ by Eqs.~\ref{eq:g},\ref{eq:dg}. 
Here, the POI is taken from the west or east subevent in calculating $\mean{v_2}$, and in calculating $\dg$, both POIs are taken from either the west subevent or the east subevent.
In addition, we use the two-particle cumulant (Eq.~\ref{eq:v22}) to compute $v_2\two$ where one particle is taken from the west subevent and the other from the east subevent, 
and the three-particle correlator~\cite{Voloshin:2004vk} to compute 
\begin{equation}
    \gamma=\mean{\cos(\phi_\alpha+\phi_\beta-2\phi_c)}/v_2\two\,,
    \label{eq:g3}
\end{equation}
where $\phi_\alpha$ and $\phi_\beta$ are the azimuthal angles of two POIs from the west (east) subevent and $\phi_c$ is that of a third particle from the other-side subevent, i.e.~the east (west) subevent.

\section{results and discussions\label{sec:results}}

\subsection{Physics Models}
Figure~\ref{fig:avfd_ess} in the appendix shows the \avfd\ simulation results of $\dg$ as functions of $\vsing$ in bins of $\qhpair^2$ analyzed by the ESS method in 30--40\% centrality Au+Au collisions at $\snn=200$~GeV. The $\qhpair^2$ is computed from the same POIs used for the $\dg$ and $\vsing$ measurements. Three values of $\ns$ are simulated. The results are consistent with those of Ref.~\cite{Xu:2023elq}.
In \avfd\ and all other physics models we studied, $\dg$ appears to be linear as a function of $\vsing$. We fit the measured range of $0<\vsing<0.2$ to obtain the intercept $\dgess$ at $\vsing=0$. 
The intercept is sensitive to CME with the $v_2$-induced background significantly reduced. 
For $\ns=0$, the intercept is consistent with zero as one would expect if the background was completely removed.
For $\ns=0.1$ and 0.2, the intercepts are found to be $1.34\times10^{-4}$ and $5.70\times10^{-4}$ with negligible statistical uncertainties, respectively. The CME signal is expected to scale with $(\ns)^2$.
The ratio of the ESS intercept for $\ns=0.2$ to that for $\ns=0.1$ is $4.25\pm0.13$, deviating from the expected ratio of 4 by about 2 standard deviations.

Figure~\ref{fig:avfd_ese} shows the \avfd\ results analyzed using the ESE method of $\dg$ as functions of $v_2$ in events classified according to $\qh^2$. The $\qh^2$ is calculated using particles from $|\eta|<0.3$, separate from the POIs from $0.3<|\eta|<2$ (upper cut of 2 instead of typically 1 to increase statistics) that are used to calculate $\dg$ and $v_2$. 
The range of dynamical variation of $v_2$ is significantly smaller than the range of variation in $\vsing$ of the ESS method, which is predominantly statistical.
The intercept $\dgese$ is obtained from a linear fit to all data points.
The intercept for the $\ns=0$ case is consistent with zero, as expected. 
The intercepts for $\ns=0.1$ and 0.2 are $(1.63\pm0.19)\times10^{-4}$ and $(6.61\pm0.19)\times10^{-4}$, respectively. 
The statistical uncertainties are larger than those from ESS because of the larger range of projection to the intercept. 
The intercept ratio between $\ns=0.2$ and 0.1 is $4.06\pm0.49$, consistent with the expected ratio of 4 although the  uncertainty is large.
It is interesting to note that the fit slopes from both ESS and ESE vary with $\ns$ significantly, whereas it is expected to measure the background strength. This is likely because the CME signal particles have altered the strength of background-contributing correlations in the final state~\cite{Li:2024pue} and thus affected the slope as well.

Figure~\ref{fig:ampt_ess} shows the \ampt\  results of  $\dg$ as functions of $\vsing$ in bins of $\qhpair^2$ analyzed by the ESS method in three centralities of Au+Au collisions at $\snn=200$~GeV and 27~GeV. The intercepts are all negative; Negative intercepts were also found by the authors of Ref.~\cite{Xu:2023elq,Milton:2021wku}. Since \ampt\ does not have CME signals, the results suggest that the ESS intercept is sensitive not only to CME signals but also to backgrounds.
The fit slope increases with decreasing centrality and is larger at 27~GeV than at 200~GeV; this is expected because the $\dg$ background is approximately inversely proportional to multiplicity.

Figure~\ref{fig:ampt_ese} shows the \ampt\ results analyzed by the ESE method of $\dg$ as functions of $v_2$ in events classified according to $\qh^2$ calculated from particles within $|\eta|<0.3$ whereas POIs are taken from $0.3<|\eta|<1$. The intercepts are consistent with zero as expected, albeit with large statistical uncertainties. 
The fit errors on the slope parameter are too large to draw firm conclusions. 
We have also used $\qh^2$ from particles at forward/backward pseudorapidities $3<|\eta|<5$ and $2<|\eta|<3$ for 200~GeV and 27~GeV collisions, respectively, keeping the same POI acceptance. 
The results are shown in Fig.~\ref{fig:ampt_ese_forward} and are found to be similar to those with midrapidity $\qh^2$ but with larger uncertainties.

Figure~\ref{fig:epos_ess} shows the \epos\ results of  $\dg$ as functions of $\vsing$ in bins of $\qhpair^2$ analyzed by the ESS method in three centralities of Au+Au collisions at $\snn=200$~GeV and 27~GeV. The POI and $\qhpair^2$ pseudorapidity acceptance is taken to be $|\eta|<2$ to increase statistics. The intercepts appear to be positive, 
opposite to those in \ampt. This again suggests that the ESS intercept may be sensitive to backgrounds since \epos\ does not have CME signals. 
There is an indication of an increasing fit slope with decreasing centrality; however, the uncertainties are large.

Figure~\ref{fig:epos_ese} shows the \epos\ results analyzed by the ESE method of $\dg$ as functions of $v_2$ in events classified according to $\qh^2$ from $|\eta|<0.3$. The POI pseudorapidity acceptance is taken to be $0.3<|\eta|<2$ to increase statistics. The intercepts are consistent with zero as expected, albeit with large statistical uncertainties. The fit errors on the slope parameter are too large to draw firm conclusions.

Figure~\ref{fig:hydjet_ess} shows the \hydjet\ results of  $\dg$ as functions of $\vsing$ in bins of $\qhpair^2$ analyzed by the ESS method in three centralities of Au+Au collisions at $\snn=200$~GeV and 27~GeV. The intercepts are consistent with zero at 200~GeV but are all positive at 27~GeV. Since \hydjet\ does not have CME signals, the results again indicate that the ESS intercept is sensitive not only to CME signals but also to backgrounds. The \hydjet\ results further suggest that the degree of the sensitivity to backgrounds depends on the collision energy, or details of the event content.
The fit slope increases with decreasing centrality and is larger at 27~GeV than at 200~GeV, as expected, similar to the \ampt\ results.

Figure~\ref{fig:hydjet_ese} shows the \hydjet\ results analyzed by the ESE method of $\dg$ as functions of $v_2$ in events classified according to $\qh^2$. The intercepts are all consistent with zero, as expected, except the 20--30\% and 40--50\% centralities at 27~GeV where the intercept is displaced from zero but only with less than 2 standard deviations. 
The fit slope increases with decreasing centrality. 

We have also used the three-particle correlator method to calculate $\dg$ and the two-particle cumulant to calculate $v_2\two$, without relying on the known EP at $\psi=0$. 
The results are thus subject to nonflow effects. 
The results are shown in Fig.~\ref{fig:hydjet_ese3} corresponding to the same \hydjet\ data as in Fig.~\ref{fig:hydjet_ese}. The results are consistent with those in Fig.~\ref{fig:hydjet_ese} calculated with the known $\psi=0$, with somewhat larger statistical uncertainties. 
This suggests that the nonflow effects are relatively small in \hydjet\ within the statistical uncertainties of the current study. 

We have also used $\qh^2$ from particles at forward/backward pseudorapidities of $3<|\eta|<5$ and $2<|\eta|<3$ for 200~GeV and 27~GeV collisions, respectively, keeping the same POI acceptance. The results are shown in Fig.~\ref{fig:hydjet_ese_forward} corresponding to the same \hydjet\ data as in Fig.~\ref{fig:hydjet_ese}. The intercepts are all consistent with zero, except the 40--50\% centrality bin at 27~GeV where the intercept is positive but within 2 standard deviations from zero.  

\subsection{Discussions on Physics Model Results}
\begin{figure*}[hbt]
    \includegraphics[width=0.4\textwidth]{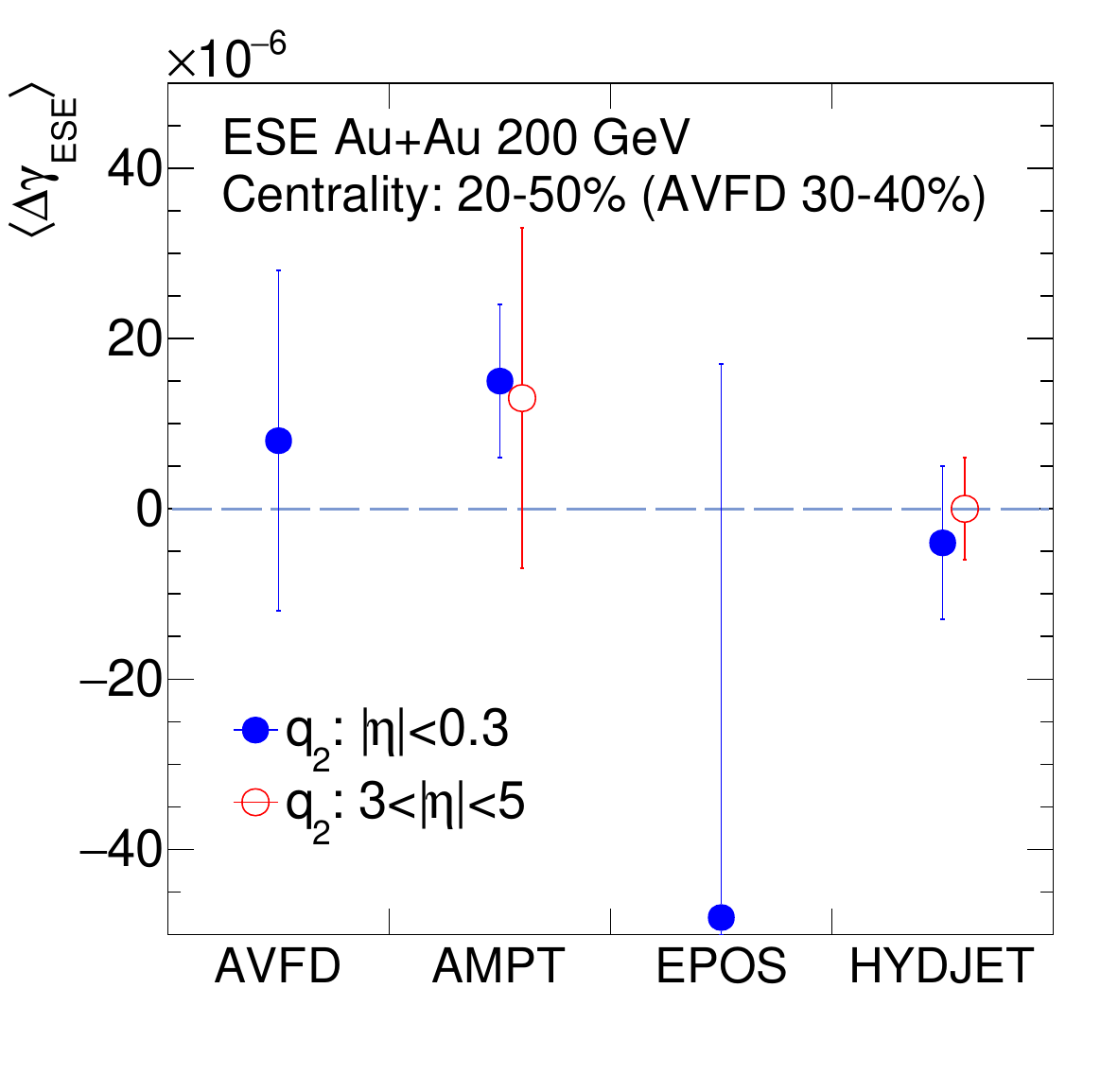}
    \includegraphics[width=0.4\textwidth]{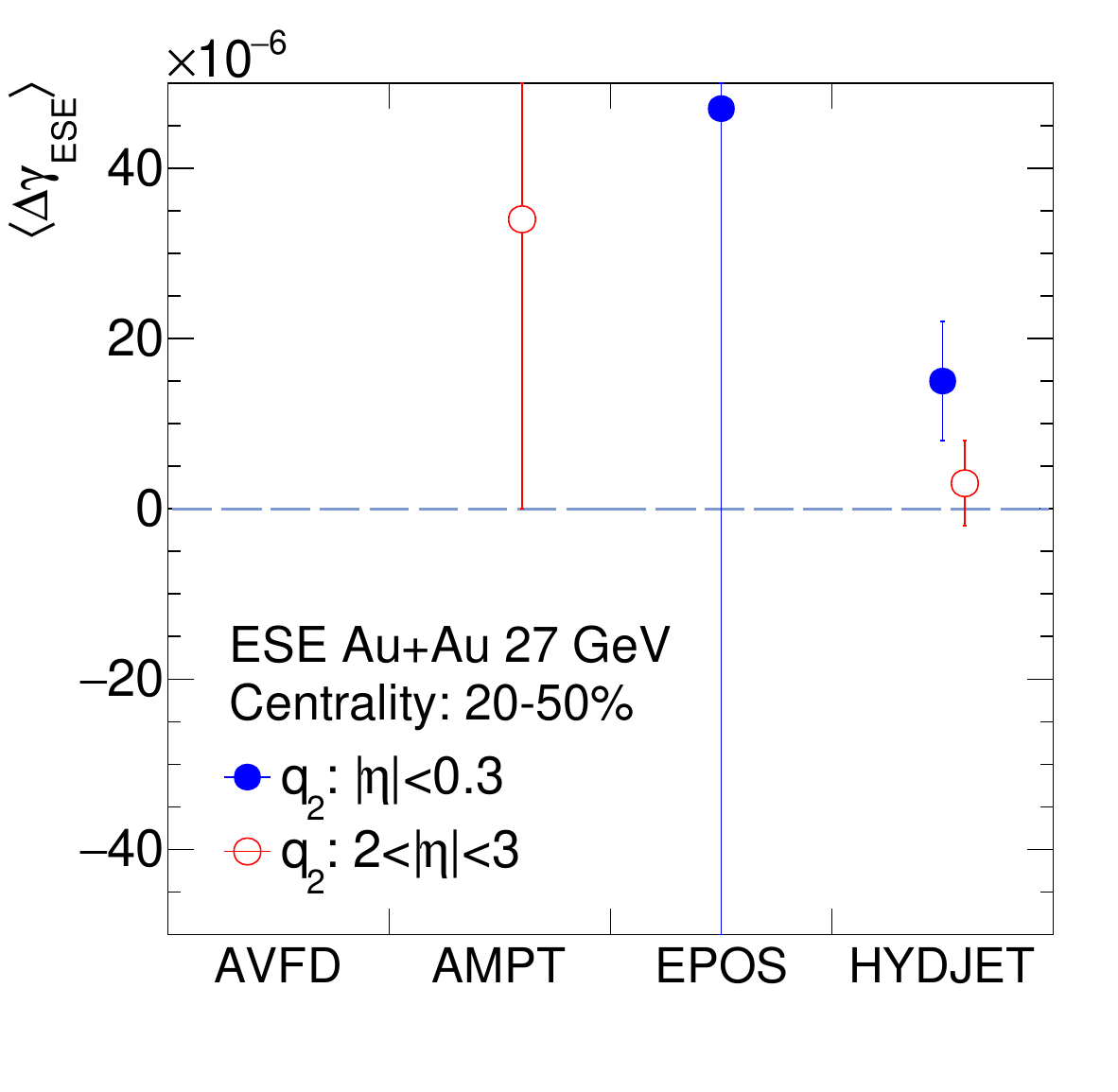}
    \vspace{-4mm}\\
    \includegraphics[width=0.4\textwidth]{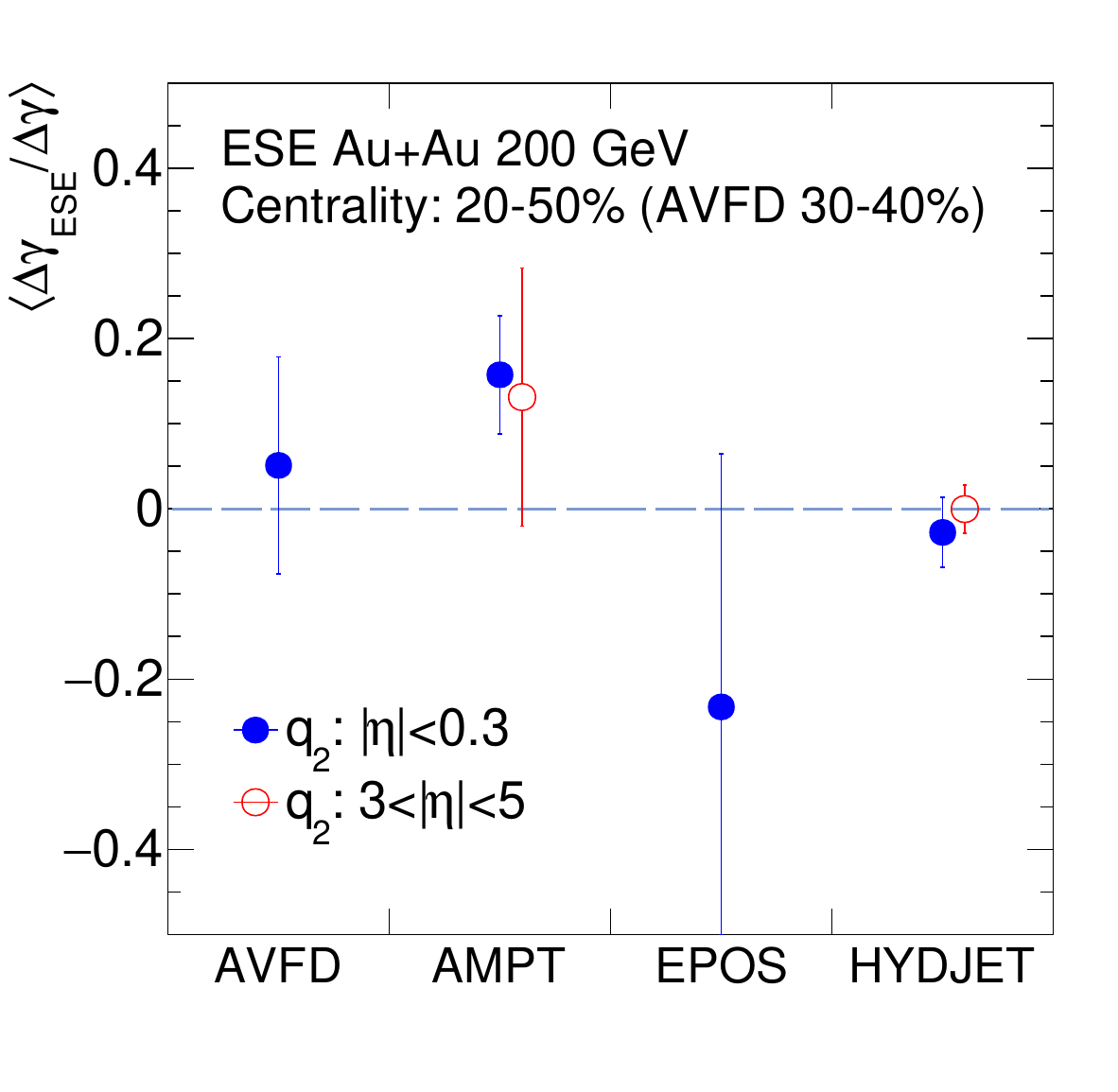}
    \includegraphics[width=0.4\textwidth]{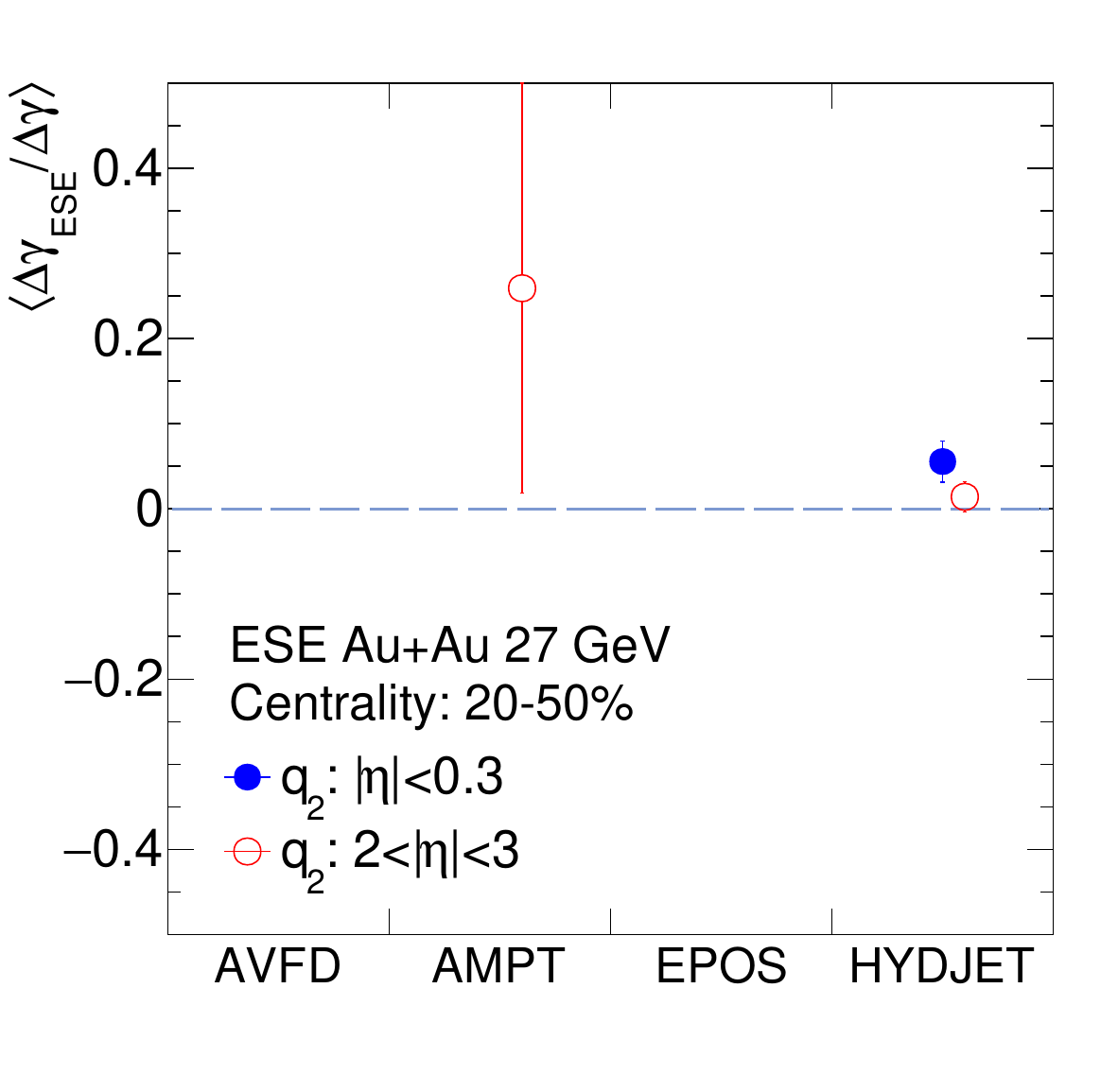}
    \caption{ESE intercepts. Shown are intercepts (upper panels) and intercepts divided by the corresponding inclusive $\dg$ values (lower panels) from the ESE method for various physics models. For \avfd\ ($\ns=0$), the result is from the 30--40\% centrality Au+Au collisions at $\snn=200$~GeV. For \ampt, \epos, and \hydjet, the results are averaged over the 20--50\% centrality range of Au+Au collisions at $\snn=200$ and 27~GeV ($\mean{\dgese/\dg}$ does not necessarily equal $\mean{\dgese}/\mean{\dg}$). The POI $\eta$ acceptance is $0.3<|\eta|<1$ for \avfd, \ampt, and \hydjet, and $0.3<|\eta|<2$ for \epos, and the $\pt$ acceptance is $0.2<\pt<2$~\gevc\ for all models. 
    The $q_2^2$ is calculated using particles from   $0.2<\pt<2$~\gevc\ and  $|\eta|<0.3$ for all models (and also forward/backward rapidities for \ampt\ and \hydjet: $3<|\eta|<5$ at 200~GeV and $2<|\eta|<3$ at 27~GeV).
    Several points are missing as they are outside the frame but consistent with zero with large uncertainties.}
    \label{fig:interESE}
\end{figure*}

\begin{figure*}[hbt]
    \includegraphics[width=0.4\textwidth]{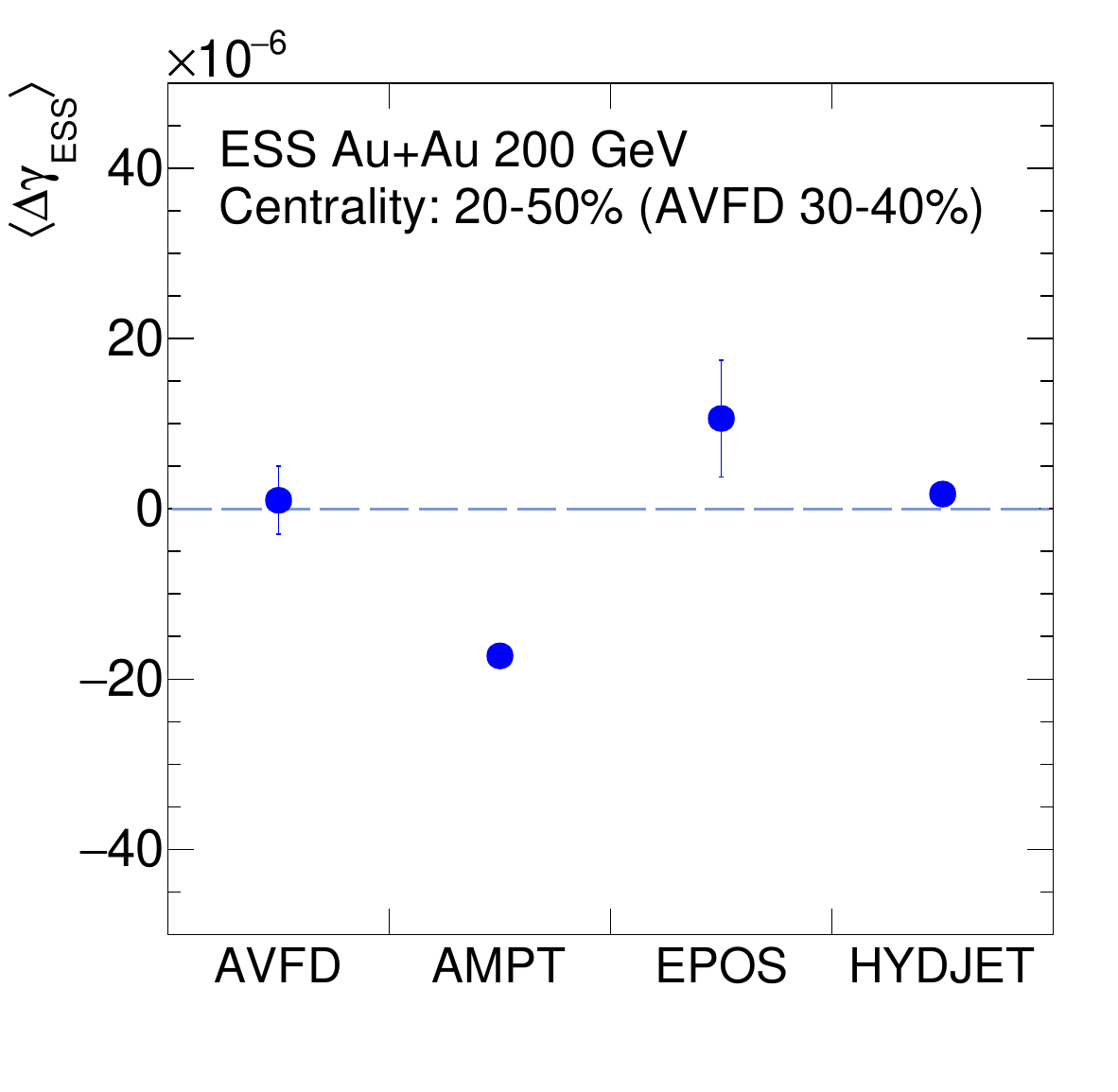}
    \includegraphics[width=0.4\textwidth]{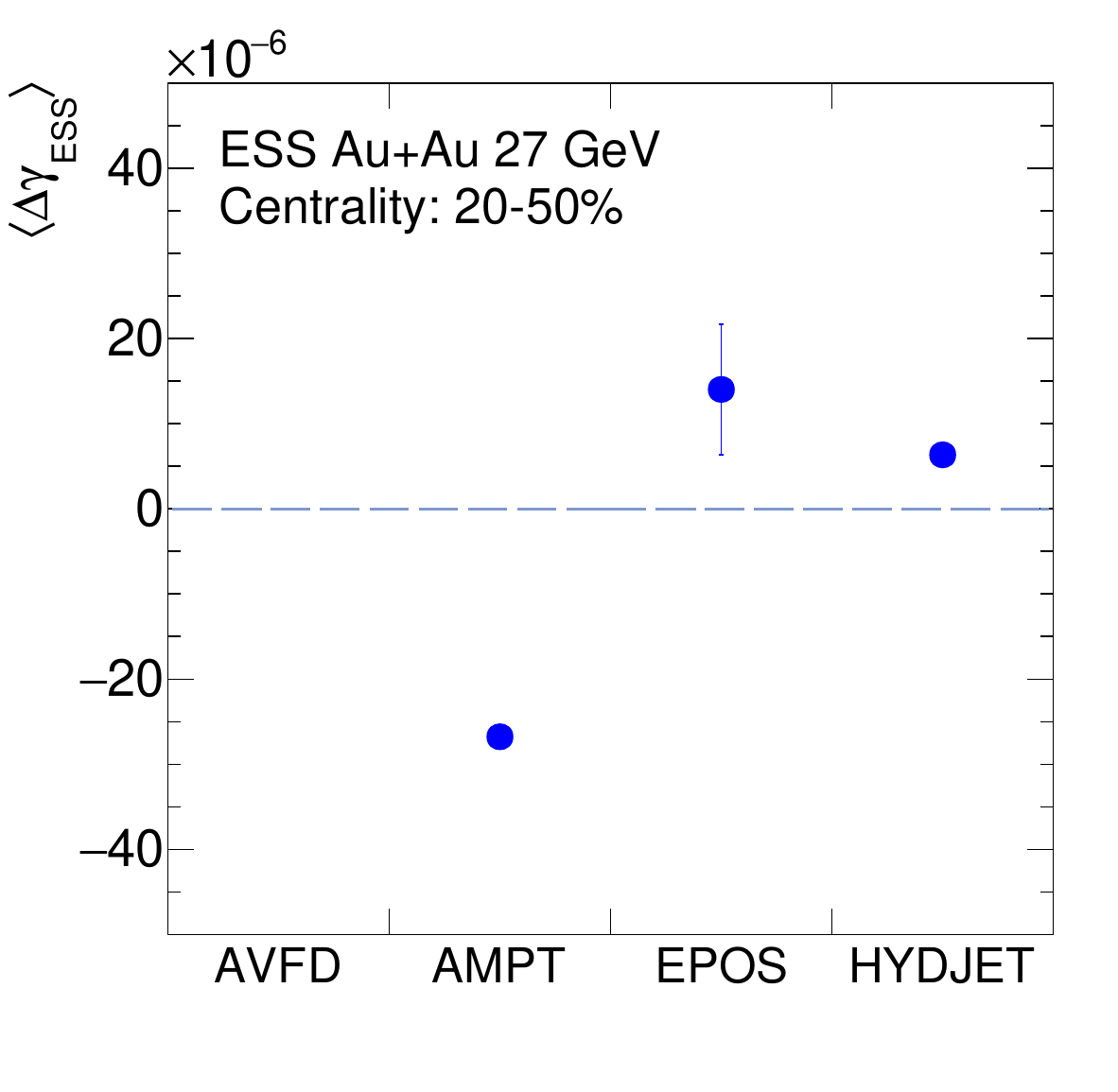} 
    \vspace{-4mm}\\
    \includegraphics[width=0.4\textwidth]{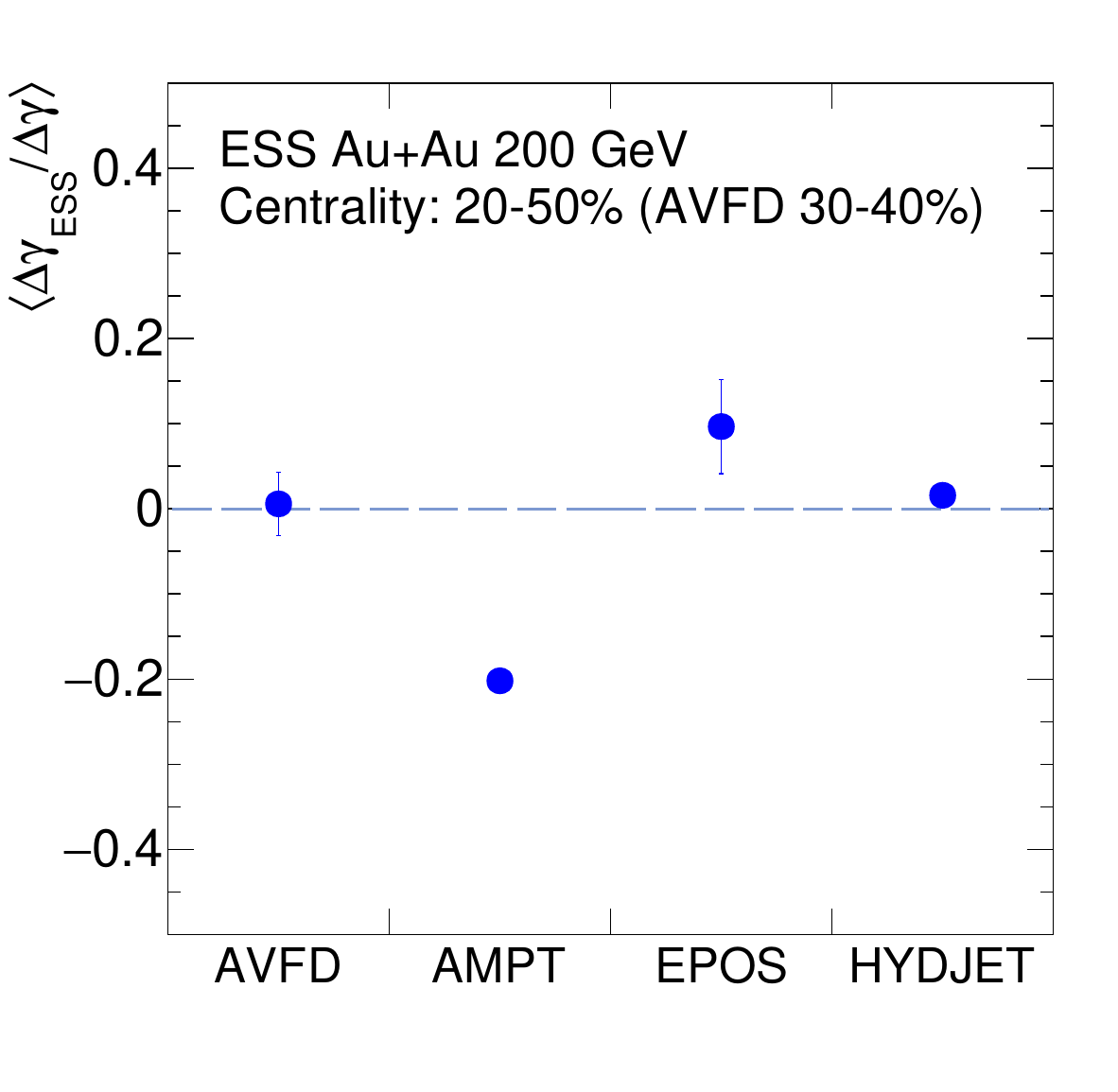}
    \includegraphics[width=0.4\textwidth]{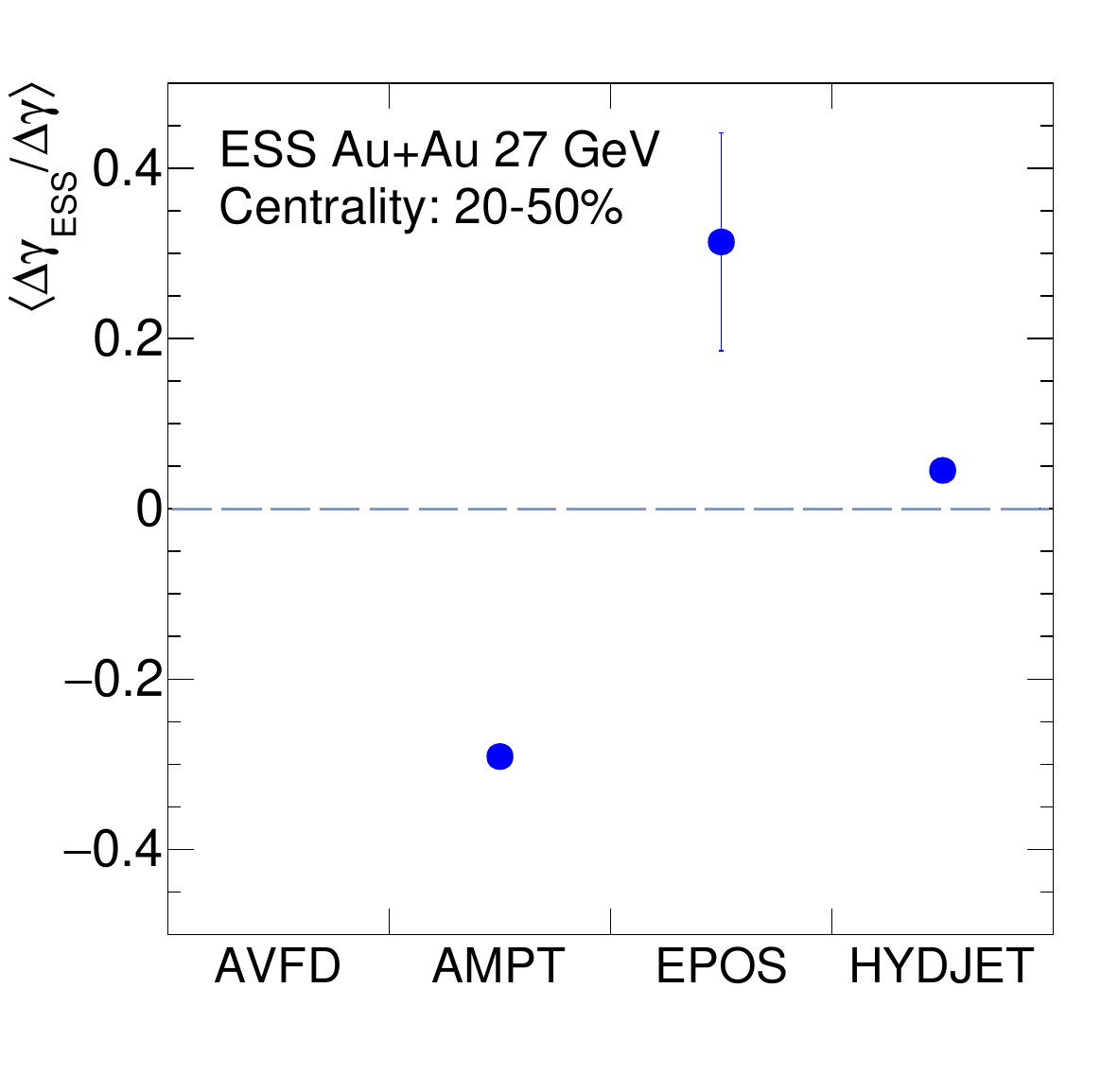}
    \caption{ESS intercepts. Shown are intercepts (upper panels) and intercepts divided by the corresponding inclusive $\dg$ values (lower panels) from the ESS method for various physics models. For \avfd\ ($\ns=0)$, the result is from the 30--40\% centrality Au+Au collisions at $\snn=200$~GeV. For \ampt, \epos, and \hydjet, the results are averaged over the 20--50\% centrality range of Au+Au collisions at $\snn=200$ and 27~GeV ($\mean{\dgess/\dg}$ does not necessarily equal $\mean{\dgess}/\mean{\dg}$). The POI $\eta$ acceptance is $|\eta|<1$ for \avfd, \ampt, and \hydjet, and $|\eta|<2$ for \epos, and the $\pt$ acceptance is $0.2<\pt<2$~\gevc\ for all models. The $\qhpair^2$ is always calculated from the same POIs.}
    \label{fig:interESS}
\end{figure*}

We summarize the intercepts $\mean{\dgese}$ and the intercepts divided by the overall $\dg$ magnitude $\mean{\dgese/\dg}$ averaged over the 20--50\% centrality range from the physics model studies by the ESE method in Fig.~\ref{fig:interESE}.
The ESE intercepts are mostly limited by large statistical uncertainties. 
The large statistical uncertainties arise from the narrow dynamical range of $\mean{v_2}$ based on the $\qh$ event selection in a given narrow centrality bin (see, particularly, Fig.~\ref{fig:ampt_ese} lower panels, Fig.~\ref{fig:ampt_ese_forward}, and Fig.~\ref{fig:epos_ese}) and from the long extrapolation lever arm to the $\mean{v_2}=0$ intercept. 
This is particularly true for 27~GeV where some of the data points fall outside the graph frame of Fig.~\ref{fig:interESE}. 
The \avfd, \ampt, and \epos\ models are computationally expensive, and it is difficult to accumulate statistics for a qualitative improvement beyond the present precision.

As discussed in Sect.~\ref{sec:tech}, the flow-induced background must be zero at the ESE intercept $\dgese$ in the absence of nonflow. In most of our model studies and for all the model studies presented in Fig.~\ref{fig:interESE}, we used the known reaction plane $\psi=0$ in computing the $v_2$ and $\dg$ variables, so nonflow is absent from these observables. 
While the statistical uncertainties are large, the \ampt\ and \hydjet\ $\dgese$ using $\qh^2$ from the $|\eta|<0.3$ range seem to be nonzero. 
The reason is likely due to nonflow effects caused by the ESE event selection using the midrapidity $\qh^2$: because the $\eta$ distance is small, there can be nonflow correlation between a POI and a particle used in the $\qh^2$ calculation. Such a correlation could bias the $\mean{v_{2,\res}}$ values of resonances contributing to CME backgrounds to be not strictly proportional to the final-state particle $\mean{v_2}$, resulting in a nonzero $\dgese$ intercept. 
To reduce this effect, we also calculate $\qh^2$ using particles at forward/backward pseudorapidities in \ampt\ and \hydjet, the results of which are shown by the red points in Fig.~\ref{fig:interESE}.
These intercepts $\dgese$ are consistent with zero, confirming our hypothesis, though the uncertainty is large for the \ampt\ result at 27~GeV.

Figure~\ref{fig:interESS} summarizes the physics model results from the ESS method on the intercepts $\mean{\dgess}$ and the intercepts divided by the overall $\dg$ magnitude $\mean{\dgess/\dg}$, averaged over the 20--50\% centrality range. 
As discussed in Sect.~\ref{sec:tech}, the ESS method uses the POIs to compute the event selection $\qhpair^2$ variable, which are automatically (self-)correlated with the $\dg$ and $\vsing$ of POIs.
Thus, the $\dg$ and $\vsing$ are affected by statistical fluctuations in $\qhpair^2$ that is used to select events. The wide range of $\vsing$ and $\dg$ are a result of these statistical fluctuations by insisting on certain values of the statistically fluctuating $\qhpair^2$. 
The $\vsing$ value can go to zero as shown in Figs.~\ref{fig:avfd_ess}, \ref{fig:ampt_ess}, \ref{fig:epos_ess}, and \ref{fig:hydjet_ess}.
Due to the long projection lever arm, the statistical precision on the intercept $\dgess$ is good. 

We observe in Fig.~\ref{fig:interESS} that the ESS intercepts are close to zero but  not always consistent with zero. 
Since these models shown in Fig.~\ref{fig:interESS} contains no CME but only physics background, our results demonstrate that the $v_2$-induced background is unnecessarily zero in the ESS intercept $\dgess$.  
Although the background may be suppressed, 
it is of {\em no} use because the amount of suppressed background is unknown, even whether it is over-suppressed or under-suppressed is unknown as shown in the lower panels of Fig.~\ref{fig:interESS}. 
It has been shown in~\cite{Xu:2023elq} that CME signals would be preserved in the ESS intercept, however, such a conclusion or further studies of CME signal response in ESS is not helpful before the background issue is fully addressed. 
The reason that the $v_2$-induced backgrounds unnecessarily must be zero in the ESS intercept $\dgess$ is because of the  statistical fluctuation nature of the observables used in the ESS analysis.
A model demonstration of this for the similar but simpler ES method (see Sect.~\ref{sec:tech}) can be found in Ref.~\cite{wang:2016iov}. There, the reason is clear, as elucidated in Sect.~\ref{sec:tech} and by the middle-panel cartoon of Fig.~\ref{cartoon}: the resonance (generally background contributing source) $v_2$ values are proportional to the overall $v_2$ of final-state particles, and when events are binned in the observed, statistically fluctuating $\vobs$ of final-state particles, the average $\mean{v_2}$ values of the background sources are positive at $\vobs=0$~\cite{wang:2016iov}.
The situation is much more complicated here in the ESS method where $\dg$ is plotted against $\vsing$ in events binned in $\qhpair^2$. 
As discussed in Sect.~\ref{subsec:ana}, the $\qhpair^2$ calculated by Eq.~\ref{eq:qhpair} is composed of several components: (i) a flow component from hydrodynamic particle pairs, which is related to the single particle flow in a non-trivial way~\cite{Li:2024pue}, (ii) a flow component of the resonances (CME background contributing sources), (iii) a nonflow component between the two pairs used to calculate the two-pair cumulant by Eq.~\ref{eq:qhpair} that contain daughters from, for example, the same resonance decay, and (iv) a self-correlation component from the sharing of a common particle between the two pairs. 
Because of these complicated nature in the event-by-event $\qhpair^2$ variable, it is unclear how the $\vsing$ and $\dg$ would behave, individually and intertwiningly, as functions of $\qhpair^2$. 
It would not be expected {\em a priori} that the $\dg$ as a function of $\vsing$, binned in $\qhpair^2$, would project to a zero intercept in the absence of CME.

\begin{figure}[hbt]
    \includegraphics[width=0.4\textwidth]{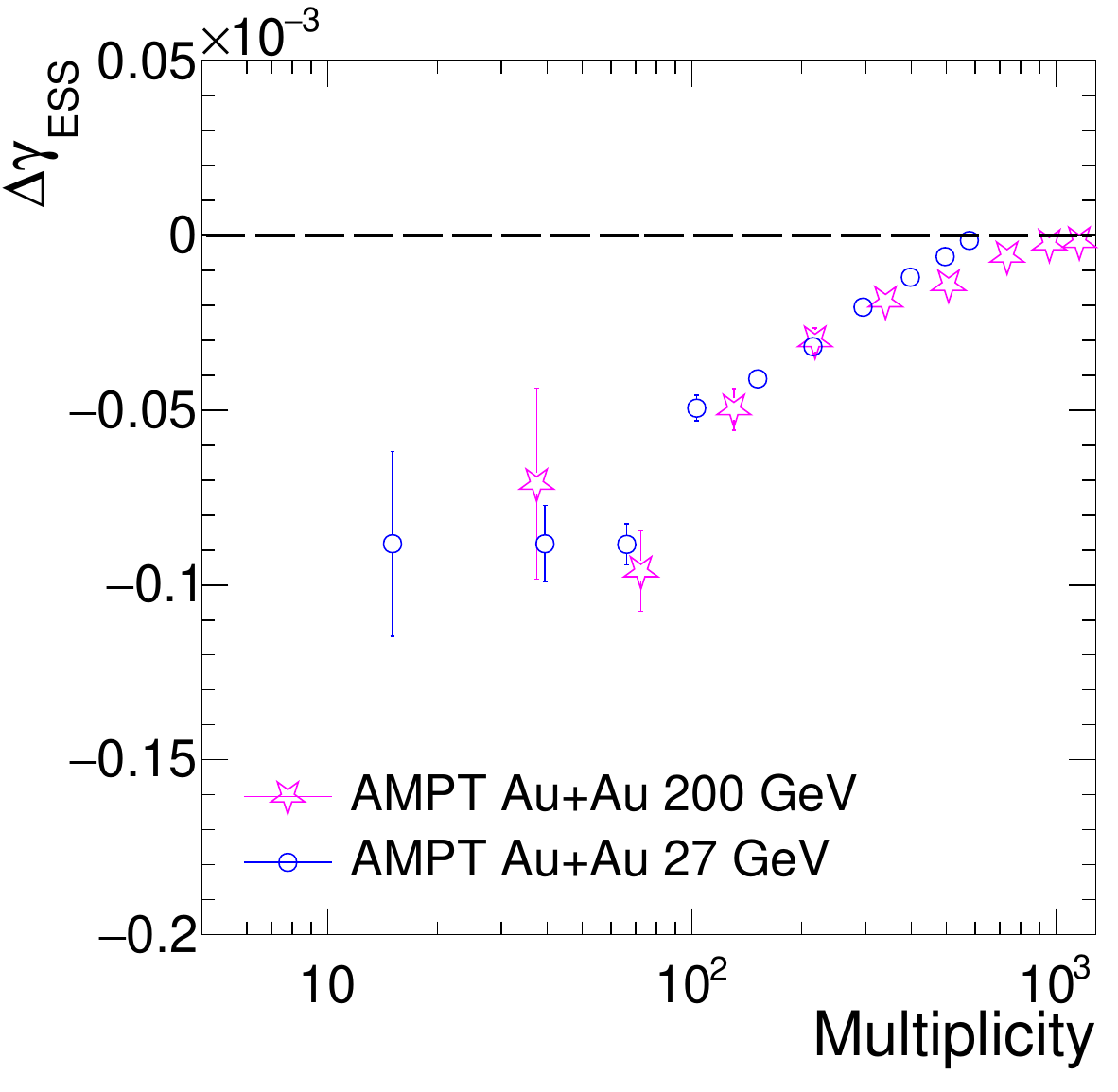}
    \includegraphics[width=0.4\textwidth]{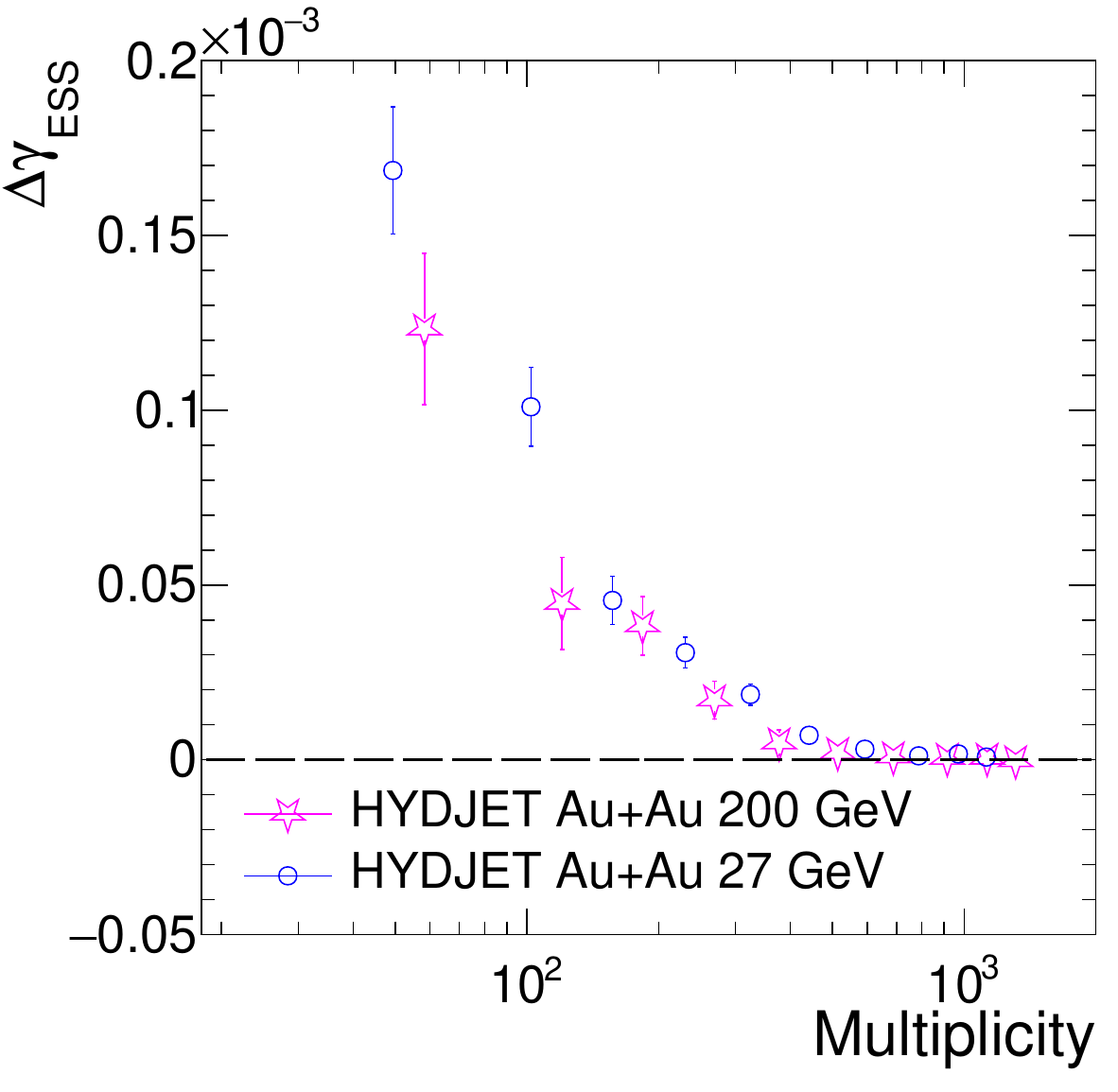}
    \caption{\label{fig:ess_cent}ESS intercept multiplicity dependence. The ESS intercept $\dgess$ as a function of the average POI multiplicity from peripheral 80--100\% to central 0--5\% Au+Au collisions at 200 and 27~GeV from \ampt\ (upper panel) and \hydjet\ (lower panel). The most peripheral point for AMPT 200~GeV is outside the frame.}
\end{figure}

It appears from the physics model studies shown in Fig.~\ref{fig:interESS} (where CME is absent) that the value of $\dgess$ can, indeed, be negative, zero, or positive, presumably dependent of model details (model descriptions can be found in Sects.~\ref{sec:avfd}--\ref{sec:hydjet}). 
Note the results in Fig.~\ref{fig:interESS} are averaged from 20--50\% centralities. One obvious difference in the simulated events, even with the same model, is the event centrality or multiplicity.
As discussed in Sect.~\ref{subsec:ana} and mentioned above, the $\qhpair$ quantity contains a self-correlation component besides being sensitive to single-particle $v_2$, multi-particle cluster $v_2$, and nonflow. The self-correlations are caused by sharing of the same POI between two pairs of POIs, and thus its effect is inversely proportional to multiplicity. To gain  insight, we show in Fig.~\ref{fig:ess_cent} the ESS intercept in \ampt\ and \hydjet\ as a function of centrality in terms of the average POI multiplicity. The centralities are defined by the charged hadron multiplicity within $|\eta|<0.5$ to be 0--5\%, 5--10\%, 10--20\%, $\cdots$ (10\% size), 70--80\%, and 80-100\% of the total MB event sample.
As seen from Fig.~\ref{fig:ess_cent}, the nonzero $\dgess$ intercept becomes stronger in more peripheral collisions. Within the same model, the $\dgess$ seems to follow the same trend in POI multiplicity. This explains why the \hydjet\ $\dgess$ in 20--50\% centrality range is consistent with zero at 200~GeV and positive at 27~GeV. 
This may suggest that the root reason for the nonzero intercept $\dgess$ is not collision energy or centrality, but POI multiplicity. 
The multiplicity dependence of the ESS intercept $\dgess$ clearly shows that it is sensitive to background correlations. The different signs of $\dgess$ between \ampt\ (negative) and \hydjet\ (positive) is presumably due to different physics implemented in these models.

To conclude this part of the discussion, our physics model results show that the ESS intercept is not an easily interpretable observable. 
Although the background in $\dgess$ is strongly suppressed, one does not know how much background still remains in $\dgess$, or even its sign. 
One does not know quantitatively what the ESS intercept exactly measures, so the ESS method is practically not useful to search for the CME.

\subsection{Toy Model Verification}
One important difference among the models we studied is the various contributions from $\Ks$. \avfd\ and \ampt\ keep the $\Ks$ stable so there is no contamination from $\Ks$ decay pions in the final state. \epos\ and \hydjet, on the other hand, decay $\Ks$ so there is maximum contamination in the final-state pions (as we have used all final-state pions of the model in our calculations). The intercepts from \ampt\ at both 200 and 27 GeV are negative, and the intercept from \avfd\ at 200 GeV is consistent with zero (\avfd\ modeling at low energy is not available). The intercepts from \epos\ at 200 and 27~GeV are  both positive. The intercept from \hydjet\ is consistent with zero at 200~GeV and positive at 27~GeV. In all models the strong-decay resonances (such as the $\rho^0$ resonance) are forced to decay. Relative particle abundances, for example, the $\Ks$ over $\rho^0$ ratio, depends on energy; the lower the energy, the larger the low- to high-mass ratio. While very much dependent on models, it seems that $\dgess$ increases with increasing $\Ks/\rho^0$ ratio. It is thus conceivable, as resonances are a main background contribution to $\dg$, that the relative resonance abundances could also be important for the ESS intercept.

To test the hypothesis that resonance mixtures may influence the robustness of the ESS method, we use toy model simulations, where we can easily alter the event content to examine the consequences. 
In real data analysis, a fraction of $\Ks$ hadrons decay into pions that are reconstructed as primordial particles, often defined experimentally as those with a reconstructed distance of closest approach ({\sc dca}) from the reconstructed primary collision point to be within a certain cut, typically a few centimeters. 
In the STAR experiment, such a fraction is on the order of 2\%~\cite{Abelev:2008ab}, and is used as default in Toy Model I (described in Sect.~\ref{sec:toy1}). 
To alter the event content, we vary the accepted fraction of $\Ks$ to 10\%, 20\%, and 30\%. 

In addition, we note in Fig.~\ref{fig:minv} that all the physics models contain a mass continuum of OS pairs in excess of SS pairs. 
We have therefore included a mass continuum in Toy Model II as described in Sect.~\ref{sec:toy2}. The accepted $\Ks$ fraction is kept at 2\%.
Adding the mass continuum is another way to alter the event content. 

Figure~\ref{fig:toy_ess} shows the ESS results from   Toy Model I (default 2\% $\Ks$), Toy Model I (10\% $\Ks$), and Toy Model II (mass continuum). It is interesting to notice that the dependence of $\dg$ on $\vsing$ in those $\qhpair^2$-binned events is not linear for  Toy Model II. This suggests that the linear dependence hypothesis, which is essential for the linear extrapolation in ESS, may not be generally valid.

Figure~\ref{fig:toy_ese} shows the ESE results from   Toy Model I (default 2\% $\Ks$), Toy Model I (10\% $\Ks$), and Toy Model II (mass continuum) with POIs from $0.3<|\eta|<1$ and $q_2^2$ from $|\eta|<0.3$. 
There seems to be an indication of nonlinear behavior in the Toy Model II result, similar to that in the ESS result in Fig.~\ref{fig:toy_ess} right panel.

\begin{figure}[hbt]
    \includegraphics[width=0.4\textwidth]{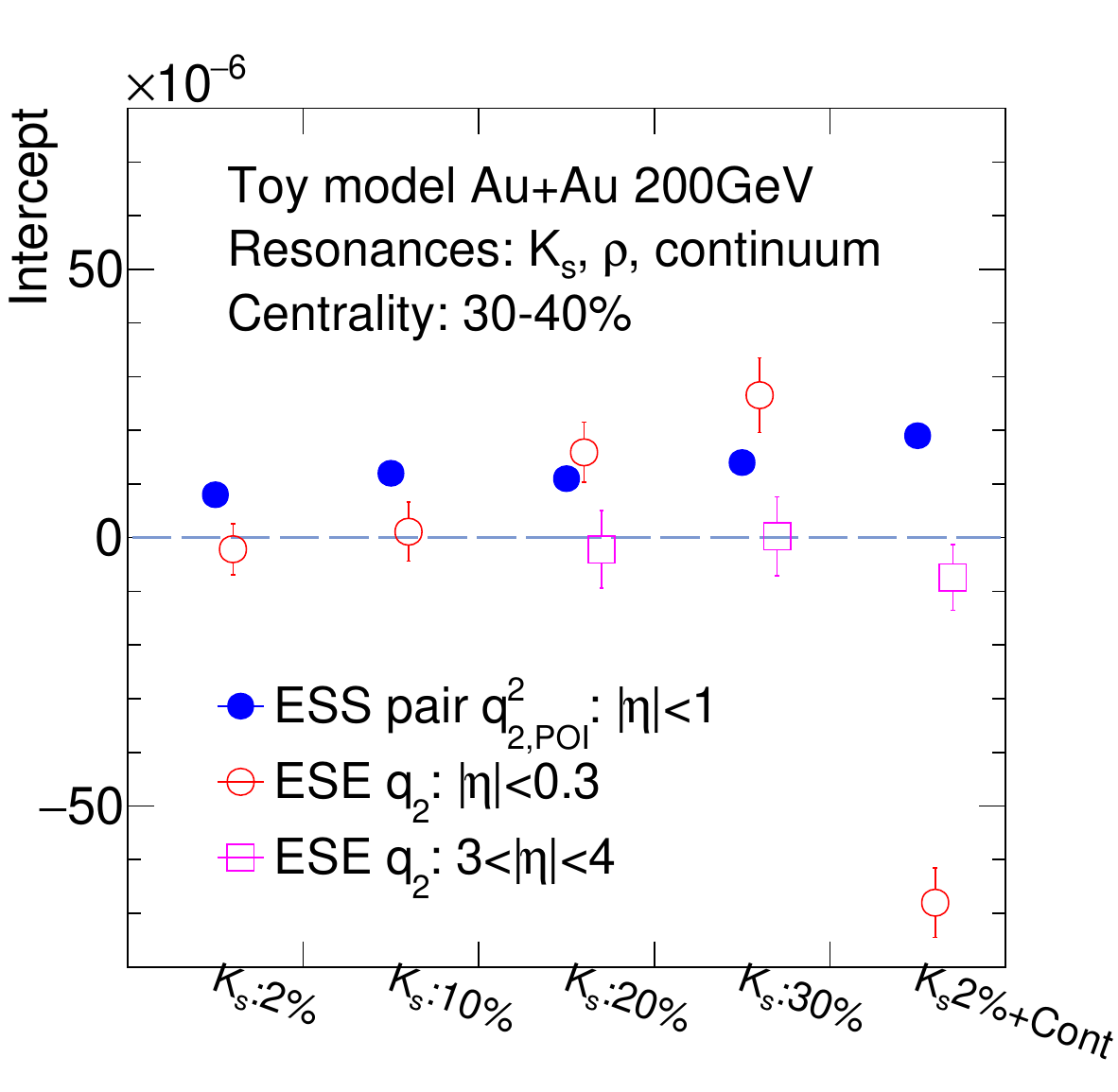}
    \includegraphics[width=0.4\textwidth]{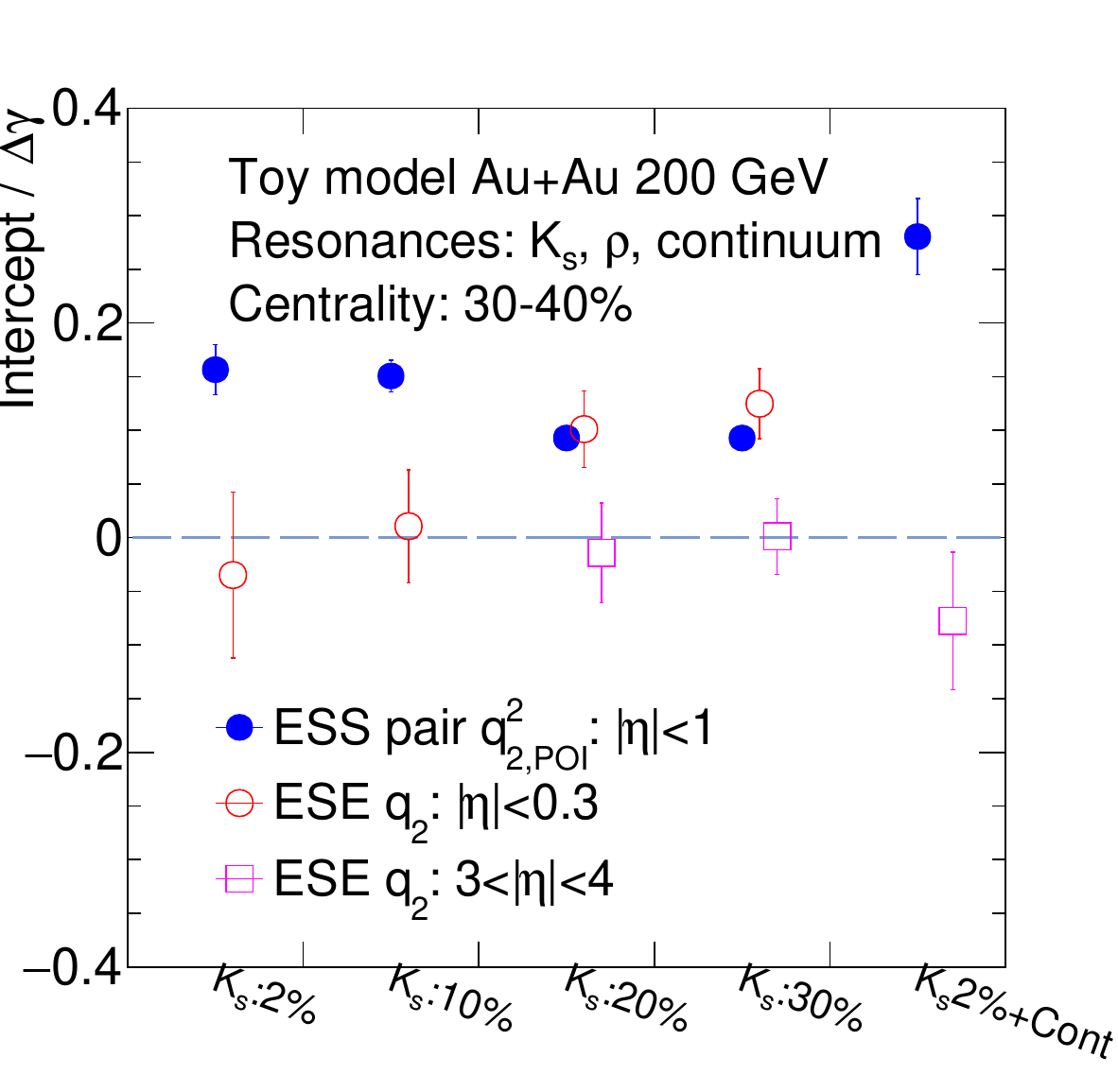}
    \caption{Toy model ESS and ESE intercepts. Shown are intercepts (upper panel) and intercepts divided by the inclusive $\dg$ value (lower panel) from the ESS (blue points) and ESE (red and pink points) methods using the toy models. Toy model inputs are from parameterizations to experimental data of  30--40\% Au+Au collisions at 200~GeV. 
    For ESS, the POI acceptance is $|\eta|<1$ and $0.2<\pt<2$~\gevc, and the $\qhpair^2$ is always calculated from the same POIs. For ESE, the POI acceptance is $0.3<|\eta|<1$, and the $\qh^2$ is calculated from particles in $|\eta|<0.3$ (red points) or $3<|\eta|<4$ for three of the studied cases (pink points); The $\pt$ range for both POIs and particles used for $\qh^2$ calculations is $0.2<\pt<2$~\gevc. The rightmost red point is outside the frame.}
    \label{fig:Ks}
\end{figure}
Figure~\ref{fig:Ks} summarizes the intercepts and the intercepts divided by the overall $\dg$ values from the toy model studies. The intercepts from the ESS method (blue points) are all positive and appear to vary over the different event contents. This may confirm our hypothesis that the ESS method is sensitive to the details of the event makeup because of the complex intertwining of the $\dg$, $\vsing$, and the event selection variable $\qhpair^2$. 

The intercepts from the ESE method are shown in the red markers. 
The statistical precision is high because the toy models are computationally inexpensive.
The ESE intercepts are consistent with zero for the two lower $\Ks$-contamination cases studied in Toy Model I. However, the ESE intercepts from the two higher $\Ks$-contamination cases of Toy Model I (positive) and the intercept from Toy Model II (negative) are inconsistent with zero. We postulate the cause to be nonflow correlations between POIs in $0.3<|\eta|<1$ and the $q_2^2$ calculated from $|\eta|<0.3$ in our toy model studies because resonance decay daughters can end up in both regions which are not far apart. 
(We have mentioned this nonflow effect in the $q_2^2$ event selection in our physics model discussion of Fig.~\ref{fig:interESE}, though the statistical uncertainties there are large.) 
In addition, the effect of adding a mass continuum is different from simply increasing the $\Ks$ contamination, presumably because of the differing decay kinematics and opening angles. 

To reduce this nonflow effect, we classify events in the ESE method using $q_2^2$ calculated from forward/backward pseudorapidity regions of $3<|\eta|<4$, far displaced from the POI's region of $0.3<|\eta|<1$. The results are shown in Fig.~\ref{fig:toy_ese2} for Toy Model I (20\% $\Ks$), Toy Model I (30\% $\Ks$), and Toy Model II (mass continuum). 
The ESE intercepts and the intercepts divided by the overall $\dg$ values are shown in the pink markers in Fig.~\ref{fig:Ks} with good statistical precision. These intercepts are consistent with zero, as one would expect in the absence of nonflow. 
These results highlight the importance to use $q_2^2$ calculated in pseudorapidity range far distanced from that of the POIs.

\section{summary\label{sec:summary}}
The charge-dependent azimuthal correlator $\dg$ is commonly used to search for the chiral magnetic effect (CME), a parity and charge-parity violating phenomenon predicted by quantum chromodynamics. 
The $\dg$ correlator is affected by a major physics background induced by elliptic flow anisotropy $v_2$ in relativistic heavy ion collisions. 
The event-shape engineering (ESE) and event-shape selection (ESS) methods have been exploited to search for the CME by projecting $\dg$ to the measured $v_2$ of vanishing value, relying on {\em dynamical} fluctuations and {\em statistical} fluctuations of $v_2$, respectively. 

We have conducted a systematic study using four physics models (\avfd, \ampt, \epos, and \hydjet) describing heavy ion collisions. 
It is found, with no CME signal, that the ESS intercept ($\dgess$) can be negative, zero, or positive depending on the models and multiplicities of the events, and the ESE intercept ($\dgese$) is mostly consistent with zero, as expected, albeit with inherently large statistical uncertainties.

It is further found, using toy models, that $\dgess$ appears to depend on the details of the event content, for example, the relative abundances of resonances and the presence of a mass continuum. 
Given that the event-shape selection variable $\qhpair^2$, the single particle $\vsing$, and the three-point correlator $\dg$ in $\qhpair^2$-classified events all depend on the particles of interest (POIs), which are made of not only primordial particles but also all decay products of electromagnetic- and strong-decay resonances and partial products of weak-decay resonances, we conjecture that the ESS intercept $\dgess$ is a complicated function of all elements that make up the event. 
These elements include resonance abundances relative to primordial pions, resonance $v_2$'s and $\pt$ spectra, as well as primordial particle $v_2$ and $\pt$ spectra. 
Depending on the detailed physics of these elements, $\dgess$ can be positive, negative, or zero. 
The issue is, unfortunately, that it is unknown, and needless to say unproven, what combinations of all those elements in the event  would give the ideal  $\dgess=0$ in the absence of CME. 
This makes it impractical to search for CME using the ESS method. 

The ESE method, on the other hand, has much more clarity: the $\dgese$ intercept of flow-induced backgrounds must be zero in the absence of nonflow. 
However, the ESE method is not without pitfalls. 
There exist nonflow correlations between particles, and when those nonflow correlations are present between POIs and the $\qh^2$ variable classifying events, then the ESE intercept can also be sensitive to non-CME physics. 
The issue is similar in nature to the ESS method--the relevant variables are now intertwined.
One way to avoid/reduce such nonflow effects is to separate POIs and particles to calculate $\qh^2$ well in pseudorapidity. 
Furthermore, the dynamical fluctuations in $v_2$ are typically not large, so a long extrapolation to the $v_2=0$ intercept is required in the ESE method. This results in a relatively large uncertainty in the $\dgese$ intercept, and large statistics are required for a precision measurement by the ESE method. 
The statistics issue, however, can easily be resolved, at least conceptually, by accumulating more data.

\section*{Acknowledgment} 
This work is supported in part by the U.S.~Department of Energy (Grant No.~DE-SC0012910).

\bibliography{refs}

\newcommand{\SNN}{$\sqrt{s_{_{NN}}}$}\newcommand{\MEAN}[1]{\langle#1\rangle}
\begin{thebibliography}{75}%
\makeatletter
\providecommand \@ifxundefined [1]{%
 \@ifx{#1\undefined}
}%
\providecommand \@ifnum [1]{%
 \ifnum #1\expandafter \@firstoftwo
 \else \expandafter \@secondoftwo
 \fi
}%
\providecommand \@ifx [1]{%
 \ifx #1\expandafter \@firstoftwo
 \else \expandafter \@secondoftwo
 \fi
}%
\providecommand \natexlab [1]{#1}%
\providecommand \enquote  [1]{``#1''}%
\providecommand \bibnamefont  [1]{#1}%
\providecommand \bibfnamefont [1]{#1}%
\providecommand \citenamefont [1]{#1}%
\providecommand \href@noop [0]{\@secondoftwo}%
\providecommand \href [0]{\begingroup \@sanitize@url \@href}%
\providecommand \@href[1]{\@@startlink{#1}\@@href}%
\providecommand \@@href[1]{\endgroup#1\@@endlink}%
\providecommand \@sanitize@url [0]{\catcode `\\12\catcode `\$12\catcode `\&12\catcode `\#12\catcode `\^12\catcode `\_12\catcode `\%12\relax}%
\providecommand \@@startlink[1]{}%
\providecommand \@@endlink[0]{}%
\providecommand \url  [0]{\begingroup\@sanitize@url \@url }%
\providecommand \@url [1]{\endgroup\@href {#1}{\urlprefix }}%
\providecommand \urlprefix  [0]{URL }%
\providecommand \Eprint [0]{\href }%
\providecommand \doibase [0]{https://doi.org/}%
\providecommand \selectlanguage [0]{\@gobble}%
\providecommand \bibinfo  [0]{\@secondoftwo}%
\providecommand \bibfield  [0]{\@secondoftwo}%
\providecommand \translation [1]{[#1]}%
\providecommand \BibitemOpen [0]{}%
\providecommand \bibitemStop [0]{}%
\providecommand \bibitemNoStop [0]{.\EOS\space}%
\providecommand \EOS [0]{\spacefactor3000\relax}%
\providecommand \BibitemShut  [1]{\csname bibitem#1\endcsname}%
\let\auto@bib@innerbib\@empty
\bibitem [{\citenamefont {Kharzeev}\ \emph {et~al.}(1998)\citenamefont {Kharzeev}, \citenamefont {Pisarski},\ and\ \citenamefont {Tytgat}}]{Kharzeev:1998kz}%
  \BibitemOpen
  \bibfield  {author} {\bibinfo {author} {\bibfnamefont {D.}~\bibnamefont {Kharzeev}}, \bibinfo {author} {\bibfnamefont {R.}~\bibnamefont {Pisarski}},\ and\ \bibinfo {author} {\bibfnamefont {M.~H.}\ \bibnamefont {Tytgat}},\ }\bibfield  {title} {\bibinfo {title} {{Possibility of spontaneous parity violation in hot QCD}},\ }\href {https://doi.org/10.1103/PhysRevLett.81.512} {\bibfield  {journal} {\bibinfo  {journal} {Phys.Rev.Lett.}\ }\textbf {\bibinfo {volume} {81}},\ \bibinfo {pages} {512} (\bibinfo {year} {1998})},\ \Eprint {https://arxiv.org/abs/hep-ph/9804221} {arXiv:hep-ph/9804221 [hep-ph]} \BibitemShut {NoStop}%
\bibitem [{\citenamefont {Kharzeev}(2006)}]{Kharzeev:2004ey}%
  \BibitemOpen
  \bibfield  {author} {\bibinfo {author} {\bibfnamefont {D.}~\bibnamefont {Kharzeev}},\ }\bibfield  {title} {\bibinfo {title} {{Parity violation in hot QCD: Why it can happen, and how to look for it}},\ }\href {https://doi.org/10.1016/j.physletb.2005.11.075} {\bibfield  {journal} {\bibinfo  {journal} {Phys.Lett.}\ }\textbf {\bibinfo {volume} {B633}},\ \bibinfo {pages} {260} (\bibinfo {year} {2006})},\ \Eprint {https://arxiv.org/abs/hep-ph/0406125} {arXiv:hep-ph/0406125 [hep-ph]} \BibitemShut {NoStop}%
\bibitem [{\citenamefont {Kharzeev}\ \emph {et~al.}(2008)\citenamefont {Kharzeev}, \citenamefont {McLerran},\ and\ \citenamefont {Warringa}}]{Kharzeev:2007jp}%
  \BibitemOpen
  \bibfield  {author} {\bibinfo {author} {\bibfnamefont {D.~E.}\ \bibnamefont {Kharzeev}}, \bibinfo {author} {\bibfnamefont {L.~D.}\ \bibnamefont {McLerran}},\ and\ \bibinfo {author} {\bibfnamefont {H.~J.}\ \bibnamefont {Warringa}},\ }\bibfield  {title} {\bibinfo {title} {{The Effects of topological charge change in heavy ion collisions: 'Event by event P and CP violation'}},\ }\href {https://doi.org/10.1016/j.nuclphysa.2008.02.298} {\bibfield  {journal} {\bibinfo  {journal} {Nucl.Phys.}\ }\textbf {\bibinfo {volume} {A803}},\ \bibinfo {pages} {227} (\bibinfo {year} {2008})},\ \Eprint {https://arxiv.org/abs/0711.0950} {arXiv:0711.0950 [hep-ph]} \BibitemShut {NoStop}%
\bibitem [{\citenamefont {Fukushima}\ \emph {et~al.}(2008)\citenamefont {Fukushima}, \citenamefont {Kharzeev},\ and\ \citenamefont {Warringa}}]{Fukushima:2008xe}%
  \BibitemOpen
  \bibfield  {author} {\bibinfo {author} {\bibfnamefont {K.}~\bibnamefont {Fukushima}}, \bibinfo {author} {\bibfnamefont {D.~E.}\ \bibnamefont {Kharzeev}},\ and\ \bibinfo {author} {\bibfnamefont {H.~J.}\ \bibnamefont {Warringa}},\ }\bibfield  {title} {\bibinfo {title} {{The chiral magnetic effect}},\ }\href {https://doi.org/10.1103/PhysRevD.78.074033} {\bibfield  {journal} {\bibinfo  {journal} {Phys.Rev.}\ }\textbf {\bibinfo {volume} {D78}},\ \bibinfo {pages} {074033} (\bibinfo {year} {2008})},\ \Eprint {https://arxiv.org/abs/0808.3382} {arXiv:0808.3382 [hep-ph]} \BibitemShut {NoStop}%
\bibitem [{\citenamefont {Skokov}\ \emph {et~al.}(2009)\citenamefont {Skokov}, \citenamefont {Illarionov},\ and\ \citenamefont {Toneev}}]{Skokov:2009qp}%
  \BibitemOpen
  \bibfield  {author} {\bibinfo {author} {\bibfnamefont {V.}~\bibnamefont {Skokov}}, \bibinfo {author} {\bibfnamefont {A.~{\relax Yu}.}\ \bibnamefont {Illarionov}},\ and\ \bibinfo {author} {\bibfnamefont {V.}~\bibnamefont {Toneev}},\ }\bibfield  {title} {\bibinfo {title} {{Estimate of the magnetic field strength in heavy-ion collisions}},\ }\href {https://doi.org/10.1142/S0217751X09047570} {\bibfield  {journal} {\bibinfo  {journal} {Int. J. Mod. Phys.}\ }\textbf {\bibinfo {volume} {A24}},\ \bibinfo {pages} {5925} (\bibinfo {year} {2009})},\ \Eprint {https://arxiv.org/abs/0907.1396} {arXiv:0907.1396 [nucl-th]} \BibitemShut {NoStop}%
\bibitem [{\citenamefont {Deng}\ and\ \citenamefont {Huang}(2012)}]{Deng:2012pc}%
  \BibitemOpen
  \bibfield  {author} {\bibinfo {author} {\bibfnamefont {W.-T.}\ \bibnamefont {Deng}}\ and\ \bibinfo {author} {\bibfnamefont {X.-G.}\ \bibnamefont {Huang}},\ }\bibfield  {title} {\bibinfo {title} {{Event-by-event generation of electromagnetic fields in heavy-ion collisions}},\ }\href {https://doi.org/10.1103/PhysRevC.85.044907} {\bibfield  {journal} {\bibinfo  {journal} {Phys. Rev.}\ }\textbf {\bibinfo {volume} {C85}},\ \bibinfo {pages} {044907} (\bibinfo {year} {2012})},\ \Eprint {https://arxiv.org/abs/1201.5108} {arXiv:1201.5108 [nucl-th]} \BibitemShut {NoStop}%
\bibitem [{\citenamefont {Voloshin}\ and\ \citenamefont {Zhang}(1996)}]{Voloshin:1994mz}%
  \BibitemOpen
  \bibfield  {author} {\bibinfo {author} {\bibfnamefont {S.}~\bibnamefont {Voloshin}}\ and\ \bibinfo {author} {\bibfnamefont {Y.}~\bibnamefont {Zhang}},\ }\bibfield  {title} {\bibinfo {title} {{Flow study in relativistic nuclear collisions by Fourier expansion of Azimuthal particle distributions}},\ }\href {https://doi.org/10.1007/s002880050141} {\bibfield  {journal} {\bibinfo  {journal} {Z.Phys.}\ }\textbf {\bibinfo {volume} {C70}},\ \bibinfo {pages} {665} (\bibinfo {year} {1996})},\ \Eprint {https://arxiv.org/abs/hep-ph/9407282} {arXiv:hep-ph/9407282 [hep-ph]} \BibitemShut {NoStop}%
\bibitem [{\citenamefont {Voloshin}(2004)}]{Voloshin:2004vk}%
  \BibitemOpen
  \bibfield  {author} {\bibinfo {author} {\bibfnamefont {S.~A.}\ \bibnamefont {Voloshin}},\ }\bibfield  {title} {\bibinfo {title} {{Parity violation in hot QCD: How to detect it}},\ }\href {https://doi.org/10.1103/PhysRevC.70.057901} {\bibfield  {journal} {\bibinfo  {journal} {Phys.Rev.}\ }\textbf {\bibinfo {volume} {C70}},\ \bibinfo {pages} {057901} (\bibinfo {year} {2004})},\ \Eprint {https://arxiv.org/abs/hep-ph/0406311} {arXiv:hep-ph/0406311 [hep-ph]} \BibitemShut {NoStop}%
\bibitem [{\citenamefont {Heinz}\ and\ \citenamefont {Snellings}(2013)}]{Heinz:2013th}%
  \BibitemOpen
  \bibfield  {author} {\bibinfo {author} {\bibfnamefont {U.}~\bibnamefont {Heinz}}\ and\ \bibinfo {author} {\bibfnamefont {R.}~\bibnamefont {Snellings}},\ }\bibfield  {title} {\bibinfo {title} {{Collective flow and viscosity in relativistic heavy-ion collisions}},\ }\href {https://doi.org/10.1146/annurev-nucl-102212-170540} {\bibfield  {journal} {\bibinfo  {journal} {Ann.Rev.Nucl.Part.Sci.}\ }\textbf {\bibinfo {volume} {63}},\ \bibinfo {pages} {123} (\bibinfo {year} {2013})},\ \Eprint {https://arxiv.org/abs/1301.2826} {arXiv:1301.2826 [nucl-th]} \BibitemShut {NoStop}%
\bibitem [{\citenamefont {Alver}\ \emph {et~al.}(2007)\citenamefont {Alver} \emph {et~al.}}]{Alver:2006wh}%
  \BibitemOpen
  \bibfield  {author} {\bibinfo {author} {\bibfnamefont {B.}~\bibnamefont {Alver}} \emph {et~al.} (\bibinfo {collaboration} {PHOBOS}),\ }\bibfield  {title} {\bibinfo {title} {{System size, energy, pseudorapidity, and centrality dependence of elliptic flow}},\ }\href {https://doi.org/10.1103/PhysRevLett.98.242302} {\bibfield  {journal} {\bibinfo  {journal} {Phys.Rev.Lett.}\ }\textbf {\bibinfo {volume} {98}},\ \bibinfo {pages} {242302} (\bibinfo {year} {2007})},\ \Eprint {https://arxiv.org/abs/nucl-ex/0610037} {arXiv:nucl-ex/0610037 [nucl-ex]} \BibitemShut {NoStop}%
\bibitem [{\citenamefont {Abelev}\ \emph {et~al.}(2009{\natexlab{a}})\citenamefont {Abelev} \emph {et~al.}}]{Abelev:2009ac}%
  \BibitemOpen
  \bibfield  {author} {\bibinfo {author} {\bibfnamefont {B.}~\bibnamefont {Abelev}} \emph {et~al.} (\bibinfo {collaboration} {STAR Collaboration}),\ }\bibfield  {title} {\bibinfo {title} {{Azimuthal Charged-Particle Correlations and Possible Local Strong Parity Violation}},\ }\href {https://doi.org/10.1103/PhysRevLett.103.251601} {\bibfield  {journal} {\bibinfo  {journal} {Phys.Rev.Lett.}\ }\textbf {\bibinfo {volume} {103}},\ \bibinfo {pages} {251601} (\bibinfo {year} {2009}{\natexlab{a}})},\ \Eprint {https://arxiv.org/abs/0909.1739} {arXiv:0909.1739 [nucl-ex]} \BibitemShut {NoStop}%
\bibitem [{\citenamefont {Abelev}\ \emph {et~al.}(2010)\citenamefont {Abelev} \emph {et~al.}}]{Abelev:2009ad}%
  \BibitemOpen
  \bibfield  {author} {\bibinfo {author} {\bibfnamefont {B.}~\bibnamefont {Abelev}} \emph {et~al.} (\bibinfo {collaboration} {STAR Collaboration}),\ }\bibfield  {title} {\bibinfo {title} {{Observation of charge-dependent azimuthal correlations and possible local strong parity violation in heavy ion collisions}},\ }\href {https://doi.org/10.1103/PhysRevC.81.054908} {\bibfield  {journal} {\bibinfo  {journal} {Phys.Rev.}\ }\textbf {\bibinfo {volume} {C81}},\ \bibinfo {pages} {054908} (\bibinfo {year} {2010})},\ \Eprint {https://arxiv.org/abs/0909.1717} {arXiv:0909.1717 [nucl-ex]} \BibitemShut {NoStop}%
\bibitem [{\citenamefont {Abelev}\ \emph {et~al.}(2013)\citenamefont {Abelev} \emph {et~al.}}]{Abelev:2012pa}%
  \BibitemOpen
  \bibfield  {author} {\bibinfo {author} {\bibfnamefont {B.}~\bibnamefont {Abelev}} \emph {et~al.} (\bibinfo {collaboration} {ALICE}),\ }\bibfield  {title} {\bibinfo {title} {{Charge separation relative to the reaction plane in Pb-Pb collisions at \SNN= 2.76 TeV}},\ }\href {https://doi.org/10.1103/PhysRevLett.110.012301} {\bibfield  {journal} {\bibinfo  {journal} {Phys.Rev.Lett.}\ }\textbf {\bibinfo {volume} {110}},\ \bibinfo {pages} {012301} (\bibinfo {year} {2013})},\ \Eprint {https://arxiv.org/abs/1207.0900} {arXiv:1207.0900 [nucl-ex]} \BibitemShut {NoStop}%
\bibitem [{\citenamefont {Kharzeev}\ \emph {et~al.}(2016)\citenamefont {Kharzeev}, \citenamefont {Liao}, \citenamefont {Voloshin},\ and\ \citenamefont {Wang}}]{Kharzeev:2015znc}%
  \BibitemOpen
  \bibfield  {author} {\bibinfo {author} {\bibfnamefont {D.~E.}\ \bibnamefont {Kharzeev}}, \bibinfo {author} {\bibfnamefont {J.}~\bibnamefont {Liao}}, \bibinfo {author} {\bibfnamefont {S.~A.}\ \bibnamefont {Voloshin}},\ and\ \bibinfo {author} {\bibfnamefont {G.}~\bibnamefont {Wang}},\ }\bibfield  {title} {\bibinfo {title} {{Chiral magnetic and vortical effects in high-energy nuclear collisions—A status report}},\ }\href {https://doi.org/10.1016/j.ppnp.2016.01.001} {\bibfield  {journal} {\bibinfo  {journal} {Prog. Part. Nucl. Phys.}\ }\textbf {\bibinfo {volume} {88}},\ \bibinfo {pages} {1} (\bibinfo {year} {2016})},\ \Eprint {https://arxiv.org/abs/1511.04050} {arXiv:1511.04050 [hep-ph]} \BibitemShut {NoStop}%
\bibitem [{\citenamefont {Zhao}(2018{\natexlab{a}})}]{Zhao:2018ixy}%
  \BibitemOpen
  \bibfield  {author} {\bibinfo {author} {\bibfnamefont {J.}~\bibnamefont {Zhao}},\ }\bibfield  {title} {\bibinfo {title} {{Search for the Chiral Magnetic Effect in Relativistic Heavy-Ion Collisions}},\ }\href {https://doi.org/10.1142/S0217751X18300107} {\bibfield  {journal} {\bibinfo  {journal} {Int. J. Mod. Phys.}\ }\textbf {\bibinfo {volume} {A33}},\ \bibinfo {pages} {1830010} (\bibinfo {year} {2018}{\natexlab{a}})},\ \Eprint {https://arxiv.org/abs/1805.02814} {arXiv:1805.02814 [nucl-ex]} \BibitemShut {NoStop}%
\bibitem [{\citenamefont {Zhao}\ and\ \citenamefont {Wang}(2019)}]{Zhao:2019hta}%
  \BibitemOpen
  \bibfield  {author} {\bibinfo {author} {\bibfnamefont {J.}~\bibnamefont {Zhao}}\ and\ \bibinfo {author} {\bibfnamefont {F.}~\bibnamefont {Wang}},\ }\bibfield  {title} {\bibinfo {title} {{Experimental searches for the chiral magnetic effect in heavy-ion collisions}},\ }\href {https://doi.org/10.1016/j.ppnp.2019.05.001} {\bibfield  {journal} {\bibinfo  {journal} {Prog. Part. Nucl. Phys.}\ }\textbf {\bibinfo {volume} {107}},\ \bibinfo {pages} {200} (\bibinfo {year} {2019})},\ \Eprint {https://arxiv.org/abs/1906.11413} {arXiv:1906.11413 [nucl-ex]} \BibitemShut {NoStop}%
\bibitem [{\citenamefont {Wang}(2010)}]{Wang:2009kd}%
  \BibitemOpen
  \bibfield  {author} {\bibinfo {author} {\bibfnamefont {F.}~\bibnamefont {Wang}},\ }\bibfield  {title} {\bibinfo {title} {{Effects of Cluster Particle Correlations on Local Parity Violation Observables}},\ }\href {https://doi.org/10.1103/PhysRevC.81.064902} {\bibfield  {journal} {\bibinfo  {journal} {Phys.Rev.}\ }\textbf {\bibinfo {volume} {C81}},\ \bibinfo {pages} {064902} (\bibinfo {year} {2010})},\ \Eprint {https://arxiv.org/abs/0911.1482} {arXiv:0911.1482 [nucl-ex]} \BibitemShut {NoStop}%
\bibitem [{\citenamefont {Liao}\ \emph {et~al.}(2010)\citenamefont {Liao}, \citenamefont {Koch},\ and\ \citenamefont {Bzdak}}]{Liao:2010nv}%
  \BibitemOpen
  \bibfield  {author} {\bibinfo {author} {\bibfnamefont {J.}~\bibnamefont {Liao}}, \bibinfo {author} {\bibfnamefont {V.}~\bibnamefont {Koch}},\ and\ \bibinfo {author} {\bibfnamefont {A.}~\bibnamefont {Bzdak}},\ }\bibfield  {title} {\bibinfo {title} {{On the Charge Separation Effect in Relativistic Heavy Ion Collisions}},\ }\href {https://doi.org/10.1103/PhysRevC.82.054902} {\bibfield  {journal} {\bibinfo  {journal} {Phys.Rev.}\ }\textbf {\bibinfo {volume} {C82}},\ \bibinfo {pages} {054902} (\bibinfo {year} {2010})},\ \Eprint {https://arxiv.org/abs/1005.5380} {arXiv:1005.5380 [nucl-th]} \BibitemShut {NoStop}%
\bibitem [{\citenamefont {Bzdak}\ \emph {et~al.}(2011)\citenamefont {Bzdak}, \citenamefont {Koch},\ and\ \citenamefont {Liao}}]{Bzdak:2010fd}%
  \BibitemOpen
  \bibfield  {author} {\bibinfo {author} {\bibfnamefont {A.}~\bibnamefont {Bzdak}}, \bibinfo {author} {\bibfnamefont {V.}~\bibnamefont {Koch}},\ and\ \bibinfo {author} {\bibfnamefont {J.}~\bibnamefont {Liao}},\ }\bibfield  {title} {\bibinfo {title} {{Azimuthal correlations from transverse momentum conservation and possible local parity violation}},\ }\href {https://doi.org/10.1103/PhysRevC.83.014905} {\bibfield  {journal} {\bibinfo  {journal} {Phys.Rev.}\ }\textbf {\bibinfo {volume} {C83}},\ \bibinfo {pages} {014905} (\bibinfo {year} {2011})},\ \Eprint {https://arxiv.org/abs/1008.4919} {arXiv:1008.4919 [nucl-th]} \BibitemShut {NoStop}%
\bibitem [{\citenamefont {Schlichting}\ and\ \citenamefont {Pratt}(2011)}]{Schlichting:2010qia}%
  \BibitemOpen
  \bibfield  {author} {\bibinfo {author} {\bibfnamefont {S.}~\bibnamefont {Schlichting}}\ and\ \bibinfo {author} {\bibfnamefont {S.}~\bibnamefont {Pratt}},\ }\bibfield  {title} {\bibinfo {title} {{Charge conservation at energies available at the BNL Relativistic Heavy Ion Collider and contributions to local parity violation observables}},\ }\href {https://doi.org/10.1103/PhysRevC.83.014913} {\bibfield  {journal} {\bibinfo  {journal} {Phys.Rev.}\ }\textbf {\bibinfo {volume} {C83}},\ \bibinfo {pages} {014913} (\bibinfo {year} {2011})},\ \Eprint {https://arxiv.org/abs/1009.4283} {arXiv:1009.4283 [nucl-th]} \BibitemShut {NoStop}%
\bibitem [{\citenamefont {Pratt}\ \emph {et~al.}(2011)\citenamefont {Pratt}, \citenamefont {Schlichting},\ and\ \citenamefont {Gavin}}]{Pratt:2010zn}%
  \BibitemOpen
  \bibfield  {author} {\bibinfo {author} {\bibfnamefont {S.}~\bibnamefont {Pratt}}, \bibinfo {author} {\bibfnamefont {S.}~\bibnamefont {Schlichting}},\ and\ \bibinfo {author} {\bibfnamefont {S.}~\bibnamefont {Gavin}},\ }\bibfield  {title} {\bibinfo {title} {{Effects of Momentum Conservation and Flow on Angular Correlations at RHIC}},\ }\href {https://doi.org/10.1103/PhysRevC.84.024909} {\bibfield  {journal} {\bibinfo  {journal} {Phys.Rev.}\ }\textbf {\bibinfo {volume} {C84}},\ \bibinfo {pages} {024909} (\bibinfo {year} {2011})},\ \Eprint {https://arxiv.org/abs/1011.6053} {arXiv:1011.6053 [nucl-th]} \BibitemShut {NoStop}%
\bibitem [{\citenamefont {Zhao}\ \emph {et~al.}(2018)\citenamefont {Zhao}, \citenamefont {Tu},\ and\ \citenamefont {Wang}}]{Zhao:2018skm}%
  \BibitemOpen
  \bibfield  {author} {\bibinfo {author} {\bibfnamefont {J.}~\bibnamefont {Zhao}}, \bibinfo {author} {\bibfnamefont {Z.}~\bibnamefont {Tu}},\ and\ \bibinfo {author} {\bibfnamefont {F.}~\bibnamefont {Wang}},\ }\bibfield  {title} {\bibinfo {title} {{Status of the Chiral Magnetic Effect Search in Relativistic Heavy-Ion Collisions}},\ }\href {https://doi.org/10.11804/NuclPhysRev.35.03.225} {\bibfield  {journal} {\bibinfo  {journal} {Nucl. Phys. Rev.}\ }\textbf {\bibinfo {volume} {35}},\ \bibinfo {pages} {225} (\bibinfo {year} {2018})},\ \Eprint {https://arxiv.org/abs/1807.05083} {arXiv:1807.05083 [nucl-ex]} \BibitemShut {NoStop}%
\bibitem [{\citenamefont {Zhao}(2018{\natexlab{b}})}]{Zhao:2018pnk}%
  \BibitemOpen
  \bibfield  {author} {\bibinfo {author} {\bibfnamefont {J.}~\bibnamefont {Zhao}} (\bibinfo {collaboration} {STAR}),\ }\bibfield  {title} {\bibinfo {title} {{Chiral magnetic effect search in p(d)+Au, Au+Au collisions at RHIC}},\ }\bibfield  {booktitle} {\emph {\bibinfo {booktitle} {{Proceedings, 21st International Conference on Particles and Nuclei (PANIC 17): Beijing, China, September 1-5, 2017}}},\ }\href {https://doi.org/10.1142/S2010194518600108} {\bibfield  {journal} {\bibinfo  {journal} {Int. J. Mod. Phys. Conf. Ser.}\ }\textbf {\bibinfo {volume} {46}},\ \bibinfo {pages} {1860010} (\bibinfo {year} {2018}{\natexlab{b}})},\ \Eprint {https://arxiv.org/abs/1802.03283} {arXiv:1802.03283 [nucl-ex]} \BibitemShut {NoStop}%
\bibitem [{\citenamefont {Adamczyk}\ \emph {et~al.}(2014{\natexlab{a}})\citenamefont {Adamczyk} \emph {et~al.}}]{Adamczyk:2013kcb}%
  \BibitemOpen
  \bibfield  {author} {\bibinfo {author} {\bibfnamefont {L.}~\bibnamefont {Adamczyk}} \emph {et~al.} (\bibinfo {collaboration} {STAR}),\ }\bibfield  {title} {\bibinfo {title} {{Measurement of charge multiplicity asymmetry correlations in high-energy nucleus-nucleus collisions at $\sqrt{{s}_{NN}} =$ 200 GeV}},\ }\href {https://doi.org/10.1103/PhysRevC.89.044908} {\bibfield  {journal} {\bibinfo  {journal} {Phys. Rev.}\ }\textbf {\bibinfo {volume} {C89}},\ \bibinfo {pages} {044908} (\bibinfo {year} {2014}{\natexlab{a}})},\ \Eprint {https://arxiv.org/abs/1303.0901} {arXiv:1303.0901 [nucl-ex]} \BibitemShut {NoStop}%
\bibitem [{\citenamefont {Adamczyk}\ \emph {et~al.}(2014{\natexlab{b}})\citenamefont {Adamczyk} \emph {et~al.}}]{Adamczyk:2014mzf}%
  \BibitemOpen
  \bibfield  {author} {\bibinfo {author} {\bibfnamefont {L.}~\bibnamefont {Adamczyk}} \emph {et~al.} (\bibinfo {collaboration} {STAR}),\ }\bibfield  {title} {\bibinfo {title} {{Beam-energy dependence of charge separation along the magnetic field in Au+Au collisions at RHIC}},\ }\href {https://doi.org/10.1103/PhysRevLett.113.052302} {\bibfield  {journal} {\bibinfo  {journal} {Phys. Rev. Lett.}\ }\textbf {\bibinfo {volume} {113}},\ \bibinfo {pages} {052302} (\bibinfo {year} {2014}{\natexlab{b}})},\ \Eprint {https://arxiv.org/abs/1404.1433} {arXiv:1404.1433 [nucl-ex]} \BibitemShut {NoStop}%
\bibitem [{\citenamefont {Khachatryan}\ \emph {et~al.}(2017)\citenamefont {Khachatryan} \emph {et~al.}}]{Khachatryan:2016got}%
  \BibitemOpen
  \bibfield  {author} {\bibinfo {author} {\bibfnamefont {V.}~\bibnamefont {Khachatryan}} \emph {et~al.} (\bibinfo {collaboration} {CMS}),\ }\bibfield  {title} {\bibinfo {title} {{Observation of charge-dependent azimuthal correlations in $p$-Pb collisions and its implication for the search for the chiral magnetic effect}},\ }\href {https://doi.org/10.1103/PhysRevLett.118.122301} {\bibfield  {journal} {\bibinfo  {journal} {Phys. Rev. Lett.}\ }\textbf {\bibinfo {volume} {118}},\ \bibinfo {pages} {122301} (\bibinfo {year} {2017})},\ \Eprint {https://arxiv.org/abs/1610.00263} {arXiv:1610.00263 [nucl-ex]} \BibitemShut {NoStop}%
\bibitem [{\citenamefont {Adam}\ \emph {et~al.}(2019)\citenamefont {Adam} \emph {et~al.}}]{STAR:2019xzd}%
  \BibitemOpen
  \bibfield  {author} {\bibinfo {author} {\bibfnamefont {J.}~\bibnamefont {Adam}} \emph {et~al.} (\bibinfo {collaboration} {STAR}),\ }\bibfield  {title} {\bibinfo {title} {{Charge-dependent pair correlations relative to a third particle in $p$ + Au and $d$+ Au collisions at RHIC}},\ }\href {https://doi.org/10.1016/j.physletb.2019.134975} {\bibfield  {journal} {\bibinfo  {journal} {Phys. Lett.}\ }\textbf {\bibinfo {volume} {B798}},\ \bibinfo {pages} {134975} (\bibinfo {year} {2019})},\ \Eprint {https://arxiv.org/abs/1906.03373} {arXiv:1906.03373 [nucl-ex]} \BibitemShut {NoStop}%
\bibitem [{\citenamefont {Acharya}\ \emph {et~al.}(2020)\citenamefont {Acharya} \emph {et~al.}}]{ALICE:2020siw}%
  \BibitemOpen
  \bibfield  {author} {\bibinfo {author} {\bibfnamefont {S.}~\bibnamefont {Acharya}} \emph {et~al.} (\bibinfo {collaboration} {ALICE}),\ }\bibfield  {title} {\bibinfo {title} {{Constraining the Chiral Magnetic Effect with charge-dependent azimuthal correlations in Pb-Pb collisions at $ \sqrt{s_{\mathrm{NN}}} $ = 2.76 and 5.02 TeV}},\ }\href {https://doi.org/10.1007/JHEP09(2020)160} {\bibfield  {journal} {\bibinfo  {journal} {JHEP}\ }\textbf {\bibinfo {volume} {09}},\ \bibinfo {pages} {160}},\ \Eprint {https://arxiv.org/abs/2005.14640} {arXiv:2005.14640 [nucl-ex]} \BibitemShut {NoStop}%
\bibitem [{\citenamefont {Acharya}\ \emph {et~al.}(2018)\citenamefont {Acharya} \emph {et~al.}}]{Acharya:2017fau}%
  \BibitemOpen
  \bibfield  {author} {\bibinfo {author} {\bibfnamefont {S.}~\bibnamefont {Acharya}} \emph {et~al.} (\bibinfo {collaboration} {ALICE}),\ }\bibfield  {title} {\bibinfo {title} {{Constraining the magnitude of the Chiral Magnetic Effect with Event Shape Engineering in Pb-Pb collisions at $\sqrt{s_\mathrm{NN}}$ = 2.76 TeV}},\ }\href {https://doi.org/10.1016/j.physletb.2017.12.021} {\bibfield  {journal} {\bibinfo  {journal} {Phys. Lett.}\ }\textbf {\bibinfo {volume} {B777}},\ \bibinfo {pages} {151} (\bibinfo {year} {2018})},\ \Eprint {https://arxiv.org/abs/1709.04723} {arXiv:1709.04723 [nucl-ex]} \BibitemShut {NoStop}%
\bibitem [{\citenamefont {Sirunyan}\ \emph {et~al.}(2018)\citenamefont {Sirunyan} \emph {et~al.}}]{Sirunyan:2017quh}%
  \BibitemOpen
  \bibfield  {author} {\bibinfo {author} {\bibfnamefont {A.~M.}\ \bibnamefont {Sirunyan}} \emph {et~al.} (\bibinfo {collaboration} {CMS}),\ }\bibfield  {title} {\bibinfo {title} {{Constraints on the chiral magnetic effect using charge-dependent azimuthal correlations in $p\mathrm{Pb}$ and PbPb collisions at the CERN Large Hadron Collider}},\ }\href {https://doi.org/10.1103/PhysRevC.97.044912} {\bibfield  {journal} {\bibinfo  {journal} {Phys. Rev.}\ }\textbf {\bibinfo {volume} {C97}},\ \bibinfo {pages} {044912} (\bibinfo {year} {2018})},\ \Eprint {https://arxiv.org/abs/1708.01602} {arXiv:1708.01602 [nucl-ex]} \BibitemShut {NoStop}%
\bibitem [{\citenamefont {Xu}\ \emph {et~al.}(2018)\citenamefont {Xu}, \citenamefont {Zhao}, \citenamefont {Wang}, \citenamefont {Li}, \citenamefont {Lin}, \citenamefont {Shen},\ and\ \citenamefont {Wang}}]{Xu:2017qfs}%
  \BibitemOpen
  \bibfield  {author} {\bibinfo {author} {\bibfnamefont {H.-J.}\ \bibnamefont {Xu}}, \bibinfo {author} {\bibfnamefont {J.}~\bibnamefont {Zhao}}, \bibinfo {author} {\bibfnamefont {X.}~\bibnamefont {Wang}}, \bibinfo {author} {\bibfnamefont {H.}~\bibnamefont {Li}}, \bibinfo {author} {\bibfnamefont {Z.-W.}\ \bibnamefont {Lin}}, \bibinfo {author} {\bibfnamefont {C.}~\bibnamefont {Shen}},\ and\ \bibinfo {author} {\bibfnamefont {F.}~\bibnamefont {Wang}},\ }\bibfield  {title} {\bibinfo {title} {{Varying the chiral magnetic effect relative to flow in a single nucleus-nucleus collision}},\ }\href {https://doi.org/10.1088/1674-1137/42/8/084103} {\bibfield  {journal} {\bibinfo  {journal} {Chin. Phys.}\ }\textbf {\bibinfo {volume} {C42}},\ \bibinfo {pages} {084103} (\bibinfo {year} {2018})},\ \Eprint {https://arxiv.org/abs/1710.07265} {arXiv:1710.07265 [nucl-th]} \BibitemShut {NoStop}%
\bibitem [{\citenamefont {Voloshin}(2018)}]{Voloshin:2018qsm}%
  \BibitemOpen
  \bibfield  {author} {\bibinfo {author} {\bibfnamefont {S.~A.}\ \bibnamefont {Voloshin}},\ }\bibfield  {title} {\bibinfo {title} {{Estimate of the signal from the chiral magnetic effect in heavy-ion collisions from measurements relative to the participant and spectator flow planes}},\ }\href {https://doi.org/10.1103/PhysRevC.98.054911} {\bibfield  {journal} {\bibinfo  {journal} {Phys. Rev. C}\ }\textbf {\bibinfo {volume} {98}},\ \bibinfo {pages} {054911} (\bibinfo {year} {2018})},\ \Eprint {https://arxiv.org/abs/1805.05300} {arXiv:1805.05300 [nucl-ex]} \BibitemShut {NoStop}%
\bibitem [{\citenamefont {Abdallah}\ \emph {et~al.}(2022)\citenamefont {Abdallah} \emph {et~al.}}]{STAR:2021pwb}%
  \BibitemOpen
  \bibfield  {author} {\bibinfo {author} {\bibfnamefont {M.}~\bibnamefont {Abdallah}} \emph {et~al.} (\bibinfo {collaboration} {STAR}),\ }\bibfield  {title} {\bibinfo {title} {{Search for the Chiral Magnetic Effect via Charge-Dependent Azimuthal Correlations Relative to Spectator and Participant Planes in Au+Au Collisions at $\sqrt{s_{NN}}$ =\, 200\,GeV}},\ }\href {https://doi.org/10.1103/PhysRevLett.128.092301} {\bibfield  {journal} {\bibinfo  {journal} {Phys. Rev. Lett.}\ }\textbf {\bibinfo {volume} {128}},\ \bibinfo {pages} {092301} (\bibinfo {year} {2022})},\ \Eprint {https://arxiv.org/abs/2106.09243} {arXiv:2106.09243 [nucl-ex]} \BibitemShut {NoStop}%
\bibitem [{\citenamefont {Choudhury}\ \emph {et~al.}(2022)\citenamefont {Choudhury} \emph {et~al.}}]{Choudhury:2021jwd}%
  \BibitemOpen
  \bibfield  {author} {\bibinfo {author} {\bibfnamefont {S.}~\bibnamefont {Choudhury}} \emph {et~al.},\ }\bibfield  {title} {\bibinfo {title} {{Investigation of experimental observables in search of the chiral magnetic effect in heavy-ion collisions in the STAR experiment}},\ }\href {https://doi.org/10.1088/1674-1137/ac2a1f} {\bibfield  {journal} {\bibinfo  {journal} {Chin. Phys. C}\ }\textbf {\bibinfo {volume} {46}},\ \bibinfo {pages} {014101} (\bibinfo {year} {2022})},\ \Eprint {https://arxiv.org/abs/2105.06044} {arXiv:2105.06044 [nucl-ex]} \BibitemShut {NoStop}%
\bibitem [{\citenamefont {Schukraft}\ \emph {et~al.}(2013)\citenamefont {Schukraft}, \citenamefont {Timmins},\ and\ \citenamefont {Voloshin}}]{Schukraft:2012ah}%
  \BibitemOpen
  \bibfield  {author} {\bibinfo {author} {\bibfnamefont {J.}~\bibnamefont {Schukraft}}, \bibinfo {author} {\bibfnamefont {A.}~\bibnamefont {Timmins}},\ and\ \bibinfo {author} {\bibfnamefont {S.~A.}\ \bibnamefont {Voloshin}},\ }\bibfield  {title} {\bibinfo {title} {{Ultra-relativistic nuclear collisions: event shape engineering}},\ }\href {https://doi.org/10.1016/j.physletb.2013.01.045} {\bibfield  {journal} {\bibinfo  {journal} {Phys. Lett.}\ }\textbf {\bibinfo {volume} {B719}},\ \bibinfo {pages} {394} (\bibinfo {year} {2013})},\ \Eprint {https://arxiv.org/abs/1208.4563} {arXiv:1208.4563 [nucl-ex]} \BibitemShut {NoStop}%
\bibitem [{\citenamefont {Xu}\ \emph {et~al.}(2024)\citenamefont {Xu}, \citenamefont {Chan}, \citenamefont {Wang}, \citenamefont {Tang},\ and\ \citenamefont {Huang}}]{Xu:2023elq}%
  \BibitemOpen
  \bibfield  {author} {\bibinfo {author} {\bibfnamefont {Z.}~\bibnamefont {Xu}}, \bibinfo {author} {\bibfnamefont {B.}~\bibnamefont {Chan}}, \bibinfo {author} {\bibfnamefont {G.}~\bibnamefont {Wang}}, \bibinfo {author} {\bibfnamefont {A.}~\bibnamefont {Tang}},\ and\ \bibinfo {author} {\bibfnamefont {H.~Z.}\ \bibnamefont {Huang}},\ }\bibfield  {title} {\bibinfo {title} {{Event shape selection method in search of the chiral magnetic effect in heavy-ion collisions}},\ }\href {https://doi.org/10.1016/j.physletb.2023.138367} {\bibfield  {journal} {\bibinfo  {journal} {Phys. Lett. B}\ }\textbf {\bibinfo {volume} {848}},\ \bibinfo {pages} {138367} (\bibinfo {year} {2024})},\ \Eprint {https://arxiv.org/abs/2307.14997} {arXiv:2307.14997 [nucl-th]} \BibitemShut {NoStop}%
\bibitem [{\citenamefont {Borghini}\ \emph {et~al.}(2000)\citenamefont {Borghini}, \citenamefont {Dinh},\ and\ \citenamefont {Ollitrault}}]{Borghini:2000cm}%
  \BibitemOpen
  \bibfield  {author} {\bibinfo {author} {\bibfnamefont {N.}~\bibnamefont {Borghini}}, \bibinfo {author} {\bibfnamefont {P.~M.}\ \bibnamefont {Dinh}},\ and\ \bibinfo {author} {\bibfnamefont {J.-Y.}\ \bibnamefont {Ollitrault}},\ }\bibfield  {title} {\bibinfo {title} {{Are flow measurements at SPS reliable?}},\ }\href {https://doi.org/10.1103/PhysRevC.62.034902} {\bibfield  {journal} {\bibinfo  {journal} {Phys.Rev.}\ }\textbf {\bibinfo {volume} {C62}},\ \bibinfo {pages} {034902} (\bibinfo {year} {2000})},\ \Eprint {https://arxiv.org/abs/nucl-th/0004026} {arXiv:nucl-th/0004026 [nucl-th]} \BibitemShut {NoStop}%
\bibitem [{\citenamefont {Borghini}(2007)}]{Borghini:2006yk}%
  \BibitemOpen
  \bibfield  {author} {\bibinfo {author} {\bibfnamefont {N.}~\bibnamefont {Borghini}},\ }\bibfield  {title} {\bibinfo {title} {{Momentum conservation and correlation analyses in heavy-ion collisions at ultrarelativistic energies}},\ }\href {https://doi.org/10.1103/PhysRevC.75.021904} {\bibfield  {journal} {\bibinfo  {journal} {Phys. Rev.}\ }\textbf {\bibinfo {volume} {C75}},\ \bibinfo {pages} {021904} (\bibinfo {year} {2007})},\ \Eprint {https://arxiv.org/abs/nucl-th/0612093} {arXiv:nucl-th/0612093 [nucl-th]} \BibitemShut {NoStop}%
\bibitem [{\citenamefont {Wang}\ and\ \citenamefont {Wang}(2010)}]{Wang:2008gp}%
  \BibitemOpen
  \bibfield  {author} {\bibinfo {author} {\bibfnamefont {Q.}~\bibnamefont {Wang}}\ and\ \bibinfo {author} {\bibfnamefont {F.}~\bibnamefont {Wang}},\ }\bibfield  {title} {\bibinfo {title} {{Non-flow correlations in a cluster model}},\ }\href {https://doi.org/10.1103/PhysRevC.81.064905} {\bibfield  {journal} {\bibinfo  {journal} {Phys.Rev.}\ }\textbf {\bibinfo {volume} {C81}},\ \bibinfo {pages} {064905} (\bibinfo {year} {2010})},\ \Eprint {https://arxiv.org/abs/0812.1176} {arXiv:0812.1176 [nucl-ex]} \BibitemShut {NoStop}%
\bibitem [{\citenamefont {Ollitrault}\ \emph {et~al.}(2009)\citenamefont {Ollitrault}, \citenamefont {Poskanzer},\ and\ \citenamefont {Voloshin}}]{Ollitrault:2009ie}%
  \BibitemOpen
  \bibfield  {author} {\bibinfo {author} {\bibfnamefont {J.-Y.}\ \bibnamefont {Ollitrault}}, \bibinfo {author} {\bibfnamefont {A.~M.}\ \bibnamefont {Poskanzer}},\ and\ \bibinfo {author} {\bibfnamefont {S.~A.}\ \bibnamefont {Voloshin}},\ }\bibfield  {title} {\bibinfo {title} {{Effect of flow fluctuations and nonflow on elliptic flow methods}},\ }\href {https://doi.org/10.1103/PhysRevC.80.014904} {\bibfield  {journal} {\bibinfo  {journal} {Phys.Rev.}\ }\textbf {\bibinfo {volume} {C80}},\ \bibinfo {pages} {014904} (\bibinfo {year} {2009})},\ \Eprint {https://arxiv.org/abs/0904.2315} {arXiv:0904.2315 [nucl-ex]} \BibitemShut {NoStop}%
\bibitem [{\citenamefont {Poskanzer}\ and\ \citenamefont {Voloshin}(1998)}]{Poskanzer:1998yz}%
  \BibitemOpen
  \bibfield  {author} {\bibinfo {author} {\bibfnamefont {A.~M.}\ \bibnamefont {Poskanzer}}\ and\ \bibinfo {author} {\bibfnamefont {S.}~\bibnamefont {Voloshin}},\ }\bibfield  {title} {\bibinfo {title} {{Methods for analyzing anisotropic flow in relativistic nuclear collisions}},\ }\href {https://doi.org/10.1103/PhysRevC.58.1671} {\bibfield  {journal} {\bibinfo  {journal} {Phys.Rev.}\ }\textbf {\bibinfo {volume} {C58}},\ \bibinfo {pages} {1671} (\bibinfo {year} {1998})},\ \Eprint {https://arxiv.org/abs/nucl-ex/9805001} {arXiv:nucl-ex/9805001 [nucl-ex]} \BibitemShut {NoStop}%
\bibitem [{\citenamefont {Bhalerao}\ and\ \citenamefont {Ollitrault}(2006)}]{Bhalerao:2006tp}%
  \BibitemOpen
  \bibfield  {author} {\bibinfo {author} {\bibfnamefont {R.~S.}\ \bibnamefont {Bhalerao}}\ and\ \bibinfo {author} {\bibfnamefont {J.-Y.}\ \bibnamefont {Ollitrault}},\ }\bibfield  {title} {\bibinfo {title} {{Eccentricity fluctuations and elliptic flow at RHIC}},\ }\href {https://doi.org/10.1016/j.physletb.2006.08.055} {\bibfield  {journal} {\bibinfo  {journal} {Phys.Lett.}\ }\textbf {\bibinfo {volume} {B641}},\ \bibinfo {pages} {260} (\bibinfo {year} {2006})},\ \Eprint {https://arxiv.org/abs/nucl-th/0607009} {arXiv:nucl-th/0607009 [nucl-th]} \BibitemShut {NoStop}%
\bibitem [{\citenamefont {Bozek}\ \emph {et~al.}(2011)\citenamefont {Bozek}, \citenamefont {Broniowski},\ and\ \citenamefont {Moreira}}]{Bozek:2010vz}%
  \BibitemOpen
  \bibfield  {author} {\bibinfo {author} {\bibfnamefont {P.}~\bibnamefont {Bozek}}, \bibinfo {author} {\bibfnamefont {W.}~\bibnamefont {Broniowski}},\ and\ \bibinfo {author} {\bibfnamefont {J.}~\bibnamefont {Moreira}},\ }\bibfield  {title} {\bibinfo {title} {{Torqued fireballs in relativistic heavy-ion collisions}},\ }\href {https://doi.org/10.1103/PhysRevC.83.034911} {\bibfield  {journal} {\bibinfo  {journal} {Phys.Rev.}\ }\textbf {\bibinfo {volume} {C83}},\ \bibinfo {pages} {034911} (\bibinfo {year} {2011})},\ \Eprint {https://arxiv.org/abs/1011.3354} {arXiv:1011.3354 [nucl-th]} \BibitemShut {NoStop}%
\bibitem [{\citenamefont {Xiao}\ \emph {et~al.}(2013)\citenamefont {Xiao}, \citenamefont {Liu},\ and\ \citenamefont {Wang}}]{Xiao:2012uw}%
  \BibitemOpen
  \bibfield  {author} {\bibinfo {author} {\bibfnamefont {K.}~\bibnamefont {Xiao}}, \bibinfo {author} {\bibfnamefont {F.}~\bibnamefont {Liu}},\ and\ \bibinfo {author} {\bibfnamefont {F.}~\bibnamefont {Wang}},\ }\bibfield  {title} {\bibinfo {title} {{Event-plane decorrelation over pseudo-rapidity and its effect on azimuthal anisotropy measurement in relativistic heavy-ion collisions}},\ }\href {https://doi.org/10.1103/PhysRevC.87.011901} {\bibfield  {journal} {\bibinfo  {journal} {Phys.Rev.}\ }\textbf {\bibinfo {volume} {C87}},\ \bibinfo {pages} {011901} (\bibinfo {year} {2013})},\ \Eprint {https://arxiv.org/abs/1208.1195} {arXiv:1208.1195 [nucl-th]} \BibitemShut {NoStop}%
\bibitem [{\citenamefont {Adams}\ \emph {et~al.}(2020)\citenamefont {Adams} \emph {et~al.}}]{Adams:2019fpo}%
  \BibitemOpen
  \bibfield  {author} {\bibinfo {author} {\bibfnamefont {J.}~\bibnamefont {Adams}} \emph {et~al.},\ }\bibfield  {title} {\bibinfo {title} {{The STAR Event Plane Detector}},\ }\href {https://doi.org/10.1016/j.nima.2020.163970} {\bibfield  {journal} {\bibinfo  {journal} {Nucl. Instrum. Meth. A}\ }\textbf {\bibinfo {volume} {968}},\ \bibinfo {pages} {163970} (\bibinfo {year} {2020})},\ \Eprint {https://arxiv.org/abs/1912.05243} {arXiv:1912.05243 [physics.ins-det]} \BibitemShut {NoStop}%
\bibitem [{\citenamefont {Yan}(2023)}]{Yan:2023ugh}%
  \BibitemOpen
  \bibfield  {author} {\bibinfo {author} {\bibfnamefont {G.}~\bibnamefont {Yan}} (\bibinfo {collaboration} {STAR}),\ }\bibfield  {title} {\bibinfo {title} {{Probing Initial- and Final-state Effects of Heavy-ion Collisions with STAR Experiment}},\ }\href {https://doi.org/10.5506/APhysPolBSupp.16.1-A137} {\bibfield  {journal} {\bibinfo  {journal} {Acta Phys. Polon. Supp.}\ }\textbf {\bibinfo {volume} {16}},\ \bibinfo {pages} {1} (\bibinfo {year} {2023})}\BibitemShut {NoStop}%
\bibitem [{\citenamefont {Feng}\ \emph {et~al.}(2022)\citenamefont {Feng}, \citenamefont {Zhao}, \citenamefont {Li}, \citenamefont {Xu},\ and\ \citenamefont {Wang}}]{Feng:2021pgf}%
  \BibitemOpen
  \bibfield  {author} {\bibinfo {author} {\bibfnamefont {Y.}~\bibnamefont {Feng}}, \bibinfo {author} {\bibfnamefont {J.}~\bibnamefont {Zhao}}, \bibinfo {author} {\bibfnamefont {H.}~\bibnamefont {Li}}, \bibinfo {author} {\bibfnamefont {H.-j.}\ \bibnamefont {Xu}},\ and\ \bibinfo {author} {\bibfnamefont {F.}~\bibnamefont {Wang}},\ }\bibfield  {title} {\bibinfo {title} {{Two- and three-particle nonflow contributions to the chiral magnetic effect measurement by spectator and participant planes in relativistic heavy ion collisions}},\ }\href {https://doi.org/10.1103/PhysRevC.105.024913} {\bibfield  {journal} {\bibinfo  {journal} {Phys. Rev. C}\ }\textbf {\bibinfo {volume} {105}},\ \bibinfo {pages} {024913} (\bibinfo {year} {2022})},\ \Eprint {https://arxiv.org/abs/2106.15595} {arXiv:2106.15595 [nucl-ex]} \BibitemShut {NoStop}%
\bibitem [{\citenamefont {Wang}\ and\ \citenamefont {Zhao}(2017)}]{wang:2016iov}%
  \BibitemOpen
  \bibfield  {author} {\bibinfo {author} {\bibfnamefont {F.}~\bibnamefont {Wang}}\ and\ \bibinfo {author} {\bibfnamefont {J.}~\bibnamefont {Zhao}},\ }\bibfield  {title} {\bibinfo {title} {{Challenges in flow background removal in search for the chiral magnetic effect}},\ }\href {https://doi.org/10.1103/PhysRevC.95.051901} {\bibfield  {journal} {\bibinfo  {journal} {Phys. Rev.}\ }\textbf {\bibinfo {volume} {C95}},\ \bibinfo {pages} {051901} (\bibinfo {year} {2017})},\ \Eprint {https://arxiv.org/abs/1608.06610} {arXiv:1608.06610 [nucl-th]} \BibitemShut {NoStop}%
\bibitem [{\citenamefont {Abelev}\ \emph {et~al.}(2009{\natexlab{b}})\citenamefont {Abelev} \emph {et~al.}}]{Abelev:2008ab}%
  \BibitemOpen
  \bibfield  {author} {\bibinfo {author} {\bibfnamefont {B.}~\bibnamefont {Abelev}} \emph {et~al.} (\bibinfo {collaboration} {STAR Collaboration}),\ }\bibfield  {title} {\bibinfo {title} {{Systematic measurements of identified particle spectra in $pp$, $d$+Au and Au+Au collisions from STAR}},\ }\href {https://doi.org/10.1103/PhysRevC.79.034909} {\bibfield  {journal} {\bibinfo  {journal} {Phys.Rev.}\ }\textbf {\bibinfo {volume} {C79}},\ \bibinfo {pages} {034909} (\bibinfo {year} {2009}{\natexlab{b}})},\ \Eprint {https://arxiv.org/abs/0808.2041} {arXiv:0808.2041 [nucl-ex]} \BibitemShut {NoStop}%
\bibitem [{\citenamefont {Shi}\ \emph {et~al.}(2018)\citenamefont {Shi}, \citenamefont {Jiang}, \citenamefont {Lilleskov},\ and\ \citenamefont {Liao}}]{Shi:2017cpu}%
  \BibitemOpen
  \bibfield  {author} {\bibinfo {author} {\bibfnamefont {S.}~\bibnamefont {Shi}}, \bibinfo {author} {\bibfnamefont {Y.}~\bibnamefont {Jiang}}, \bibinfo {author} {\bibfnamefont {E.}~\bibnamefont {Lilleskov}},\ and\ \bibinfo {author} {\bibfnamefont {J.}~\bibnamefont {Liao}},\ }\bibfield  {title} {\bibinfo {title} {{Anomalous Chiral Transport in Heavy Ion Collisions from Anomalous-Viscous Fluid Dynamics}},\ }\href {https://doi.org/10.1016/j.aop.2018.04.026} {\bibfield  {journal} {\bibinfo  {journal} {Annals Phys.}\ }\textbf {\bibinfo {volume} {394}},\ \bibinfo {pages} {50} (\bibinfo {year} {2018})},\ \Eprint {https://arxiv.org/abs/1711.02496} {arXiv:1711.02496 [nucl-th]} \BibitemShut {NoStop}%
\bibitem [{\citenamefont {Jiang}\ \emph {et~al.}(2018)\citenamefont {Jiang}, \citenamefont {Shi}, \citenamefont {Yin},\ and\ \citenamefont {Liao}}]{Jiang:2016wve}%
  \BibitemOpen
  \bibfield  {author} {\bibinfo {author} {\bibfnamefont {Y.}~\bibnamefont {Jiang}}, \bibinfo {author} {\bibfnamefont {S.}~\bibnamefont {Shi}}, \bibinfo {author} {\bibfnamefont {Y.}~\bibnamefont {Yin}},\ and\ \bibinfo {author} {\bibfnamefont {J.}~\bibnamefont {Liao}},\ }\bibfield  {title} {\bibinfo {title} {{Quantifying the chiral magnetic effect from anomalous-viscous fluid dynamics}},\ }\href {https://doi.org/10.1088/1674-1137/42/1/011001} {\bibfield  {journal} {\bibinfo  {journal} {Chin. Phys. C}\ }\textbf {\bibinfo {volume} {42}},\ \bibinfo {pages} {011001} (\bibinfo {year} {2018})},\ \Eprint {https://arxiv.org/abs/1611.04586} {arXiv:1611.04586 [nucl-th]} \BibitemShut {NoStop}%
\bibitem [{\citenamefont {Shi}\ \emph {et~al.}(2020)\citenamefont {Shi}, \citenamefont {Zhang}, \citenamefont {Hou},\ and\ \citenamefont {Liao}}]{Shi:2019wzi}%
  \BibitemOpen
  \bibfield  {author} {\bibinfo {author} {\bibfnamefont {S.}~\bibnamefont {Shi}}, \bibinfo {author} {\bibfnamefont {H.}~\bibnamefont {Zhang}}, \bibinfo {author} {\bibfnamefont {D.}~\bibnamefont {Hou}},\ and\ \bibinfo {author} {\bibfnamefont {J.}~\bibnamefont {Liao}},\ }\bibfield  {title} {\bibinfo {title} {{Signatures of Chiral Magnetic Effect in the Collisions of Isobars}},\ }\href {https://doi.org/10.1103/PhysRevLett.125.242301} {\bibfield  {journal} {\bibinfo  {journal} {Phys. Rev. Lett.}\ }\textbf {\bibinfo {volume} {125}},\ \bibinfo {pages} {242301} (\bibinfo {year} {2020})},\ \Eprint {https://arxiv.org/abs/1910.14010} {arXiv:1910.14010 [nucl-th]} \BibitemShut {NoStop}%
\bibitem [{\citenamefont {Song}\ \emph {et~al.}(2011)\citenamefont {Song}, \citenamefont {Bass},\ and\ \citenamefont {Heinz}}]{Song:2010aq}%
  \BibitemOpen
  \bibfield  {author} {\bibinfo {author} {\bibfnamefont {H.}~\bibnamefont {Song}}, \bibinfo {author} {\bibfnamefont {S.~A.}\ \bibnamefont {Bass}},\ and\ \bibinfo {author} {\bibfnamefont {U.}~\bibnamefont {Heinz}},\ }\bibfield  {title} {\bibinfo {title} {{Viscous QCD matter in a hybrid hydrodynamic+Boltzmann approach}},\ }\href {https://doi.org/10.1103/PhysRevC.83.024912} {\bibfield  {journal} {\bibinfo  {journal} {Phys. Rev. C}\ }\textbf {\bibinfo {volume} {83}},\ \bibinfo {pages} {024912} (\bibinfo {year} {2011})},\ \Eprint {https://arxiv.org/abs/1012.0555} {arXiv:1012.0555 [nucl-th]} \BibitemShut {NoStop}%
\bibitem [{\citenamefont {Bloczynski}\ \emph {et~al.}(2013)\citenamefont {Bloczynski}, \citenamefont {Huang}, \citenamefont {Zhang},\ and\ \citenamefont {Liao}}]{Bloczynski:2012en}%
  \BibitemOpen
  \bibfield  {author} {\bibinfo {author} {\bibfnamefont {J.}~\bibnamefont {Bloczynski}}, \bibinfo {author} {\bibfnamefont {X.-G.}\ \bibnamefont {Huang}}, \bibinfo {author} {\bibfnamefont {X.}~\bibnamefont {Zhang}},\ and\ \bibinfo {author} {\bibfnamefont {J.}~\bibnamefont {Liao}},\ }\bibfield  {title} {\bibinfo {title} {{Azimuthally fluctuating magnetic field and its impacts on observables in heavy-ion collisions}},\ }\href {https://doi.org/10.1016/j.physletb.2012.12.030} {\bibfield  {journal} {\bibinfo  {journal} {Phys.Lett.}\ }\textbf {\bibinfo {volume} {B718}},\ \bibinfo {pages} {1529} (\bibinfo {year} {2013})},\ \Eprint {https://arxiv.org/abs/1209.6594} {arXiv:1209.6594 [nucl-th]} \BibitemShut {NoStop}%
\bibitem [{\citenamefont {Lin}\ \emph {et~al.}(2005)\citenamefont {Lin}, \citenamefont {Ko}, \citenamefont {Li}, \citenamefont {Zhang},\ and\ \citenamefont {Pal}}]{Lin:2004en}%
  \BibitemOpen
  \bibfield  {author} {\bibinfo {author} {\bibfnamefont {Z.-W.}\ \bibnamefont {Lin}}, \bibinfo {author} {\bibfnamefont {C.~M.}\ \bibnamefont {Ko}}, \bibinfo {author} {\bibfnamefont {B.-A.}\ \bibnamefont {Li}}, \bibinfo {author} {\bibfnamefont {B.}~\bibnamefont {Zhang}},\ and\ \bibinfo {author} {\bibfnamefont {S.}~\bibnamefont {Pal}},\ }\bibfield  {title} {\bibinfo {title} {{A Multi-phase transport model for relativistic heavy ion collisions}},\ }\href {https://doi.org/10.1103/PhysRevC.72.064901} {\bibfield  {journal} {\bibinfo  {journal} {Phys.Rev.}\ }\textbf {\bibinfo {volume} {C72}},\ \bibinfo {pages} {064901} (\bibinfo {year} {2005})},\ \Eprint {https://arxiv.org/abs/nucl-th/0411110} {arXiv:nucl-th/0411110 [nucl-th]} \BibitemShut {NoStop}%
\bibitem [{\citenamefont {Gyulassy}\ and\ \citenamefont {Wang}(1994)}]{Gyulassy:1994ew}%
  \BibitemOpen
  \bibfield  {author} {\bibinfo {author} {\bibfnamefont {M.}~\bibnamefont {Gyulassy}}\ and\ \bibinfo {author} {\bibfnamefont {X.-N.}\ \bibnamefont {Wang}},\ }\bibfield  {title} {\bibinfo {title} {{HIJING 1.0: A Monte Carlo program for parton and particle production in high-energy hadronic and nuclear collisions}},\ }\href {https://doi.org/10.1016/0010-4655(94)90057-4} {\bibfield  {journal} {\bibinfo  {journal} {Comput.Phys.Commun.}\ }\textbf {\bibinfo {volume} {83}},\ \bibinfo {pages} {307} (\bibinfo {year} {1994})},\ \Eprint {https://arxiv.org/abs/nucl-th/9502021} {arXiv:nucl-th/9502021 [nucl-th]} \BibitemShut {NoStop}%
\bibitem [{\citenamefont {Lin}\ and\ \citenamefont {Ko}(2002)}]{Lin:2001zk}%
  \BibitemOpen
  \bibfield  {author} {\bibinfo {author} {\bibfnamefont {Z.-W.}\ \bibnamefont {Lin}}\ and\ \bibinfo {author} {\bibfnamefont {C.}~\bibnamefont {Ko}},\ }\bibfield  {title} {\bibinfo {title} {{Partonic effects on the elliptic flow at RHIC}},\ }\href {https://doi.org/10.1103/PhysRevC.65.034904} {\bibfield  {journal} {\bibinfo  {journal} {Phys.Rev.}\ }\textbf {\bibinfo {volume} {C65}},\ \bibinfo {pages} {034904} (\bibinfo {year} {2002})},\ \Eprint {https://arxiv.org/abs/nucl-th/0108039} {arXiv:nucl-th/0108039 [nucl-th]} \BibitemShut {NoStop}%
\bibitem [{\citenamefont {Zhang}(1998)}]{Zhang:1997ej}%
  \BibitemOpen
  \bibfield  {author} {\bibinfo {author} {\bibfnamefont {B.}~\bibnamefont {Zhang}},\ }\bibfield  {title} {\bibinfo {title} {{ZPC 1.0.1: A Parton cascade for ultrarelativistic heavy ion collisions}},\ }\href {https://doi.org/10.1016/S0010-4655(98)00010-1} {\bibfield  {journal} {\bibinfo  {journal} {Comput.Phys.Commun.}\ }\textbf {\bibinfo {volume} {109}},\ \bibinfo {pages} {193} (\bibinfo {year} {1998})},\ \Eprint {https://arxiv.org/abs/nucl-th/9709009} {arXiv:nucl-th/9709009 [nucl-th]} \BibitemShut {NoStop}%
\bibitem [{\citenamefont {Werner}(2023)}]{Werner:2023zvo}%
  \BibitemOpen
  \bibfield  {author} {\bibinfo {author} {\bibfnamefont {K.}~\bibnamefont {Werner}},\ }\bibfield  {title} {\bibinfo {title} {{Revealing a deep connection between factorization and saturation: New insight into modeling high-energy proton-proton and nucleus-nucleus scattering in the EPOS4 framework}},\ }\href {https://doi.org/10.1103/PhysRevC.108.064903} {\bibfield  {journal} {\bibinfo  {journal} {Phys. Rev. C}\ }\textbf {\bibinfo {volume} {108}},\ \bibinfo {pages} {064903} (\bibinfo {year} {2023})},\ \Eprint {https://arxiv.org/abs/2301.12517} {arXiv:2301.12517 [hep-ph]} \BibitemShut {NoStop}%
\bibitem [{\citenamefont {Werner}\ and\ \citenamefont {Guiot}(2023)}]{Werner:2023fne}%
  \BibitemOpen
  \bibfield  {author} {\bibinfo {author} {\bibfnamefont {K.}~\bibnamefont {Werner}}\ and\ \bibinfo {author} {\bibfnamefont {B.}~\bibnamefont {Guiot}},\ }\bibfield  {title} {\bibinfo {title} {{Perturbative QCD concerning light and heavy flavor in the EPOS4 framework}},\ }\href {https://doi.org/10.1103/PhysRevC.108.034904} {\bibfield  {journal} {\bibinfo  {journal} {Phys. Rev. C}\ }\textbf {\bibinfo {volume} {108}},\ \bibinfo {pages} {034904} (\bibinfo {year} {2023})},\ \Eprint {https://arxiv.org/abs/2306.02396} {arXiv:2306.02396 [hep-ph]} \BibitemShut {NoStop}%
\bibitem [{\citenamefont {Werner}(2024{\natexlab{a}})}]{Werner:2023mod}%
  \BibitemOpen
  \bibfield  {author} {\bibinfo {author} {\bibfnamefont {K.}~\bibnamefont {Werner}},\ }\bibfield  {title} {\bibinfo {title} {{Parallel scattering, saturation, and generalized Abramovskii-Gribov-Kancheli (AGK) theorem in the EPOS4 framework, with applications for heavy-ion collisions at sNN of 5.02 TeV and 200 GeV}},\ }\href {https://doi.org/10.1103/PhysRevC.109.034918} {\bibfield  {journal} {\bibinfo  {journal} {Phys. Rev. C}\ }\textbf {\bibinfo {volume} {109}},\ \bibinfo {pages} {034918} (\bibinfo {year} {2024}{\natexlab{a}})},\ \Eprint {https://arxiv.org/abs/2310.09380} {arXiv:2310.09380 [hep-ph]} \BibitemShut {NoStop}%
\bibitem [{\citenamefont {Werner}(2024{\natexlab{b}})}]{Werner:2023jps}%
  \BibitemOpen
  \bibfield  {author} {\bibinfo {author} {\bibfnamefont {K.}~\bibnamefont {Werner}},\ }\bibfield  {title} {\bibinfo {title} {{Core-corona procedure and microcanonical hadronization to understand strangeness enhancement in proton-proton and heavy ion collisions in the EPOS4 framework}},\ }\href {https://doi.org/10.1103/PhysRevC.109.014910} {\bibfield  {journal} {\bibinfo  {journal} {Phys. Rev. C}\ }\textbf {\bibinfo {volume} {109}},\ \bibinfo {pages} {014910} (\bibinfo {year} {2024}{\natexlab{b}})},\ \Eprint {https://arxiv.org/abs/2306.10277} {arXiv:2306.10277 [hep-ph]} \BibitemShut {NoStop}%
\bibitem [{\citenamefont {Lokhtin}\ \emph {et~al.}(2012)\citenamefont {Lokhtin}, \citenamefont {Belyaev}, \citenamefont {Malinina}, \citenamefont {Petrushanko}, \citenamefont {Rogochaya},\ and\ \citenamefont {Snigirev}}]{Lokhtin:2012re}%
  \BibitemOpen
  \bibfield  {author} {\bibinfo {author} {\bibfnamefont {I.~P.}\ \bibnamefont {Lokhtin}}, \bibinfo {author} {\bibfnamefont {A.~V.}\ \bibnamefont {Belyaev}}, \bibinfo {author} {\bibfnamefont {L.~V.}\ \bibnamefont {Malinina}}, \bibinfo {author} {\bibfnamefont {S.~V.}\ \bibnamefont {Petrushanko}}, \bibinfo {author} {\bibfnamefont {E.~P.}\ \bibnamefont {Rogochaya}},\ and\ \bibinfo {author} {\bibfnamefont {A.~M.}\ \bibnamefont {Snigirev}},\ }\bibfield  {title} {\bibinfo {title} {{Hadron spectra, flow and correlations in PbPb collisions at the LHC: interplay between soft and hard physics}},\ }\href {https://doi.org/10.1140/epjc/s10052-012-2045-7} {\bibfield  {journal} {\bibinfo  {journal} {Eur. Phys. J. C}\ }\textbf {\bibinfo {volume} {72}},\ \bibinfo {pages} {2045} (\bibinfo {year} {2012})},\ \Eprint {https://arxiv.org/abs/1204.4820} {arXiv:1204.4820 [hep-ph]} \BibitemShut {NoStop}%
\bibitem [{\citenamefont {Bravina}\ \emph {et~al.}(2014)\citenamefont {Bravina}, \citenamefont {Brusheim~Johansson}, \citenamefont {Eyyubova}, \citenamefont {Korotkikh}, \citenamefont {Lokhtin}, \citenamefont {Malinina}, \citenamefont {Petrushanko}, \citenamefont {Snigirev},\ and\ \citenamefont {Zabrodin}}]{Bravina:2013xla}%
  \BibitemOpen
  \bibfield  {author} {\bibinfo {author} {\bibfnamefont {L.~V.}\ \bibnamefont {Bravina}}, \bibinfo {author} {\bibfnamefont {B.~H.}\ \bibnamefont {Brusheim~Johansson}}, \bibinfo {author} {\bibfnamefont {G.~K.}\ \bibnamefont {Eyyubova}}, \bibinfo {author} {\bibfnamefont {V.~L.}\ \bibnamefont {Korotkikh}}, \bibinfo {author} {\bibfnamefont {I.~P.}\ \bibnamefont {Lokhtin}}, \bibinfo {author} {\bibfnamefont {L.~V.}\ \bibnamefont {Malinina}}, \bibinfo {author} {\bibfnamefont {S.~V.}\ \bibnamefont {Petrushanko}}, \bibinfo {author} {\bibfnamefont {A.~M.}\ \bibnamefont {Snigirev}},\ and\ \bibinfo {author} {\bibfnamefont {E.~E.}\ \bibnamefont {Zabrodin}},\ }\bibfield  {title} {\bibinfo {title} {{Higher harmonics of azimuthal anisotropy in relativistic heavy ion collisions in HYDJET++ model}},\ }\href {https://doi.org/10.1140/epjc/s10052-014-2807-5} {\bibfield  {journal} {\bibinfo  {journal} {Eur. Phys. J. C}\ }\textbf {\bibinfo {volume} {74}},\ \bibinfo {pages} {2807} (\bibinfo {year} {2014})},\ \Eprint
  {https://arxiv.org/abs/1311.7054} {arXiv:1311.7054 [nucl-th]} \BibitemShut {NoStop}%
\bibitem [{\citenamefont {Sjostrand}\ \emph {et~al.}(2001)\citenamefont {Sjostrand}, \citenamefont {Eden}, \citenamefont {Friberg}, \citenamefont {Lonnblad}, \citenamefont {Miu} \emph {et~al.}}]{Sjostrand:2000wi}%
  \BibitemOpen
  \bibfield  {author} {\bibinfo {author} {\bibfnamefont {T.}~\bibnamefont {Sjostrand}}, \bibinfo {author} {\bibfnamefont {P.}~\bibnamefont {Eden}}, \bibinfo {author} {\bibfnamefont {C.}~\bibnamefont {Friberg}}, \bibinfo {author} {\bibfnamefont {L.}~\bibnamefont {Lonnblad}}, \bibinfo {author} {\bibfnamefont {G.}~\bibnamefont {Miu}}, \emph {et~al.},\ }\bibfield  {title} {\bibinfo {title} {{High-energy physics event generation with PYTHIA 6.1}},\ }\href {https://doi.org/10.1016/S0010-4655(00)00236-8} {\bibfield  {journal} {\bibinfo  {journal} {Comput.Phys.Commun.}\ }\textbf {\bibinfo {volume} {135}},\ \bibinfo {pages} {238} (\bibinfo {year} {2001})},\ \Eprint {https://arxiv.org/abs/hep-ph/0010017} {arXiv:hep-ph/0010017 [hep-ph]} \BibitemShut {NoStop}%
\bibitem [{\citenamefont {Olive}\ \emph {et~al.}(2014)\citenamefont {Olive} \emph {et~al.}}]{ParticleDataGroup:2014cgo}%
  \BibitemOpen
  \bibfield  {author} {\bibinfo {author} {\bibfnamefont {K.~A.}\ \bibnamefont {Olive}} \emph {et~al.} (\bibinfo {collaboration} {Particle Data Group}),\ }\bibfield  {title} {\bibinfo {title} {{Review of Particle Physics}},\ }\href {https://doi.org/10.1088/1674-1137/38/9/090001} {\bibfield  {journal} {\bibinfo  {journal} {Chin. Phys. C}\ }\textbf {\bibinfo {volume} {38}},\ \bibinfo {pages} {090001} (\bibinfo {year} {2014})}\BibitemShut {NoStop}%
\bibitem [{\citenamefont {Adams}\ \emph {et~al.}(2004{\natexlab{a}})\citenamefont {Adams} \emph {et~al.}}]{STAR:2003vqj}%
  \BibitemOpen
  \bibfield  {author} {\bibinfo {author} {\bibfnamefont {J.}~\bibnamefont {Adams}} \emph {et~al.} (\bibinfo {collaboration} {STAR}),\ }\bibfield  {title} {\bibinfo {title} {{Rho0 production and possible modification in Au+Au and p+p collisions at S(NN)**1/2 = 200-GeV}},\ }\href {https://doi.org/10.1103/PhysRevLett.92.092301} {\bibfield  {journal} {\bibinfo  {journal} {Phys. Rev. Lett.}\ }\textbf {\bibinfo {volume} {92}},\ \bibinfo {pages} {092301} (\bibinfo {year} {2004}{\natexlab{a}})},\ \Eprint {https://arxiv.org/abs/nucl-ex/0307023} {arXiv:nucl-ex/0307023} \BibitemShut {NoStop}%
\bibitem [{\citenamefont {Adamczyk}\ \emph {et~al.}(2015)\citenamefont {Adamczyk} \emph {et~al.}}]{STAR:2015tnn}%
  \BibitemOpen
  \bibfield  {author} {\bibinfo {author} {\bibfnamefont {L.}~\bibnamefont {Adamczyk}} \emph {et~al.} (\bibinfo {collaboration} {STAR}),\ }\bibfield  {title} {\bibinfo {title} {{Measurements of Dielectron Production in Au$+$Au Collisions at $\sqrt{s_{\rm NN}}$ = 200 GeV from the STAR Experiment}},\ }\href {https://doi.org/10.1103/PhysRevC.92.024912} {\bibfield  {journal} {\bibinfo  {journal} {Phys. Rev. C}\ }\textbf {\bibinfo {volume} {92}},\ \bibinfo {pages} {024912} (\bibinfo {year} {2015})},\ \Eprint {https://arxiv.org/abs/1504.01317} {arXiv:1504.01317 [hep-ex]} \BibitemShut {NoStop}%
\bibitem [{\citenamefont {Adler}\ \emph {et~al.}(2003{\natexlab{a}})\citenamefont {Adler} \emph {et~al.}}]{PHENIX:2003qdj}%
  \BibitemOpen
  \bibfield  {author} {\bibinfo {author} {\bibfnamefont {S.~S.}\ \bibnamefont {Adler}} \emph {et~al.} (\bibinfo {collaboration} {PHENIX}),\ }\bibfield  {title} {\bibinfo {title} {{Suppressed $\pi^0$ production at large transverse momentum in central Au+ Au collisions at $\sqrt{S_{NN}}$ = 200 GeV}},\ }\href {https://doi.org/10.1103/PhysRevLett.91.072301} {\bibfield  {journal} {\bibinfo  {journal} {Phys. Rev. Lett.}\ }\textbf {\bibinfo {volume} {91}},\ \bibinfo {pages} {072301} (\bibinfo {year} {2003}{\natexlab{a}})},\ \Eprint {https://arxiv.org/abs/nucl-ex/0304022} {arXiv:nucl-ex/0304022} \BibitemShut {NoStop}%
\bibitem [{\citenamefont {Adams}\ \emph {et~al.}(2004{\natexlab{b}})\citenamefont {Adams} \emph {et~al.}}]{STAR:2003jwm}%
  \BibitemOpen
  \bibfield  {author} {\bibinfo {author} {\bibfnamefont {J.}~\bibnamefont {Adams}} \emph {et~al.} (\bibinfo {collaboration} {STAR}),\ }\bibfield  {title} {\bibinfo {title} {{Identified particle distributions in pp and Au+Au collisions at s(NN)**(1/2) = 200 GeV}},\ }\href {https://doi.org/10.1103/PhysRevLett.92.112301} {\bibfield  {journal} {\bibinfo  {journal} {Phys. Rev. Lett.}\ }\textbf {\bibinfo {volume} {92}},\ \bibinfo {pages} {112301} (\bibinfo {year} {2004}{\natexlab{b}})},\ \Eprint {https://arxiv.org/abs/nucl-ex/0310004} {arXiv:nucl-ex/0310004} \BibitemShut {NoStop}%
\bibitem [{\citenamefont {Dong}\ \emph {et~al.}(2004)\citenamefont {Dong}, \citenamefont {Esumi}, \citenamefont {Sorensen}, \citenamefont {Xu},\ and\ \citenamefont {Xu}}]{Dong:2004ve}%
  \BibitemOpen
  \bibfield  {author} {\bibinfo {author} {\bibfnamefont {X.}~\bibnamefont {Dong}}, \bibinfo {author} {\bibfnamefont {S.}~\bibnamefont {Esumi}}, \bibinfo {author} {\bibfnamefont {P.}~\bibnamefont {Sorensen}}, \bibinfo {author} {\bibfnamefont {N.}~\bibnamefont {Xu}},\ and\ \bibinfo {author} {\bibfnamefont {Z.}~\bibnamefont {Xu}},\ }\bibfield  {title} {\bibinfo {title} {{Resonance decay effects on anisotropy parameters}},\ }\href {https://doi.org/10.1016/j.physletb.2004.06.110} {\bibfield  {journal} {\bibinfo  {journal} {Phys. Lett. B}\ }\textbf {\bibinfo {volume} {597}},\ \bibinfo {pages} {328} (\bibinfo {year} {2004})},\ \Eprint {https://arxiv.org/abs/nucl-th/0403030} {arXiv:nucl-th/0403030} \BibitemShut {NoStop}%
\bibitem [{\citenamefont {Ackermann}\ \emph {et~al.}(2003)\citenamefont {Ackermann} \emph {et~al.}}]{Ackermann:2002ad}%
  \BibitemOpen
  \bibfield  {author} {\bibinfo {author} {\bibfnamefont {K.}~\bibnamefont {Ackermann}} \emph {et~al.} (\bibinfo {collaboration} {STAR Collaboration}),\ }\bibfield  {title} {\bibinfo {title} {{STAR detector overview}},\ }\href {https://doi.org/10.1016/S0168-9002(02)01960-5} {\bibfield  {journal} {\bibinfo  {journal} {Nucl.Instrum.Meth.}\ }\textbf {\bibinfo {volume} {A499}},\ \bibinfo {pages} {624} (\bibinfo {year} {2003})}\BibitemShut {NoStop}%
\bibitem [{\citenamefont {Li}\ \emph {et~al.}(2024)\citenamefont {Li}, \citenamefont {Feng},\ and\ \citenamefont {Wang}}]{Li:2024pue}%
  \BibitemOpen
  \bibfield  {author} {\bibinfo {author} {\bibfnamefont {H.-S.}\ \bibnamefont {Li}}, \bibinfo {author} {\bibfnamefont {Y.}~\bibnamefont {Feng}},\ and\ \bibinfo {author} {\bibfnamefont {F.}~\bibnamefont {Wang}},\ }\bibfield  {title} {\bibinfo {title} {{Influence of the chiral magnetic effect on particle-pair elliptic anisotropy}},\ }\href@noop {} {\  (\bibinfo {year} {2024})},\ \bibinfo {note} {to appear in Phys.Rev.C.},\ \Eprint {https://arxiv.org/abs/2404.05032} {arXiv:2404.05032 [hep-ph]} \BibitemShut {NoStop}%
\bibitem [{\citenamefont {Adler}\ \emph {et~al.}(2003{\natexlab{b}})\citenamefont {Adler} \emph {et~al.}}]{Adler:2003sp}%
  \BibitemOpen
  \bibfield  {author} {\bibinfo {author} {\bibfnamefont {C.}~\bibnamefont {Adler}} \emph {et~al.},\ }\bibfield  {title} {\bibinfo {title} {{The RHIC zero-degree calorimeters}},\ }\href {https://doi.org/10.1016/j.nima.2003.08.112} {\bibfield  {journal} {\bibinfo  {journal} {Nucl.Instrum.Meth.}\ }\textbf {\bibinfo {volume} {A499}},\ \bibinfo {pages} {433} (\bibinfo {year} {2003}{\natexlab{b}})}\BibitemShut {NoStop}%
\bibitem [{\citenamefont {Milton}\ \emph {et~al.}(2021)\citenamefont {Milton}, \citenamefont {Wang}, \citenamefont {Sergeeva}, \citenamefont {Shi}, \citenamefont {Liao},\ and\ \citenamefont {Huang}}]{Milton:2021wku}%
  \BibitemOpen
  \bibfield  {author} {\bibinfo {author} {\bibfnamefont {R.}~\bibnamefont {Milton}}, \bibinfo {author} {\bibfnamefont {G.}~\bibnamefont {Wang}}, \bibinfo {author} {\bibfnamefont {M.}~\bibnamefont {Sergeeva}}, \bibinfo {author} {\bibfnamefont {S.}~\bibnamefont {Shi}}, \bibinfo {author} {\bibfnamefont {J.}~\bibnamefont {Liao}},\ and\ \bibinfo {author} {\bibfnamefont {H.~Z.}\ \bibnamefont {Huang}},\ }\bibfield  {title} {\bibinfo {title} {{Utilization of event shape in search of the chiral magnetic effect in heavy-ion collisions}},\ }\href {https://doi.org/10.1103/PhysRevC.104.064906} {\bibfield  {journal} {\bibinfo  {journal} {Phys. Rev. C}\ }\textbf {\bibinfo {volume} {104}},\ \bibinfo {pages} {064906} (\bibinfo {year} {2021})},\ \Eprint {https://arxiv.org/abs/2110.01435} {arXiv:2110.01435 [nucl-th]} \BibitemShut {NoStop}%
\end{thebibliography}%
\onecolumngrid
\appendix

\renewcommand\thefigure{\thesection.\arabic{figure}}    
\setcounter{figure}{0}    
\section{Simulation Plots}
\begin{figure*}[hbt]
    \includegraphics[width=0.33\textwidth]{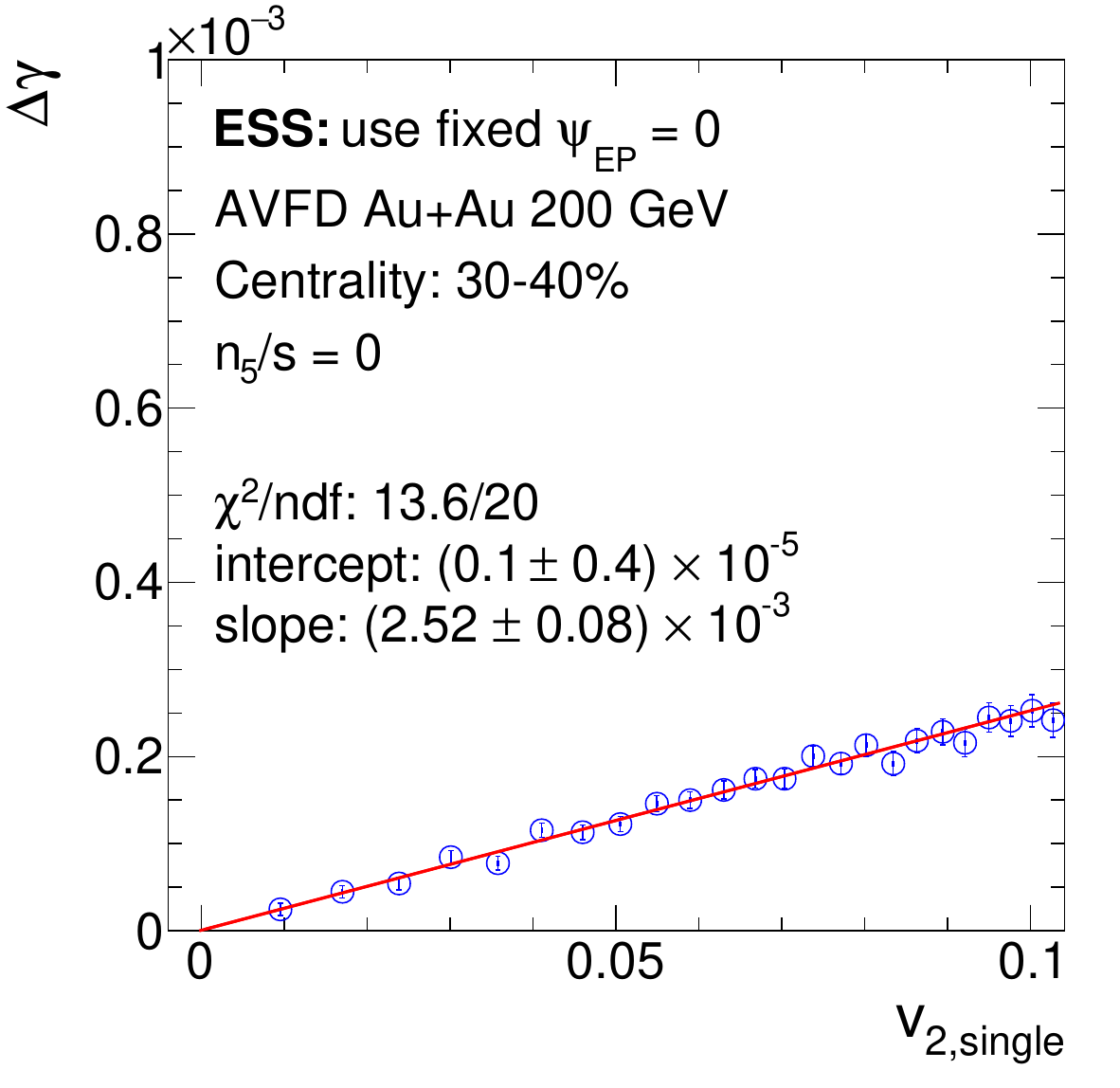}\hfill
    \includegraphics[width=0.33\textwidth]{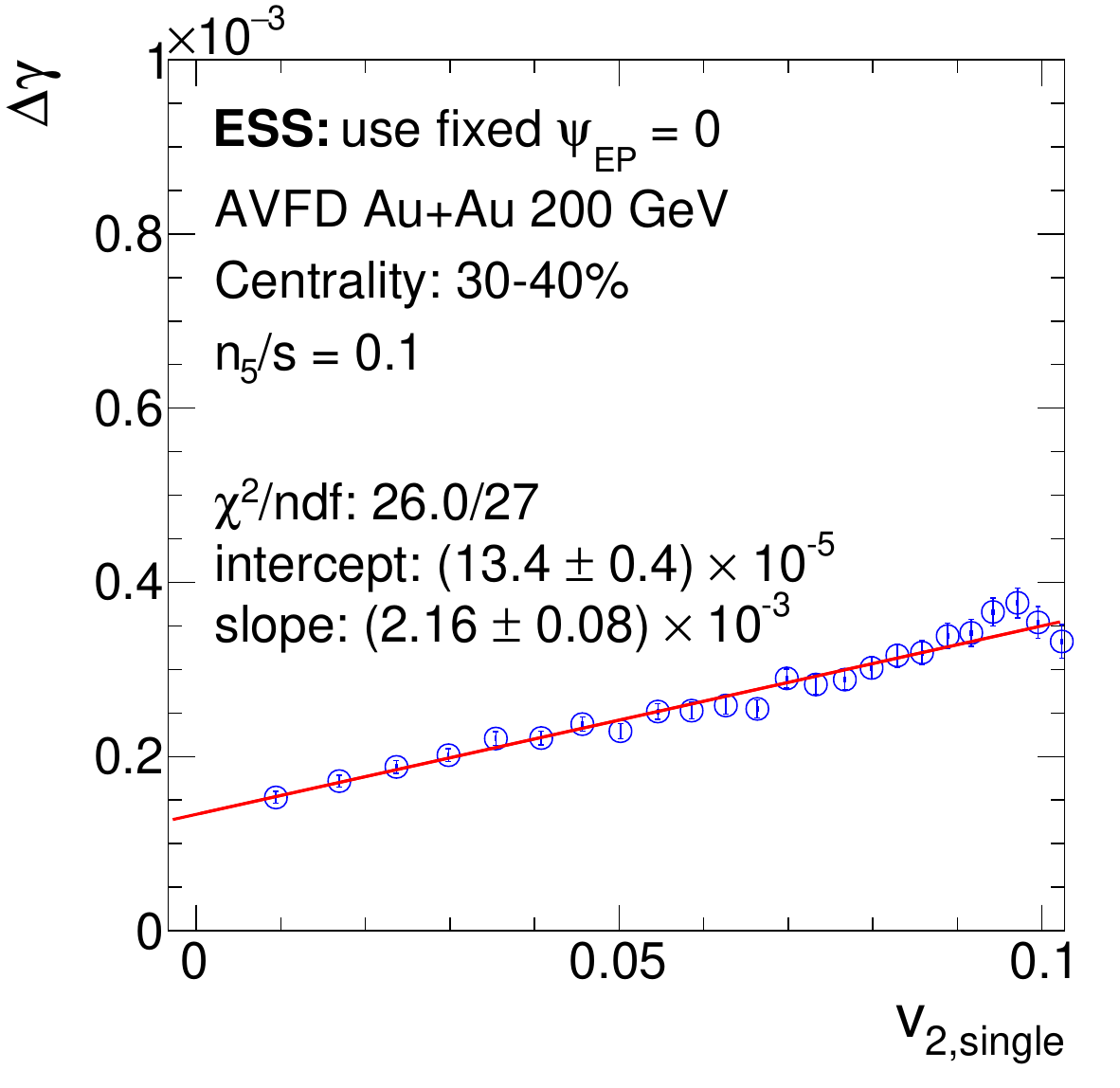}\hfill
    \includegraphics[width=0.33\textwidth]{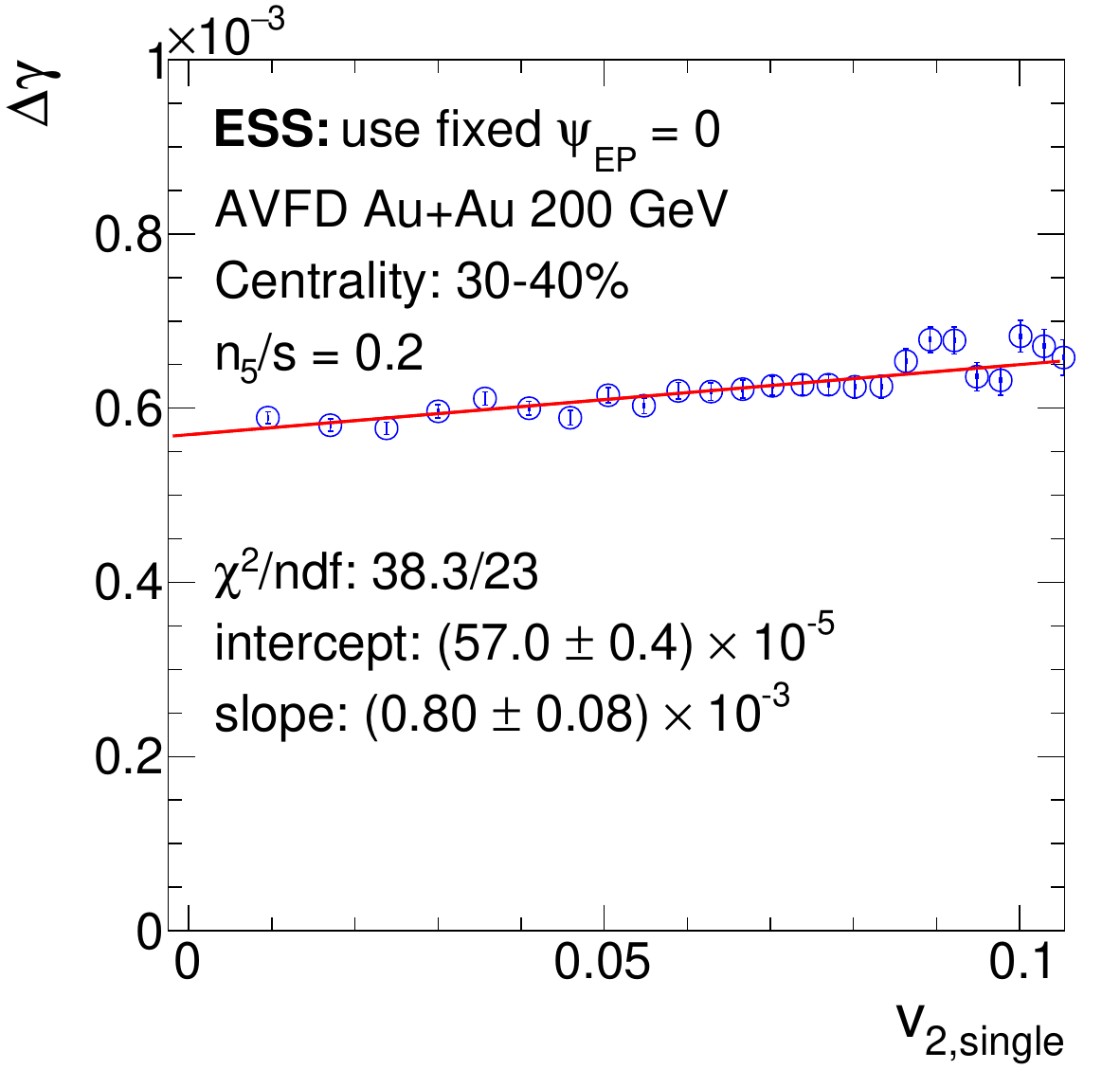}\hfill
    \vspace{-4mm}\\
    \caption{\label{fig:avfd_ess}\avfd\ ESS results. Three values of $\ns=0$ (left panel), 0.1 (center panel), and 0.2 (right panel) are implemented in the \avfd\ simulations of Au+Au collisions at $\snn=200$~GeV, with approximately $2\times10^7$ events each for the 30--40\% centrality. The $\dg$ is plotted as a function of $\vsing$ in events binned in $\qhpair^2\two$ (Eqs.~\ref{eq:q2},\ref{eq:qhpair}). POIs are from acceptance $|\eta|<1$ and $0.2 < \pt < 2$~\gevc, and the event selection variable $\qhpair^2\two$ is computed from the same POIs. 
    The model's known impact parameter direction $\psi=0$ is taken as the EP in calculating $\dg$  (Eqs.~\ref{eq:g},\ref{eq:dg}) and $\vsing$ (Eq.~\ref{eq:v2}). The red line is a first-order polynomial fit in the range of $0<\vsing<0.2$.}
    \includegraphics[width=0.33\textwidth]{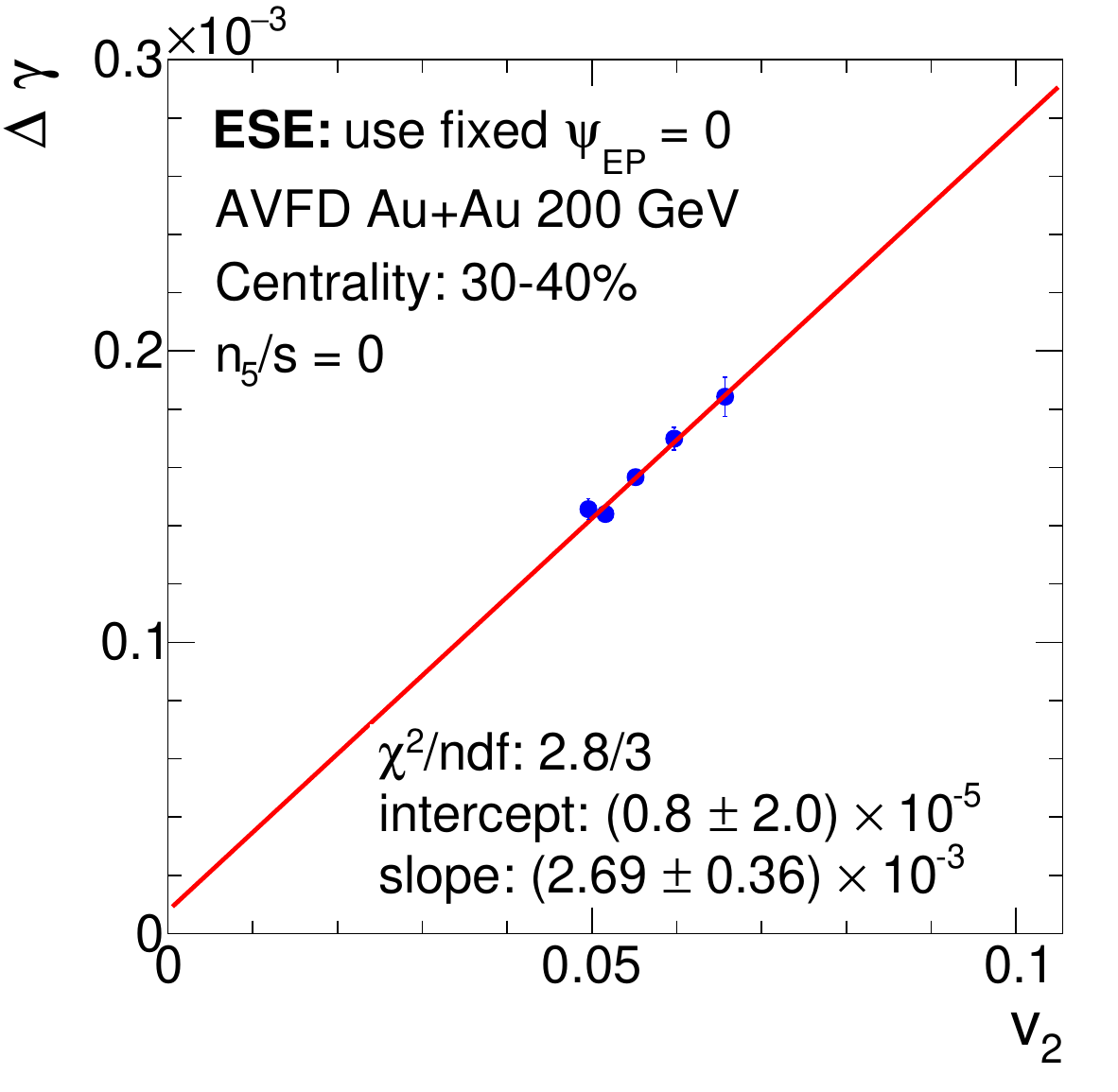}\hfill
    \includegraphics[width=0.33\textwidth]{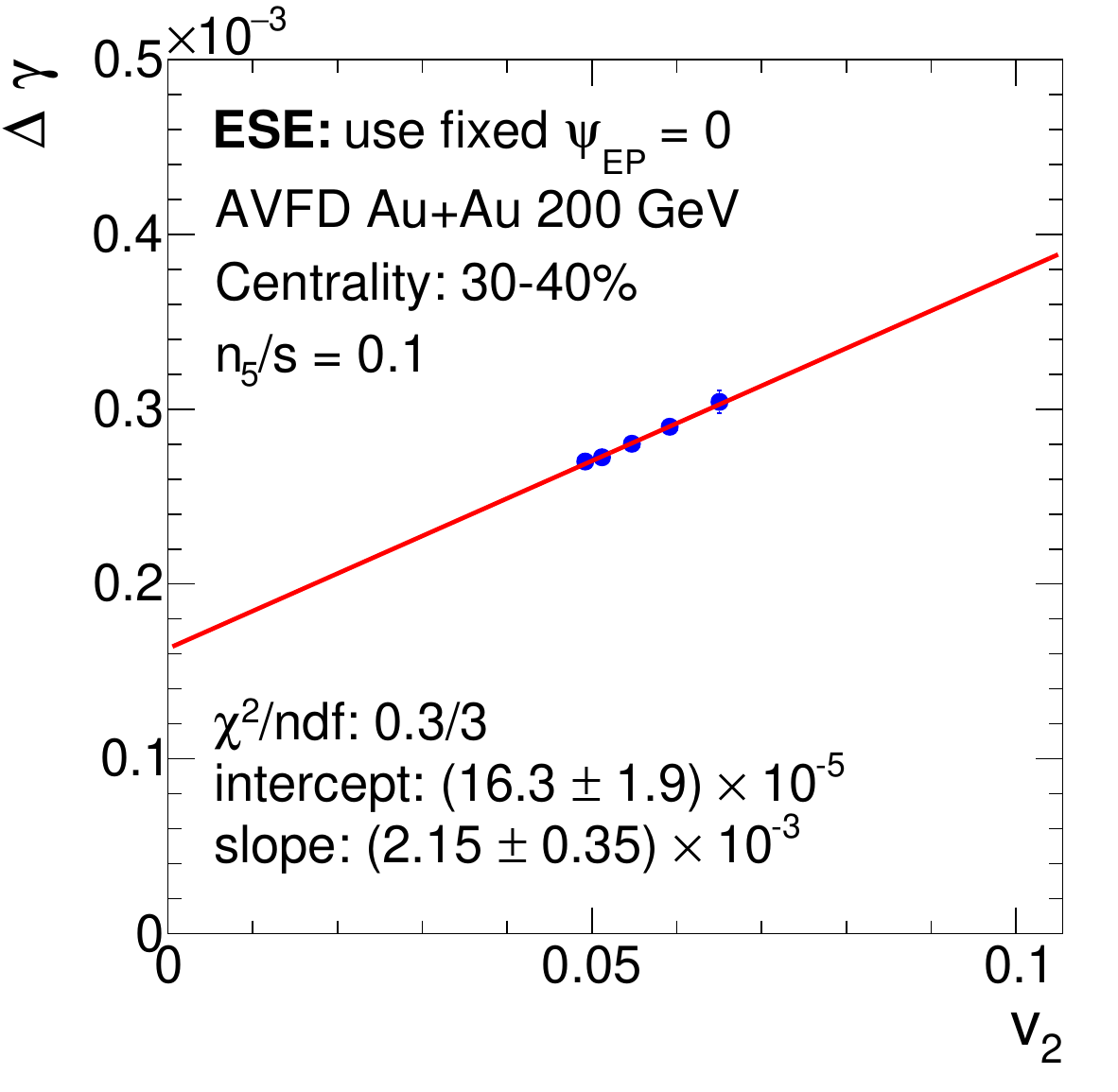}\hfill
    \includegraphics[width=0.33\textwidth]{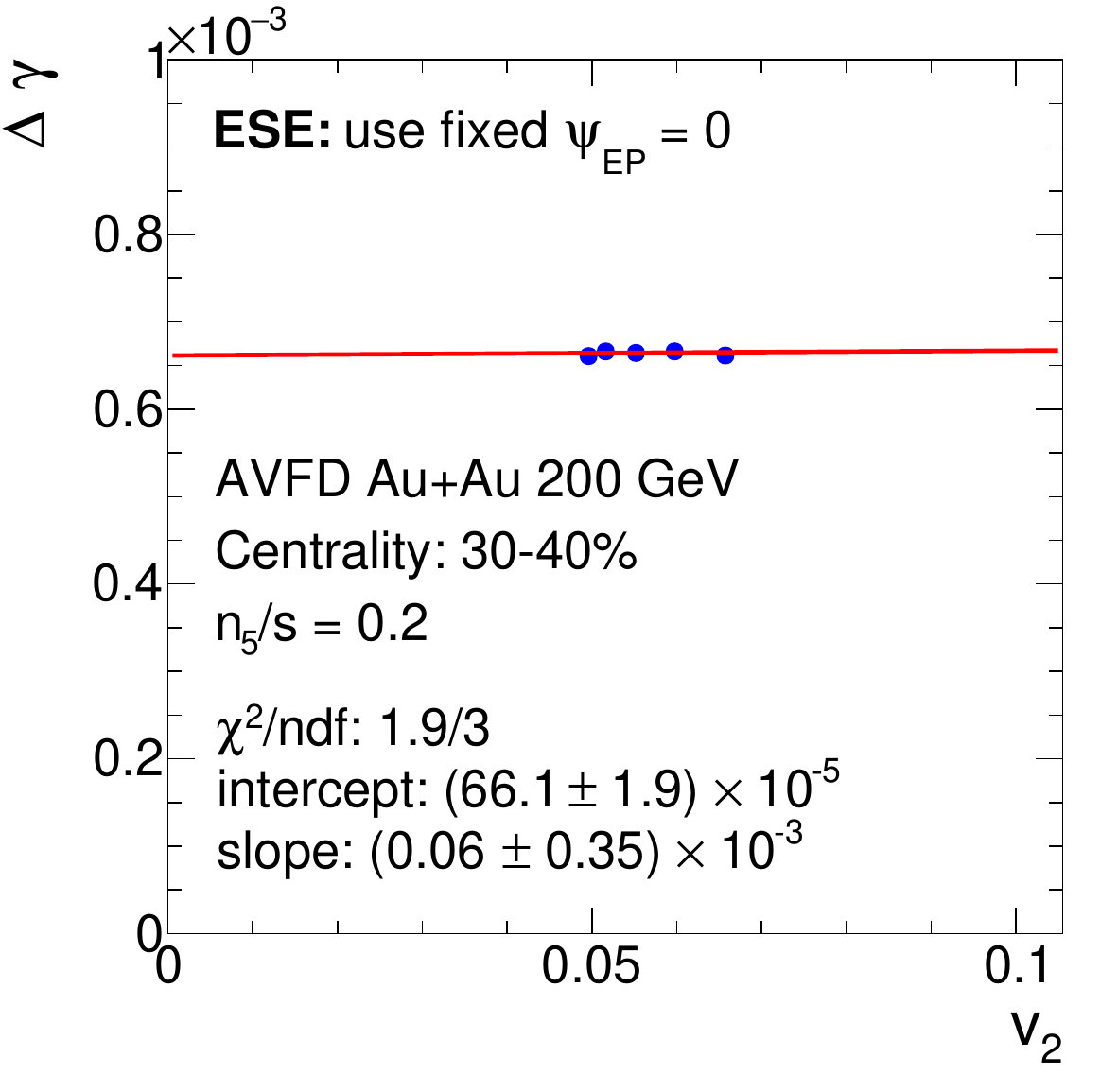}\hfill
    \vspace{-4mm}\\
    \caption{\label{fig:avfd_ese}\avfd\ ESE results. The same \avfd\ data as in Fig.~\ref{fig:avfd_ess} is used. 
    The $\dg$ is plotted as a function of $\mean{v_2}$ in events binned in $\qh^2\two$ (Eqs.~\ref{eq:q2},\ref{eq:qh}). POIs are from acceptance $0.3<|\eta|<2$, and the event selection variable $\qh^2\two$ is computed from particles in $|\eta|<0.3$, both with $0.2<\pt<2$~\gevc. 
    The model's known impact parameter direction $\psi=0$ is taken as the EP in calculating $\dg$ (Eqs.~\ref{eq:g},\ref{eq:dg}) and $\mean{v_2}$ (Eq.~\ref{eq:v2}). The red line is a first-order polynomial fit to all data points.}
\end{figure*}

\begin{figure*}[hbt]
    \includegraphics[width=0.33\textwidth]{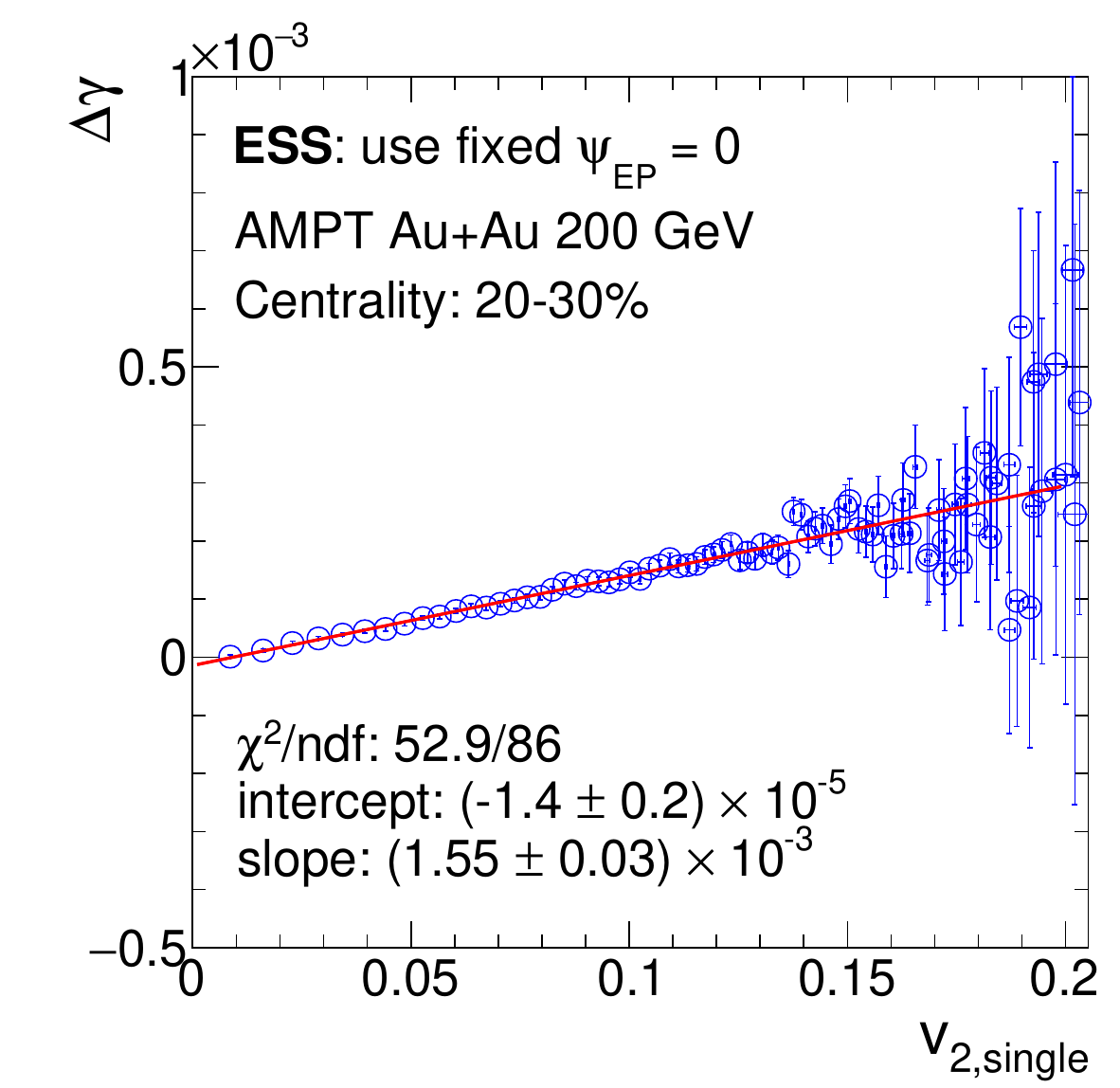}\hfill
    \includegraphics[width=0.33\textwidth]{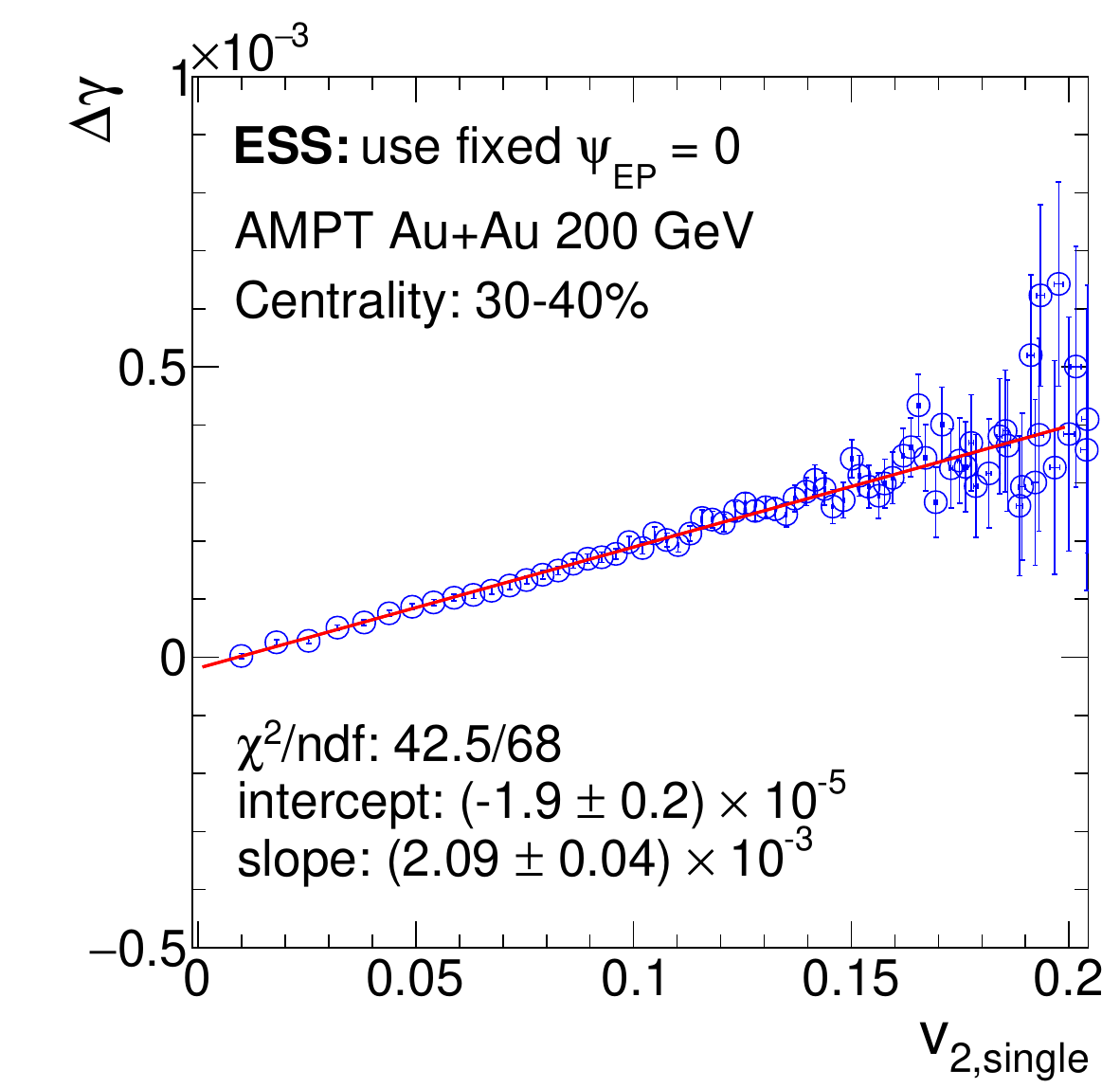}\hfill
    \includegraphics[width=0.33\textwidth]{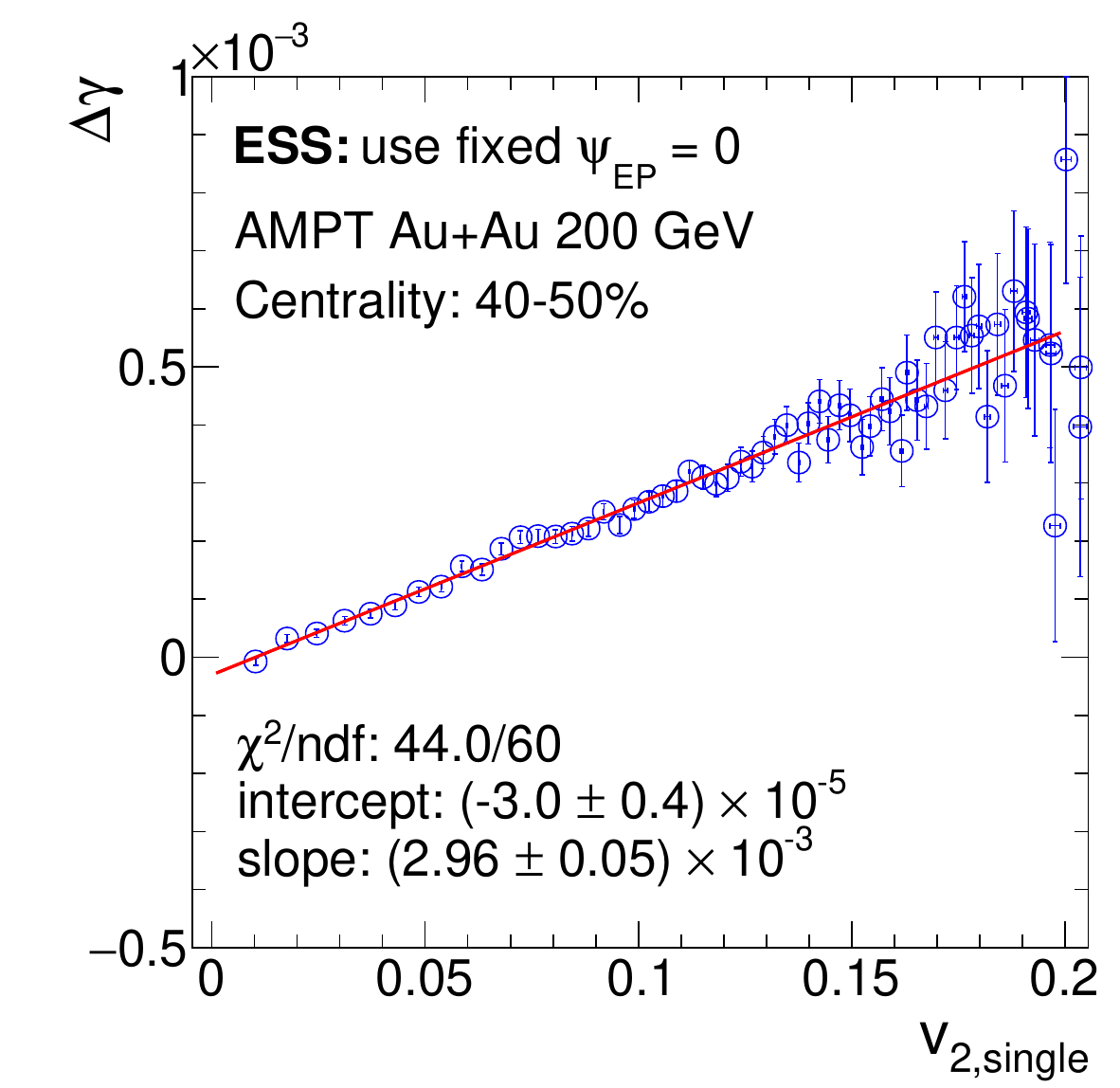}\hfill
    \vspace{-5mm}\\
    \includegraphics[width=0.33\textwidth]{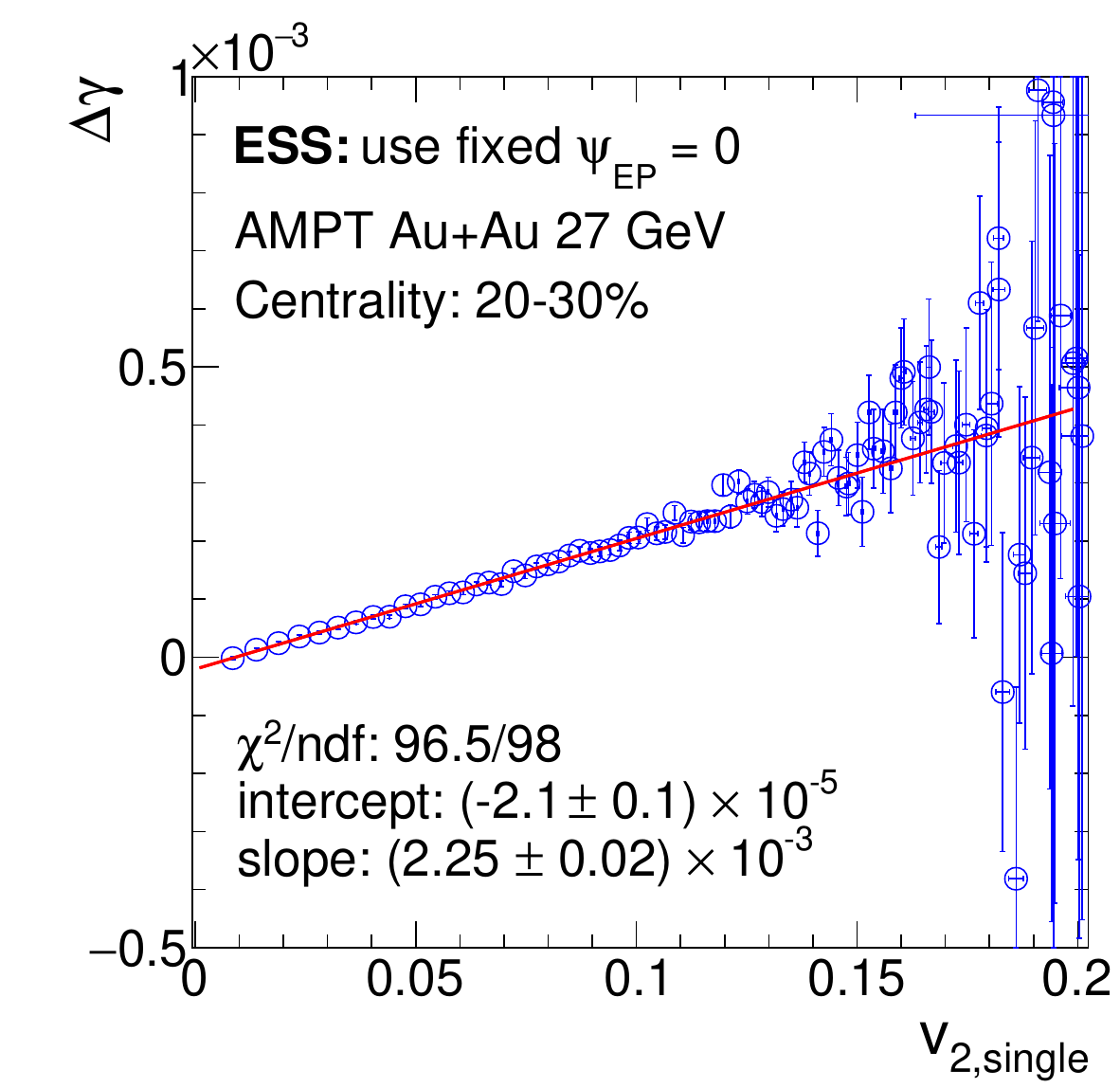}\hfill
    \includegraphics[width=0.33\textwidth]{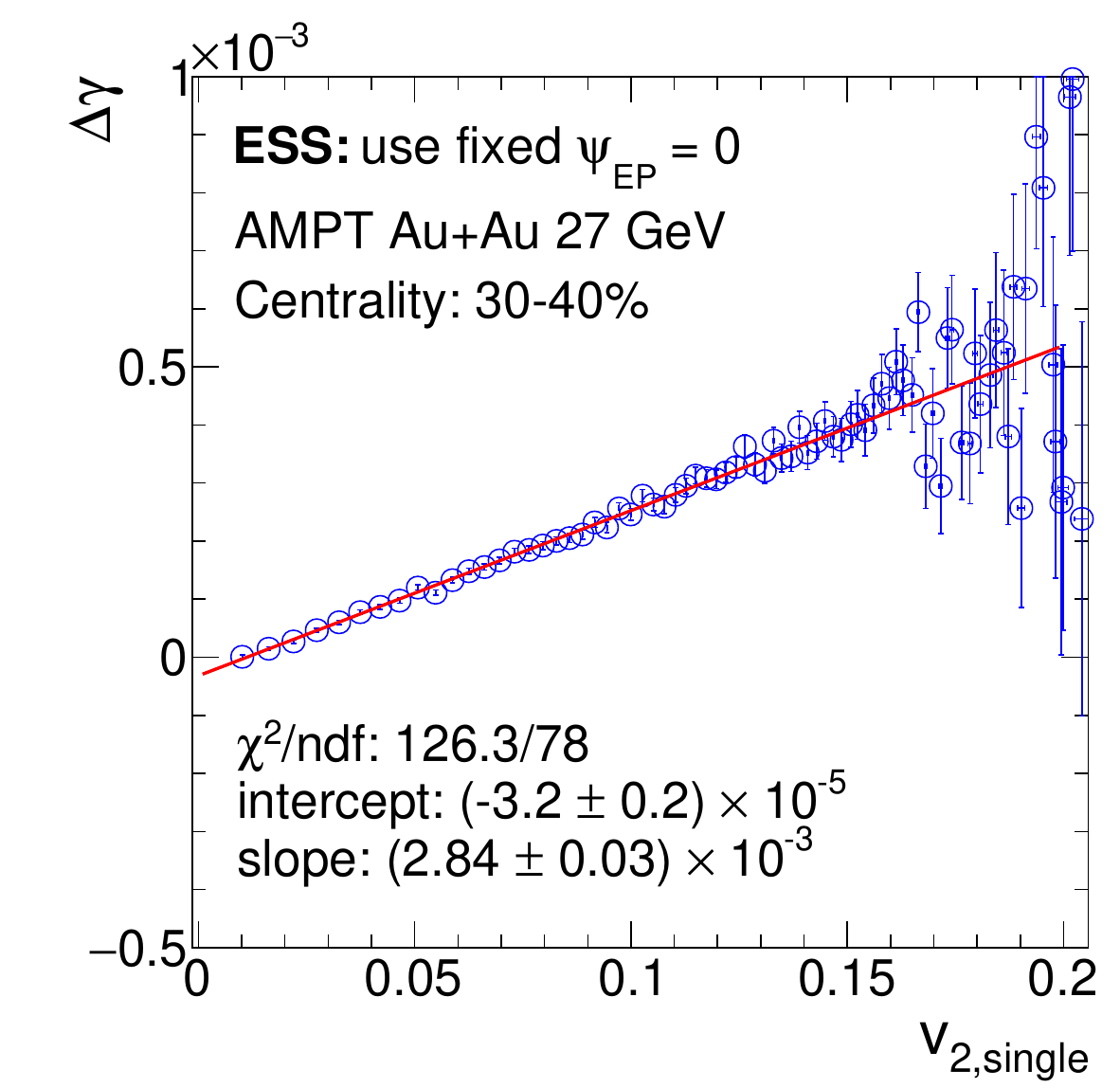}\hfill
    \includegraphics[width=0.33\textwidth]{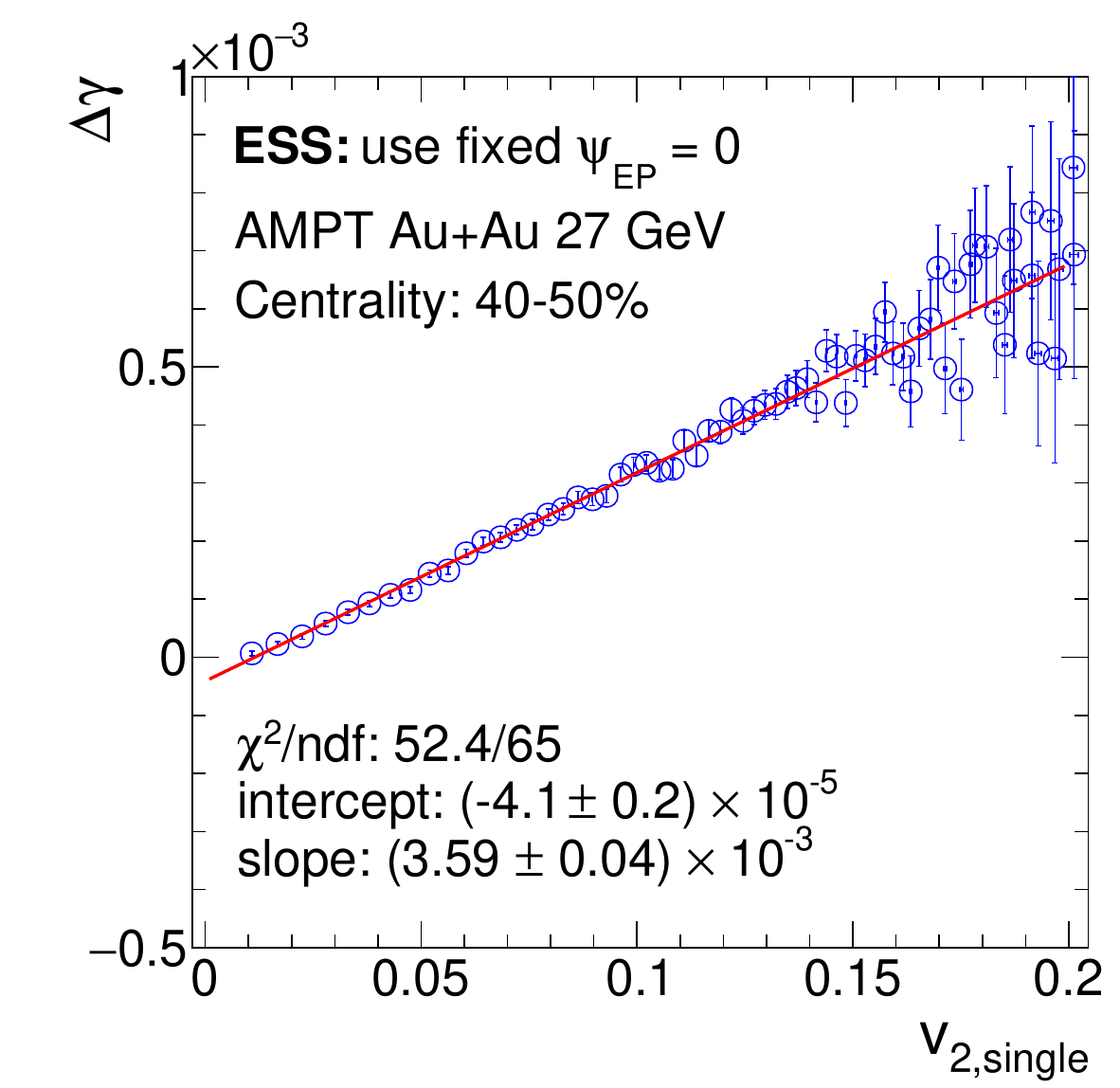}\hfill
    \vspace{-4mm}\\
    \caption{\label{fig:ampt_ess}\ampt\ ESS results. Shown are three centralities of Au+Au collisions at $\snn=200$~GeV (upper panels) and at 27~GeV (lower panels) simulated by \ampt, with approximately $4.2\times10^8$ and $1.8\times10^9$ events for each centrality at each energy. The $\dg$ is plotted as a function of $\vsing$ in events binned in $\qhpair^2\two$ (Eqs.~\ref{eq:q2},\ref{eq:qhpair}). POIs are from acceptance $|\eta|<1$ and $0.2 < \pt < 2$~\gevc, and the event selection variable $\qhpair^2\two$ is computed from the same POIs. The model's known impact parameter direction $\psi=0$ is taken as the EP in calculating $\dg$ (Eqs.~\ref{eq:g},\ref{eq:dg}) and $\vsing$ (Eq.~\ref{eq:v2}). The red line is a first-order polynomial fit in the range of $0<\vsing<0.2$}
\end{figure*}

\begin{figure*}[hbt]
    \includegraphics[width=0.33\textwidth]{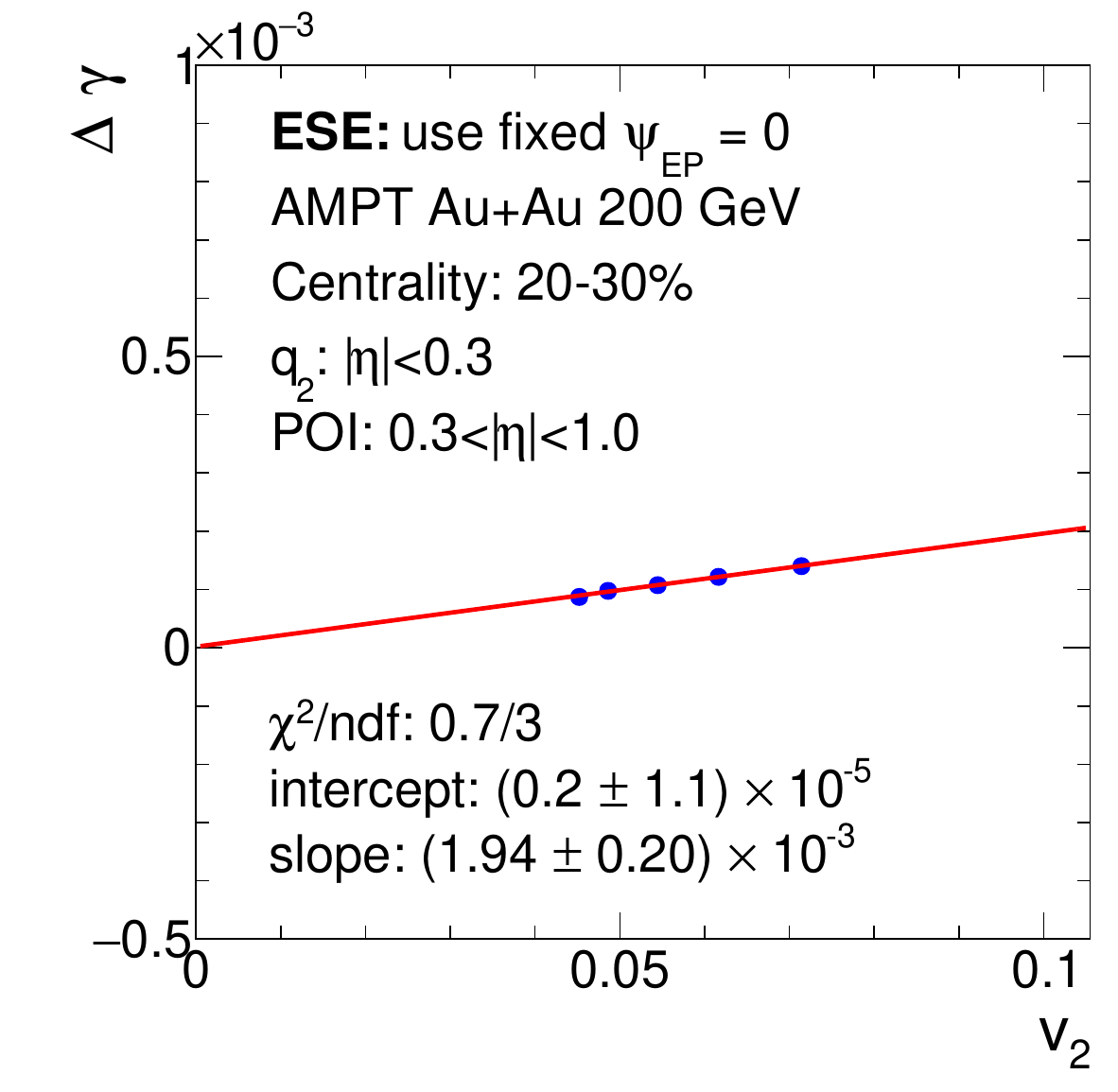}\hfill
    \includegraphics[width=0.33\textwidth]{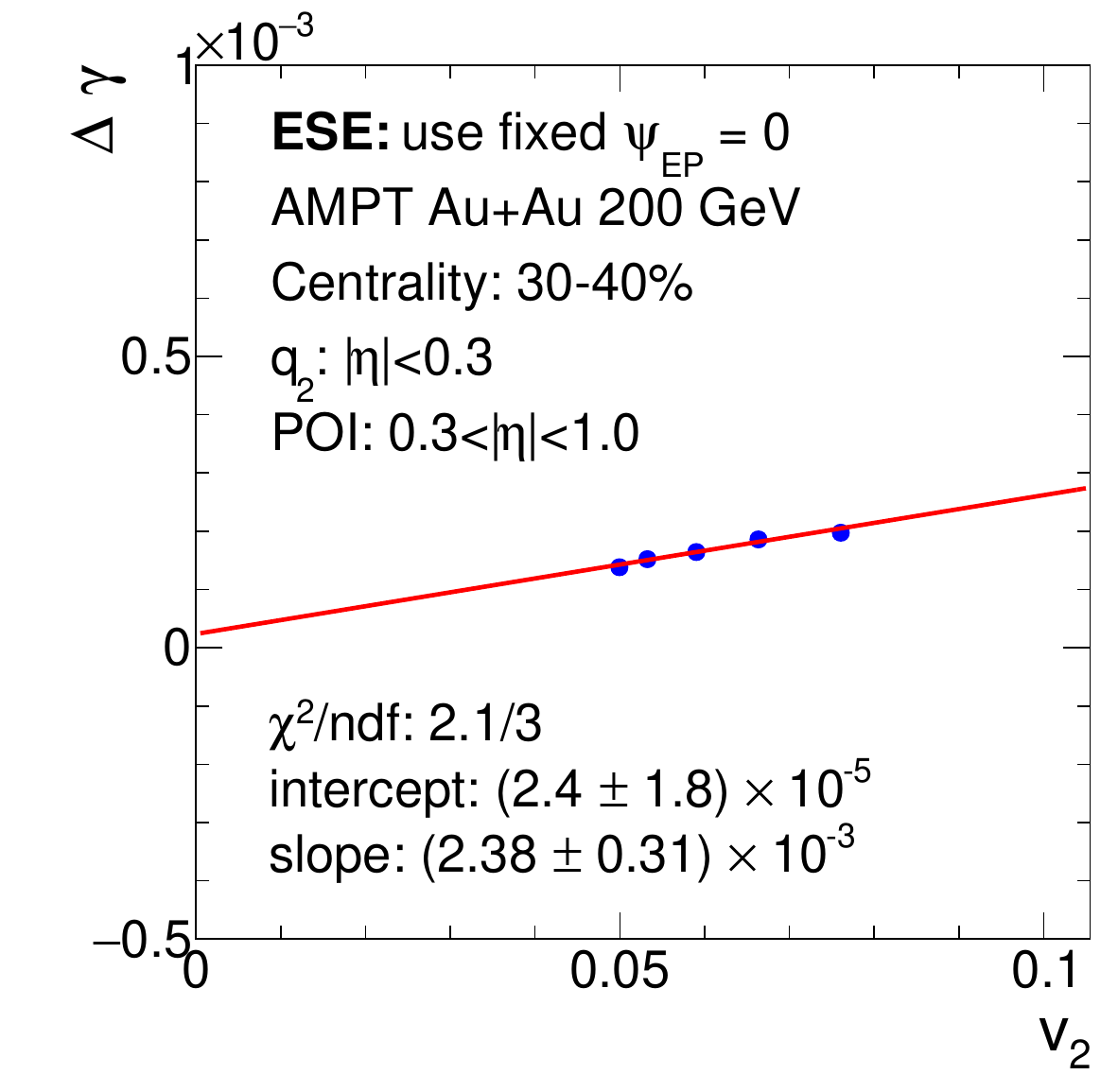}\hfill
    \includegraphics[width=0.33\textwidth]{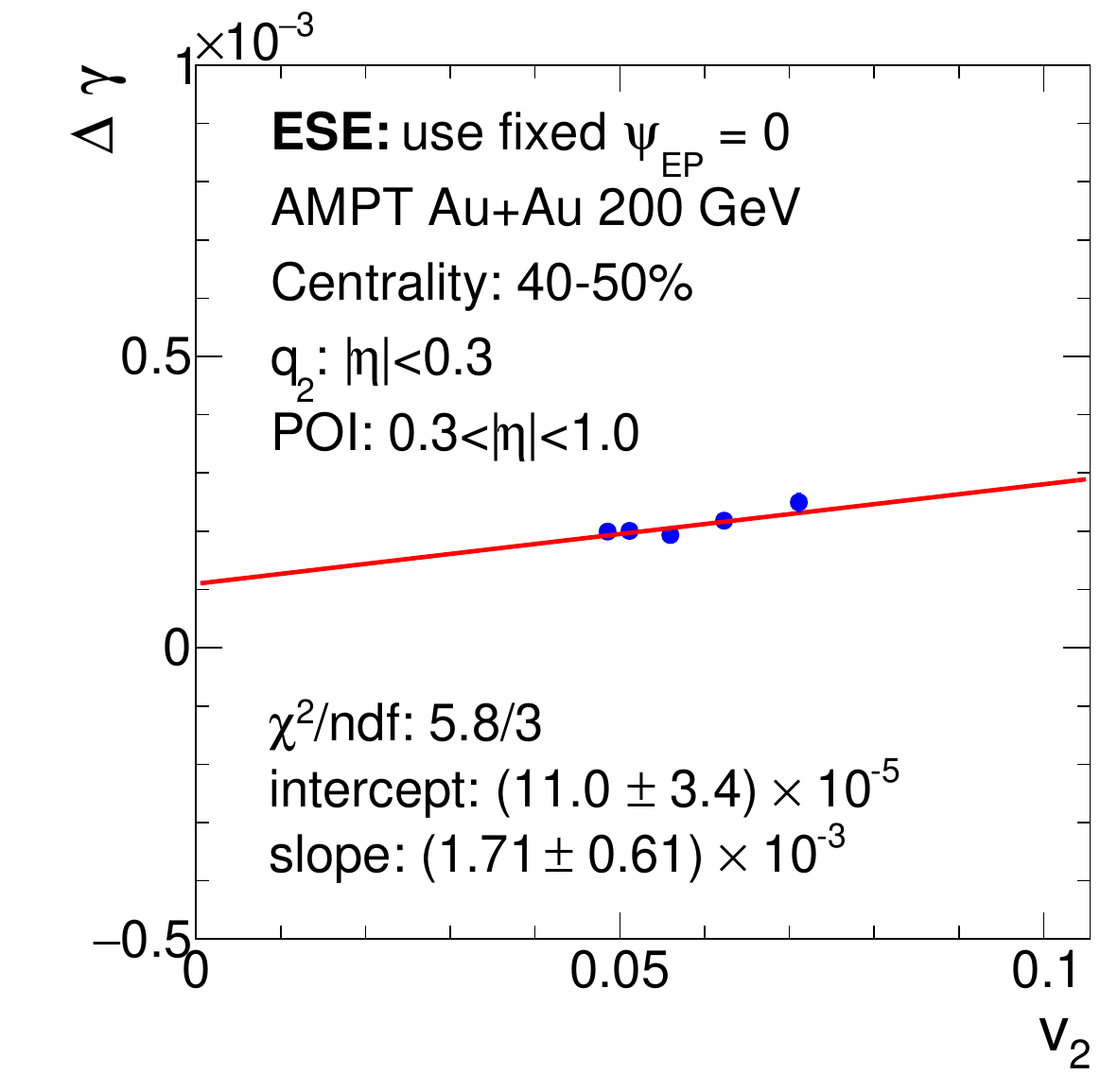}\hfill
    \vspace{-5mm}\\
    \includegraphics[width=0.33\textwidth]{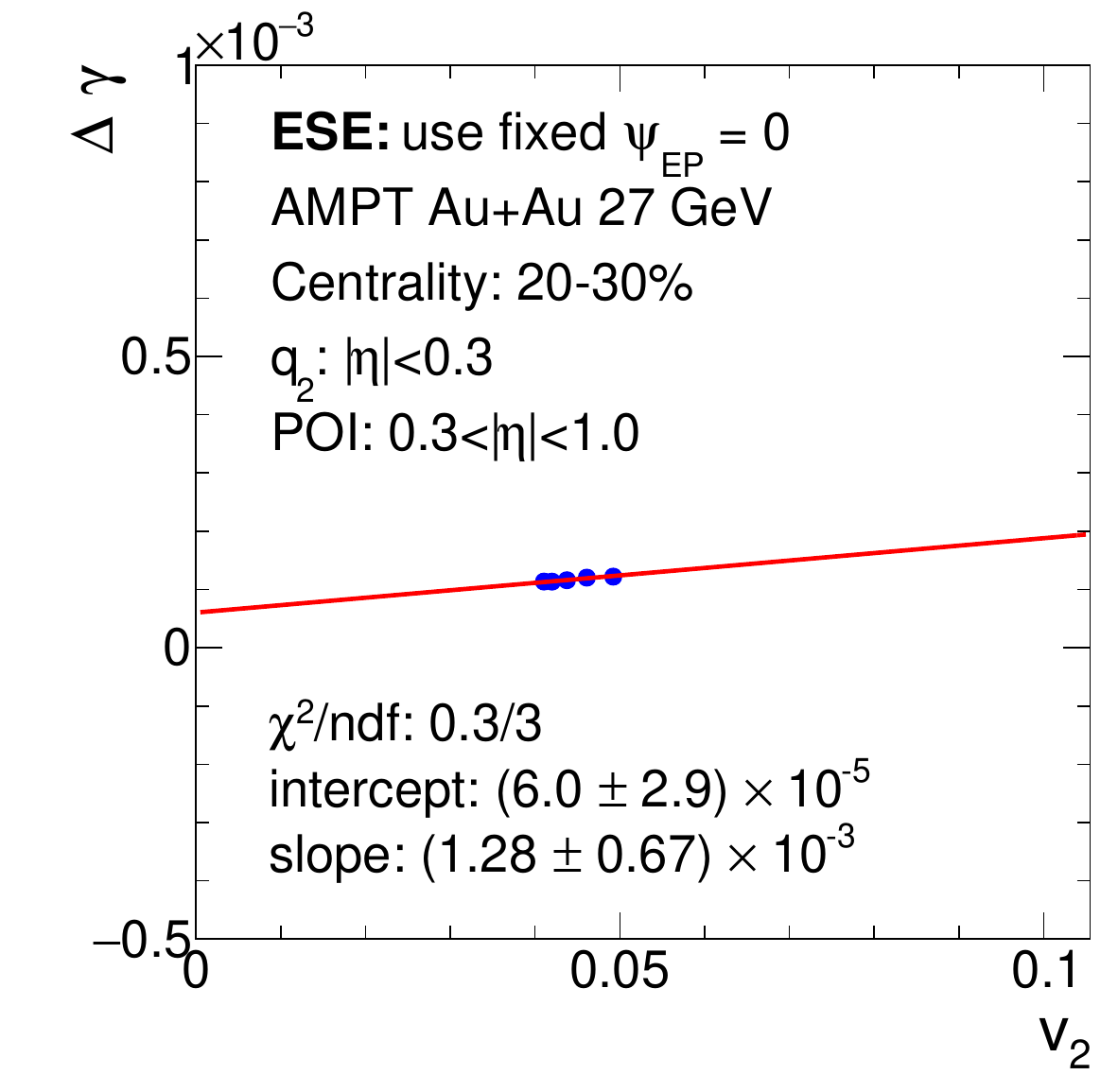}\hfill
    \includegraphics[width=0.33\textwidth]{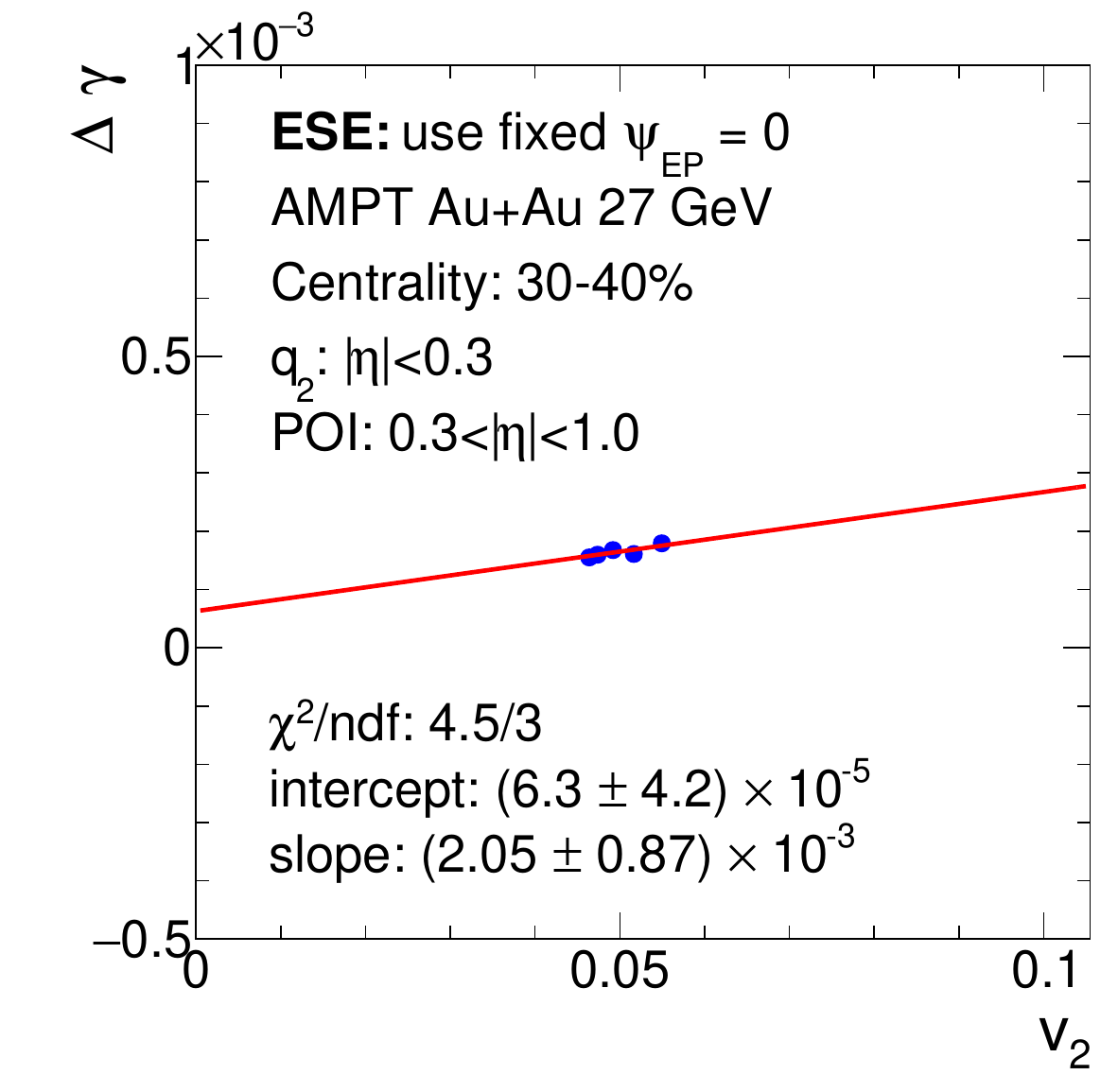}\hfill
    \includegraphics[width=0.33\textwidth]{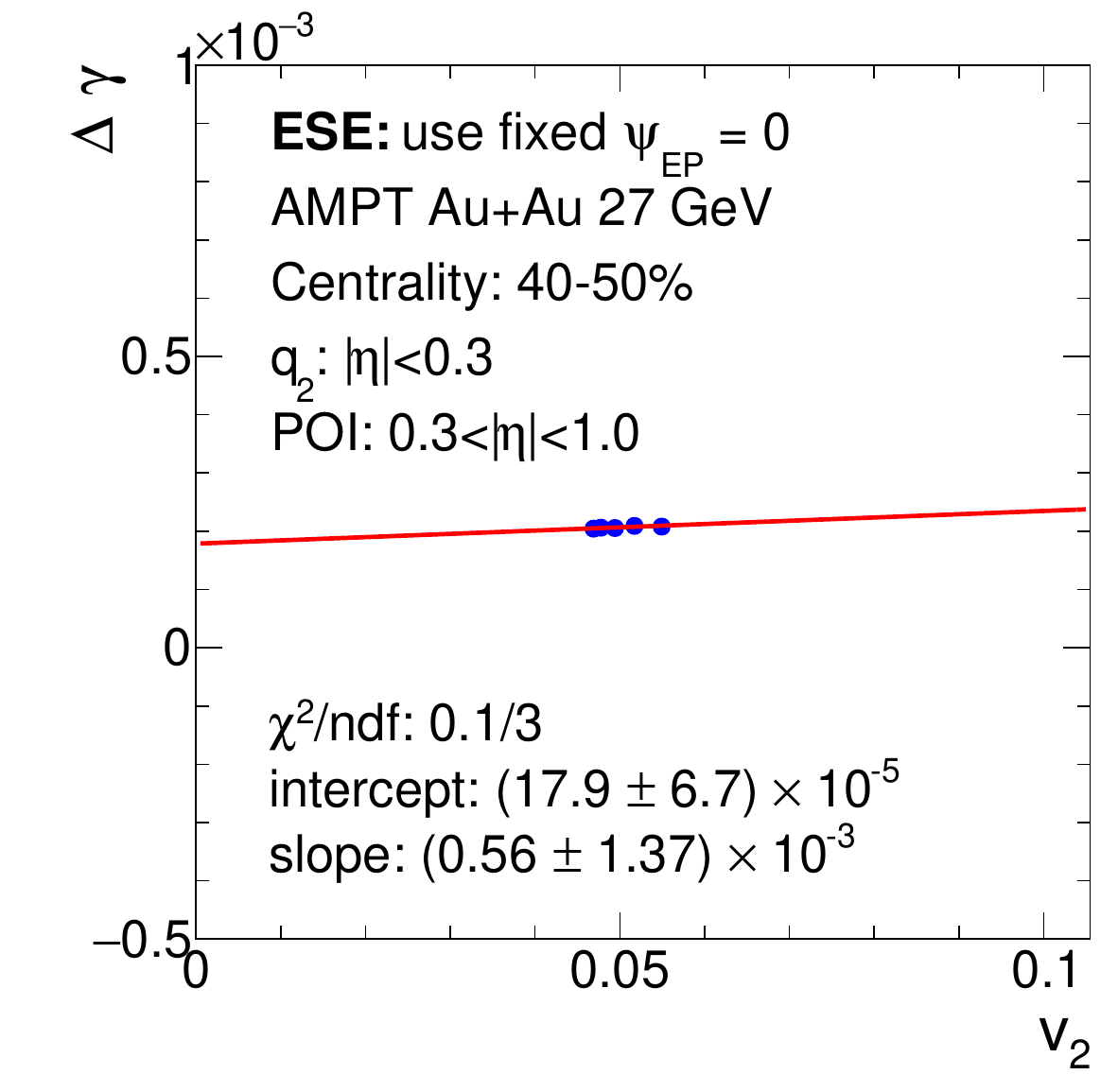}\hfill
    \vspace{-4mm}\\
    \caption{\label{fig:ampt_ese}\ampt\ ESE results. 
    The same \ampt\ data as in Fig.~\ref{fig:ampt_ess} is used.  
    The $\dg$ is plotted as a function of $\mean{v_2}$ in events binned in $\qh^2\two$ (Eqs.~\ref{eq:q2},\ref{eq:qh}). POIs are from acceptance $0.3<|\eta|<1$, and the event selection variable $\qh^2\two$ is computed from particles in $|\eta|<0.3$, both with $0.2<\pt<2$~\gevc. The model's known impact parameter direction $\psi=0$ is taken as the EP in calculating $\dg$ (Eqs.~\ref{eq:g},\ref{eq:dg}) and $\mean{v_2}$ (Eq.~\ref{eq:v2}). The red line is a first-order polynomial fit to all data points.}
    \includegraphics[width=0.33\textwidth]{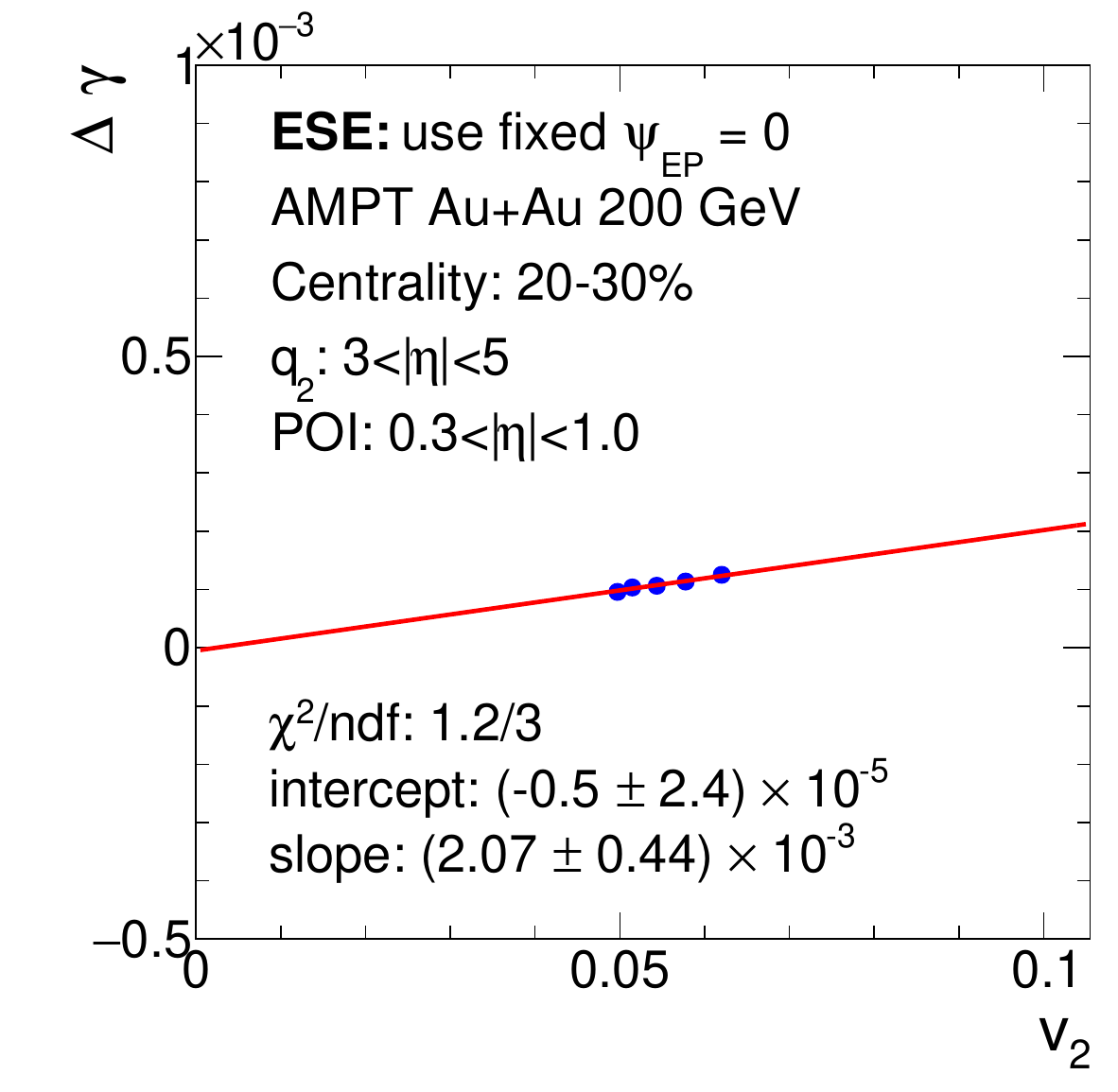}\hfill
    \includegraphics[width=0.33\textwidth]{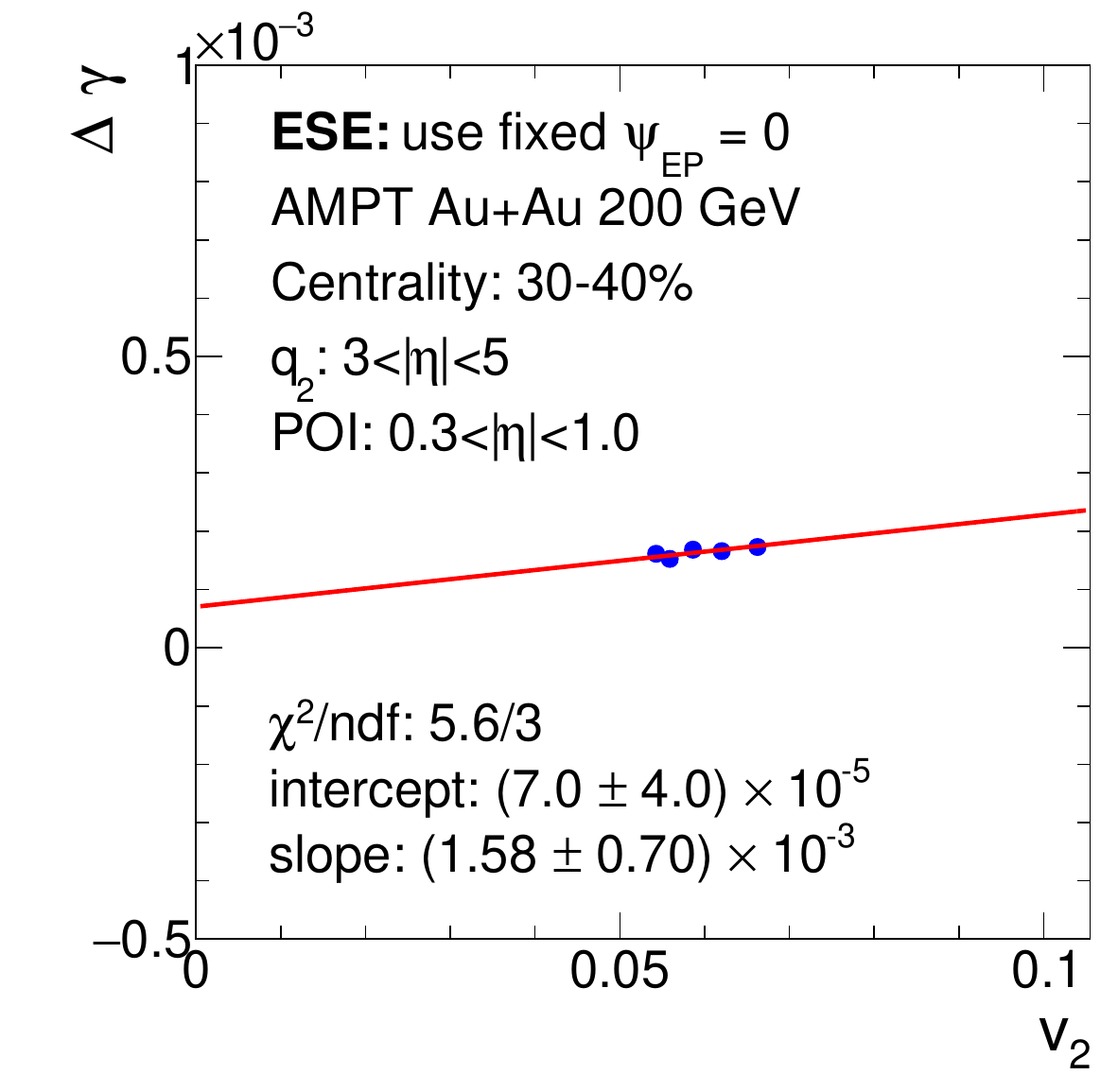}\hfill
    \includegraphics[width=0.33\textwidth]{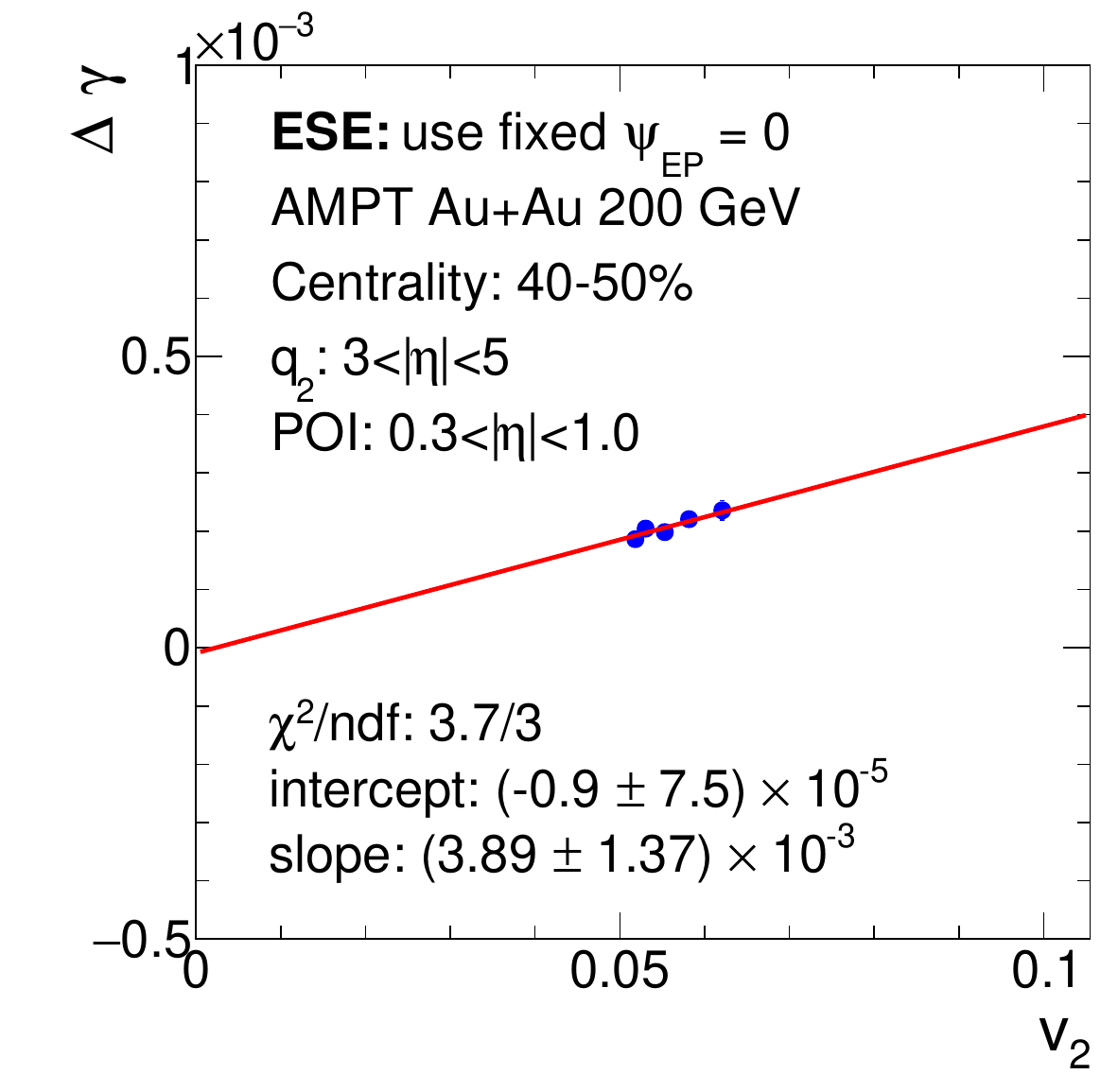}\hfill
    \vspace{-5mm}\\
    \includegraphics[width=0.33\textwidth]{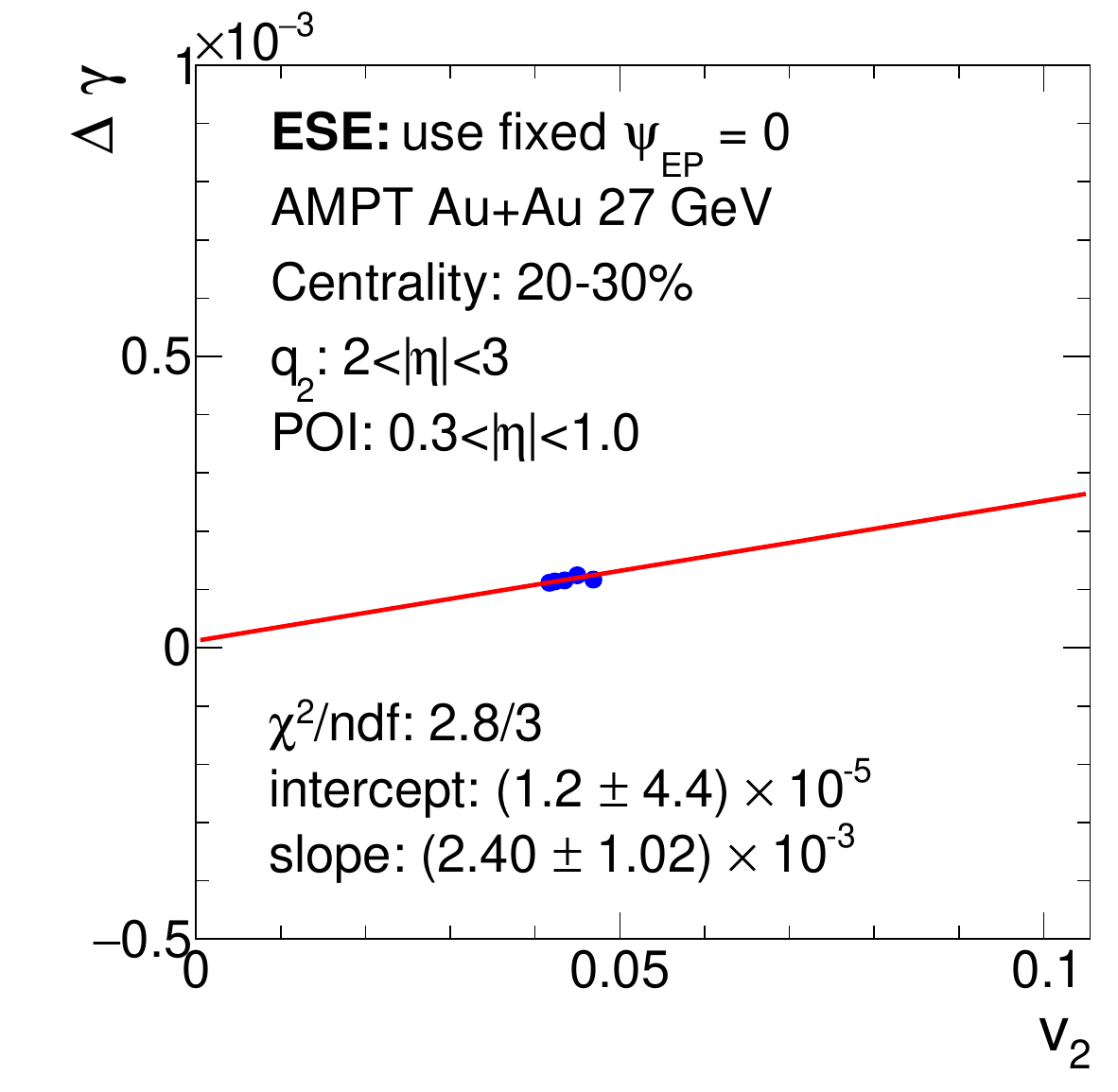}\hfill
    \includegraphics[width=0.33\textwidth]{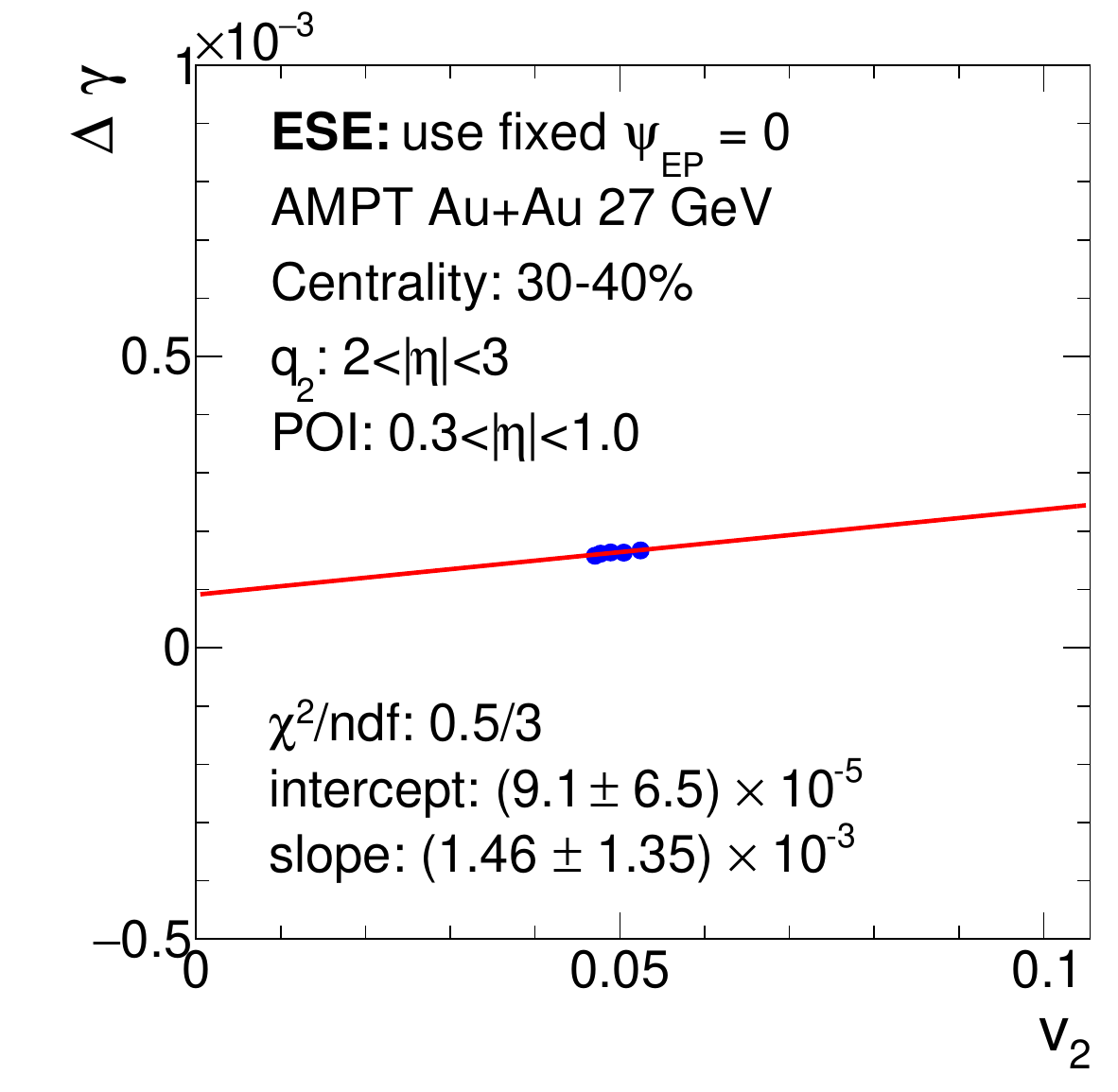}\hfill
    \includegraphics[width=0.33\textwidth]{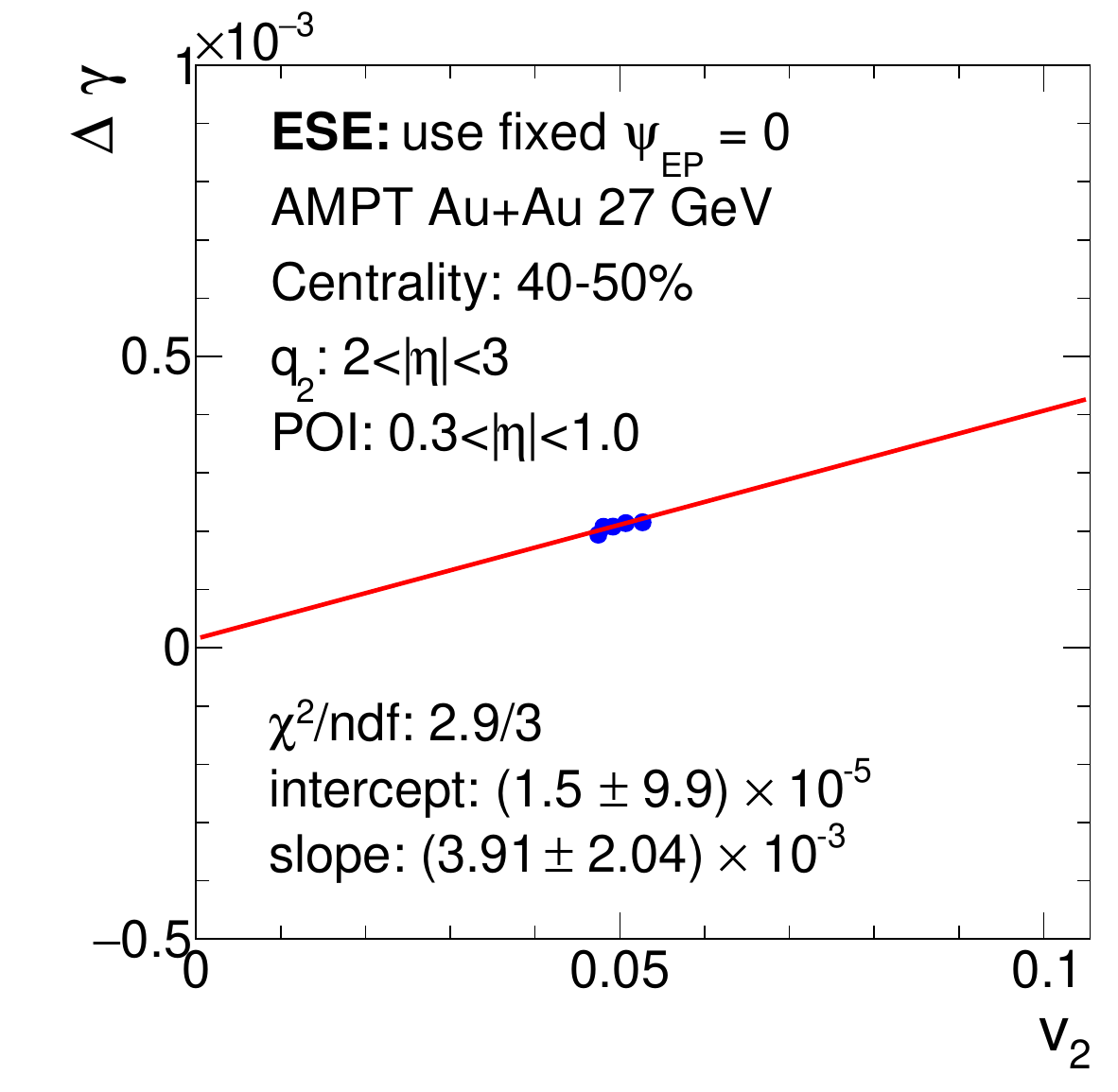}\hfill
    \vspace{-4mm}\\
    \caption{\label{fig:ampt_ese_forward}\ampt\ ESE results with forward/backward $\qh^2\two$. As same as Fig.~\ref{fig:ampt_ese} except that the event selection variable $\qh^2\two$ is computed from particles in the forward/backward pseudorapidity region of $3<|\eta|<5$ for 200~GeV and $2<|\eta|<3$ for 27~GeV.} 
\end{figure*}

\begin{figure*}[hbt]
    \includegraphics[width=0.33\textwidth]{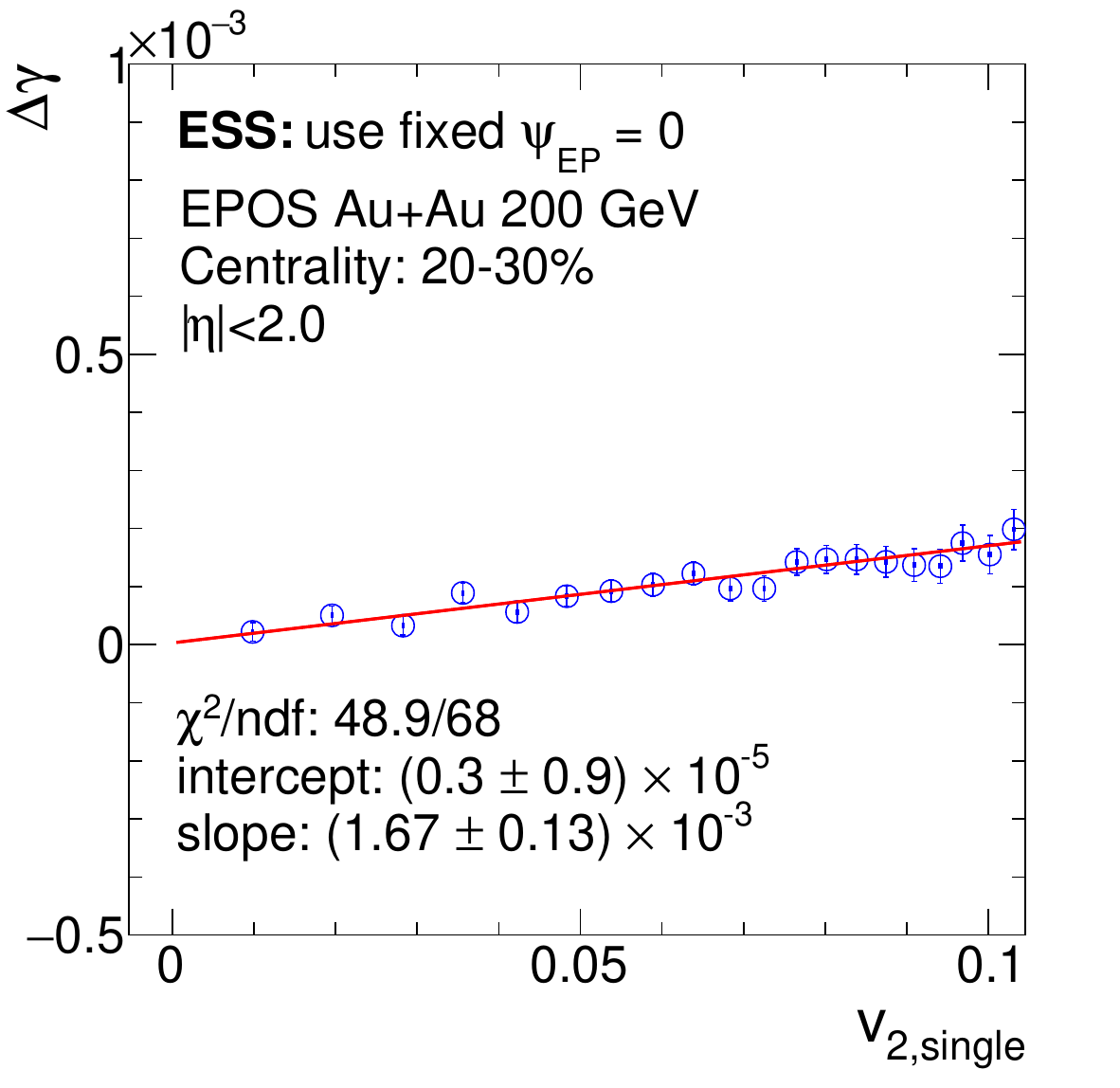}\hfill
    \includegraphics[width=0.33\textwidth]{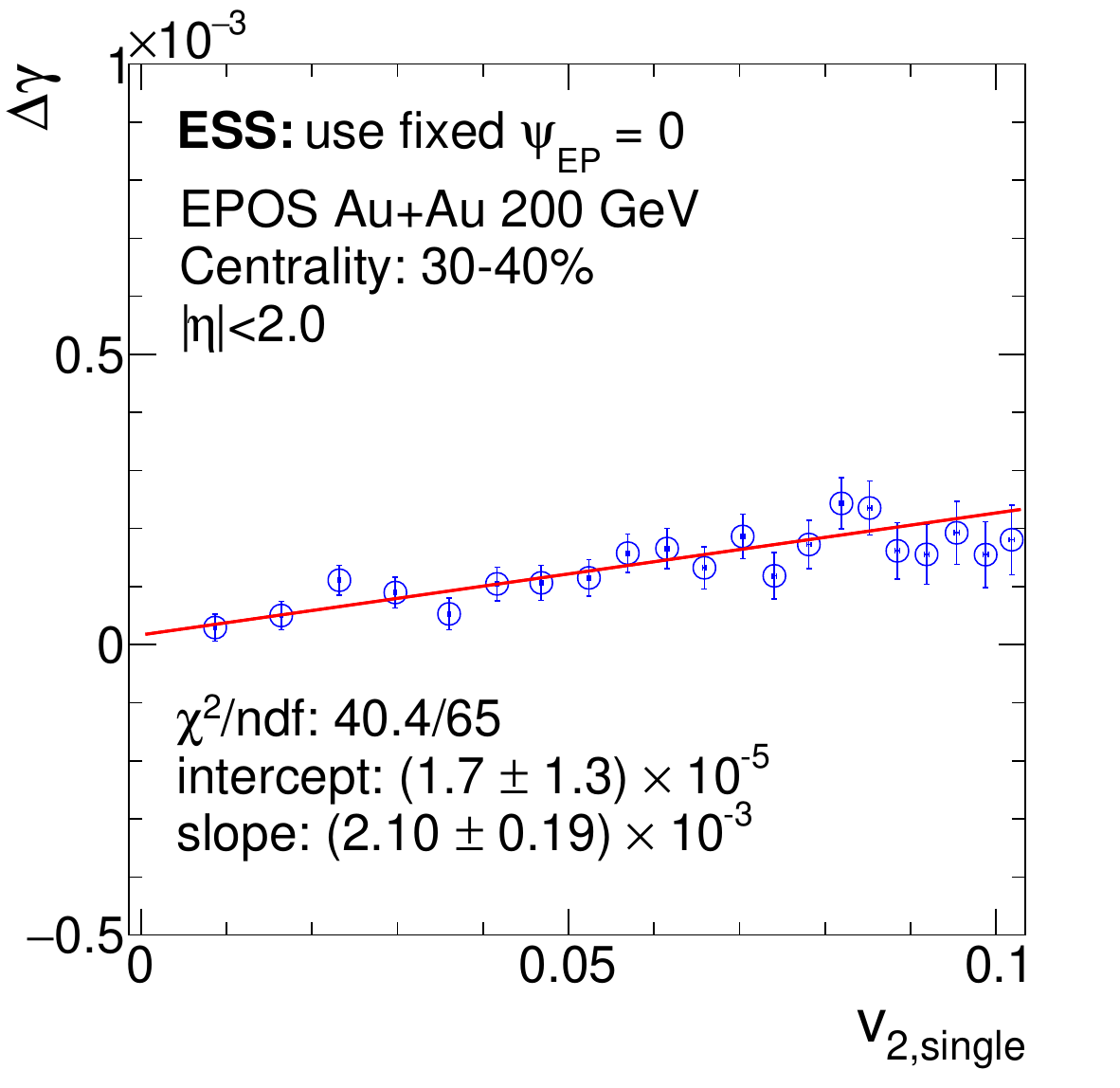}\hfill
    \includegraphics[width=0.33\textwidth]{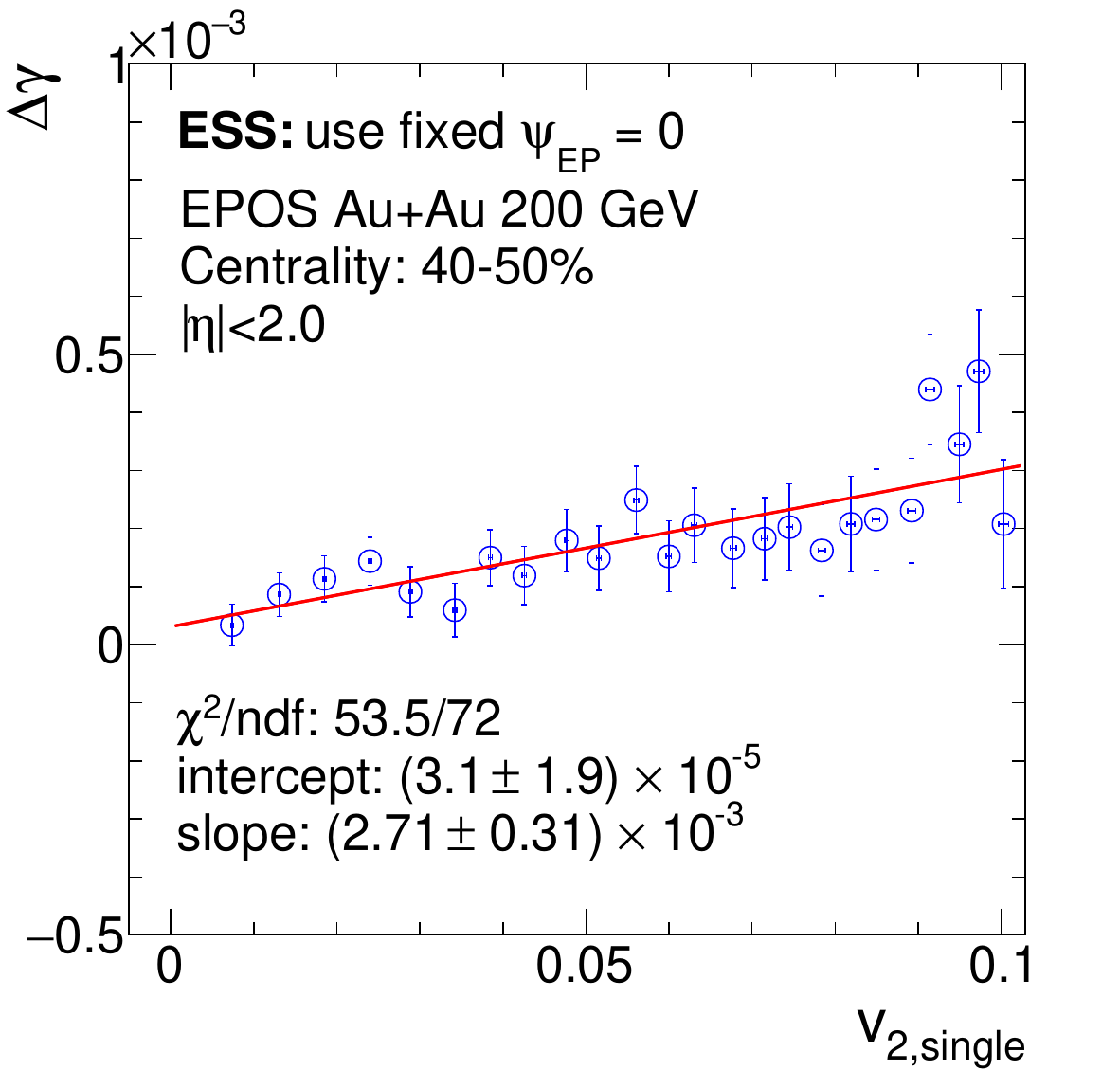}\hfill
    \vspace{-5mm}\\
    \includegraphics[width=0.33\textwidth]{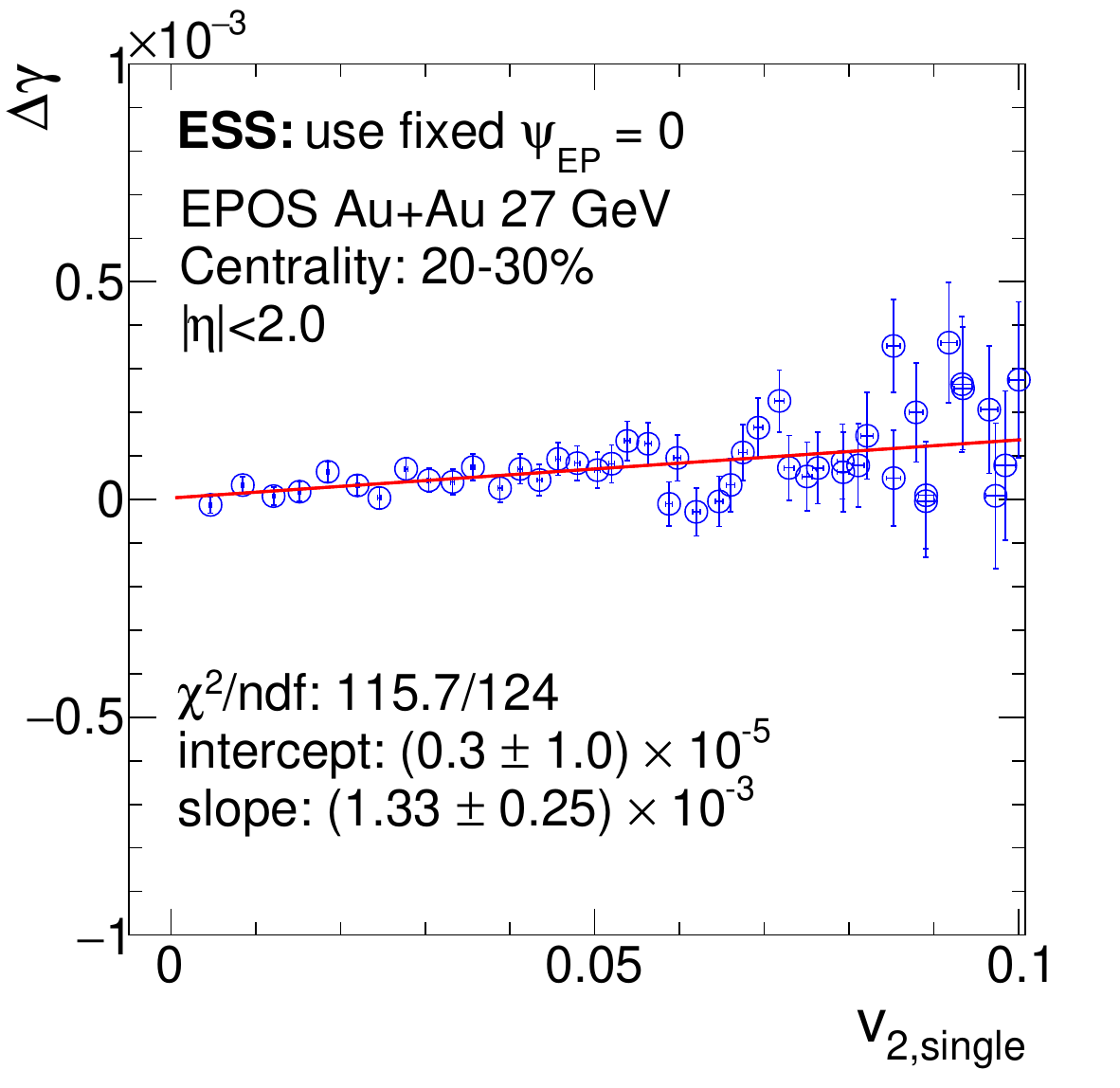}\hfill
    \includegraphics[width=0.33\textwidth]{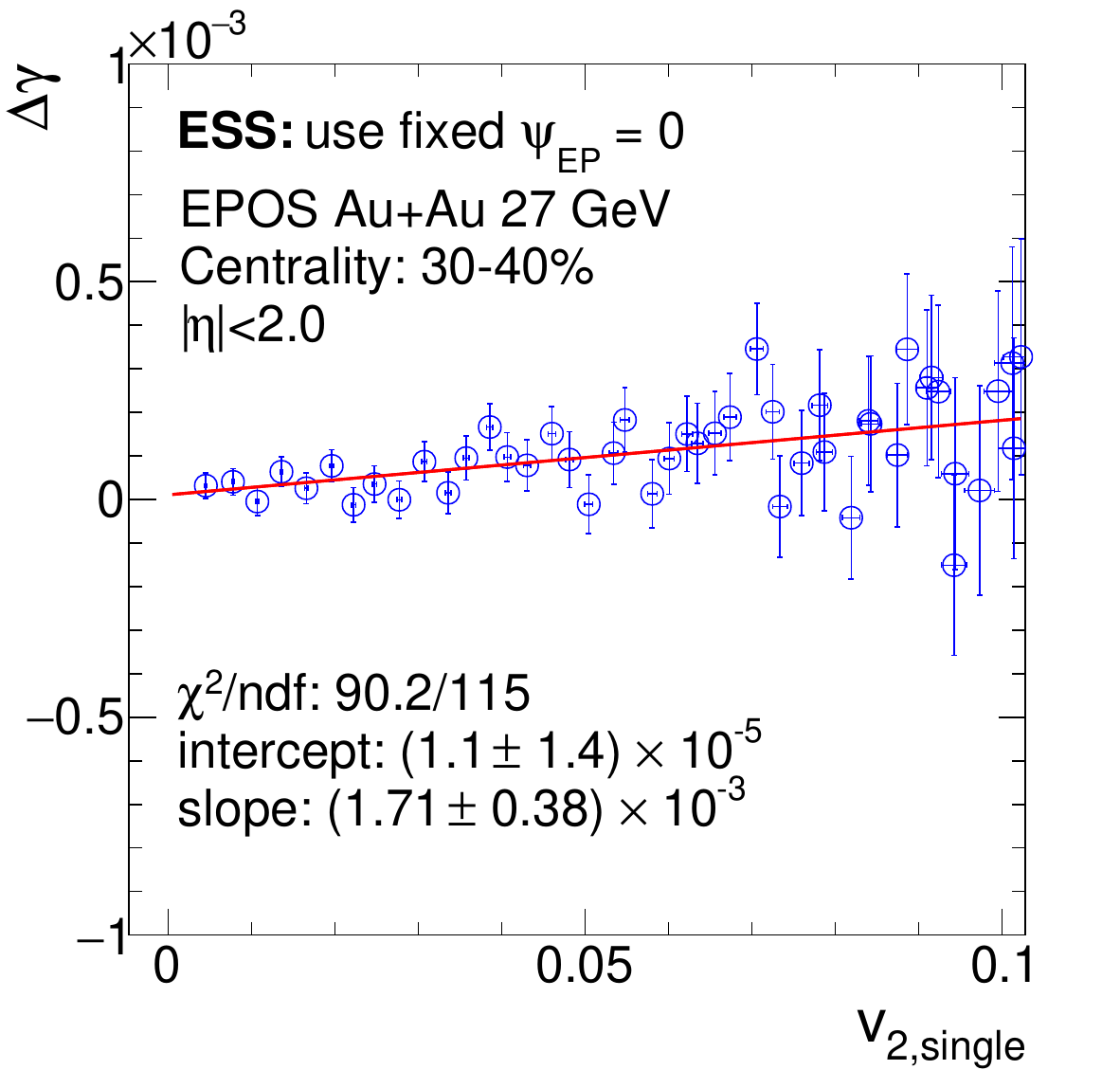}\hfill
    \includegraphics[width=0.33\textwidth]{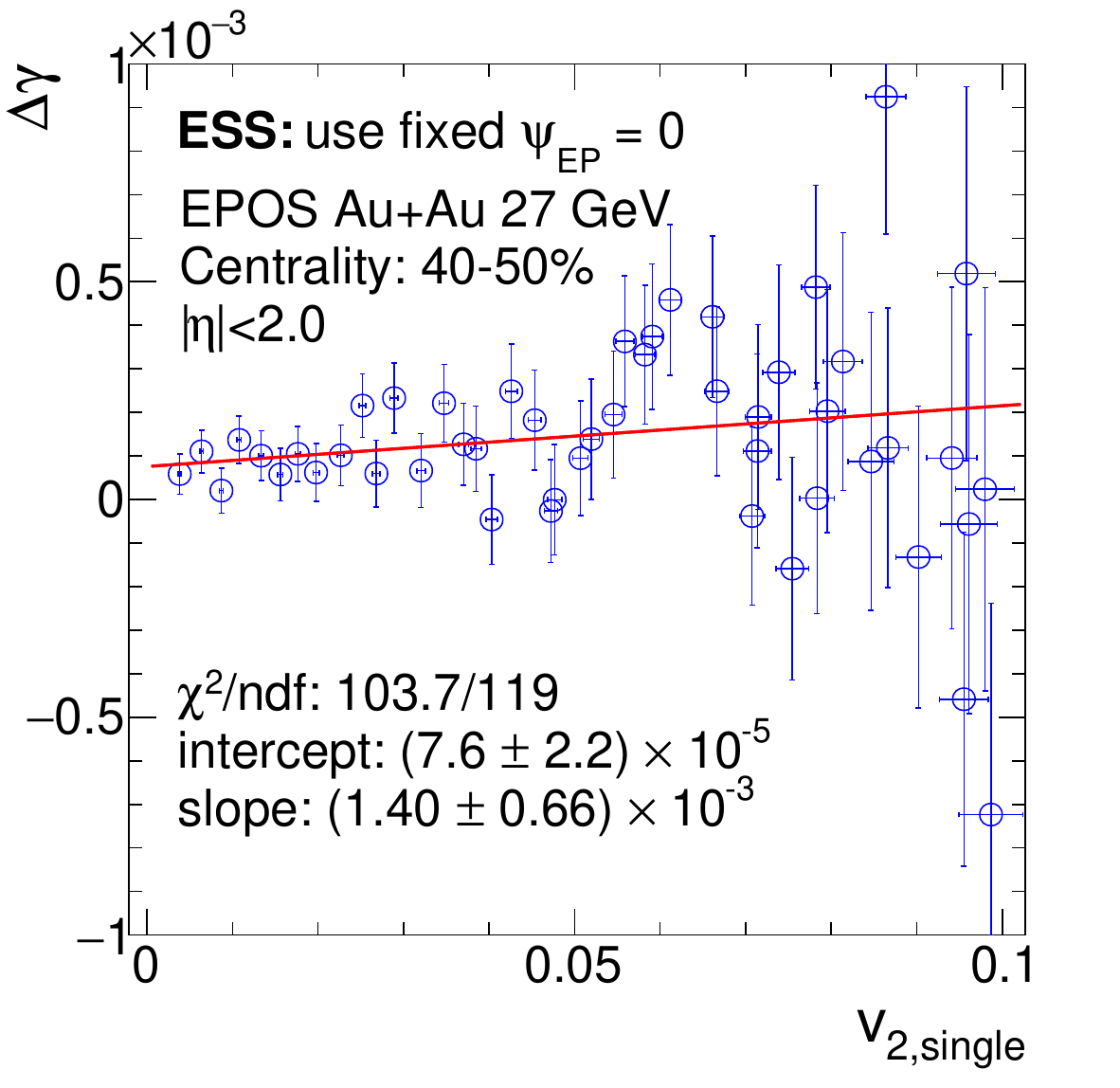}\hfill
    \vspace{-4mm}\\
    \caption{\label{fig:epos_ess}\epos\ ESS results. Shown are three centralities of Au+Au collisions at $\snn=200$~GeV (upper panels) and at 27~GeV (lower panels) simulated by \epos, with approximately $1.6\times10^7$ and $8.5\times10^6$ events for each centrality, respectively. The $\dg$ is plotted as a function of $\vsing$ in events binned in $\qhpair^2\two$ (Eqs.~\ref{eq:q2},\ref{eq:qhpair}). POIs are from acceptance $|\eta|<2$ and $0.2 < \pt < 2$~\gevc, and the event selection variable $\qhpair^2\two$ is computed from the same POIs. The model's known impact parameter direction $\psi=0$ is taken as the EP in calculating $\dg$ (Eqs.~\ref{eq:g},\ref{eq:dg}) and $\vsing$ (Eq.~\ref{eq:v2}).}
    \includegraphics[width=0.33\textwidth]{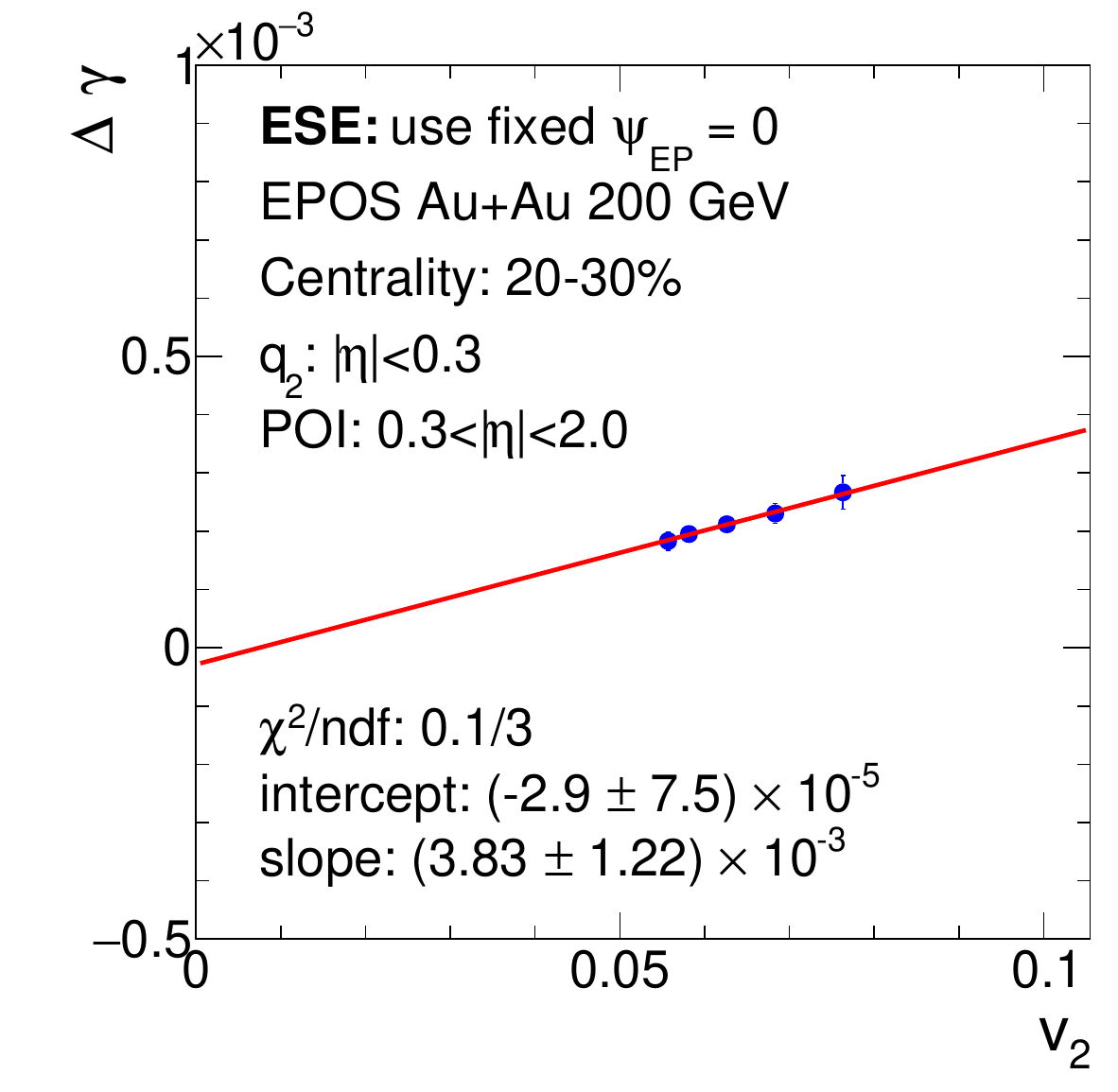}\hfill
    \includegraphics[width=0.33\textwidth]{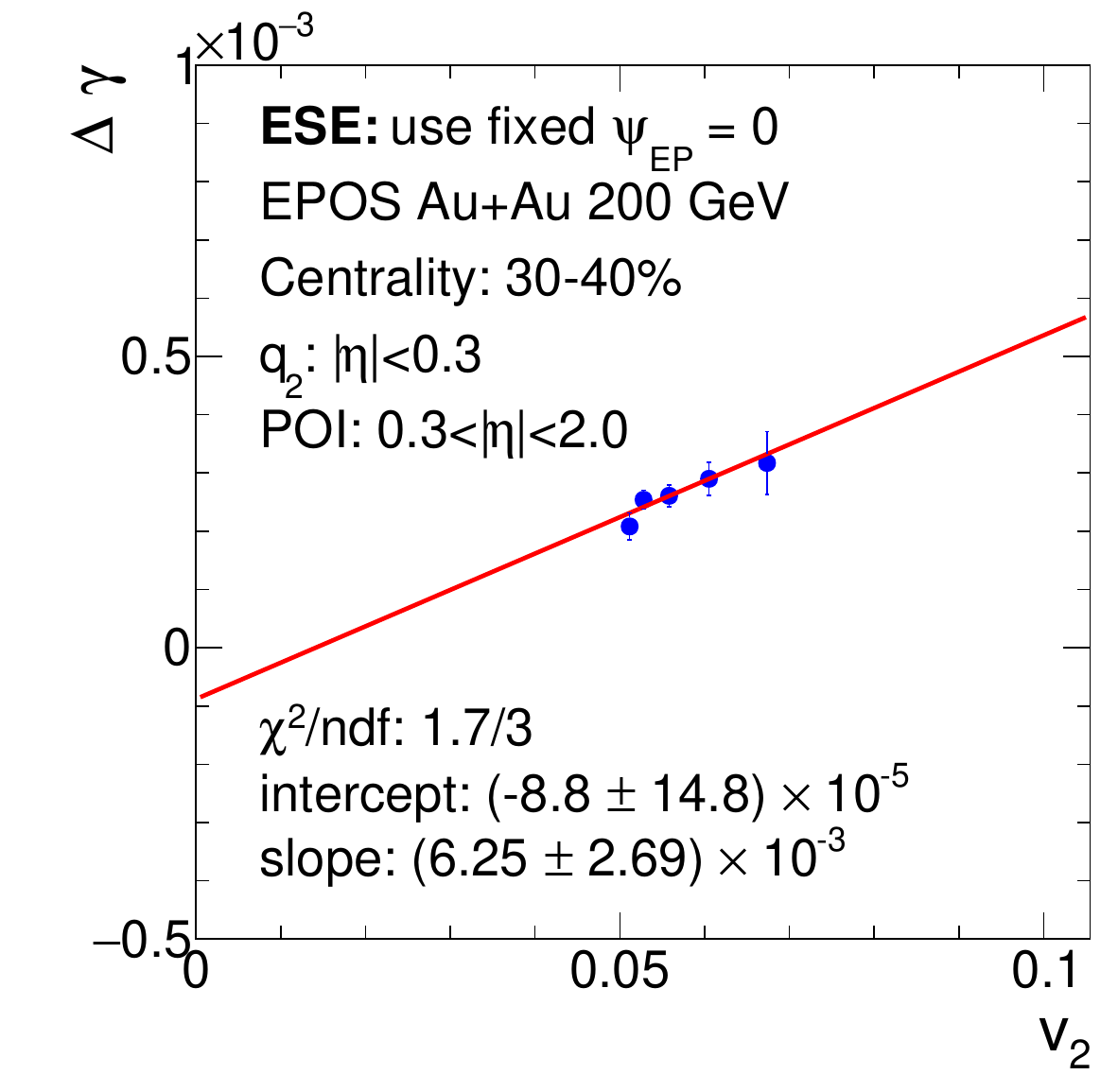}\hfill
    \includegraphics[width=0.33\textwidth]{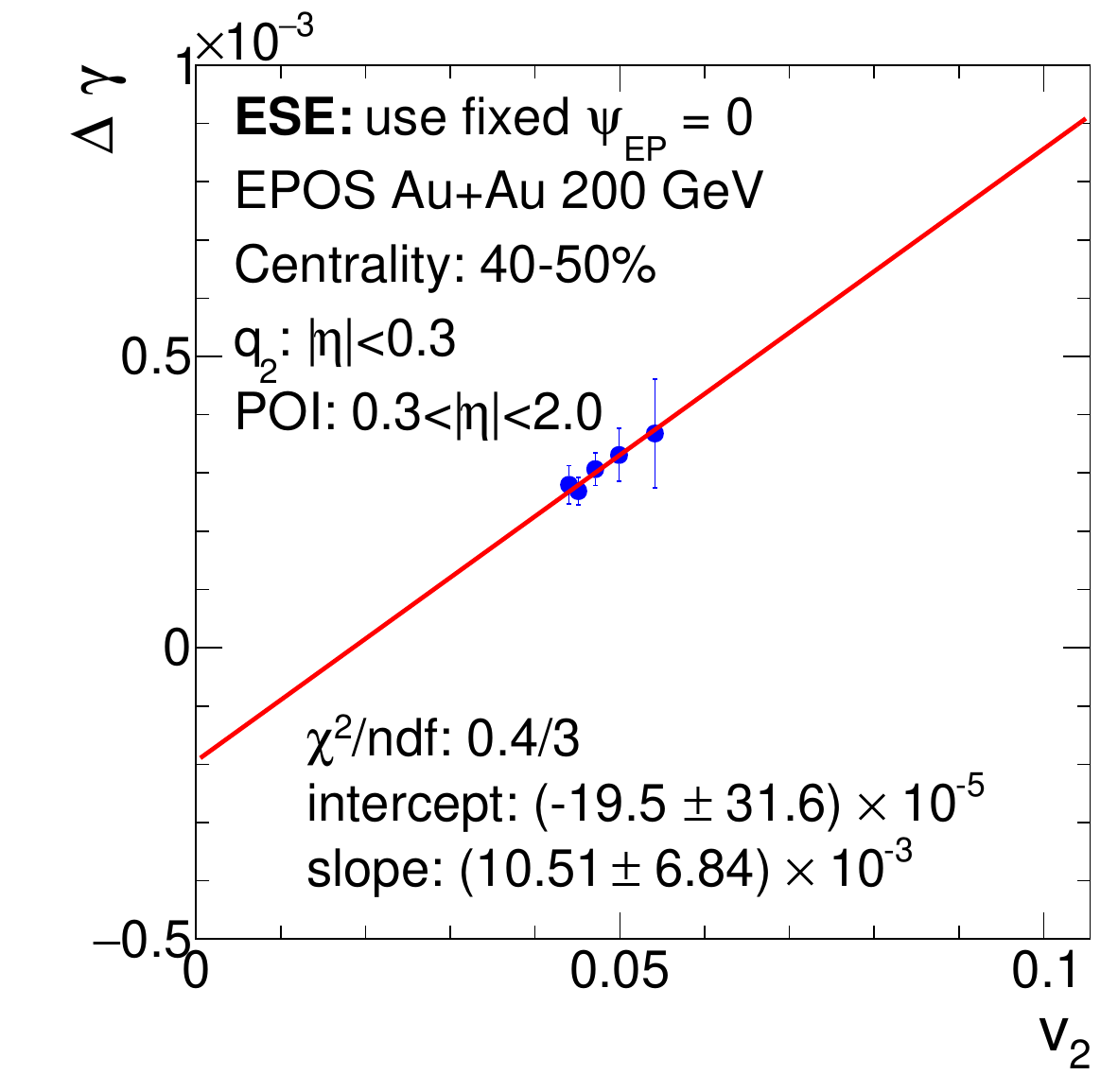}\hfill
    \vspace{-5mm}\\
    \includegraphics[width=0.33\textwidth]{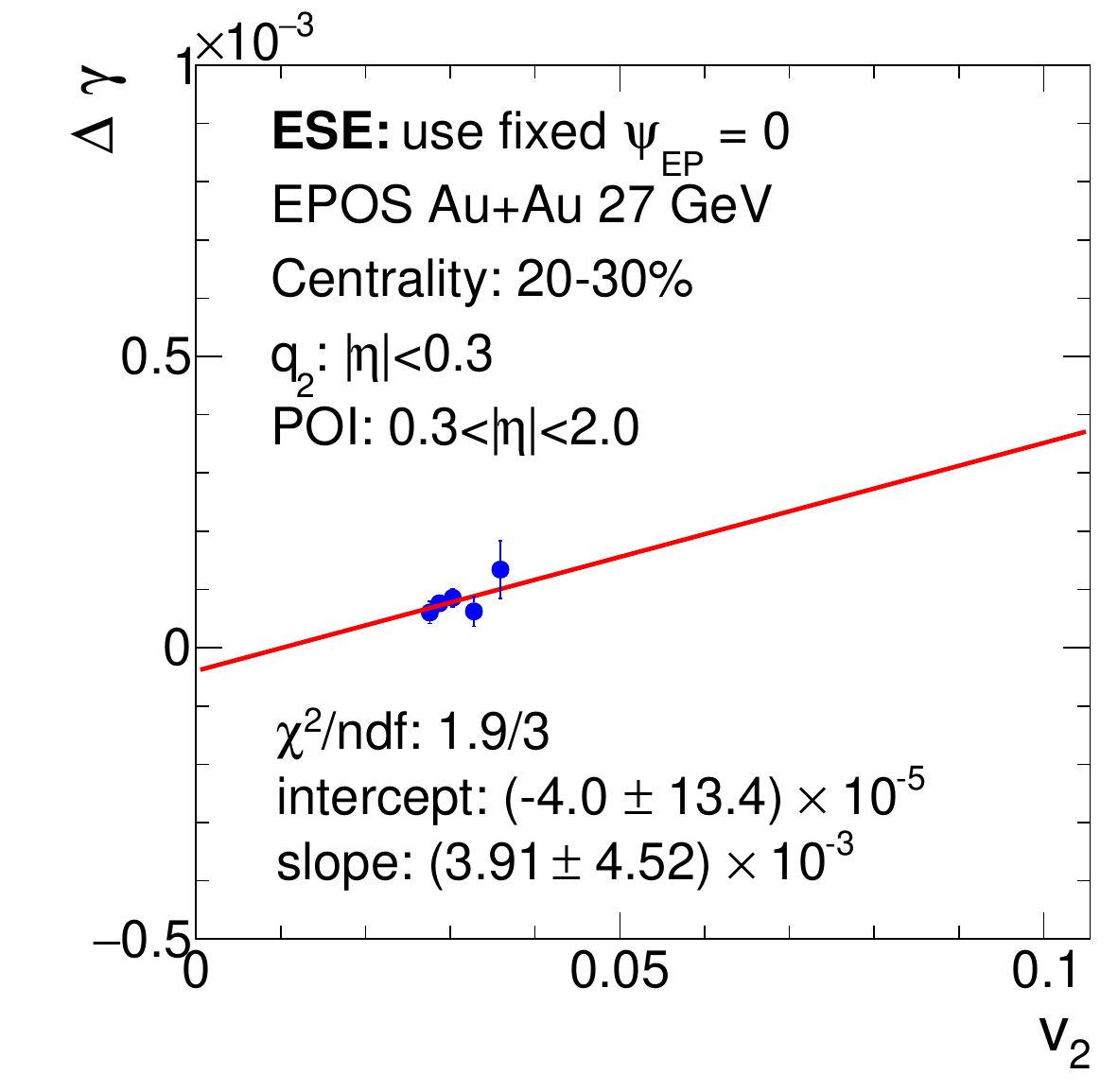}\hfill
    \includegraphics[width=0.33\textwidth]{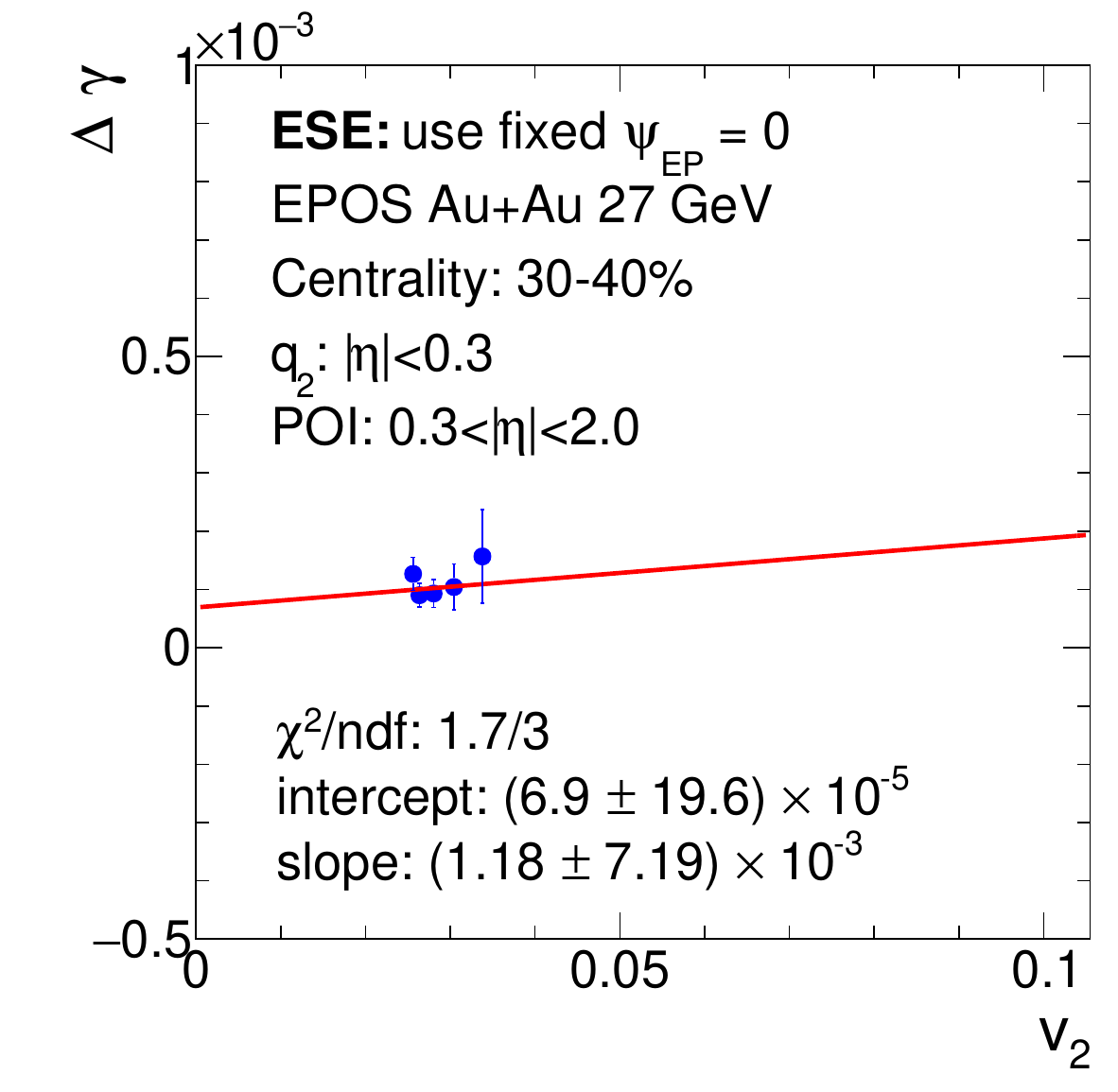}\hfill
    \includegraphics[width=0.33\textwidth]{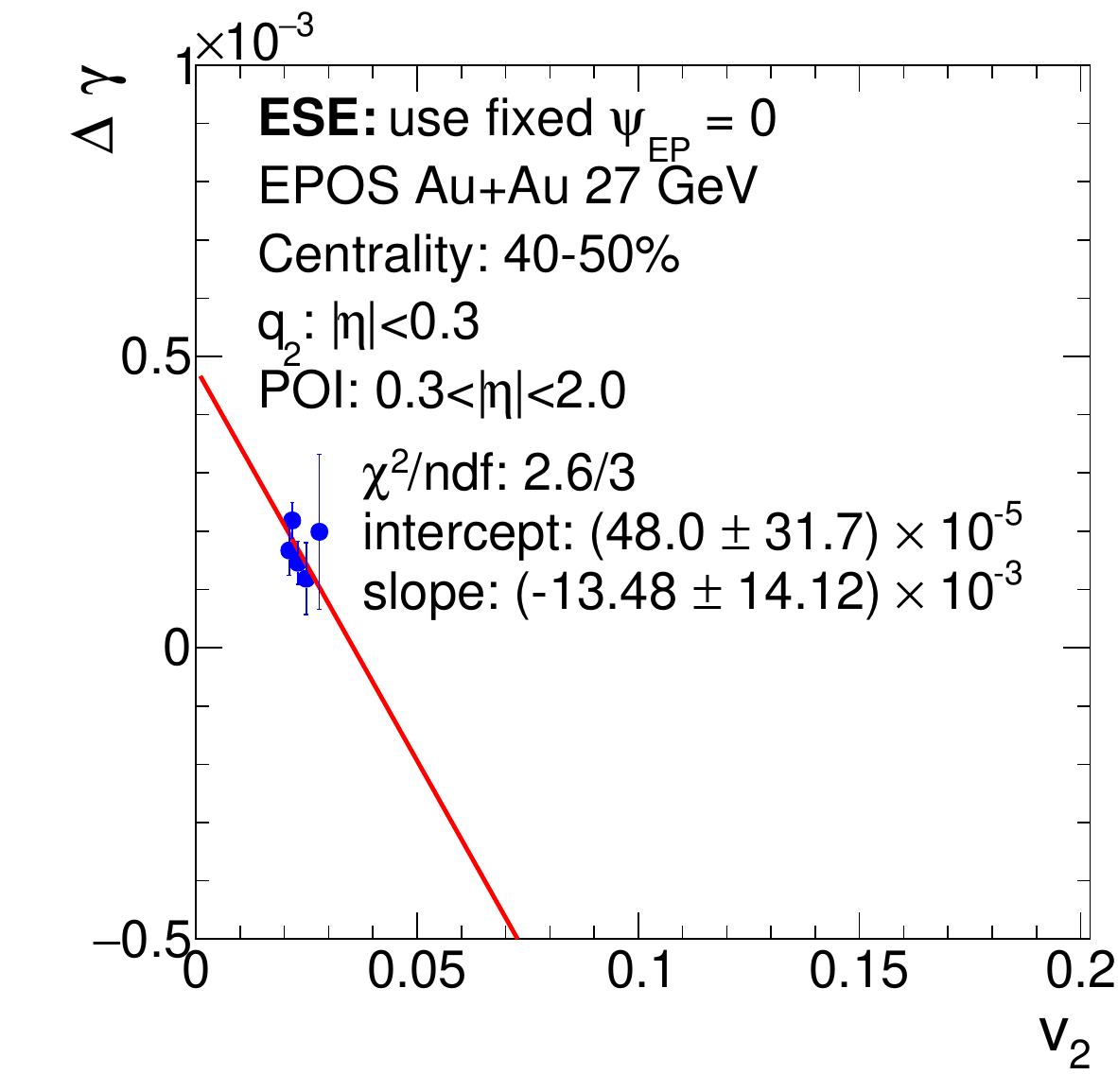}\hfill
    \vspace{-4mm}\\
    \caption{\label{fig:epos_ese}\epos\ ESE results. The same \epos\ data as in Fig.~\ref{fig:epos_ess} is used. 
    The $\dg$ is plotted as a function of $\mean{v_2}$ in events binned in $\qh^2\two$ (Eqs.~\ref{eq:q2},\ref{eq:qh}). POIs are from acceptance $0.3<|\eta|<2$, and the event selection variable $\qh^2\two$ is computed from particles in $|\eta|<0.3$, both with $0.2<\pt<2$~\gevc. The model's known impact parameter direction $\psi=0$ is taken as the EP in calculating $\dg$ (Eqs.~\ref{eq:g},\ref{eq:dg}) and $\mean{v_2}$ (Eq.~\ref{eq:v2}).}
\end{figure*}

\begin{figure*}[hbt]
    \includegraphics[width=0.33\textwidth]{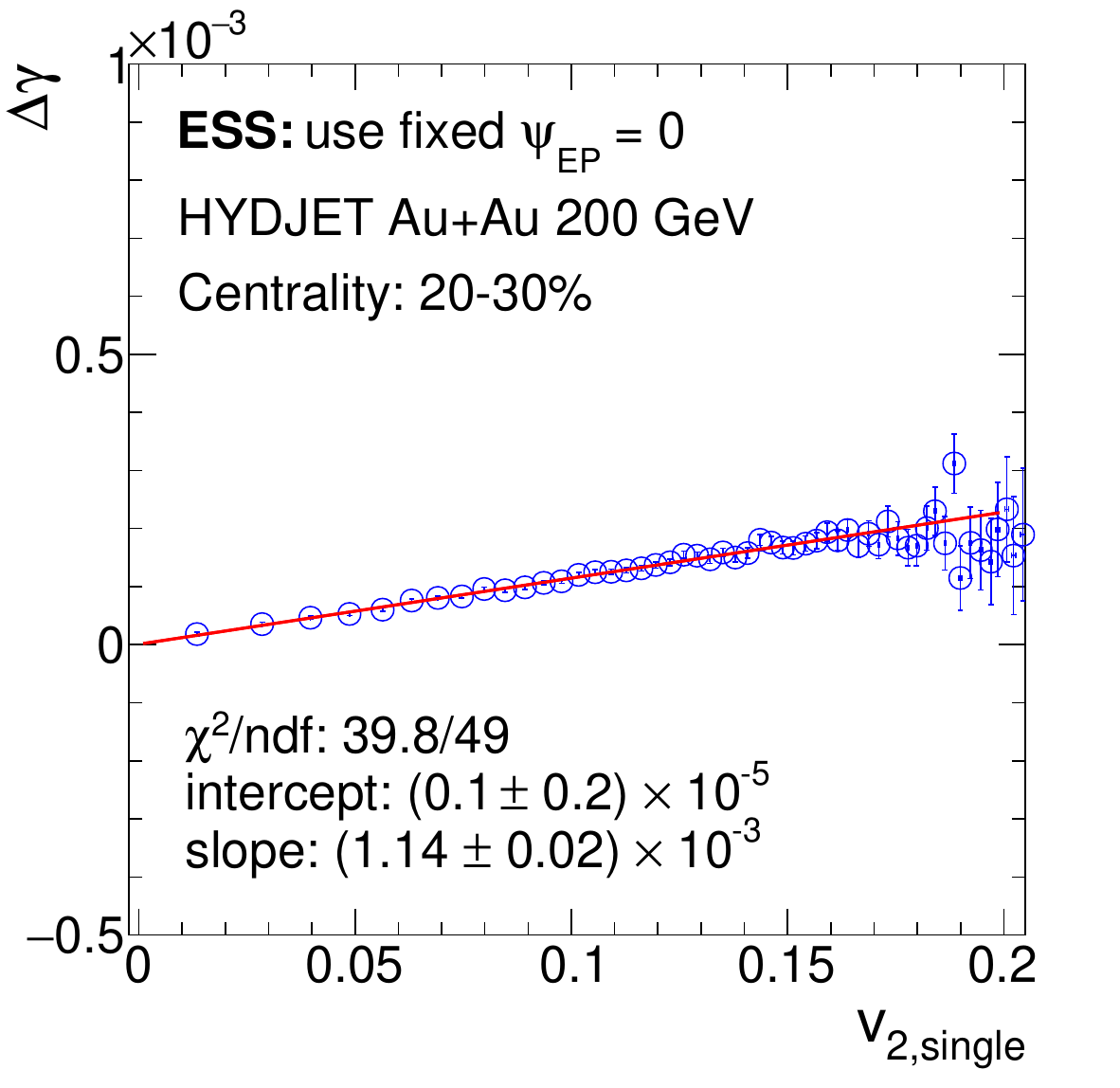}\hfill
    \includegraphics[width=0.33\textwidth]{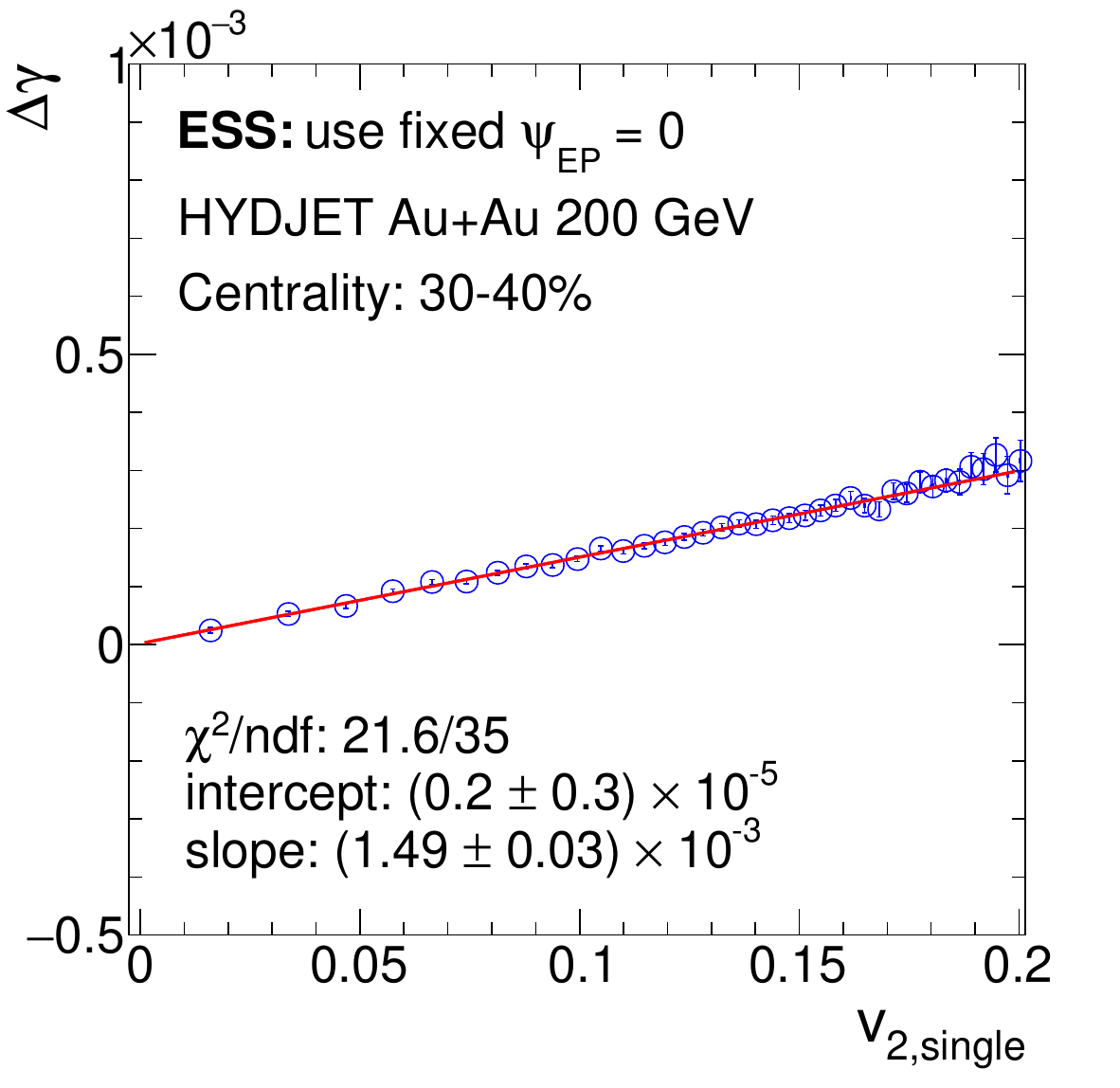}\hfill
    \includegraphics[width=0.33\textwidth]{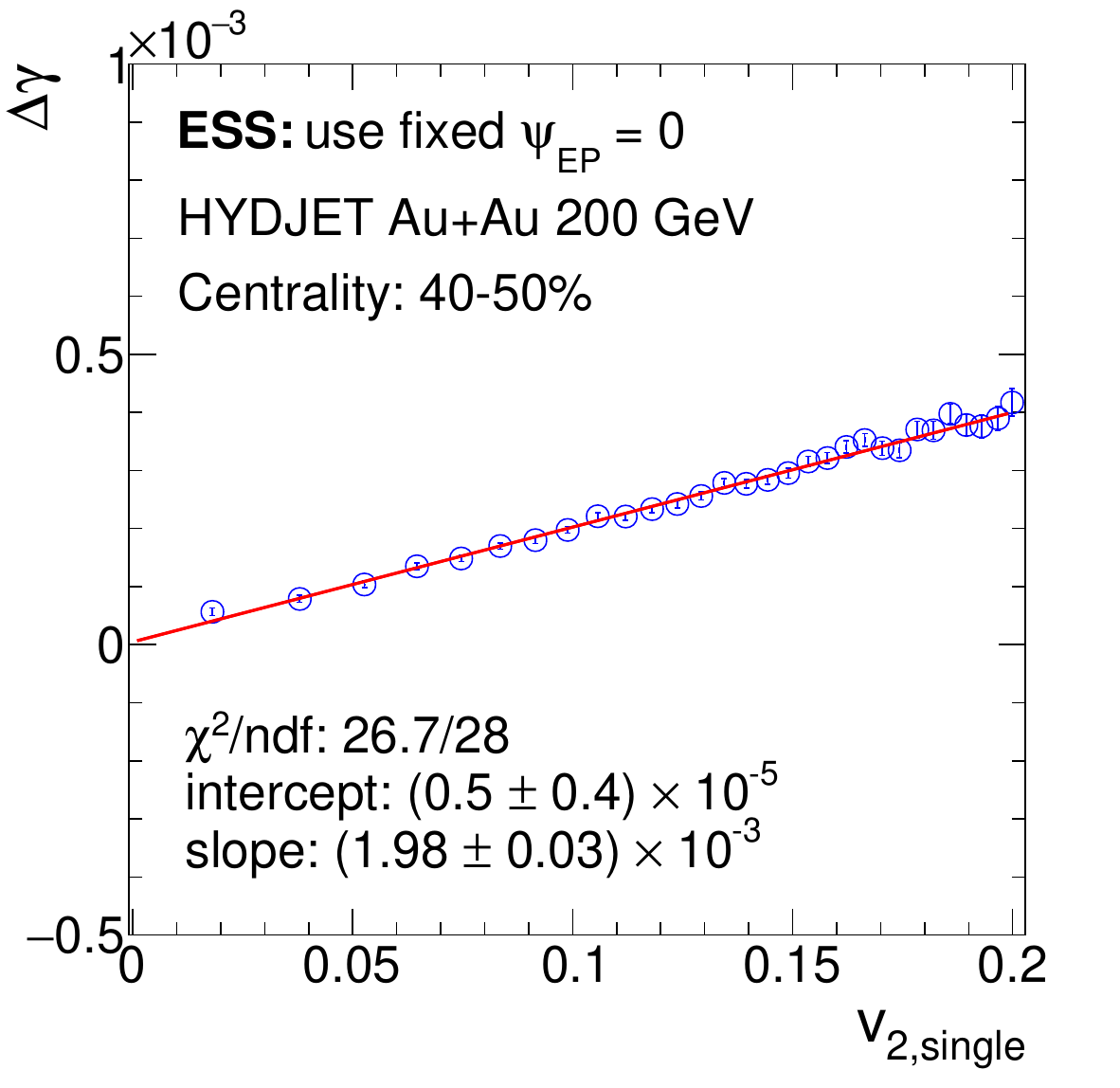}\hfill
    \vspace{-5mm}\\
    \includegraphics[width=0.33\textwidth]{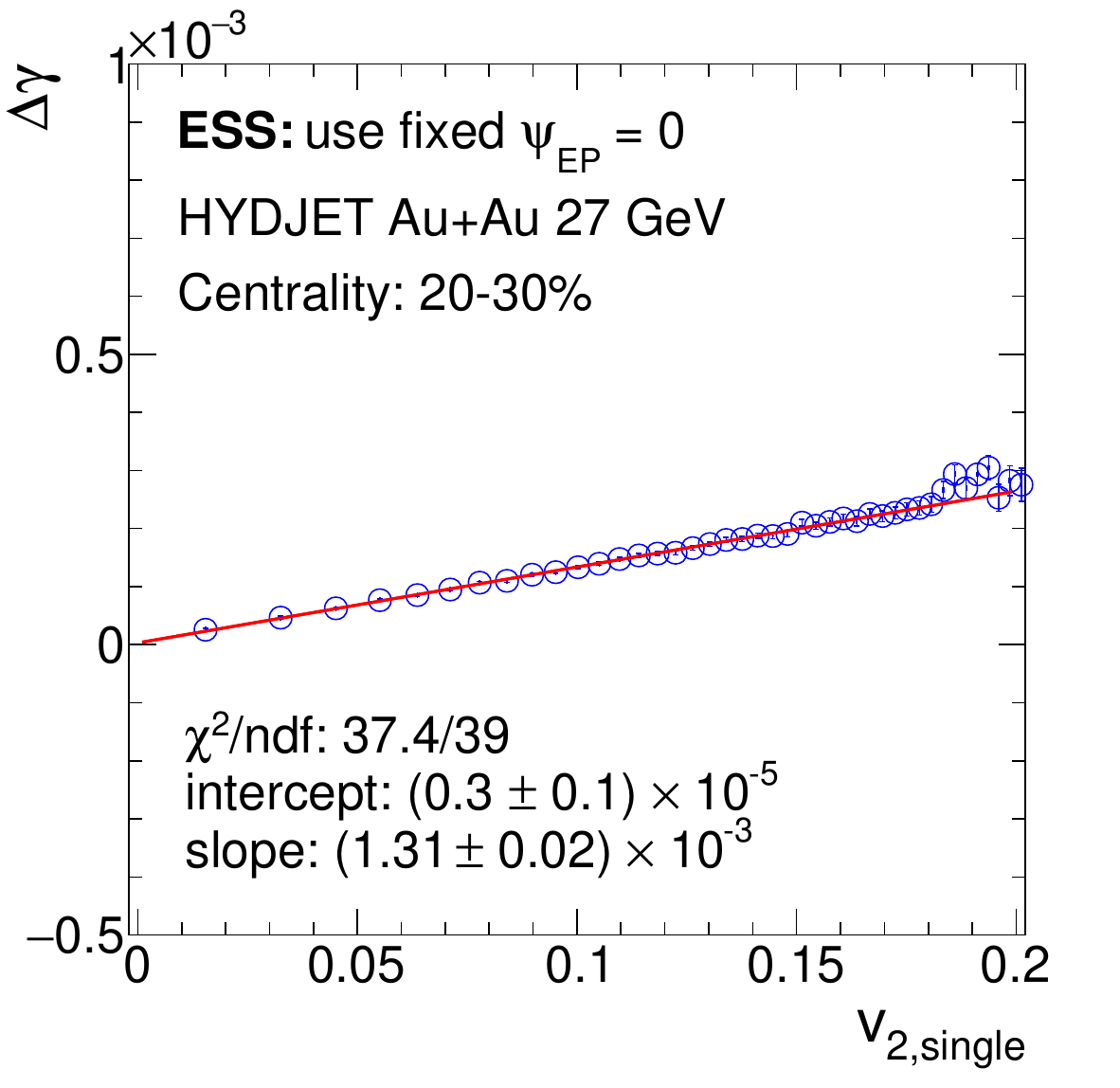}\hfill
    \includegraphics[width=0.33\textwidth]{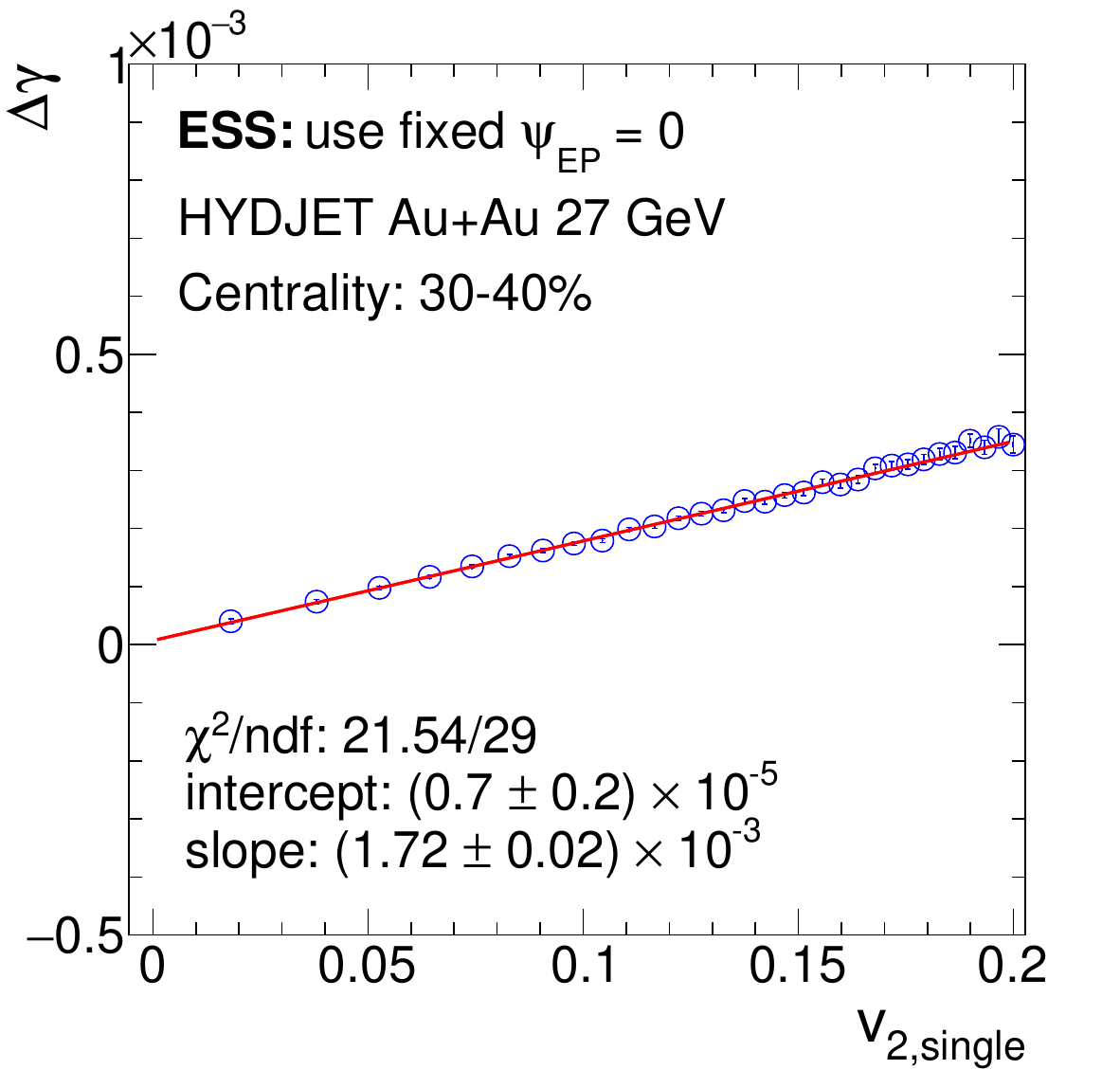}\hfill
    \includegraphics[width=0.33\textwidth]{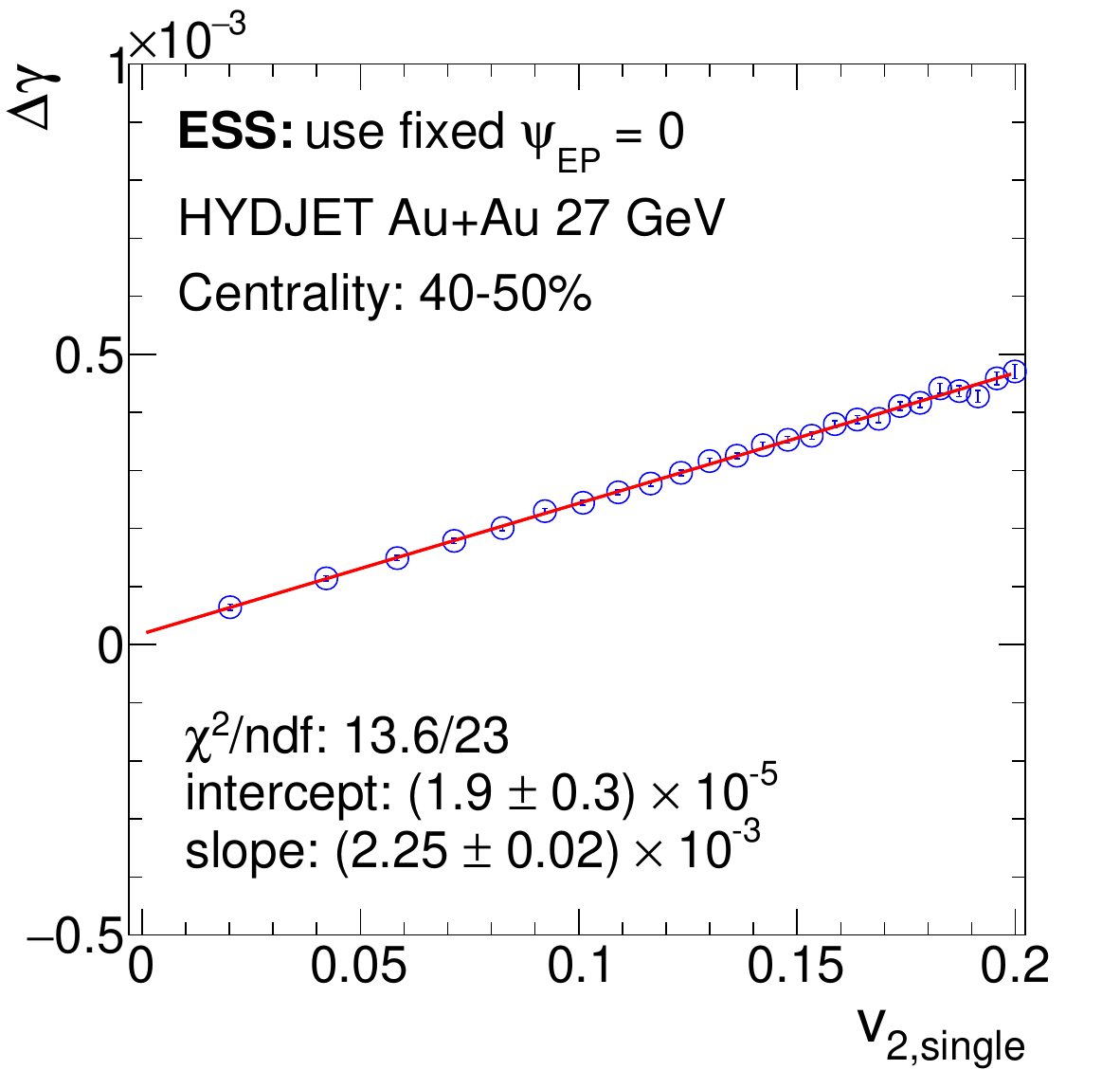}\hfill
    \vspace{-4mm}\\
    \caption{\label{fig:hydjet_ess}\hydjet\ ESS results. Shown are three centralities of Au+Au collisions at $\snn=200$~GeV (upper panels) and at 27~GeV (lower panels) simulated by \hydjet, with approximately $2.5\times10^8$ and $5.7\times10^8$ events for each centrality, respectively. The $\dg$ is plotted as a function of $\vsing$ in events binned in $\qhpair^2\two$ (Eqs.~\ref{eq:q2},\ref{eq:qhpair}). POIs are from acceptance $|\eta|<1$ and $0.2 < \pt < 2$~\gevc, and the event selection variable $\qhpair^2\two$ is computed from the same POIs. The model's known impact parameter direction $\psi=0$ is taken as the EP in calculating $\dg$ (Eqs.~\ref{eq:g},\ref{eq:dg}) and $\vsing$ (Eq.~\ref{eq:v2}).}
    \includegraphics[width=0.33\textwidth]{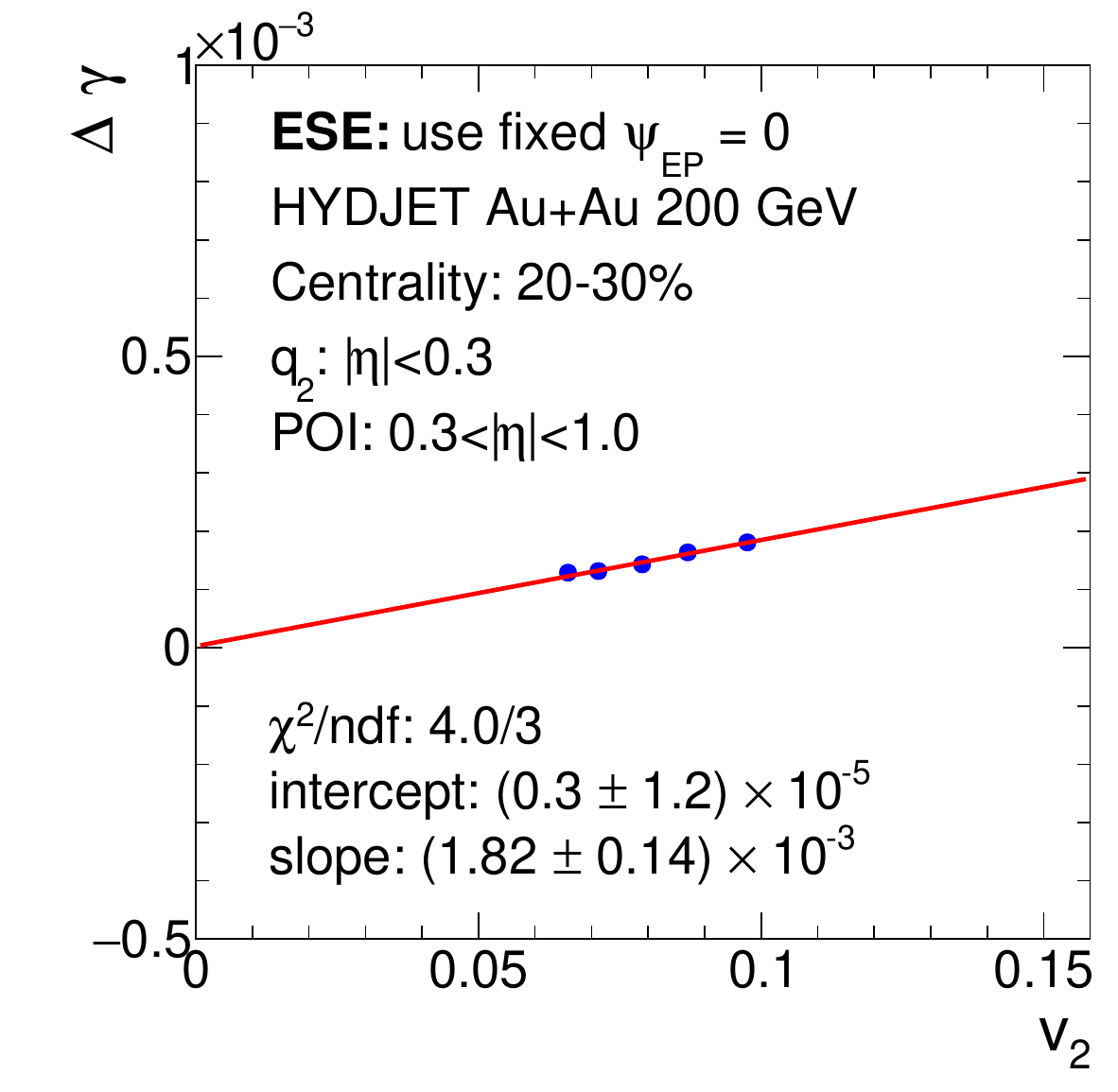}\hfill
    \includegraphics[width=0.33\textwidth]{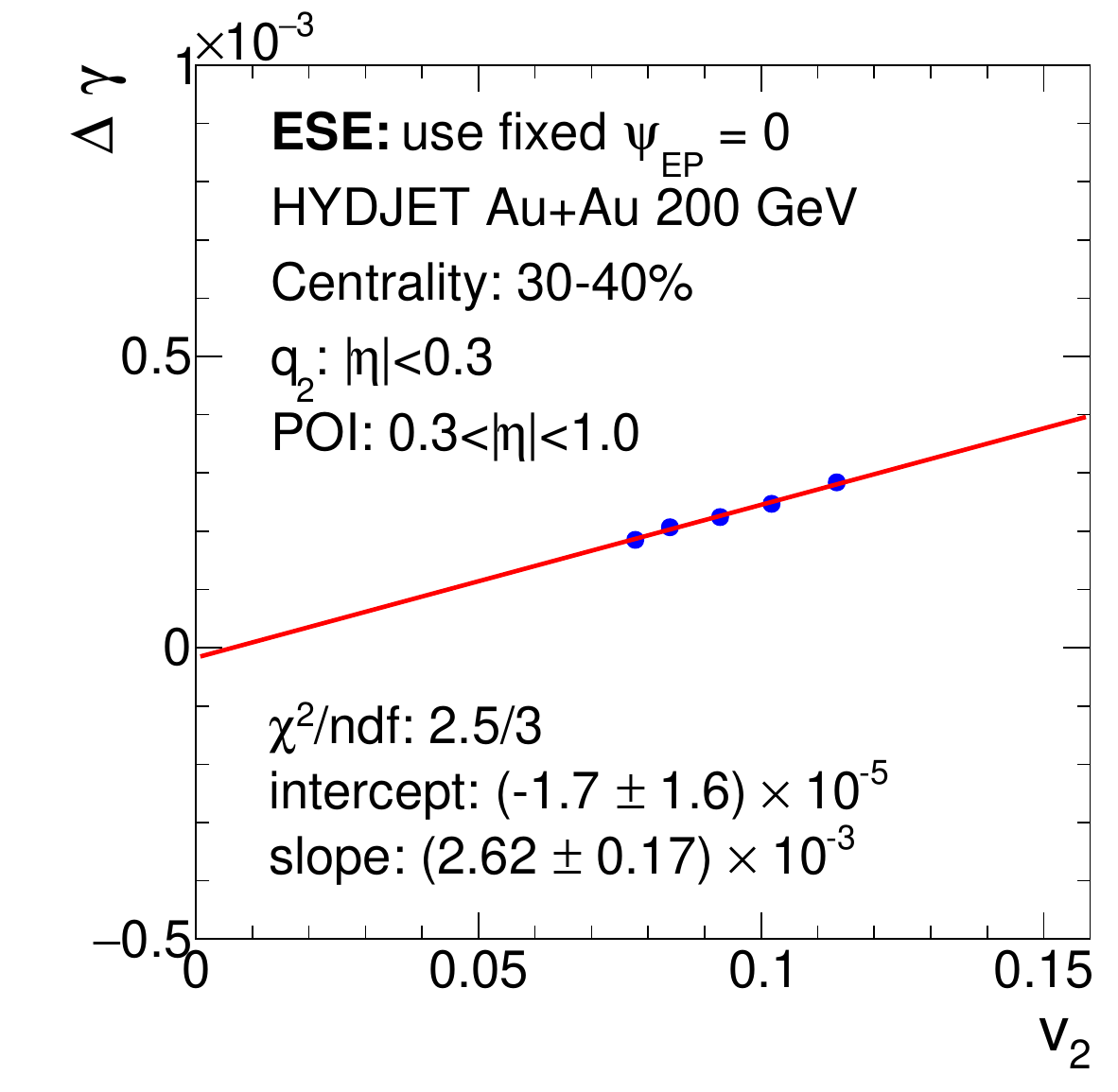}\hfill
    \includegraphics[width=0.33\textwidth]{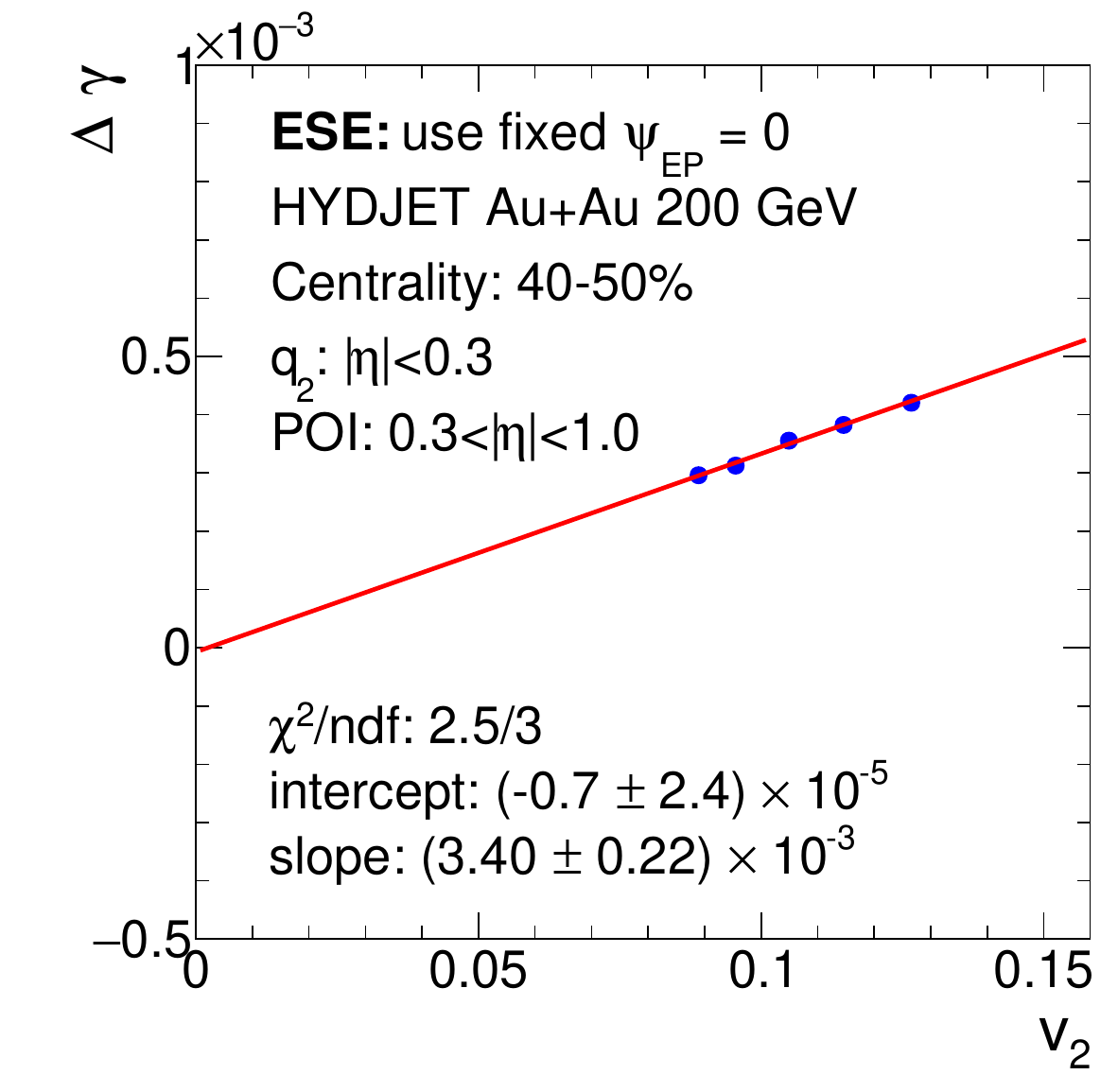}\hfill
    \vspace{-5mm}\\
    \includegraphics[width=0.33\textwidth]{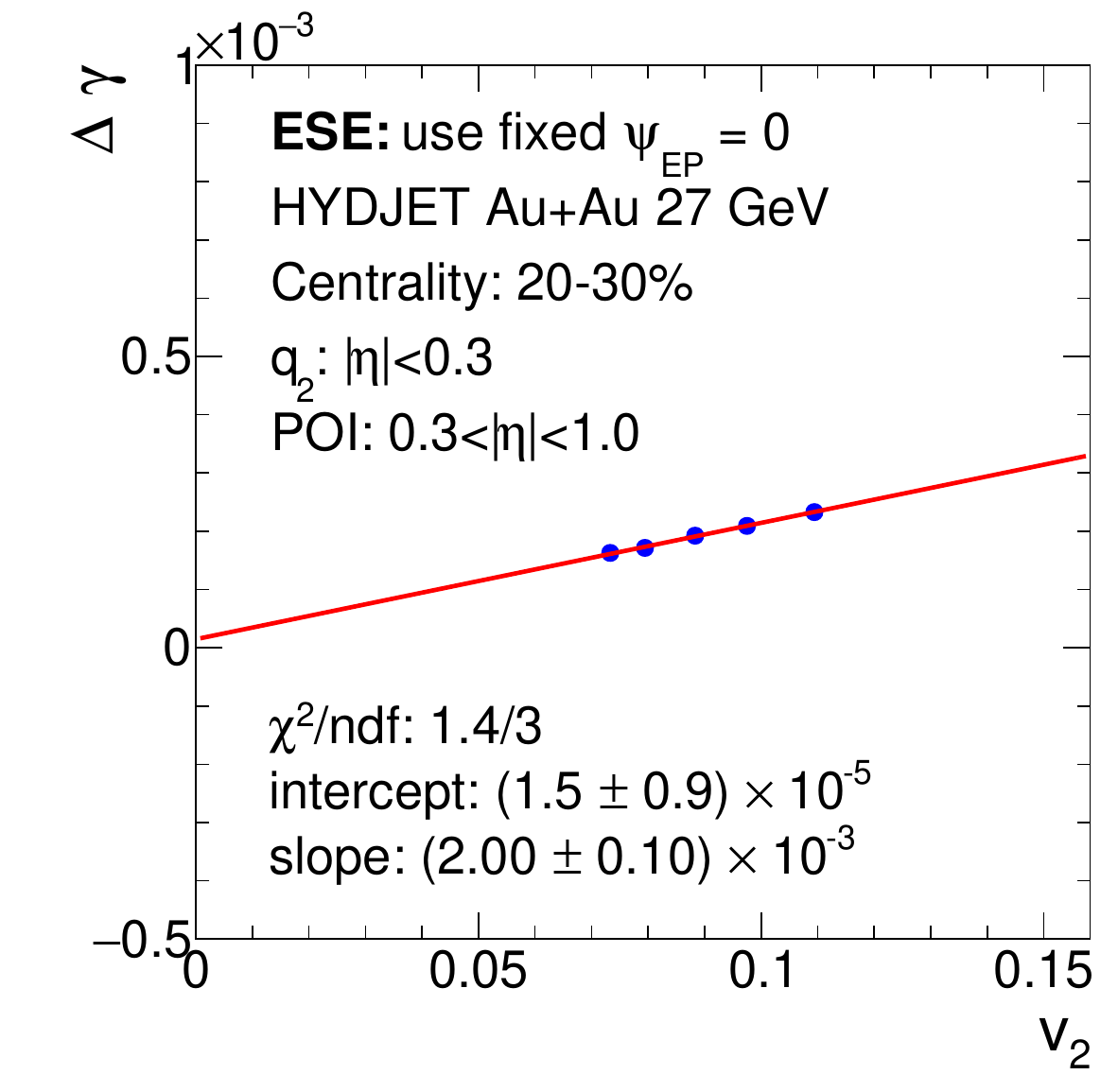}\hfill
    \includegraphics[width=0.33\textwidth]{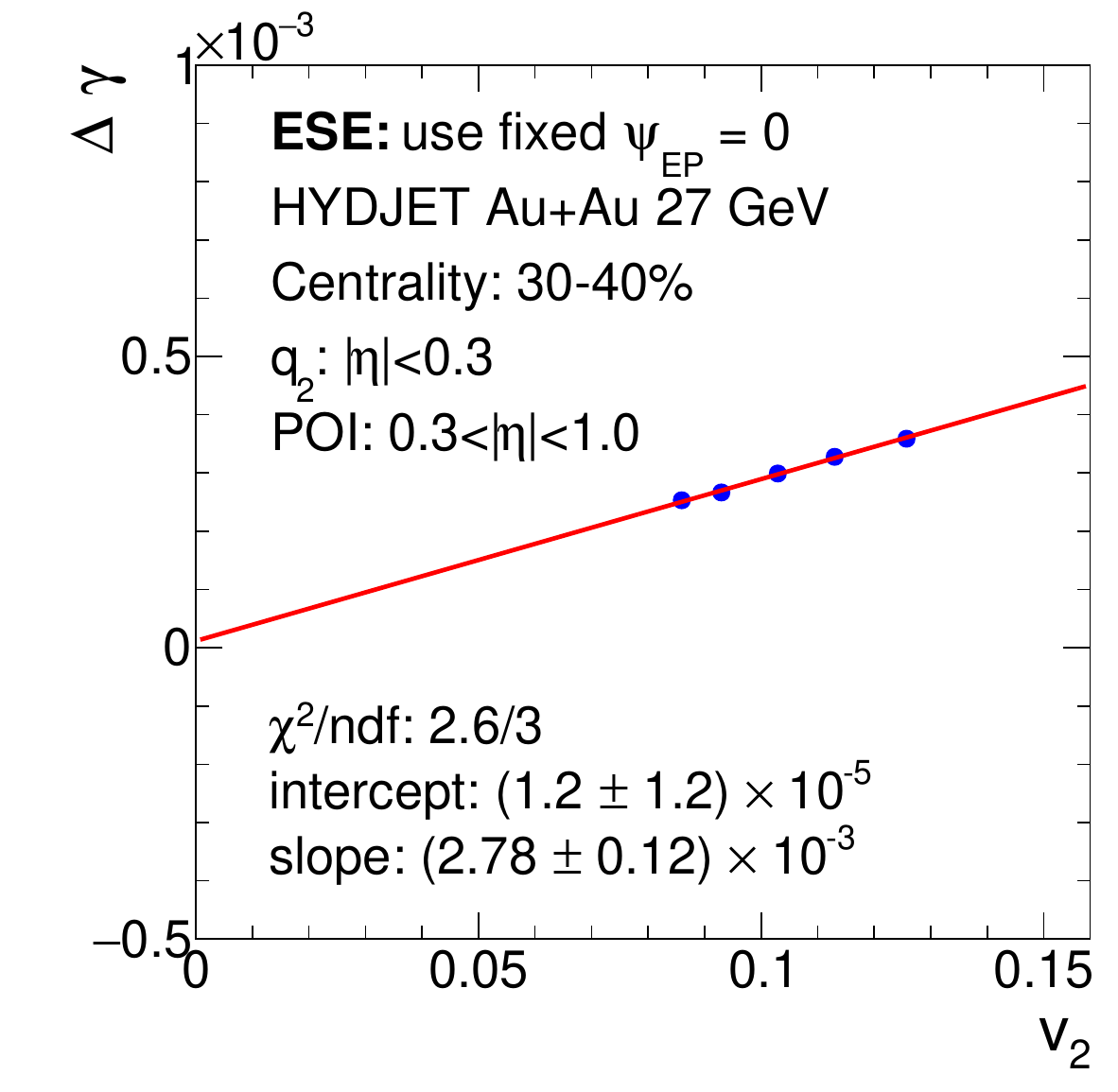}\hfill
    \includegraphics[width=0.33\textwidth]{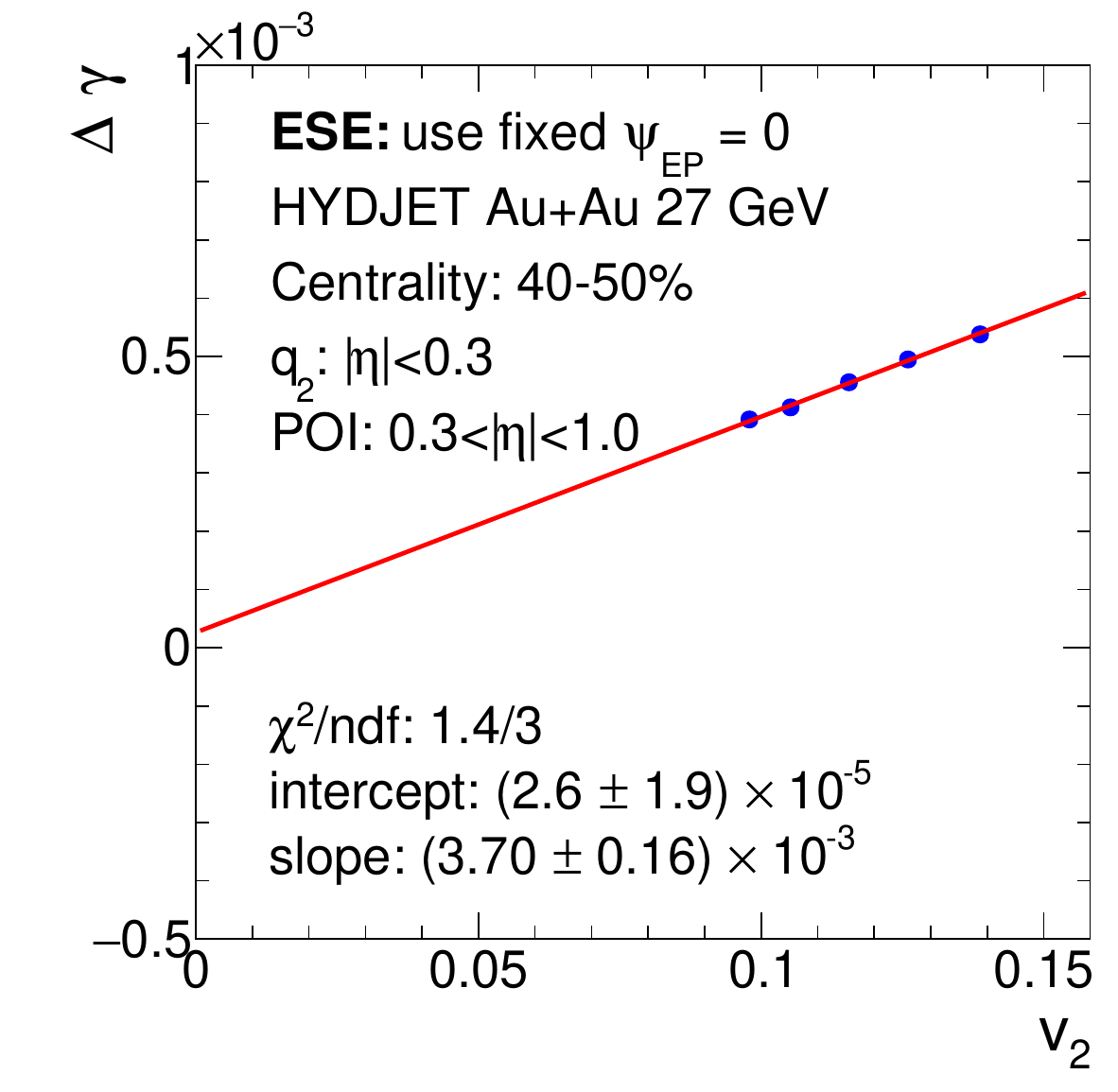}\hfill
    \vspace{-4mm}\\
    \caption{\label{fig:hydjet_ese}\hydjet\ ESE results. 
    The same \hydjet\ data as in Fig.~\ref{fig:hydjet_ess} is used. 
    The $\dg$ is plotted as a function of $\mean{v_2}$ in events binned in $\qh^2\two$ (Eqs.~\ref{eq:q2},\ref{eq:qh}). POIs are from acceptance $0.3<|\eta|<1$, and the event selection variable $\qh^2\two$ is computed from particles in $|\eta|<0.3$, both with $0.2<\pt<2$~\gevc. The model's known impact parameter direction $\psi=0$ is taken as the EP in calculating $\dg$ (Eqs.~\ref{eq:g},\ref{eq:dg}) and $\mean{v_2}$ (Eq.~\ref{eq:v2}).}
\end{figure*}

\begin{figure*}[hbt]
    \includegraphics[width=0.33\textwidth]{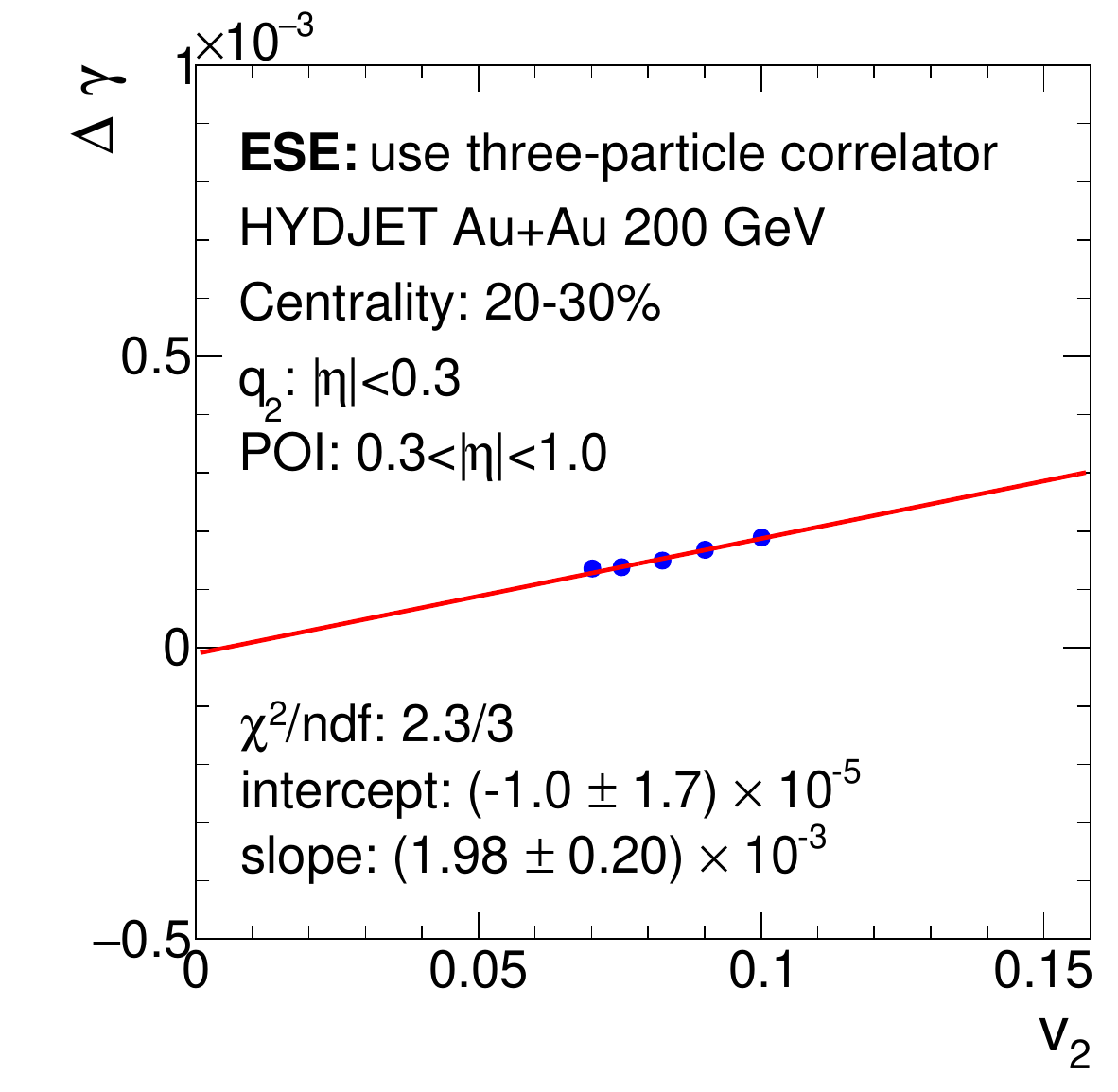}\hfill
    \includegraphics[width=0.33\textwidth]{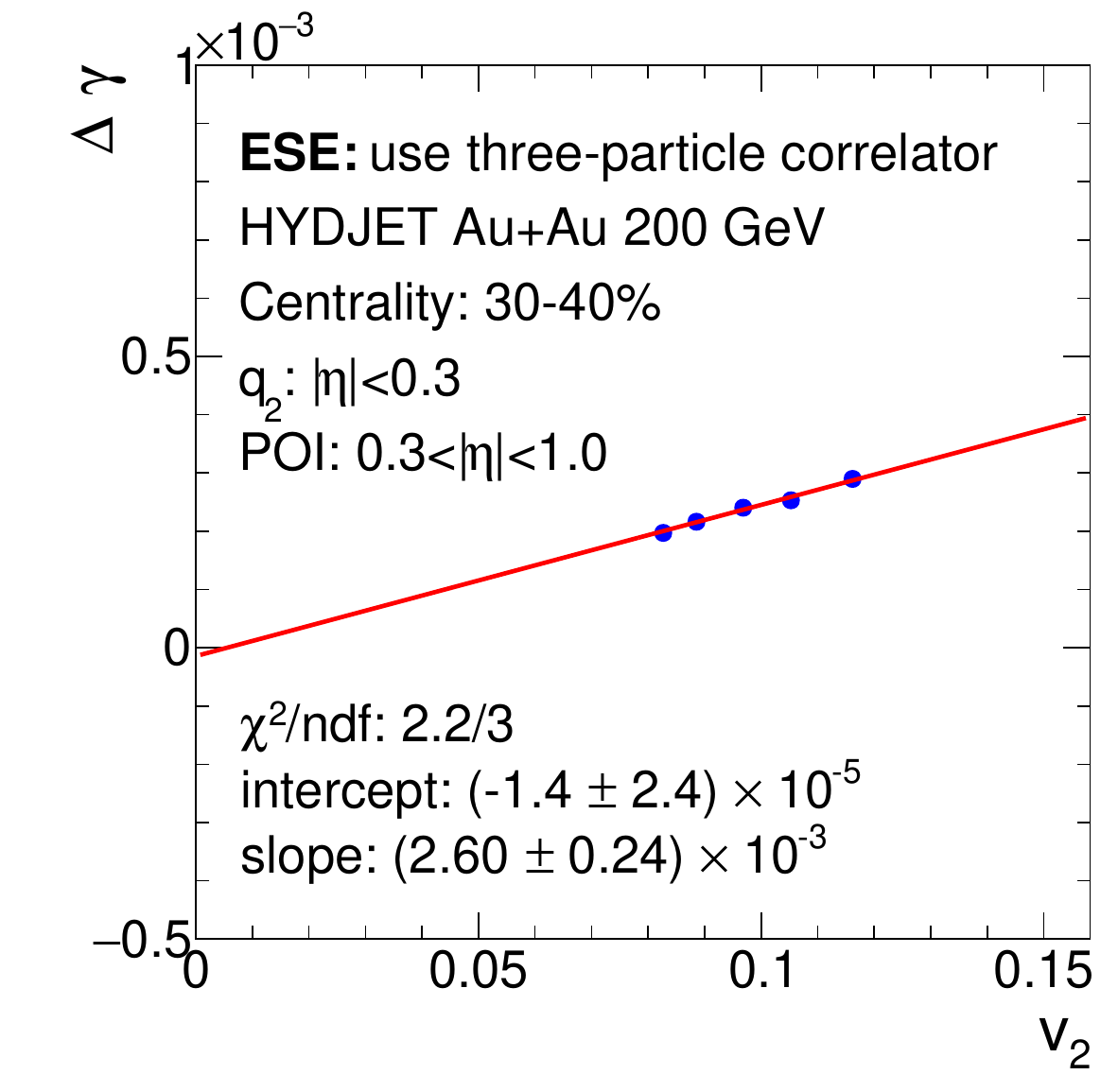}\hfill
    \includegraphics[width=0.33\textwidth]{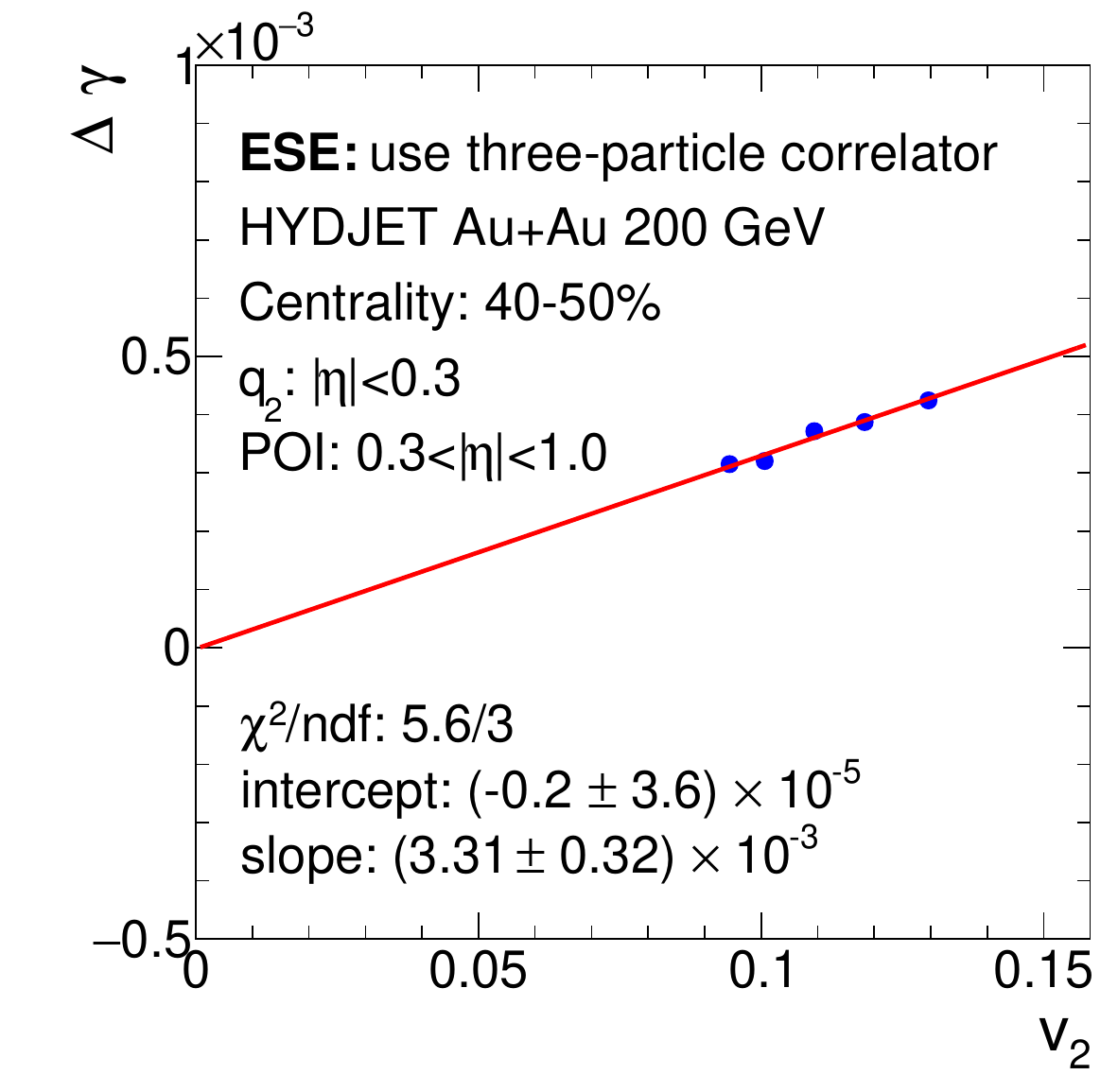}\hfill
    \vspace{-5mm}\\
    \includegraphics[width=0.33\textwidth]{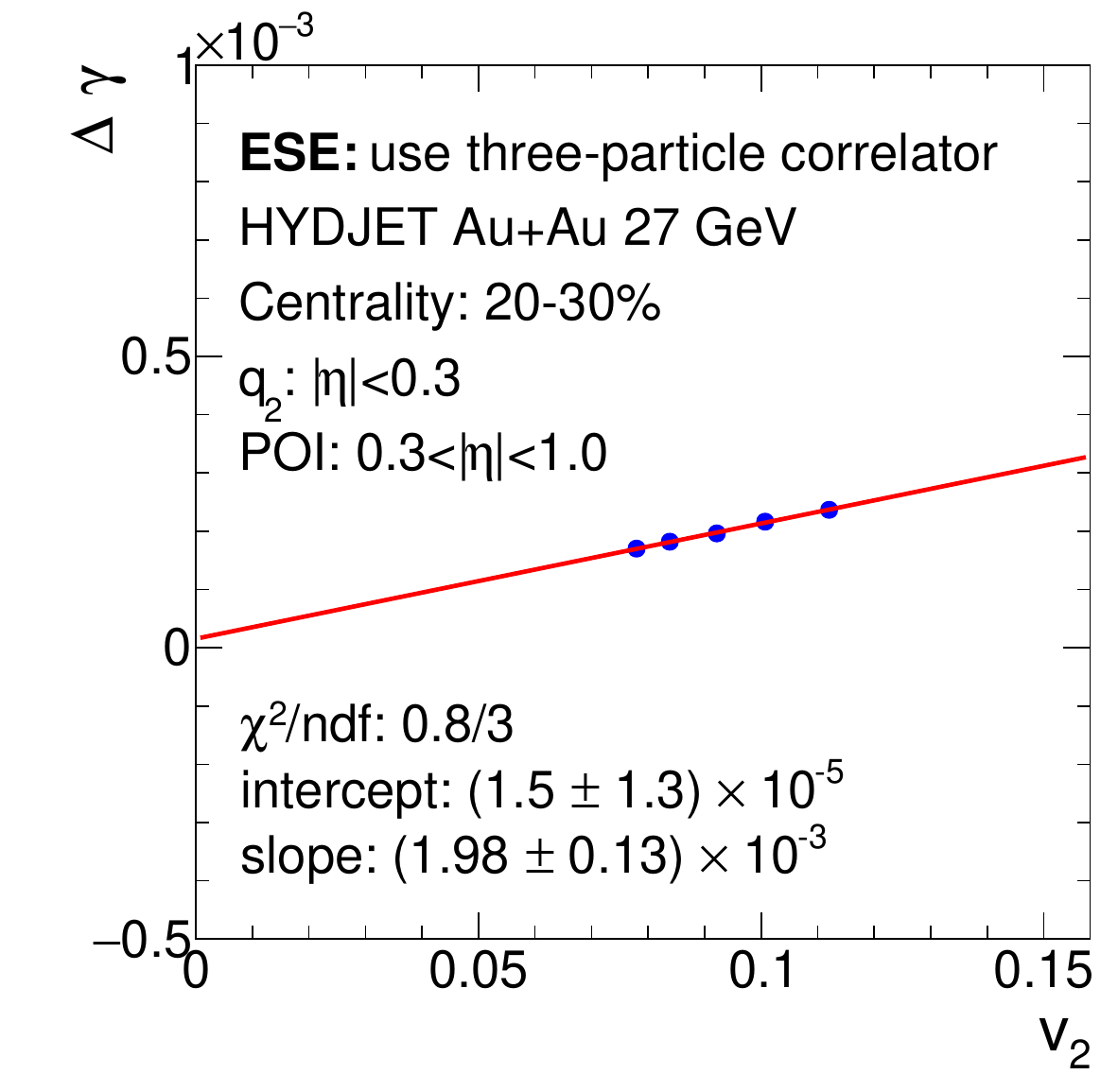}\hfill
    \includegraphics[width=0.33\textwidth]{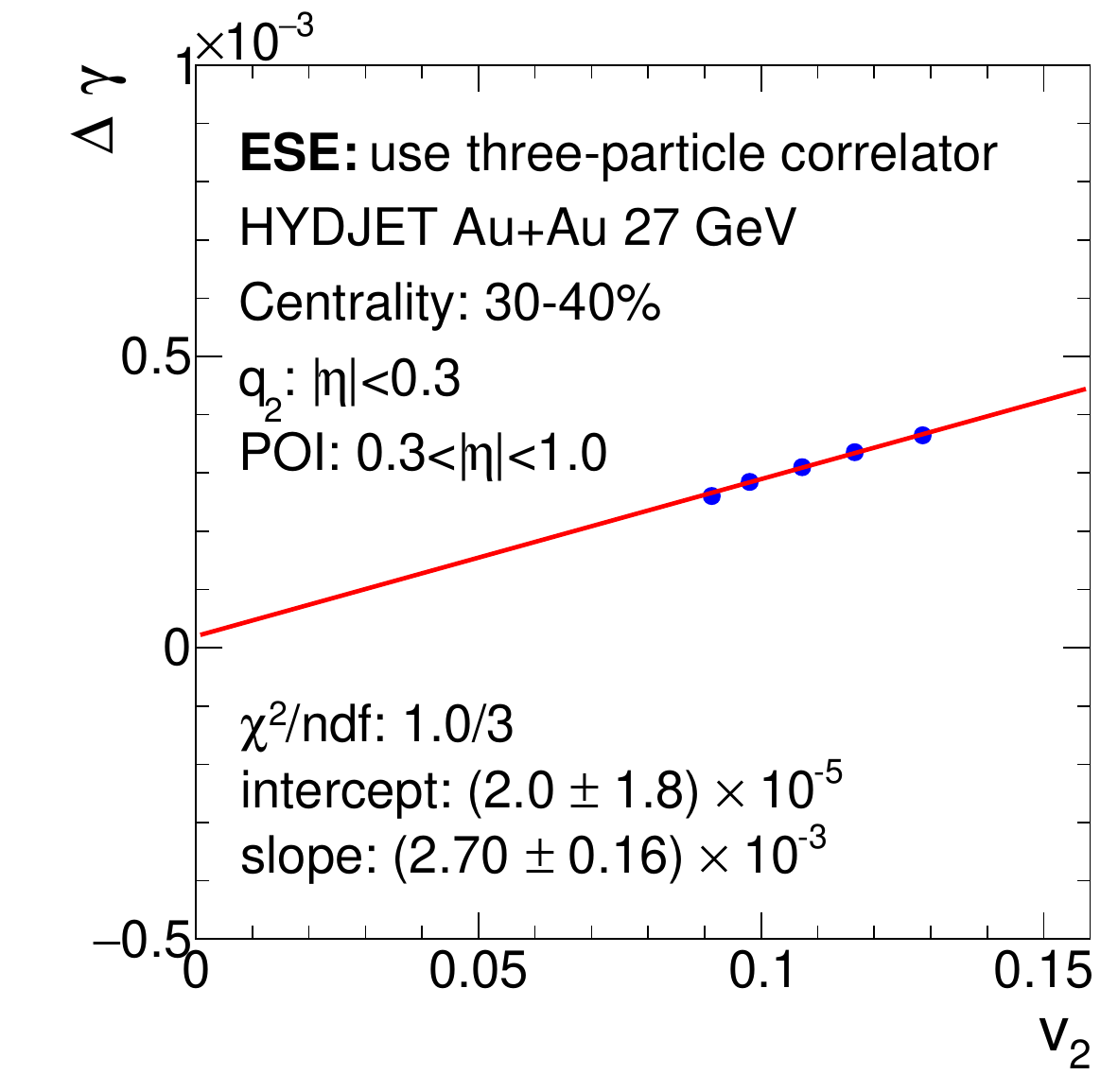}\hfill
    \includegraphics[width=0.33\textwidth]{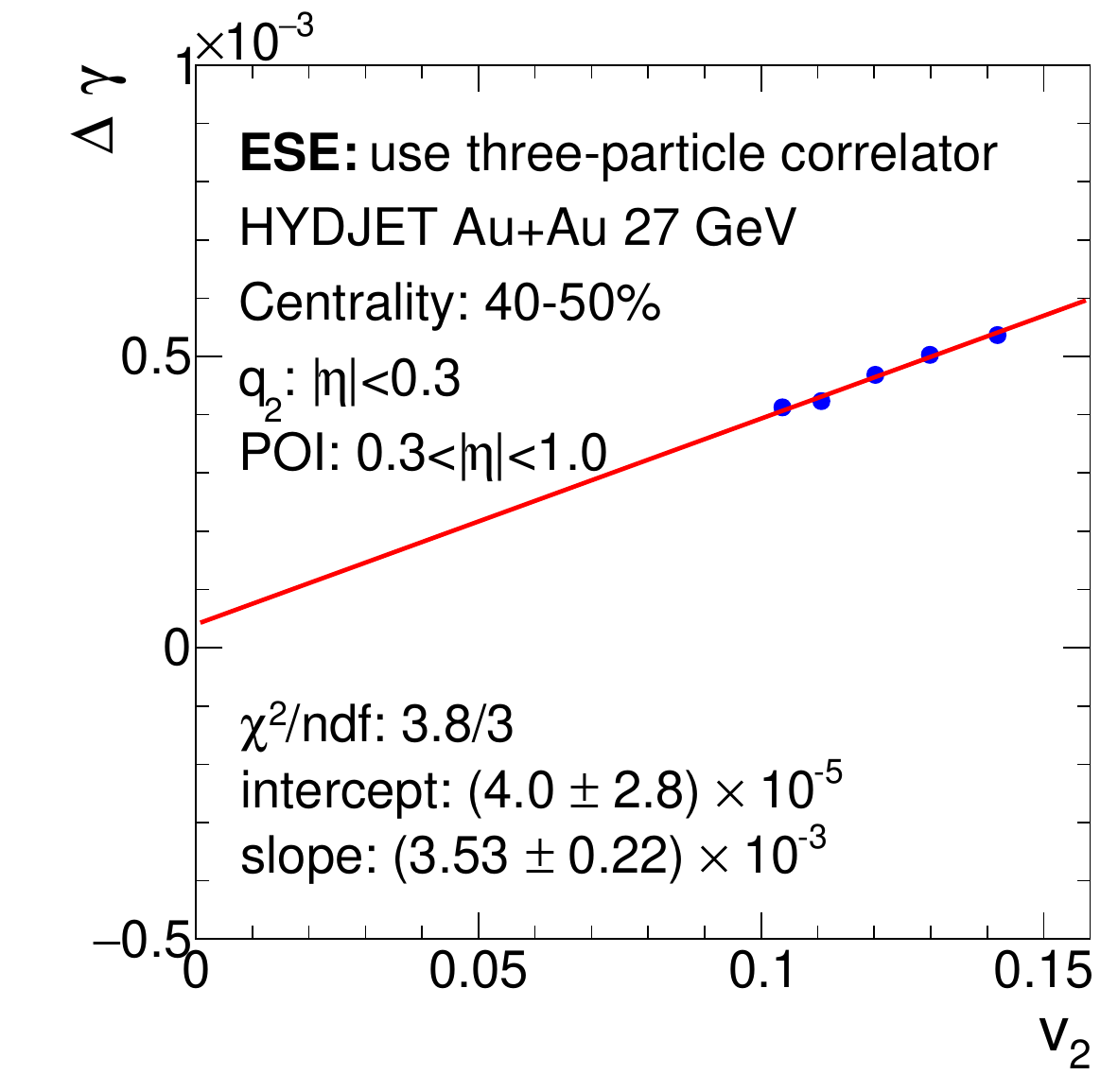}\hfill
    \vspace{-4mm}\\
    \caption{\label{fig:hydjet_ese3}\hydjet\ ESE results (three-particle correlator). As same as Fig.~\ref{fig:hydjet_ese} except that the three-particle correlator method is used to calculate $\dg$ (Eq.~\ref{eq:g3}) and the two-particle cumulant is used to calculate $\mean{v_2}$ (Eq.~\ref{eq:v22}), instead of using the known impact parameter direction in these calculations.}
    \includegraphics[width=0.33\textwidth]{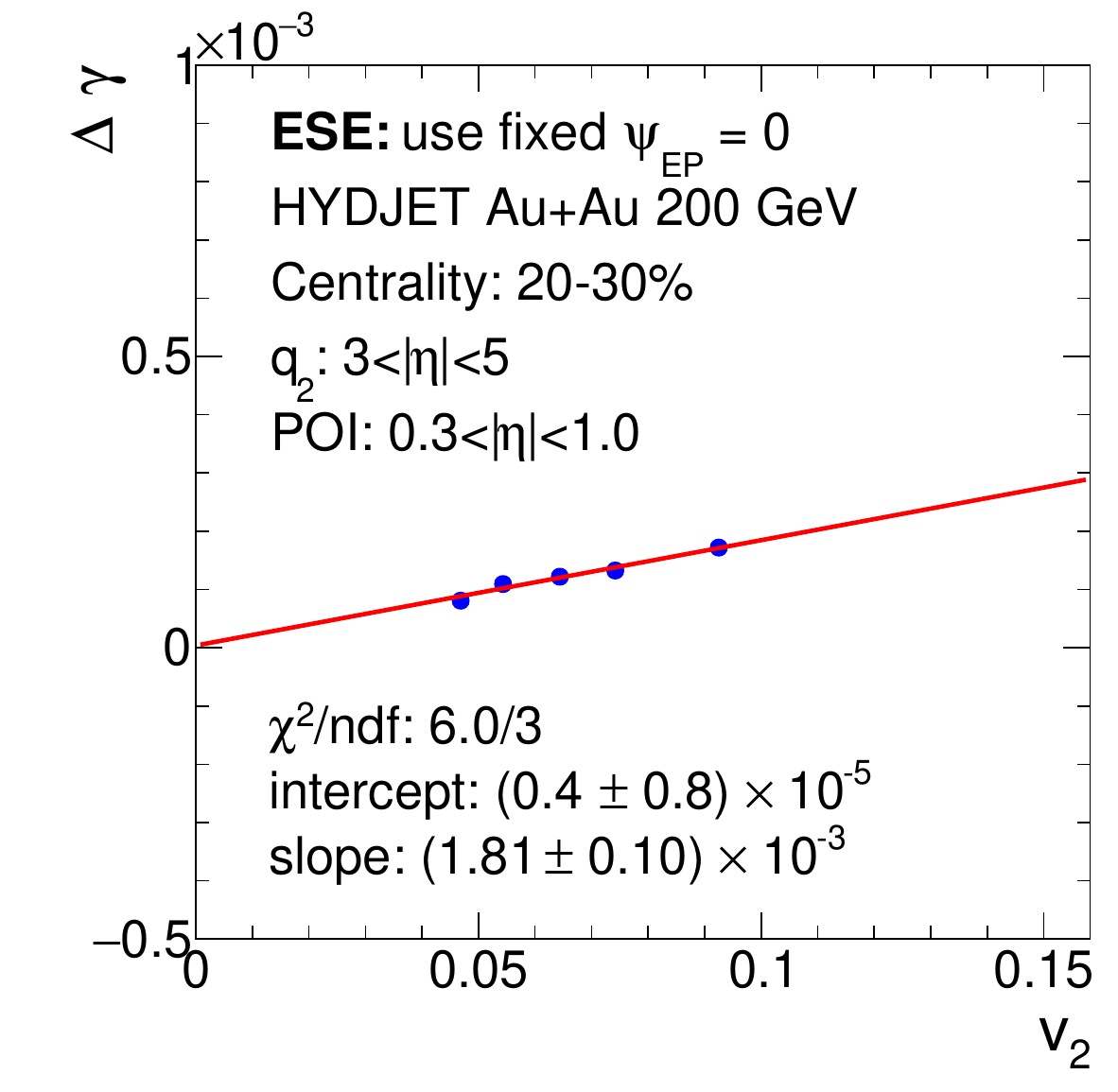}\hfill
    \includegraphics[width=0.33\textwidth]{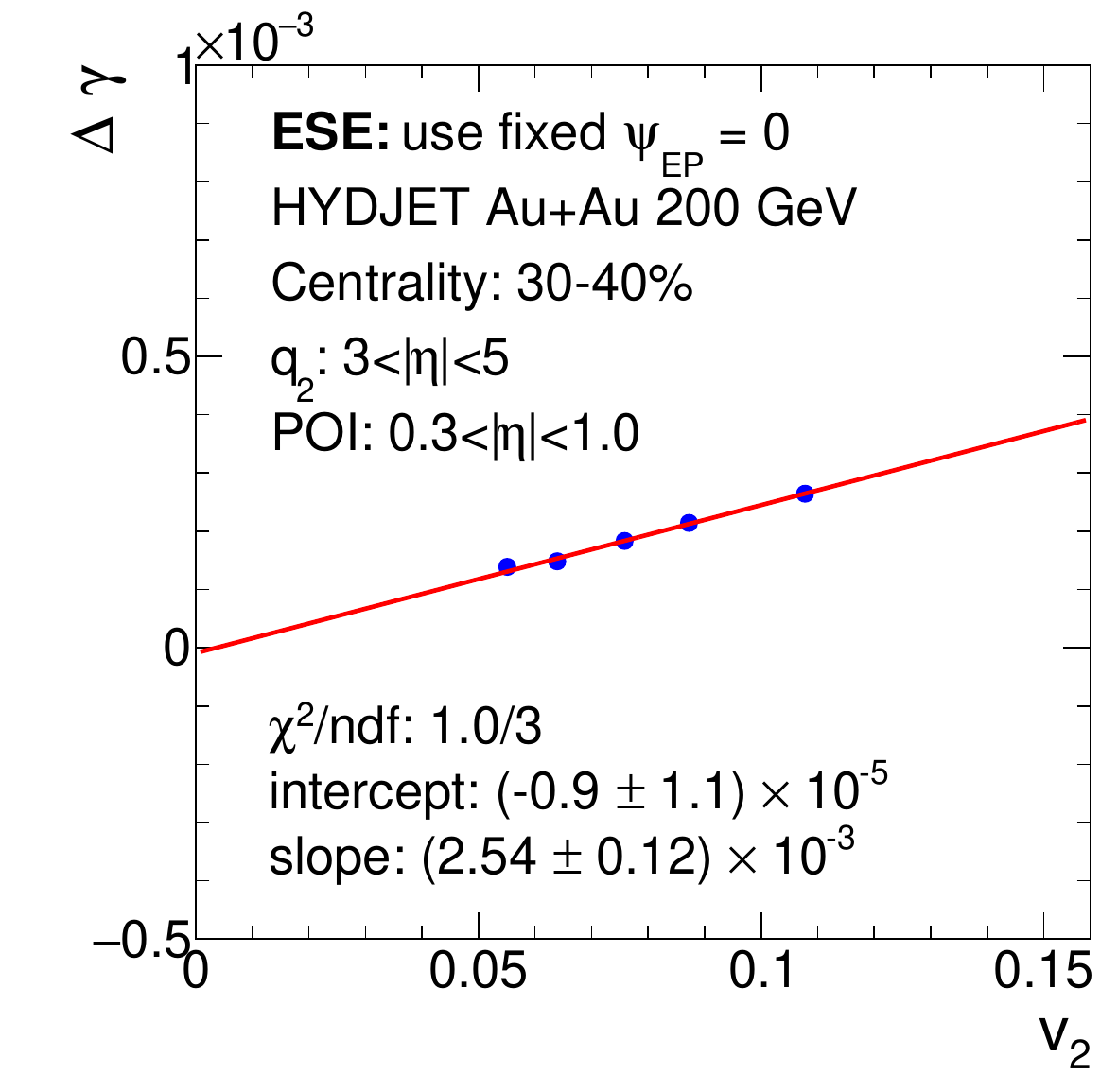}\hfill
    \includegraphics[width=0.33\textwidth]{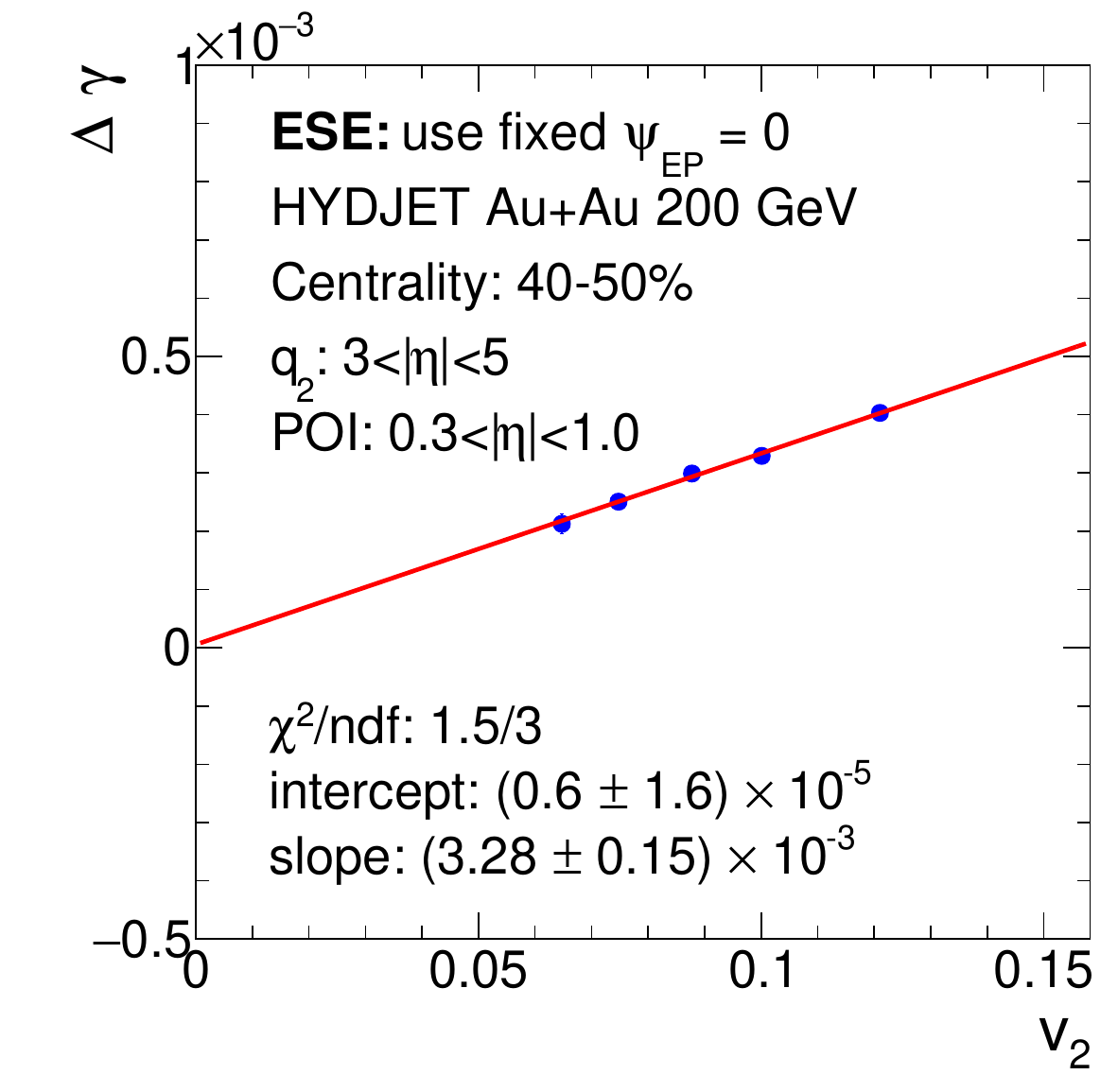}\hfill
    \vspace{-5mm}\\
    \includegraphics[width=0.33\textwidth]{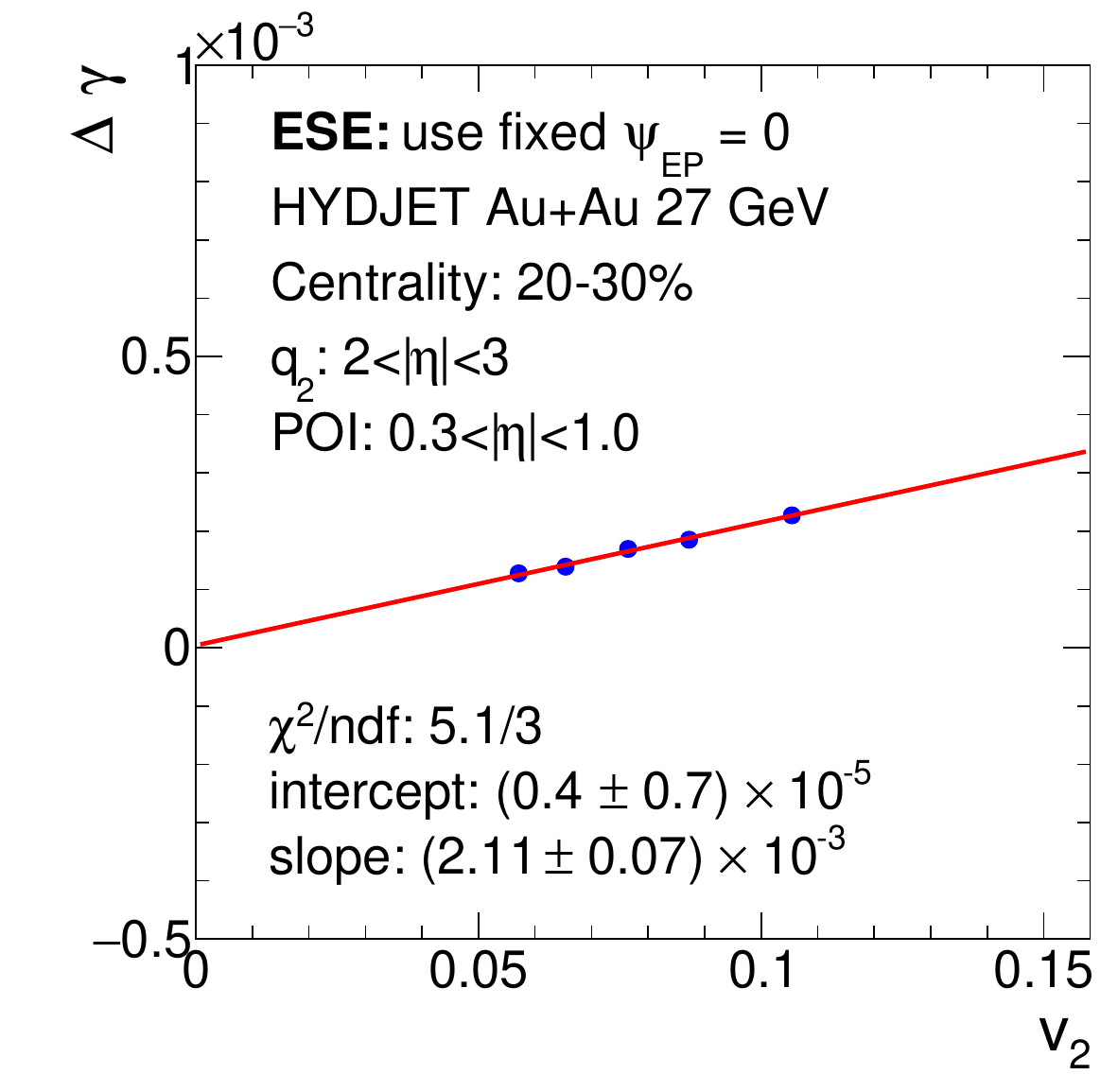}\hfill
    \includegraphics[width=0.33\textwidth]{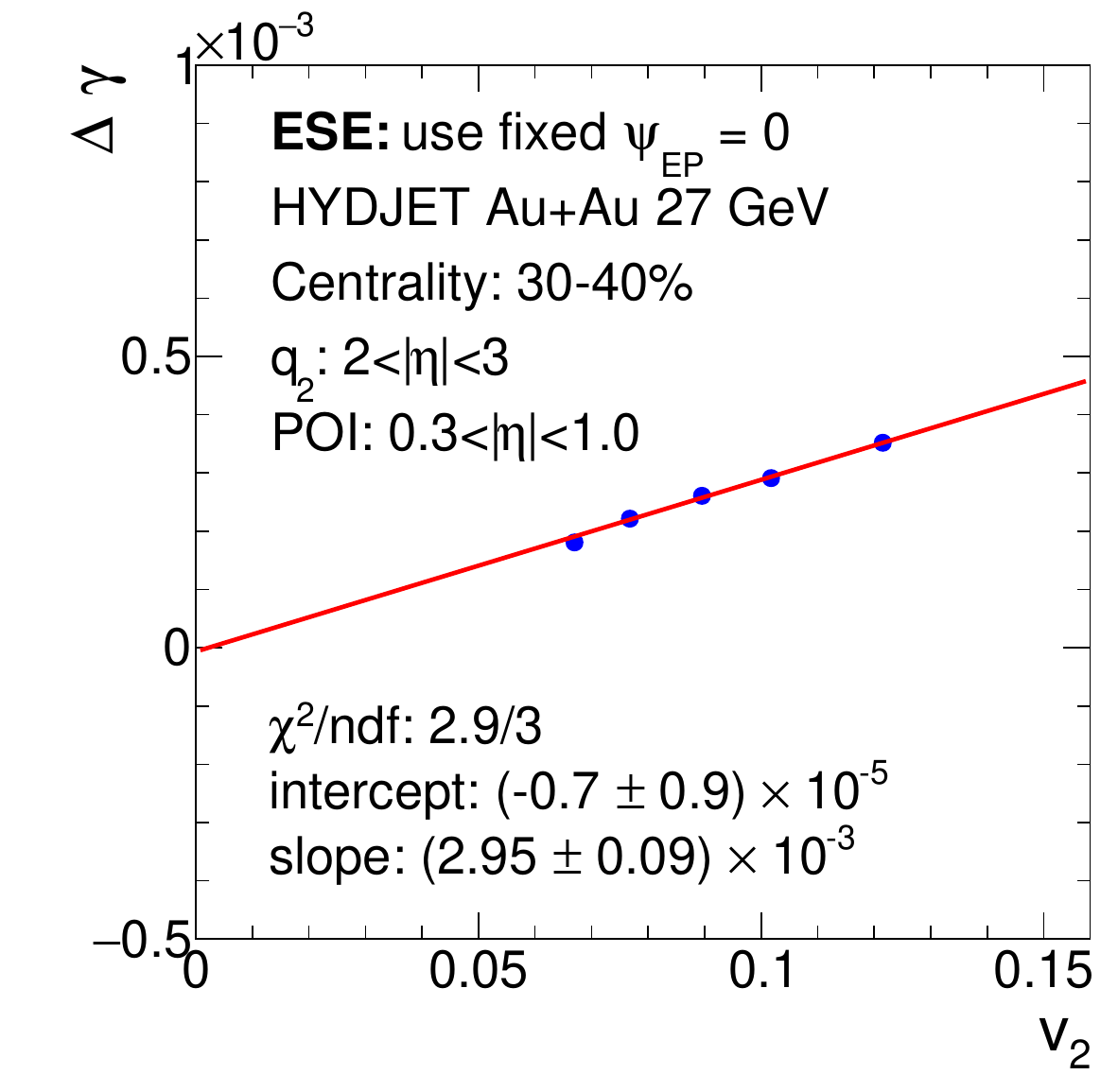}\hfill
    \includegraphics[width=0.33\textwidth]{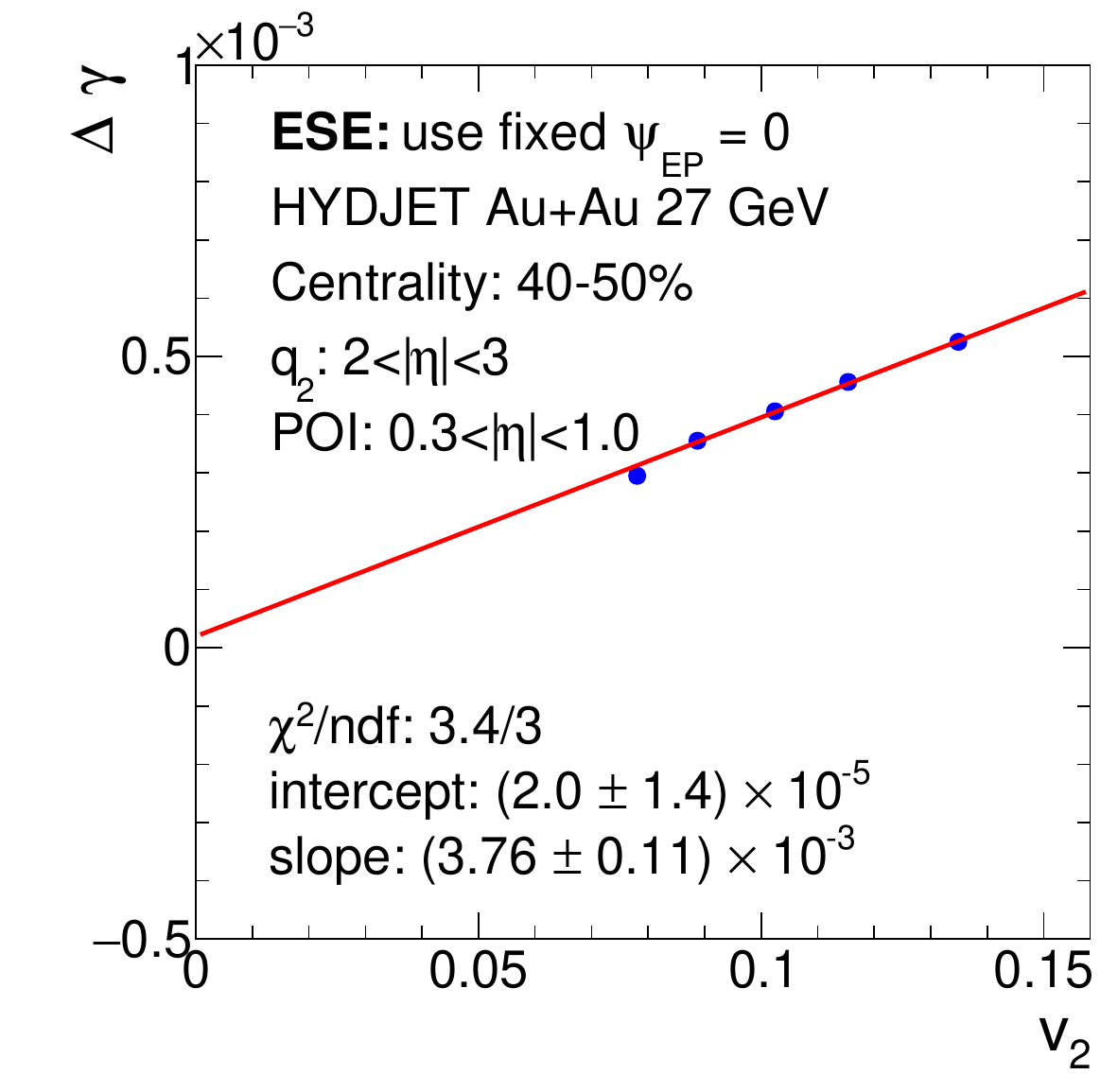}\hfill
    \vspace{-4mm}\\
    \caption{\label{fig:hydjet_ese_forward}\hydjet\ ESE results (forward/backward rapidity $\qh^2$). As same as Fig.~\ref{fig:hydjet_ese} except that the event selection variable $\qh^2\two$ (Eqs.~\ref{eq:q2},\ref{eq:qh}) is computed from particles in forward/backward pseudorapidity regions of $3<|\eta|<5$ for 200~GeV and $2<|\eta|<3$ for 27~GeV.}
\end{figure*}

\begin{figure*}[hbt]
    \includegraphics[width=0.33\textwidth]{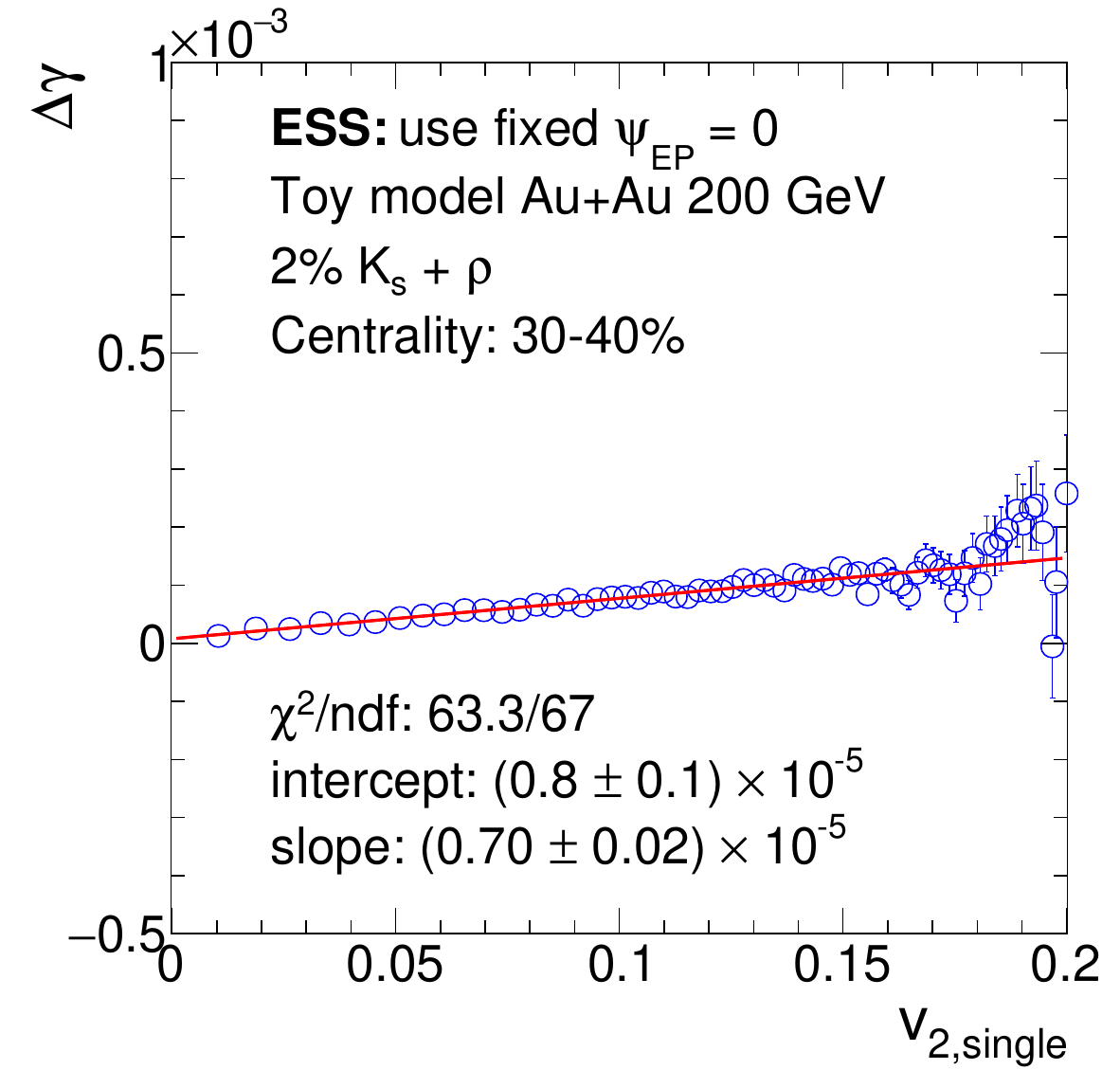}\hfill
    \includegraphics[width=0.33\textwidth]{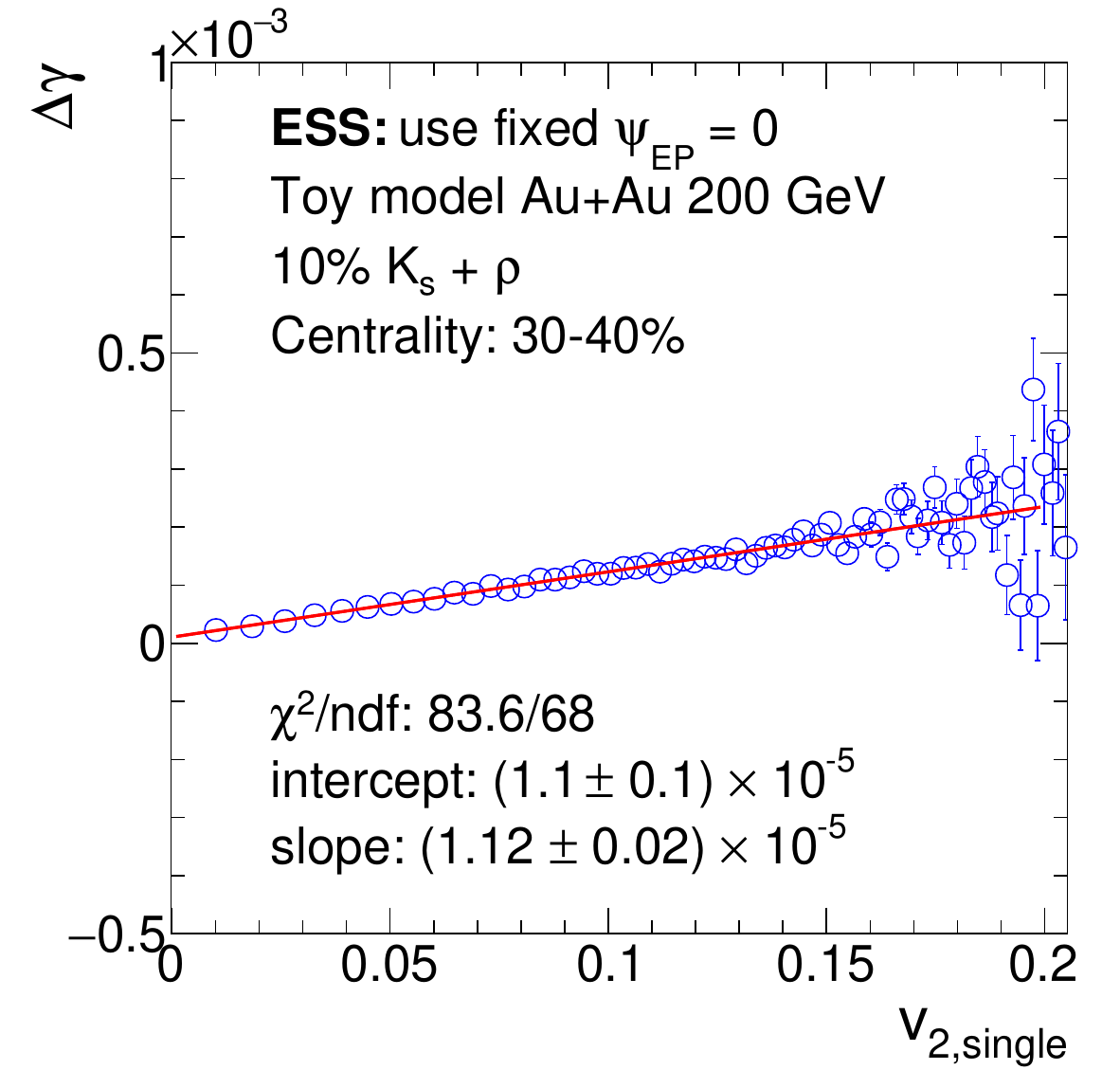}\hfill
    \includegraphics[width=0.33\textwidth]{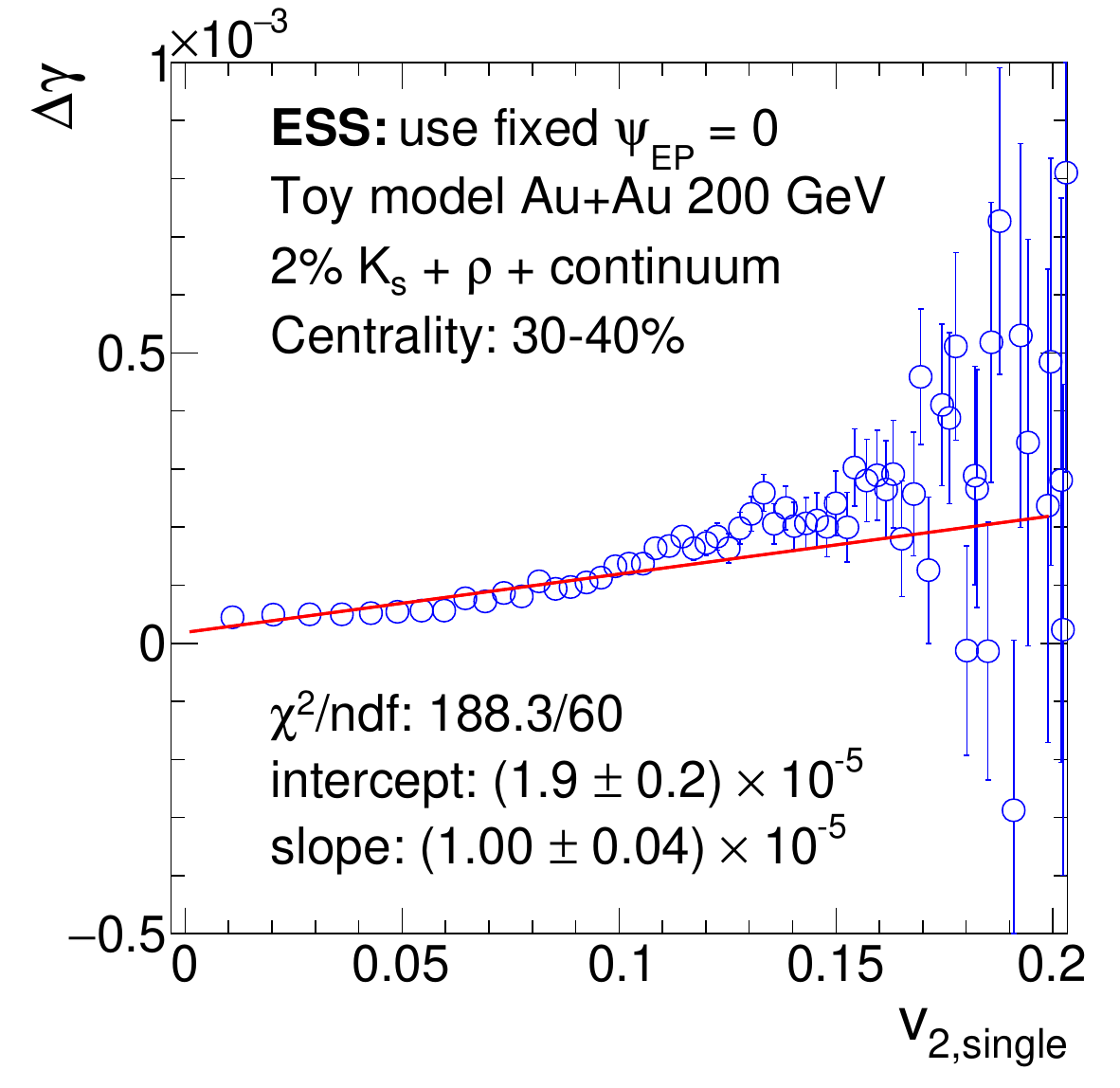}\hfill
    \vspace{-4mm}\\
    \caption{\label{fig:toy_ess}Toy model ESS results. Toy Model I with the accepted $\Ks$ fraction of the default 2\% (left panel) and of 10\% (middle panel), and Toy Model II with mass continuum (right panel). The $\Ks$, $\rho^0$, and primordial pion inputs are from parameterizations to experimental data of 30--40\% centrality Au+Au collisions at $\snn=200$~GeV. The numbers of events are $2\times10^8$ for each panel. The $\dg$ is plotted as a function of $\vsing$ in events binned in $\qhpair^2\two$ (Eqs.~\ref{eq:q2},\ref{eq:qhpair}). POIs are from acceptance $|\eta|<1$ and $0.2 < \pt < 2$~\gevc, and the event selection variable $\qhpair^2\two$ is computed from the same POIs. 
    The model's known impact parameter direction $\psi=0$ is taken as the EP in calculating $\dg$  (Eqs.~\ref{eq:g},\ref{eq:dg}) and $\vsing$ (Eq.~\ref{eq:v2}). The red line is a first-order polynomial fit in the range of $0<\vsing<0.2$.}
    \includegraphics[width=0.33\textwidth]{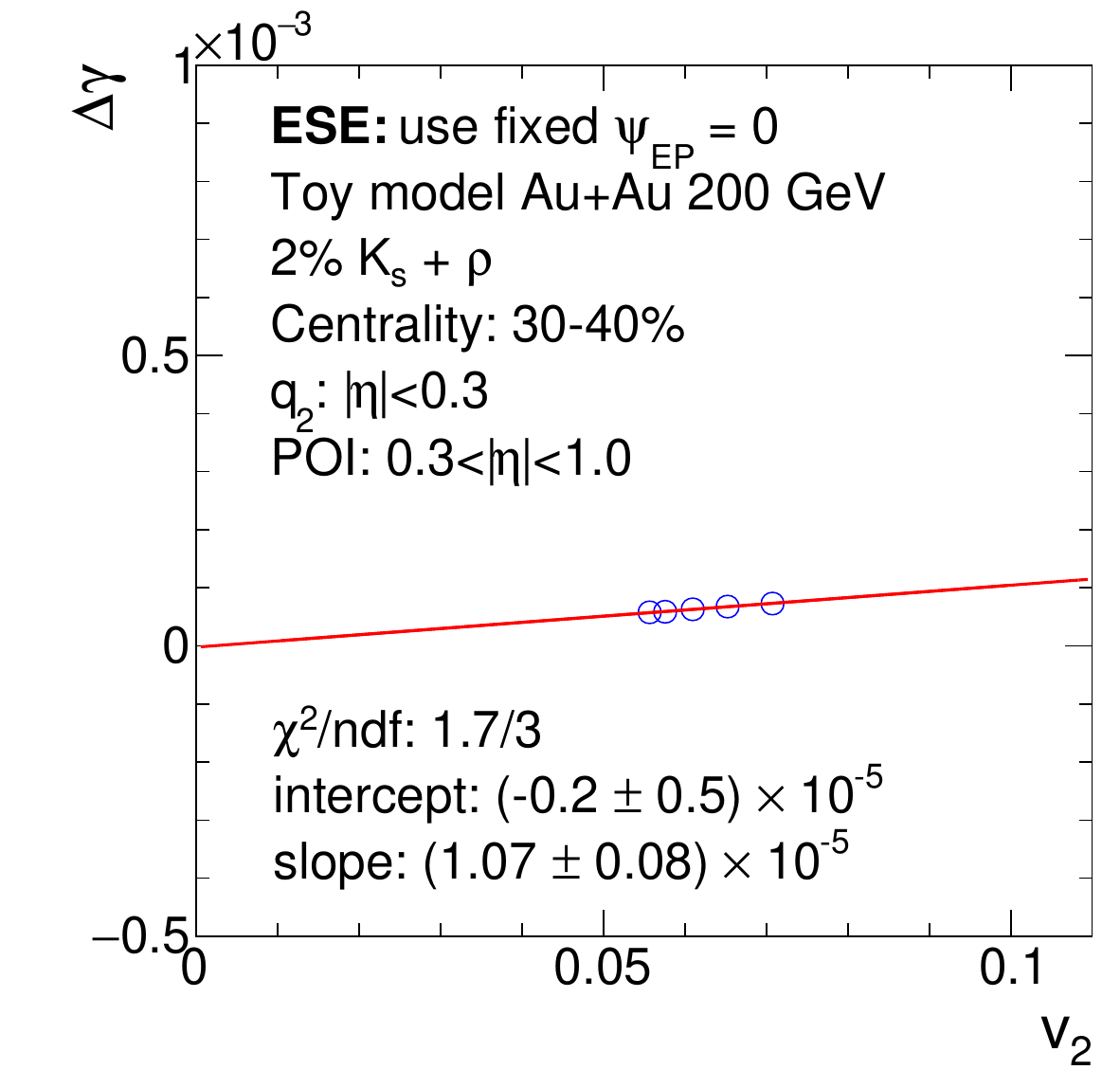}\hfill
    \includegraphics[width=0.33\textwidth]{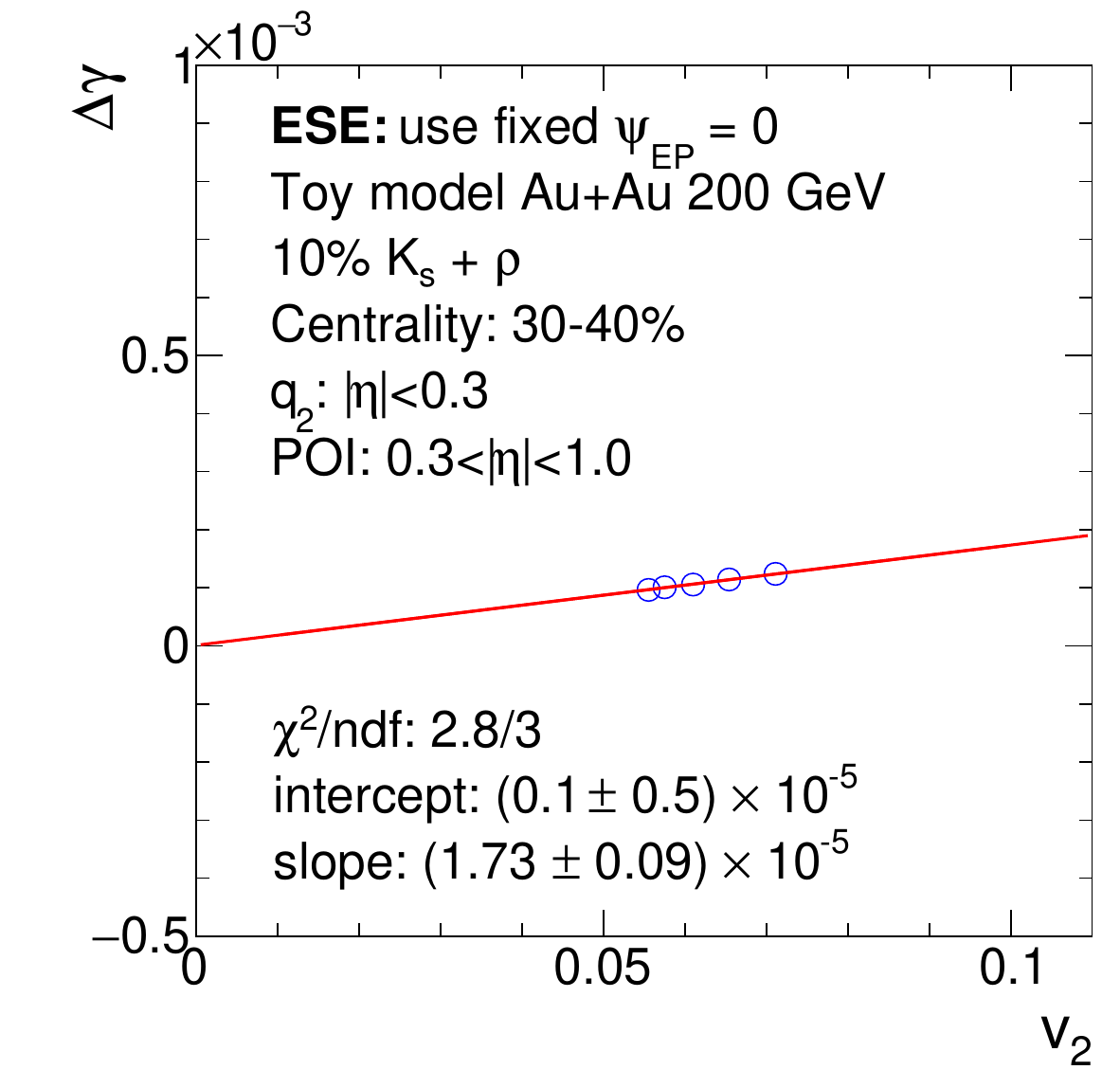}\hfill
    \includegraphics[width=0.33\textwidth]{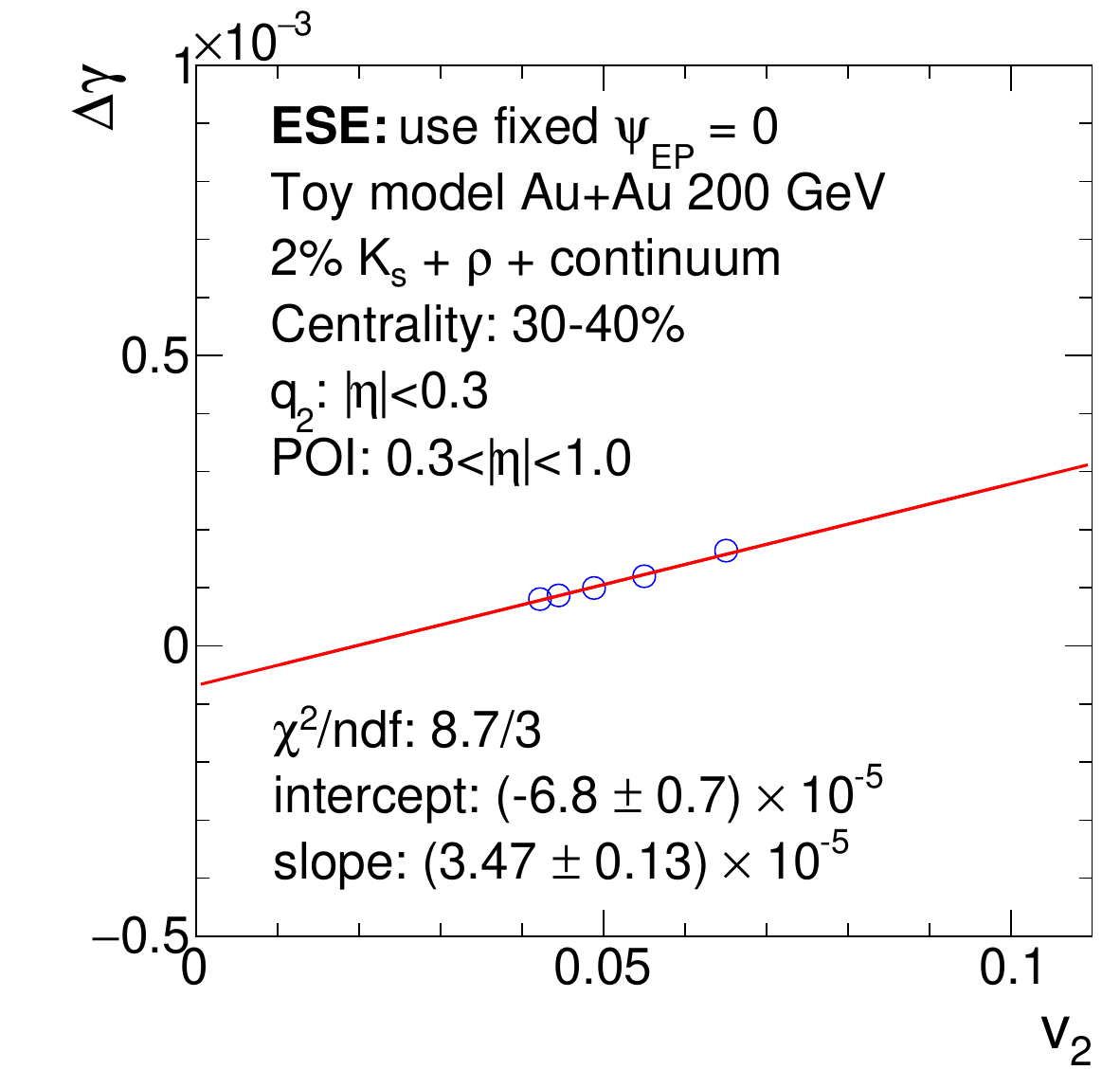}\hfill
    \vspace{-4mm}\\
    \caption{\label{fig:toy_ese}Toy model ESE results. Toy Model I with the accepted $\Ks$ fraction of the default 2\% (left panel, $2.8\times10^{9}$ events) and of 10\% (middle panel, $2.0\times10^9$ events), and Toy Model II with mass continuum (right panel, $2.2\times10^8$ events). 
    The $\dg$ is plotted as a function of $\mean{v_2}$ in events binned in $\qh^2\two$ (Eqs.~\ref{eq:q2},\ref{eq:qh}). POIs are from acceptance $0.3<|\eta|<1$, and the event selection variable $\qh^2\two$ is computed from particles in $|\eta|<0.3$, both with $0.2<\pt<2$~\gevc. 
    The model's known impact parameter direction $\psi=0$ is taken as the EP in calculating $\dg$ (Eqs.~\ref{eq:g},\ref{eq:dg}) and $\mean{v_2}$ (Eq.~\ref{eq:v2}). 
    Other information is as same as those in Fig.~\ref{fig:toy_ess}. The red line is a first-order polynomial fit to all data points.}
    \includegraphics[width=0.33\textwidth]{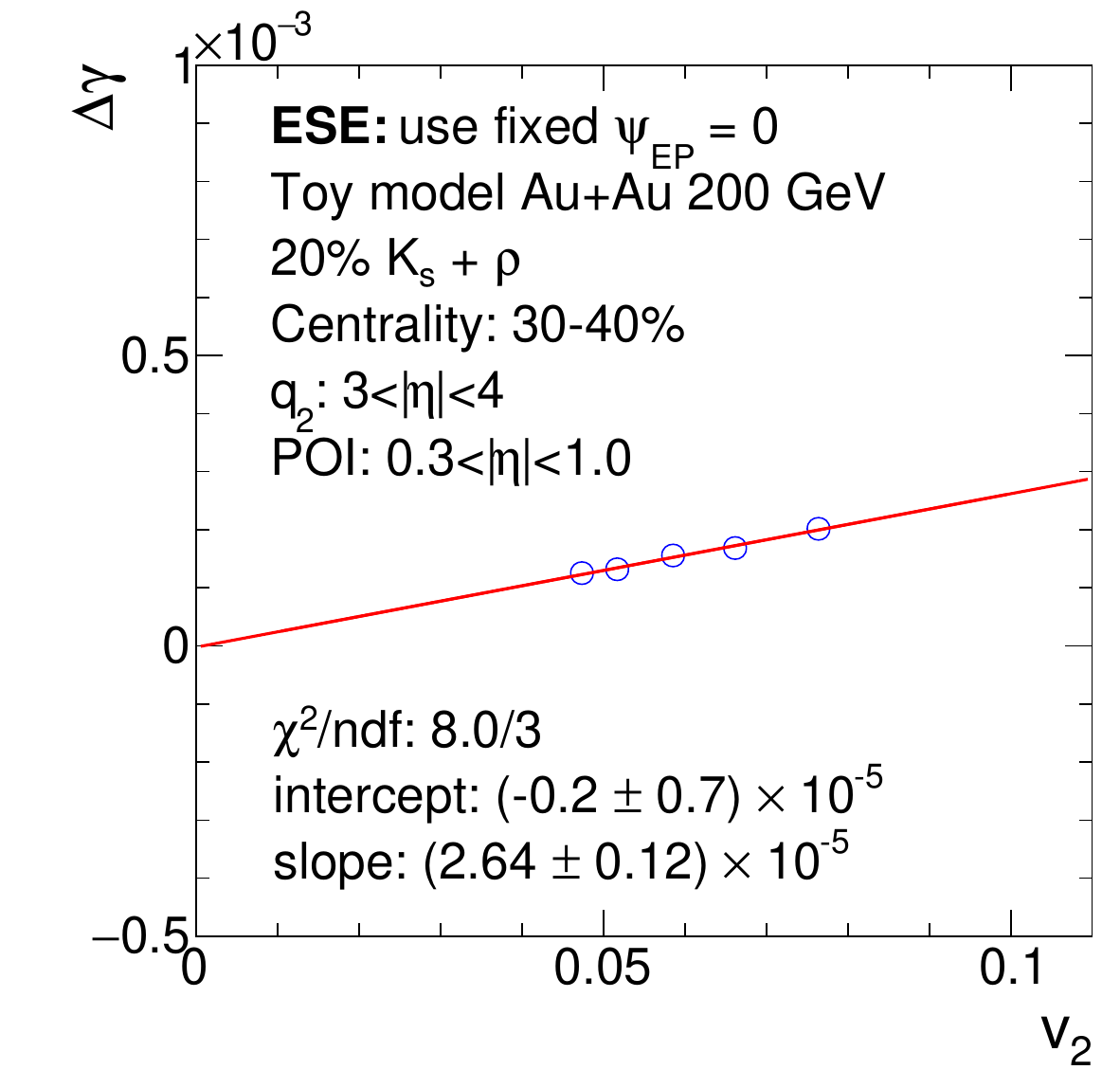}\hfill
    \includegraphics[width=0.33\textwidth]{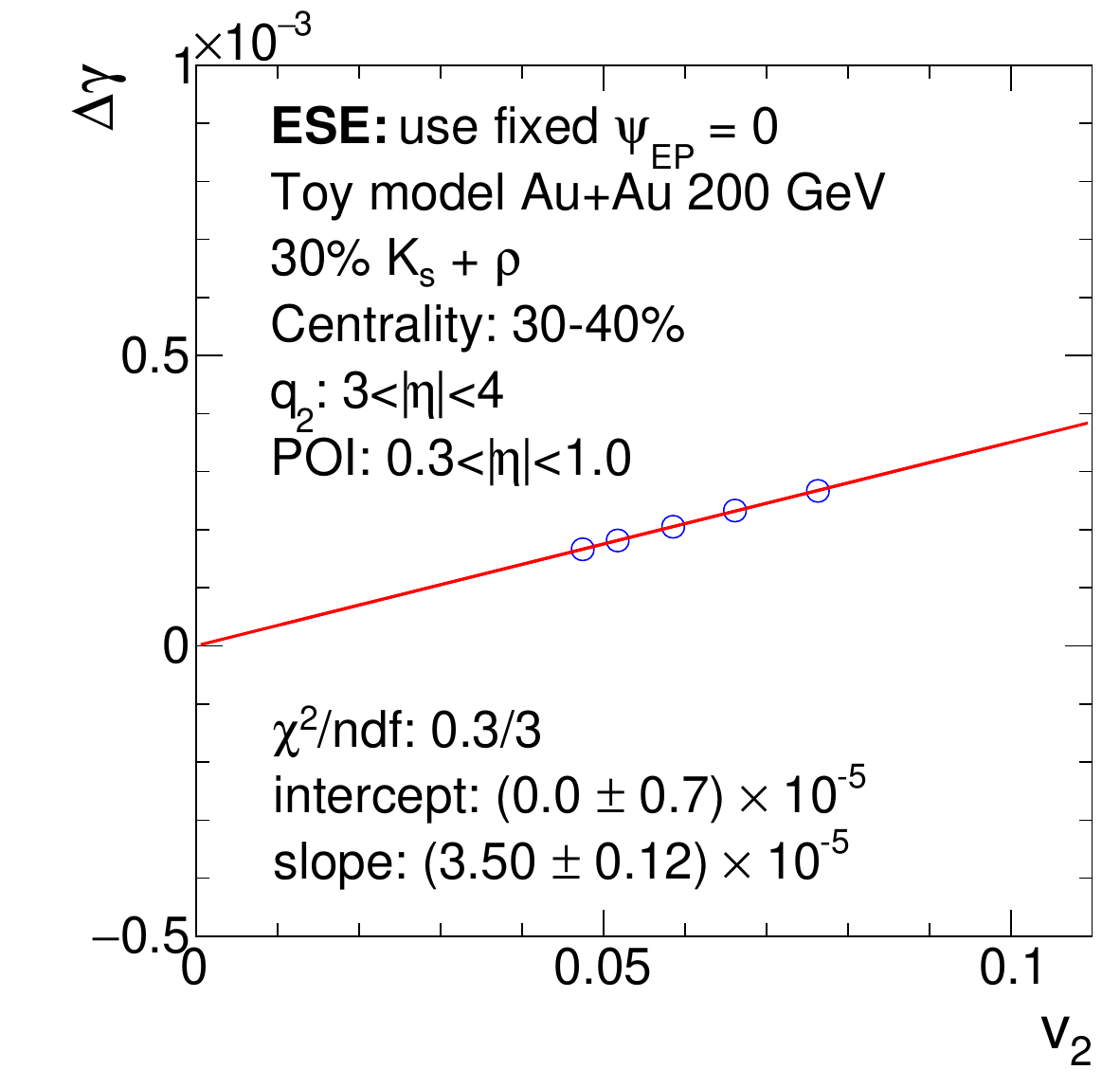}\hfill
    \includegraphics[width=0.33\textwidth]{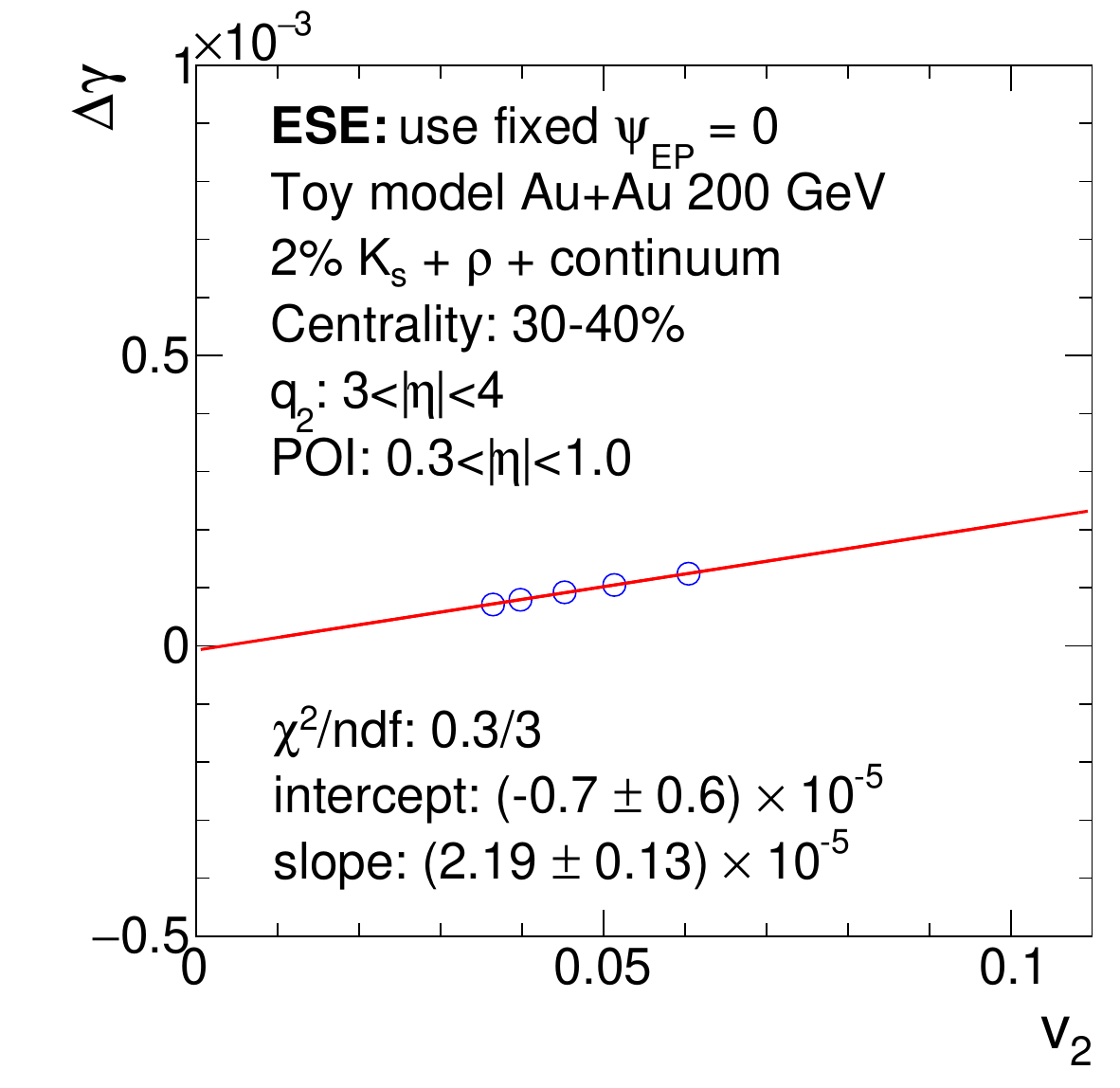}\hfill
    \vspace{-4mm}\\
    \caption{\label{fig:toy_ese2}Toy model ESE results (forward/backward rapidity $\qh^2$). 
    Toy Model I with the accepted $\Ks$ fraction of 20\% (left panel, $2.2\times10^{8}$ events) and of 30\% (middle panel, $2.2\times10^8$ events), and Toy Model II with mass continuum (right panel, $10^8$ events). 
    The event selection variable $\qh^2\two$ is computed from particles in $3<|\eta|<4$. Other information is as same as those in Fig.~\ref{fig:toy_ese}.}
\end{figure*}

This appendix compiles, in Figs.~\ref{fig:avfd_ess}-\ref{fig:toy_ese2}, all simulation plots of $\dg$ as functions of $v_2$ analyzed by the ESS and ESE methods from physics models as well as from toy models.

\end{document}